\RequirePackage{fix-cm}
\documentclass[pdflatex]{svjour3}                     
\smartqed  
\usepackage{graphicx}
\usepackage{natbib}
\usepackage{epsfig}
\usepackage{makeidx}
\makeindex
\usepackage{color}
\usepackage{xcolor}
\usepackage{amssymb}
\usepackage{amsmath}
\usepackage{mathtools}
\usepackage{mathrsfs}
\usepackage{bm}
\usepackage{siunitx} 
\usepackage{longtable}
\usepackage{hyperref}
\usepackage{academicons}
\usepackage{orcidlink}

    \newcommand{\Bm}{B} 
    \newcommand{\B}{\bm{B}} 
    \newcommand{\va}{V_{\rm A}} 
    
    \newcommand\bmm{{$\beta$-meteoroid }}
    \newcommand\bms{{$\beta$-meteoroids }}
    \newcommand\amsn{{$\alpha$-meteoroids}}
    
    \newcommand\ams{{$\alpha$-meteoroids }}
    \newcommand\bmsn{{$\beta$-meteoroids}}
    \newcommand\bs{{$\beta$-stream }}
    \newcommand\bsn{{$\beta$-stream}}

     \newcommand\rs{R_\odot}
    \newcommand\rst{$R_\odot$}

\newcommand{\orcid}[1]{\href{https://orcid.org/#1}{\textcolor[HTML]{A6CE39}{\aiOrcid}}}

\begin{document}

\title{Parker Solar Probe: Four Years of Discoveries at Solar Cycle Minimum
}


\author{N. E. Raouafi$^1$\orcidlink{0000-0003-2409-3742} \and
L. Matteini$^2$\orcidlink{0000-0002-6276-7771} \and
J. Squire$^3$\orcidlink{0000-0001-8479-962X} \and
S. T. Badman$^{4,5}$\orcidlink{0000-0002-6145-436X} \and
M. Velli$^6$\orcidlink{0000-0002-2381-3106} \and
K.G. Klein$^7$\orcidlink{0000-0001-6038-1923} \and
C. H. K. Chen$^8$\orcidlink{0000-0003-4529-3620} \and
W. H. Matthaeus$^9$\orcidlink{0000-0001-7224-6024} \and
A. Szabo$^{10}$\orcidlink{0000-0003-3255-9071} \and
M. Linton$^{11}$ \and
R. C. Allen$^1$\orcidlink{0000-0003-2079-5683} \and
J. R. Szalay$^{12}$\orcidlink{0000-0003-2685-9801} \and
R. Bruno$^{13}$\orcidlink{0000-0002-2152-0115} \and
R. B. Decker$^{1}$ \and
M. Akhavan-Tafti$^{14}$\orcidlink{0000-0003-3721-2114} \and
O. V. Agapitov$^5$\orcidlink{0000-0001-6427-1596} \and
S. D. Bale$^{5,15}$\orcidlink{0000-0002-1989-3596} \and
R. Bandyopadhyay$^{12}$\orcidlink{0000-0002-6962-0959} \and
K. Battams$^{11}$\orcidlink{0000-0002-8692-6925} \and
L. Ber\v{c}i\v{c}$^{16}$\orcidlink{0000-0002-6075-1813} \and
S. Bourouaine$^1$\orcidlink{0000-0002-2358-6628} \and
T. Bowen$^5$\orcidlink{0000-0002-4625-3332} \and
C. Cattell$^{17}$\orcidlink{0000-0002-3805-320X} \and
B. D. G. Chandran$^{18,19}$\orcidlink{0000-0003-4177-3328} \and
R. Chhiber$^{9,10}$\orcidlink{0000-0002-7174-6948} \and
C. M. S. Cohen$^{20}$\orcidlink{0000-0002-0978-8127} \and
R. D'Amicis$^{13}$\orcidlink{0000-0003-2647-117X} \and
J. Giacalone$^7$\orcidlink{0000-0002-0850-4233} \and
P. Hess$^{11}$\orcidlink{0000-0003-1377-6353} \and
R.A. Howard$^1$\orcidlink{0000-0001-9027-8249} \and
T. S. Horbury$^2$\orcidlink{0000-0002-7572-4690} \and
V. K. Jagarlamudi$^1$\orcidlink{0000-0001-6287-6479} \and
C. J. Joyce$^{21}$\orcidlink{0000-0002-3841-5020} \and
J. C. Kasper$^{14,22}$\orcidlink{0000-0002-7077-930X} \and
J. Kinnison$^{1}$ \and
R. Laker$^2$\orcidlink{0000-0002-6577-5515} \and
P. Liewer$^{23}$\orcidlink{0000-0002-5068-4637} \and
D. M. Malaspina$^{24,25}$\orcidlink{0000-0003-1191-1558} \and 
I. Mann$^{26}$\orcidlink{0000-0002-2805-3265} \and
D. J. McComas$^{12}$\orcidlink{0000-0001-6160-1158} \and
T. Niembro-Hernandez$^{4}$\orcidlink{0000-0001-6692-9187} \and
O. Panasenco$^{27}$\orcidlink{0000-0002-4440-7166} \and
P. Pokorn\'{y}$^{28,29,30}$\orcidlink{0000-0002-5667-9337} \and
A. Pusack$^{25}$\orcidlink{0000-0002-3081-8597} \and
M. Pulupa$^5$\orcidlink{0000-0002-1573-7457} \and
J. C. Perez$^{31}$\orcidlink{0000-0002-8841-6443} \and
P. Riley$^{32}$\orcidlink{0000-0002-1859-456X} \and
A. P. Rouillard$^{33}$\orcidlink{0000-0003-4039-5767} \and
C. Shi$^6$\orcidlink{0000-0002-2582-7085} \and
G. Stenborg$^1$\orcidlink{0000-0001-8480-947X} \and
A. Tenerani$^{34}$\orcidlink{0000-0003-2880-6084} \and
J. L. Verniero$^{10}$\orcidlink{0000-0003-1138-652X} \and
N. Viall$^{10}$\orcidlink{0000-0003-1692-1704} \and
A. Vourlidas$^1$\orcidlink{0000-0002-8164-5948} \and
B. E. Wood$^{11}$\orcidlink{0000-0002-4998-0893} \and
L. D. Woodham$^2$\orcidlink{0000-0003-2845-4250} \and
T. Woolley$^2$\orcidlink{0000-0002-9202-619X}
        }



\institute{Corresponding author: Nour E. Raouafi --- \href{mailto:Nour.Raouafi@jhuapl.edu}{Nour.Raouafi@jhuapl.edu} \\
$^1$Johns Hopkins Applied Physics Laboratory, Laurel, MD 20723, USA \\
$^2$Department of Physics, Imperial College London, South Kensington Campus, London SW7 2AZ, UK \\
$^3$Physics Department, University of Otago, Dunedin 9010, New Zealand  \\
$^4$Smithsonian Astrophysical Observatory, Cambridge, MA 02138 USA\\
$^5$Space Sciences Laboratory, University of California, Berkeley, CA 94720-7450, USA  \\
$^6$Earth Planetary and Space Sciences, UCLA, CA 90095, USA  \\
$^7$Lunar \& Planetary Laboratory, University of Arizona, Tucson, AZ 85721, USA  \\
$^8$Department of Physics and Astronomy, Queen Mary University of London, London E1 4NS, UK  \\
$^9$University of Delaware, Department of Physics and Astronomy, Newark, DE 19716, USA  \\
$^{10}$NASA Goddard Space Flight Center, Greenbelt MD 20771, USA  \\
$^{11}$Space Science Division, Naval Research Laboratory, Washington, DC 20375, USA  \\
$^{12}$Department of Astrophysical Sciences, Princeton University, Princeton, NJ 08540, USA  \\
$^{13}$National Institute for Astrophysics (INAF) – Institute for Space Astrophysics and Planetology (IAPS), 00133 Rome, Italy  \\
$^{14}$Climate and Space Sciences and Engineering, University of Michigan, Ann Arbor, MI 48109, USA  \\
$^15$Physics Department, University of California, Berkeley, CA 94720-7300, USA  \\
$^{16}$Mullard Space Science Laboratory, University College London, Dorking RH5 6NT, UK  \\ 
$^{17}$School of Physics and Astronomy, University of Minnesota, Minneapolis, USA  \\
$^{18}$Department of Physics \& Astronomy, University of New Hampshire, Durham, NH, USA  \\ 
$^{19}$Space Science Center, University of New Hampshire, Durham, NH 03824, USA  \\
$^{20}$California Institute of Technology, Pasadena, CA 91125, USA  \\
$^{21}$University of New Hampshire, Durham, NH 03824, USA \\
$^{22}$BWX Technologies, Inc., Washington DC 20002, USA  \\
$^{23}$Jet Propulsion Laboratory, California Institute of Technology, Pasadena, CA 91109, USA\\
$^{24}$Department of Astrophysical and Planetary Sciences, University of Colorado Boulder, Boulder, CO 80309, USA  \\
$^{25}$Laboratory for Atmospheric and Space Physics, University of Colorado, Boulder, CO 80303, USA  \\
$^{26}$Department of Physics and Technology, Postboks 6050 Langnes, 9037 Tromso, Norway  \\
$^{27}$Advanced Heliophysics, Pasadena, CA 91106, USA  \\
$^{28}$Astrophysics Science Division, NASA Goddard Spaceflight Center, Greenbelt, MD 20771, USA  \\
$^{29}$Department of Physics, The Catholic University of America, Washington, DC 20064, USA  \\
$^{30}$Center for Research and Exploration in Space Science and Technology, NASA/GSFC, Greenbelt, MD 20771, USA
$^{31}$Department of Aerospace, Physics and Space Sciences, Florida Institute of Technology, Melbourne, Florida 32901, USA  \\
$^{32}$Predictive Science Inc., San Diego, California, USA  \\
$^{33}$IRAP, Universit\'e de Toulouse, CNRS, CNES, UPS, Toulouse, France  \\
$^{34}$Department of Physics, University of Texas at Austin, TX 78712, USA 
}

\date{Received: date / Accepted: date}

\maketitle

\begin{abstract}
Launched on 12 Aug. 2018, NASA's Parker Solar Probe had completed 13 of its scheduled 24 orbits around the Sun by Nov. 2022. The mission's primary science goal is to determine the structure and dynamics of the Sun's coronal magnetic field, understand how the solar corona and wind are heated and accelerated, and determine what processes accelerate energetic particles. Parker Solar Probe returned a treasure trove of science data that far exceeded quality, significance, and quantity expectations, leading to a significant number of discoveries reported in nearly 700 peer-reviewed publications. The first four years of the 7-year primary mission duration have been mostly during solar minimum conditions with few major solar events. Starting with orbit 8 (i.e., 28 Apr. 2021), Parker flew through the magnetically dominated corona, i.e., sub-Alfv\'enic solar wind, which is one of the mission's primary objectives. In this paper, we present an overview of the scientific advances made mainly during the first four years of the Parker Solar Probe mission, which go well beyond the three science objectives that are: (1) Trace the flow of energy that heats and accelerates the solar corona and solar wind; (2) Determine the structure and dynamics of the plasma and magnetic fields at the sources of the solar wind; and (3) Explore mechanisms that accelerate and transport energetic particles.



\end{abstract}

\tableofcontents

\section{Introduction}
\label{PSPINTRO}

Parker Solar Probe \cite[{\emph{PSP}};][]{2016SSRv..204....7F,doi:10.1063/PT.3.5120} is flying closer to the Sun than any previous spacecraft (S/C). Launched on 12 Aug. 2018, on 6 Dec. 2022 {\emph{PSP}} had completed 14 of its 24 scheduled perihelion encounters (Encs.)\footnote{The PSP solar Enc. is defined as the orbit section where the S/C is below 0.25~AU from the Sun's center.} around the Sun over the 7-year nominal mission duration. The S/C flew by Venus for the fifth time on 16 Oct. 2021, followed by the closest perihelion of 13.28 solar radii (\rst) on 21 Nov. 2021. The S/C will remain on the same orbit for a total of seven solar Encs. After Enc.~16, {\emph{PSP}} is scheduled to fly by Venus for the sixth time to lower the perihelion to 11.44~$R_\odot$ for another five orbits. The seventh and last Venus gravity assist (VGA) of the primary mission is scheduled for 6 Nov. 2024. This gravity assist will set {\emph{PSP}} for its last three orbits of the primary mission. The perihelia of orbits 22, 23, and 24 of 9.86~$R_\odot$ will be on 24 Dec. 2024, 22 Mar. 2025, and 19 Jun. 2025, respectively. 

The mission’s overarching science objective is to determine the structure and dynamics of the Sun’s coronal magnetic field and to understand how the corona is heated, the solar wind accelerated, and how energetic particles are produced and their distributions evolve. The {\emph{PSP}} mission targets processes and dynamics that characterize the Sun’s expanding corona and solar wind, enabling the measurement of coronal conditions leading to the nascent solar wind and eruptive transients that create space weather. {\emph{PSP}} is sampling the solar corona and solar wind to reveal how it is heated and the solar wind and solar energetic particles (SEPs) are accelerated. To achieve this, {\emph{PSP}} measurements will be used to address the following three science goals: (1) Trace the flow of energy that heats the solar corona and accelerates the solar wind; (2) Determine the structure and dynamics of the plasma and magnetic fields at the sources of the solar wind; and (3) Explore mechanisms that accelerate and transport energetic particles. Understanding these phenomena has been a top science goal for over six decades. {\emph{PSP}} is primarily an exploration mission that is flying through one of the last unvisited and most challenging regions of space within our solar system, and the potential for discovery is huge. 

The returned science data is a treasure trove yielding insights into the nature of the young solar wind and its evolution as it propagates away from the Sun. Numerous discoveries have been made over the first four years of the prime mission, most notably the ubiquitous magnetic field switchbacks closer to the Sun, the dust-free zone (DFZ), novel kinetic aspects in the young solar wind, excessive tangential flows beyond the Alfv\'en critical point, dust $\beta$-streams resulting from collisions of the Geminids meteoroid stream with the zodiacal dust cloud (ZDC), and the shortest wavelength thermal emission from the Venusian surface. Since 28 Apr. 2021 ({\emph{i.e.}}, perihelion of Enc.~8), the S/C has been sampling the solar wind plasma within the magnetically-dominated corona, {\emph{i.e.}}, sub-Alfv\'enic solar wind, marking the beginning of a critical phase of the {\emph{PSP}} mission. In this region solar wind physics changes because of the multi-directionality of wave propagation (waves moving sunward and anti-sunward can affect the local dynamics including the turbulent evolution, heating and acceleration of the plasma). This is also the region where velocity gradients between the fast and slow speed streams develop, forming the initial conditions for the formation, further out, of corotating interaction regions (CIRs).

The science data return ({\emph{i.e.}}, data volume) from {\emph{PSP}} exceeded the pre-launch estimates by a factor of over four. Since the second orbit, orbital coverage extended from the nominal perihelion Enc. to over 70\% of the following orbit duration. We expect this to continue throughout the mission. The {\emph{PSP}} team is also looking into ways to extend the orbital coverage to the whole orbit duration. This will allow sampling the solar wind and SEPs over a large range of heliodistances. 

The {\emph{PSP}} science payload comprises four instrument suites:
\begin{enumerate}
    \item{}FIELDS investigation makes measurements of the electric and magnetic fields and waves, S/C floating potential, density fluctuations, and radio emissions over 20 MHz of bandwidth and 140 dB of dynamic range. It comprises 
    \begin{itemize}
        \item Four electric antennas (V1-V4) mounted at the base of the S/C thermal protection system (TPS). The electric preamplifiers connected to each antenna provide outputs to the Radio Frequency
Spectrometer (RFS), the Time Domain Sampler (TDS), and the Antenna Electronics Board (AEB) and Digital Fields Board (DFB). The V1-V4 antennas are exposed to the full solar environment.
        \item A fifth antenna (V5) provides low (LF) and medium (MF) frequency outputs.
        \item Two fluxgate magnetometers (MAGs) provide data with bandwidth of $\sim140$~Hz and at 292.97 Samples/sec over a dynamic range of $\pm65,536$~nT with a resolution of 16 bits.
        \item A search coil magnetometer (SCM) measures the AC magnetic signature of solar wind fluctuations, from 10 Hz up to 1 MHz.
    \end{itemize}
V5, the MAGs, and the SCM are all mounted on the boom in the shade of the TPS. For further details, see \citet{2016SSRv..204...49B}.
    \item{}The Solar Wind Electrons Alphas and Protons (SWEAP) Investigation measures the thermal solar wind, {\emph{i.e.}}, electrons, protons and alpha particles. SWEAP measures the velocity distribution functions (VDFs) of ions and electrons with high energy and angular resolution. It consists of the Solar Probe Cup (SPC) and the Solar Probe Analyzers (SPAN-A and SPAN-B), and the SWEAP Electronics Module (SWEM):
    \begin{itemize}
    \item SPC is fully exposed to the solar environment as it looks directly at the Sun and measures ion and electron fluxes and flow angles as a function of energy.
    \item SPAN-A is mounted on the ram side and comprises an ion and electron electrostatic analyzers (SPAN-i and SPAN-e, respectively).
    \item SPAN-B is an electron electrostatic analyzer on the anti-ram side of the S/C.
    \item The SWEM manages the suite by distributing power, formatting onboard data products, and serving as a single electrical interface to the S/C.
    \end{itemize}
The SPANs and the SWEM reside on the S/C bus behind the TPS. See \citet{2016SSRv..204..131K} for more information.
    
    \item{} The Integrated Science Investigation of the Sun (IS$\odot$IS) investigation measures energetic particles over a very broad energy range (10s of keV to 100 MeV). IS$\odot$IS is mounted on the ram side of the S/C bus. It comprises two Energetic Particle Instruments (EPI) to measure low (EPI-Lo) and high (EPI-Hi) energy:
    \begin{itemize}
    \item EPI-Lo is time-of-flight (TOF) mass spectrometer that measures electrons from $\sim25–1000$~keV, protons from $\sim0.04–7$~MeV, and heavy ions from $\sim0.02–2$~MeV/nuc. EPI-Lo has 80 apertures distributed over eight wedges. Their combined fields-of-view (FOVs) cover nearly an entire hemisphere.
    \item EPI-Hi measures electrons from $\sim0.5–6$~MeV and ions from $\sim1–200$ MeV/nuc. EPI-Hi consists of three telescopes: a high energy telescope (HET; double ended) and two low energy telescopes LET1 (double ended) and LET2 (single ended).
    \end{itemize}
See \citep{2016SSRv..204..187M} for a full description of the IS$\odot$IS investigation.
    \item{} The Wide-Field Imager for Solar PRobe (WISPR) is the only remote-sensing instrument suite on the S/C. WISPR is a white-light imager providing observations of flows and transients in the solar wind over a $95^\circ\times58^\circ$ (radial and transverse, respectively) FOV covering elongation angles from $13.5^\circ$ to $108^\circ$. It comprises two telescopes:
    \begin{itemize}
    \item WISPR-i covers the inner part of the FOV ($40^\circ\times40^\circ$).
    \item WISPR-o covers the outer part of the FOV ($58^\circ\times58^\circ$).
    \end{itemize}
See \citet{2016SSRv..204...83V} for further details. 
\end{enumerate}

Before tackling the {\emph{PSP}} achievements during the first four years of the prime mission, a brief historical context is given in \S\ref{HistoricalContext}. \S\ref{PSPMSTAT} provides a brief summary of the {\emph{PSP}} mission status. \S\S\ref{MagSBs}-\ref{PSPVENUS} describe the {\emph{PSP}} discoveries during the first four years of operations: switchbacks, solar wind sources, kinetic physics, turbulence, large-scale structures, energetic particles, dust, and Venus, respectively. The conclusions and discussion are given in \S\ref{SUMCONC}.

{\emph{Although sections 3-12 may have some overlap and cross-referencing, each section can be read independently from the rest of the paper.}}

\section{Historical Context: {\emph{Mariner~2}}, {\emph{Helios}}, and {\emph{Ulysses}}}
\label{HistoricalContext}

Before {\emph{PSP}}, several space missions shaped our understanding of the solar wind for decades. Three stand out as trailblazers, {\emph{i.e.}}, {\emph{Mariner~2}}, {\emph{Helios}} \citep{Marsch1990} , and {\emph{Ulysses}} \citep{1992AAS...92..207W}.

{\emph{Mariner~2}}, launched on 27 Aug. 1962, was the first successful mission to a planet other than the Earth ({\emph{i.e.}}, Venus). Its measurements of the solar wind are a first and among the most significant discoveries of the mission \citep[see ][]{1962Sci...138.1095N}. Although the mission returned data for only a few months, the measurements showed the highly variable nature and complexity of the plasma flow expanding anti-sunward \citep{1965ASSL....3...67S}. However, before the launch of {\emph{PSP}},  almost everything we knew about the inner interplanetary medium was due to the double {\emph{Helios}} mission. This mission set the stage for an initial understanding of the major physical processes occurring in the inner heliosphere. It greatly helped the development and tailoring of instruments onboard subsequent missions.

The two {\emph{Helios}} probes were launched on 10 Dec. 1974 and 15 Jan. 1976 and placed in orbit in the ecliptic plane. Their distance from the Sun varied between 0.29 and 1 astronomical unit (AU) with an orbital period of about six months. The payload of the two {\emph{Helios}} comprised several instruments:
\begin{itemize}
\item Proton, electron, and alpha particle analyzers;
\item Two DC magnetometers;
\item A search coil magnetometer;
\item A radio wave experiment;
\item Two cosmic ray experiments;
\item Three photometers for the zodiacal light; and
\item A dust particle detector
\end{itemize}
Here we provide a very brief overview of some of the scientific goals achieved by {\emph{Helios}} to make the reader aware of the importance that this mission has had in the study of the solar wind and beyond.

{\emph{Helios}} established the mechanisms which generate dust particles at the origin of the zodiacal light (ZL), their relationship with micrometeorites and comets, and the radial dependence of dust density \citep{1976BAAS....8R.457L}. Micrometeorite impacts of the dust particle sensors allowed to study asymmetries with respect to (hereafter w.r.t.) the ecliptic plane and the different origins related to stone meteorites or iron meteorites and suggested that many particles run on hyperbolic orbits aiming out of the solar system \citep{1980P&SS...28..333G}.

{\emph{Helios}}' plasma wave experiment firstly confirmed that the generation of type~III radio bursts is a two-step process, as theoretically predicted by \citet{1958SvA.....2..653G} and revealed enhanced levels of ion acoustic wave turbulence in the solar wind. In addition, the radial excursion of {\emph{Helios}} allowed proving that the frequency of the radio emission increases with decreasing the distance from the Sun and the consequent increase of plasma density \citep{1979JGR....84.2029G,1986A&A...169..329K}. The radio astronomy experiments onboard both S/C were the first to provide ``three-dimensional (3D) direction finding" in space, allowing to follow the source of type~III radio bursts during its propagation along the interplanetary magnetic field lines. In practice, they provided a significant extension of our knowledge of the large-scale structure of the interplanetary medium via remote sensing \citep{1984GeCAS......111K}.

The galactic and cosmic ray experiment studied the energy spectra, charge composition, and flow patterns of both solar and galactic cosmic rays (GCRs). {\emph{Helios}} was highly relevant as part of a large network of cosmic ray experiments onboard S/C located between 0.3 and 10~AU. It contributed significantly to confirming the role of the solar wind suprathermal particles as seed particles injected into interplanetary shocks to be eventually accelerated \citep{1976ApJ...203L.149M}. Coupling observations by {\emph{Helios}} and other S/C at 1~AU allowed studying the problem of transport performing measurements in different conditions relative to magnetic connectivity and radial distance from the source region. Moreover, joint measurements between {\emph{Helios}} and {\emph{Pioneer}}~10 gave important results about the modulation of cosmic rays in the heliosphere \citep{1978cosm.conf...73K,1984GeCAS......124K}.

The solar wind plasma experiment and the magnetic field experiments allowed us to investigate the interplanetary medium from the large-scale structure to spatial scales of the order of the proton Larmor radius for more than an entire 11-year solar cycle. The varying vantage point due to a highly elliptic orbit allowed us to reach an unprecedented description of the solar wind's large-scale structure and the dynamical processes that take place during the expansion into the interplanetary medium \citep{1981sowi.conf..126S}. {\emph{Helios}}' plasma and magnetic field continuous observations allowed new insights into the study of magneto-hydrodynamic (MHD) turbulence opening Pandora's box in our understanding of this phenomenon of astrophysical relevance \citep[see reviews by][]{1995Sci...269.1124T,2013LRSP...10....2B}. Similarly, detailed observations of the 3D velocity distribution of protons, alphas, and electrons not only revealed the presence of anisotropies w.r.t. the local magnetic field but also the presence of proton and alpha beams as well as electron strahl. Moreover, these observations allowed us to study the variability and evolution of these kinetic features with heliocentric distance and different Alfv\'enic turbulence levels at fluid scales \citep[see the review by][]{1995Sci...269.1124T}.

Up to the launch of {\emph{Ulysses}} on 6 Oct. 1990, the solar wind exploration was limited to measurements within the ecliptic plane. Like {\emph{PSP}}, the idea of flying a mission to explore the solar polar region dates back to the 1959 Simpson's Committee report. Using a Jupiter gravity assist, {\emph{Ulysses}} slang shot out of the ecliptic to fly above the solar poles and provide unique measurements. During its three solar passes in 1994-95, 2000-01, and 2005, {\emph{Ulysses}} covered two minima and one maximum of the solar sunspot cycle, revealing phenomena unknown to us before \citep[see][]{2008GeoRL..3518103M}. All measurements were, however, at heliodistances beyond 1~AU and only {\emph{in~situ}}, as there were no remote-sensing instruments onboard.

\section{Mission Status}
\label{PSPMSTAT}
\begin{figure}[!ht]
\begin{center}
\includegraphics[width=1\columnwidth]{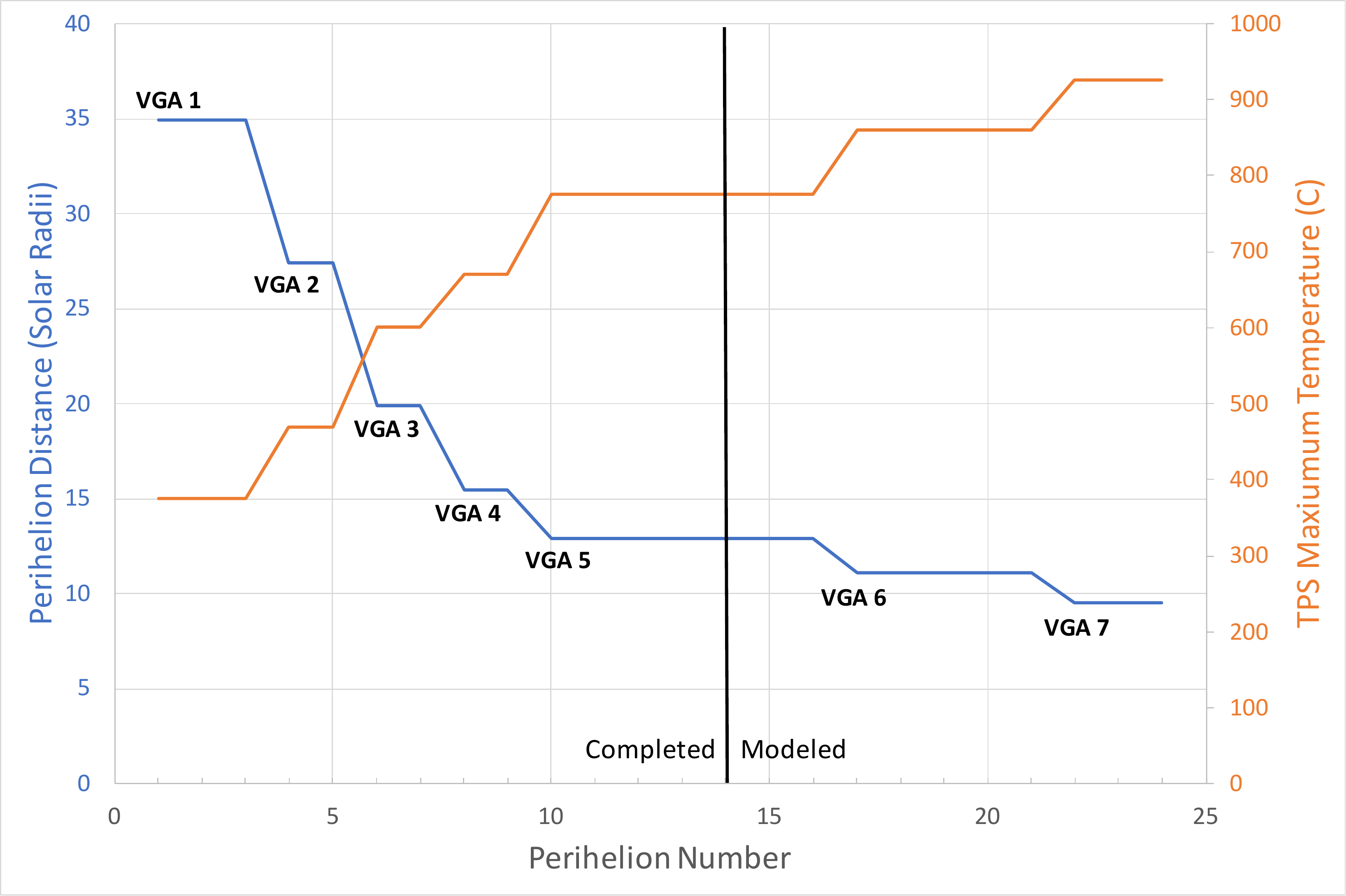}
\includegraphics[width=1\columnwidth]{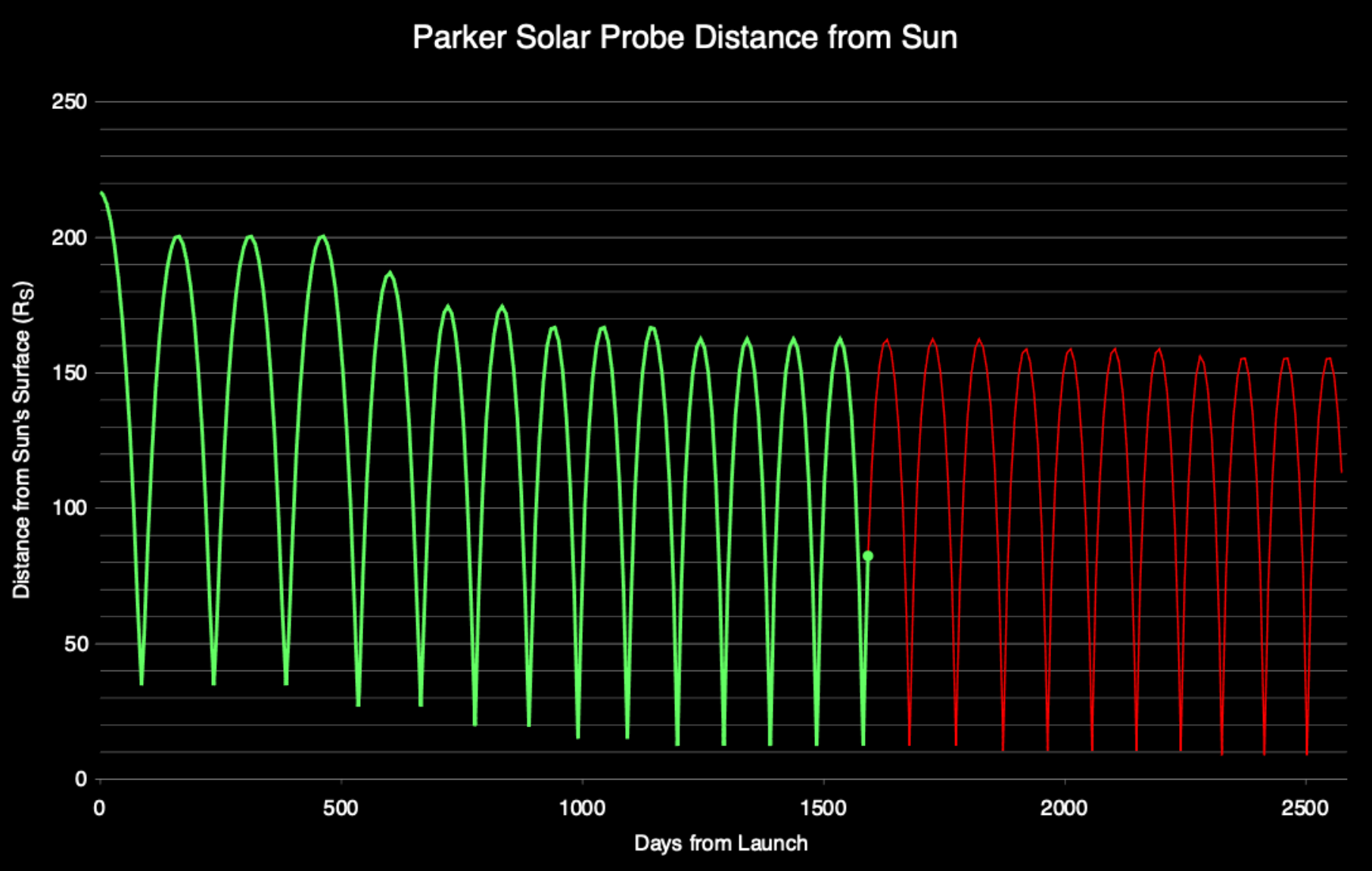}
\caption{(Top-blue) PSP's perihelion distance is decreased by performing gravity assists using Venus (VGAs). After seven close flybys of Venus, the final perihelion is anticipated to be 9.86~$R_\odot$ from the Sun's center. (Top-orange) The modeled temperature of the TPS sunward face at each perihelion. The thermal sensors on the S/C (behind the TPS) confirm the TPS thermal model. It is noteworthy that there are no thermal sensors on the TPS itself. (Bottom) The trajectory of PSP during the 7-year primary mission phase as a function of days after the launch on 12 Aug. 2018. The green (red) color indicates the completed (future) part of the PSP orbit. The green dot shows the PSP heliodistance on 21 Dec. 2022.}
\label{Fig_PSPStatus}
\end{center}
\end{figure}

After a decade in the making, {\emph{PSP}} began its 7-year journey into the Sun’s corona on 12 Aug. 2018 \citep{9172703}. Following the launch, about six weeks later, the S/C flew by Venus for the first of seven gravity assists to target the initial perihelion of 35.6~$R_\odot$. As the S/C continues to perform VGAs, the perihelion has been decreased to 13.28~$R_\odot$ after the fifth VGA, with the anticipation of a final perihelion of 9.86~$R_\odot$ in the last three orbits. Fig.~\ref{Fig_PSPStatus} shows the change in perihelia as the S/C has successfully completed the VGAs and the anticipated performance in future orbits. Following the seventh VGA, the aphelion is below Venus’ orbit. So, no more VGAs will be possible, and the orbit perihelion will remain the same for a potential extended mission. 

As shown in Fig.~\ref{Fig_PSPStatus}, the S/C had completed 13 orbits by Oct. of 2022, with an additional 11 orbits remaining in the primary mission. As designed, these orbits are separated into a solar Enc. phase and a cruise phase. Solar Encs. are dedicated to taking the data that characterize the near-Sun environment and the corona. The cruise phase of each orbit is devoted to a mix of science data downlink, S/C operations, and maintenance, and science in regions further away from the Sun. 

The major engineering challenge for the mission before launch was to design and build a TPS that would keep the bulk of the S/C at comfortable temperatures during each solar Enc. period. Fig.~\ref{Fig_PSPStatus} also shows the temperature of the TPS' sunward face at each perihelion, the maximum temperature in each orbit. Given the anticipated temperature at the final perihelion of nearly 1000$^\circ$C, the TPS does not include sensors for the direct measurement of the TPS temperature. However, the S/C has other sensors, such as the barrier blanket temperature sensor and monitoring of the cooling system, with which the S/C's overall thermal model has been validated. Through orbit 13, the thermal model and measured temperatures agree very well, though actual temperatures are slightly lower as the model included conservative assumptions for inputs such as surface properties. This good agreement holds throughout the orbits, including aphelion. For the early orbits, the reduced solar illumination when the S/C is further away from the Sun raised concerns before launch that the cooling system might freeze unless extra energy was provided to the S/C by tilting the system to expose more surface area to the Sun near aphelion. This design worked as expected, and temperatures near aphelion have been comfortably above the point where freezing might occur. 

The mission was designed to collect science data during solar Encs. ({\emph{i.e.}}, inside 0.25~AU) and return that data to Earth during the cruise phase, when the S/C is further away from the Sun. The system was designed to do this using a Ka-band telecommunications link, one of the first uses of this technology for APL\footnote{The Johns Hopkins University Applied Physics Laboratory, Laurel, Maryland.}, with the requirement of returning an average of 85 Gbits of science data per orbit. While the pre-launch operations plan comfortably exceeded this, the mission has returned over three times the planned data volume through the first 13 orbits, with increased data return expected through the remaining orbits. The increased data return is mainly due to better than expected performance of the Ka-band telecommunications system. It has resulted in the ability to measure and return data throughout the orbit, not just in solar Encs., to characterize the solar environment fully. 

Another major engineering challenge before launch was the ability of the system to detect and recover from faults and to maintain attitude control of the S/C to prevent unintended exposure to solar illumination. The fault management is, by necessity, autonomous, since the S/C spends a significant amount of time out of communication with Mission Operations during the solar Enc. periods in each orbit. A more detailed discussion of the design and operation of the S/C autonomy system is found in \citet{9172703}. Through 13 orbits, the S/C has successfully executed each orbit and operated as expected in this harsh environment. We have seen some unanticipated issues, associated mainly with the larger-than-expected dust environment, that have affected the S/C. However, the autonomy system has successfully detected and recovered from all of these events. The robust design of the autonomy system has kept the S/C safe, and we expect this to continue through the primary mission. 

Generally, the S/C has performed well within the expectations of the engineering team, who used conservative design and robust, redundant systems to build the highly capable {\emph{PSP}}. Along with this, a major factor in the mission's success so far is the tight coupling between the engineering and operations teams and the science team. Before launch, this interaction gave the engineering team insight into this unexplored near-Sun environment, resulting in designs that were conservative. After launch, the operations and science teams have worked together to exploit this conservatism to achieve results far beyond expectations. 

\section{Magnetic Field Switchbacks}
\label{MagSBs}

Abrupt and significant changes of the interplanetary magnetic field direction were reported as early as the mid-1960's \citep[see][]{1966JGR....71.3315M}. The cosmic ray anisotropy remained well aligned with the field. \citet{1967JGR....72.1917M} also reported increases in the radial solar wind speed accompanying the magnetic field deviations from the Parker spiral. Using {\emph{Ulysses}}’ data recorded above the solar poles at heliodistances $\ge1$~AU, \citep{1999GeoRL..26..631B} analyzed the propagation direction of waves to show that these rotations in the magnetic field of $90^\circ$ w.r.t. the Parker spiral are magnetic field line folds rather than opposite polarity flux tubes originating at the Sun. Magnetic field inversions were observed at 1~AU by the International Sun-Earth Explorer-3 ({\emph{ISEE}}-3 [\citealt{1979NCimC...2..722D}]; \citealt{1996JGR...10124373K}) and the  Advanced Composition ({\emph{ACE}} [\citealt{1998SSRv...86....1S}]; \citealt{2009ApJ...695L.213G,2016JGRA..12110728L}). Inside 1~AU, the magnetic field reversals were also observed in the {\emph{Helios}} \citep{1981ESASP.164...43P} solar wind measurements as close as 0.3~AU from the Sun's center \citep{2016JGRA..121.5055B,2018MNRAS.478.1980H}. 

The magnetic field switchbacks took center stage recently owing to their prominence and ubiquitousness in the {\emph{PSP}} measurements inside 0.2~AU.

\subsection{What is a switchback?}
Switchbacks are short magnetic field rotations that are ubiquitously observed in the solar wind. They are consistent with local folds in the magnetic field rather than changes in the magnetic connectivity to solar source regions. This interpretation is supported by the observation of suprathermal electrons \citep{1996JGR...10124373K}, the differential streaming of alpha particles \citep{2004JGRA..109.3104Y} and proton beams \citep{2013AIPC.1539...46N}, and the directionality of Alfv\'en waves (AWs) \citep{1999GeoRL..26..631B}. Because of the intrinsic Alfv\'enic nature of these structures $-$ implying a high degree of correlation between magnetic and velocity fluctuations in all field components $-$ the magnetic field fold has a distinct velocity counterpart. Moreover, the so called \emph{one-sided} aspect of solar wind fluctuations during Alfv\'enic streams \citep{2009ApJ...695L.213G}, which is a consequence of the approximate constancy of the magnetic field strength $\Bm=|\B|$ during these intervals, has a direct impact on the distribution of $B_R$ and $V_R$ in switchbacks. Under such conditions (constant $B$ and Alfv\'enic fluctuations), large magnetic fields rotations, and switchbacks in particular, always lead to bulk speed enhancements \citep{2014GeoRL..41..259M}, resulting in a spiky solar wind velocity profile during Alfv\'enic periods. Since the amplitude of the velocity spikes associated to switchbacks is proportional to the local Alfv\'en speed $\va$, the speed modulation is particularly intense in fast-solar-wind streams observed inside 1~AU, where $\va$ is larger, and it was suggested that velocity spikes could be even larger closer-in \citep{2018MNRAS.478.1980H}.

Despite the previous knowledge of switchbacks in the solar wind community and some expectations that they could have played some role closer to the Sun, our consideration of these structures has been totally changed by {\emph{PSP}}, since its first observations inside 0.3~AU  \citep{2019Natur.576..228K,2019Natur.576..237B}. The switchback occurrence rate, morphology, and amplitude as observed by {\emph{PSP}}, as well as the fact that they are ubiquitously observed also in slow, though mostly Alfv\'enic, solar wind, made them one of the most interesting and intriguing aspects of the first {\emph{PSP}} Encs.

In this section, we summarize recent findings about switchbacks from the first {\emph{PSP}} orbits. In Section \ref{SB_obs} we provide an overview of the main observational properties of these structures in terms of size, shape, radial evolution, and internal and boundary properties; in Section \ref{sec: theory switchbacks} we present current theories for the generation and evolution of switchbacks, presenting different types of models, based on their generation at the solar surface or {\emph{in situ}} in the wind. \S\ref{SB_discussion} contains a final discussion of the state-of-art of switchbacks' observational and theoretical studies and a list of current open questions to be answered by {\emph{PSP}} in future Encs.

\subsection{Observational properties of switchbacks}\label{SB_obs}

\subsubsection{Velocity increase inside switchbacks}\label{sub:obs: velocity increase}
\begin{figure}
\begin{center}
\includegraphics[width=1\columnwidth]{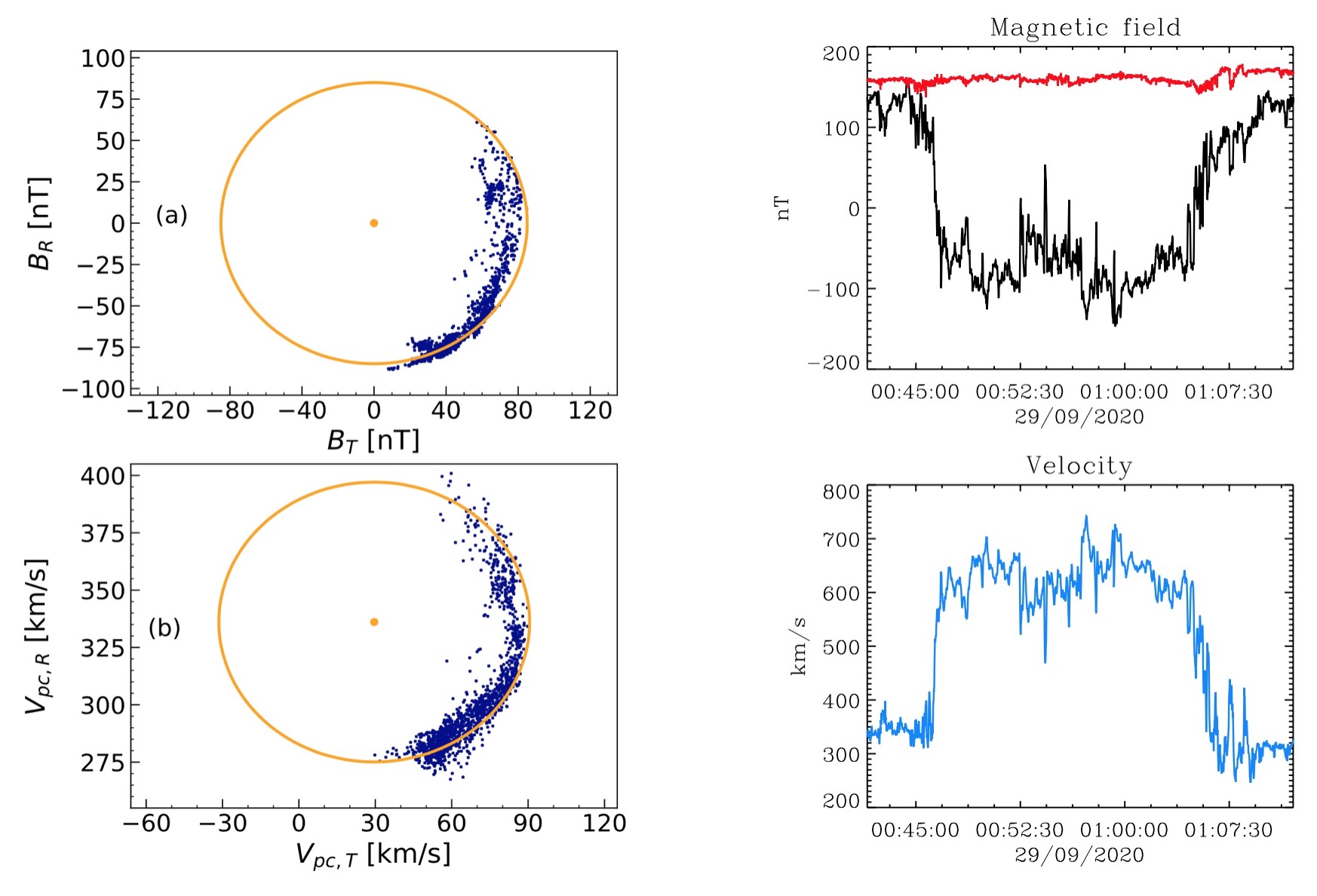}
\caption{
{\it Left:} Magnetic field and velocity vector rotations during a large amplitude switchback during {\emph{PSP}} Enc.~1 \citep{2020MNRAS.498.5524W}. {\it Right:} An example of switchback observed by {\emph{PSP}} during Enc.~6. Top panel shows the almost complete magnetic field reversal of $B_R$ (black), while the magnetic field intensity $|B|$ (red) remains almost constant through the whole structure. The bottom panel shows the associated jump in the radial velocity $V_R$. In a full switchback the bulk speed of the solar wind protons can increase by up to twice the Alfv\'en speed $\va$; as a consequence we observe a jump from $\sim300$~km~s$^{-1}$ to $\sim600$~km~s$^{-1}$ in the speed during this interval ($\va\sim150$~km~s$^{-1}$).}
\label{fig:big_sb}
\end{center}
\end{figure}

At first order, switchbacks can be considered as strong rotations of the magnetic-field vector $\B$, with no change in the magnetic field intensity $\Bm=|\B|$. Geometrically, this corresponds to a rotation of $\B$ with its tip constrained on a sphere of constant radius $\Bm$. Such excursions are well represented by following the $\Bm$ direction in the RT plane, during the time series of a large amplitude switchback, like in the left panels of Fig.~\ref{fig:big_sb}. The top left panel represents the typical $\B$ pattern observed since  Enc.~1 \citep{2020MNRAS.498.5524W}: the background magnetic field, initially almost aligned with the radial ($B_R<0$) in the near-Sun regions observed by {\emph{PSP}}, makes a significant rotation in the RT plane, locally inverting its polarity ($B_R>0$). All this occurs keeping $\Bm\sim\mathrm{const.}$ and points follow a circle of approximately constant radius during the rotation; as a consequence this increases significantly the transverse component of $\B$ and $B_T\gg B_R$ when approaching $90^\circ$. Due to the high Alfv\'enicity of the fluctuations in near-Sun streams sampled by {\emph{PSP}}, the same pattern is observed for the velocity vector, with similar and proportional variations in $V_R$ and $V_T$ (bottom left panel). While the magnetic field is frame-invariant, the circular pattern seen for the velocity vector is not and its center identifies the so-called de Hoffman-Teller frame (dHT): the frame in which the motional electric field associated to the fluctuations is zero and where the switchbacks magnetic structure can be considered at rest. This frame is typically traveling at the local Alfv\'en speed ahead of the solar wind protons, along the magnetic field. This is consistent with the velocity measurements in the bottom left panel of Fig.~\ref{fig:big_sb}, where the local $\va$ is of the order of $\sim50$~km~s$^{-1}$ and agrees well with the local of the centre of the circle, which is roughly 50~km~s$^{-1}$ ahead of the minimum $V_R$ seen at the beginning of the interval. 

Because of the geometrical property above, there is a direct relation between the $\B$ excursion and the resulting modulation of the flow speed in switchbacks. Remarkably, switchbacks always lead to speed increases, characterized by a spiky, one-sided profile of $V_R$, independent of the underlying magnetic field polarity; {\emph{i.e.}}, regardless $\B$ rotates from $0^\circ$ towards $180^\circ$, or vice-versa \citep{2014GeoRL..41..259M}. As a consequence, it is possible to derive a simple phenomenological relation that links the instantaneous proton radial velocity $V_R$ to the magnetic field angle w.r.t. the radial $\theta_{BR}$, where $\cos\theta_{BR}=B_R$/\Bm. Moreover, since the solar wind speed is typically  dominated by its radial component, this can be considered an approximate expression for the proton bulk speed within switchbacks \citep{2015ApJ...802...11M}:
\begin{equation}
    V_p=V_0+\va[1\pm\cos{\theta_{BR}}],\label{eq_v_in_sb}
\end{equation}
where $V_0$ is the background solar wind speed and the sign in front of the cosine takes into account the underlying Parker spiral polarity ($-\cos{\theta_{BR}}$ if $B_R>0$, $+\cos{\theta_{BR}}$ otherwise). As apparent from Eq.~(\ref{eq_v_in_sb}), the speed increase inside a switchback with constant $\Bm$ has a maximum amplitude of $2\times\va$. This corresponds to magnetic field rotations that are full reversals;  for moderate deflections, the speed increase is smaller, typically of the order of $\sim{\va}$ for a $90^\circ$ deflection. Also, because the increase in $V_p$ is proportional to the local Alfv\'en speed, larger enhancements are expected closer to the Sun.

The right panels of Fig.~\ref{fig:big_sb} show  one of the most striking examples of switchbacks observed by {\emph{PSP}} during Enc.~6. This corresponds to an almost full reversal of $B_R$, from approximately $100$ to ${-100}$ nT, maintaining the magnetic field intensity remarkably constant during the vector $\B$ rotation.
As a consequence, the background bulk flow proton velocity ($\sim300$~km~s$^{-1}$) goes up by almost $2~\va$, leading to a speed enhancement up to 600~km~s$^{-1}$ inside the structure ($\va\sim150$~km~s$^{-1}$). This has the impressive effect of turning the ambient slow solar wind into fast for the duration of the crossing, without a change in the connection to the source. It is an open question if even larger velocity jumps could be observed closer in, when $\va$ approaches $200-300$~km~s$^{-1}$ and becomes comparable to the bulk flow itself, and what would be the consequences on the overall flow energy and dynamics.

Finally, it is worth emphazising that the velocity enhancements  discussed above relate only to the main proton core population in the solar wind plasma. Other species, like proton beams and alpha particles, react differently to switchbacks and may or may not partake in the Alfv\'enic motion associated to these structures, depending on their relative drift w.r.t. the proton core. In fact, alpha particles typically stream faster than protons along the magnetic field in Alfv\'enic streams, with a drift speed that in the inner heliosphere can be quite close to $\va$. As a consequence they sit close to the zero electric field reference frame (dHT) and display much smaller oscillations and speed variations in switchbacks (in the case they stream exactly at the same speed as the phase velocity of the switchback, they are totally unaffected and do not feel any fold in the field \citep[see {\emph{e.g.}},][]{2015ApJ...802...11M}. Similarly, proton beams have drift speeds that exceed the local Alfv\'en speed close to the Sun and therefore, because they stream faster than the dHT, they are observed to oscillate out of phase with the main proton core \citep[{\emph{i.e.}}, they get slower inside switchbacks and the core-beam structure of the proton VDF is locally reversed where $B_R$ flips sign;][]{2013AIPC.1539...46N}. The same happens for the electron strahl, leading to an inversion in the electron heat-flux.

\subsubsection{Characteristic Scales, Size and Shape}\label{sub:obs: shapes and sizes}
Ideally, switchbacks would be imaged from a range of angles, providing a straightforward method to visualize their shape. However, as mentioned above, these structures are Alfv\'enic and have little change in plasma density, which is essential for line of sight (LOS) images from remote sensing instruments. We must instead rely on the {\emph{in~situ}} observations from a single S/C, which are fundamentally local measurements. Therefore, it is important to understand the relationship between the true physical structure of a switchback and the data measured by a S/C, as this can influence the way in which we think about and study them. For example, a small duration switchback in the {\emph{PSP}} time series may be due to a physically smaller switchback, or because {\emph{PSP}} clipped the edge of a larger switchback. This ambiguity also applies to a series of multiple switchbacks, which may truly be several closely spaced switchbacks or in fact one larger, more degraded switchback \citep{2021ApJ...915...68F}.

\citet{2020ApJS..246...39D} provided the first detailed statistics on switchbacks for {\emph{PSP}}'s first Enc. They showed that switchback duration could vary from a few seconds to over an hour, with no characteristic timescale. Through studying the waiting time (the time between each switchback) statistics, they found that switchbacks exhibited long term memory, and tended to aggregate, which they take as evidence for similar coronal origin. Many authors define switchbacks as deflections, above some threshold, away from the Parker spiral. The direction of this deflection, {\emph{i.e.}} towards +T, is also interesting as it could act as a testable prediction of switchback origin theories \cite{2021ApJ...909...95S}. For Enc.~1 at least, \citet{2020ApJS..246...39D} showed that deflections were isotropic about the Parker spiral direction, although they did note that the longest switchbacks displayed a weak preference for deflections in +T. \citet{2020ApJS..246...45H} also found that switchbacks displayed a slight preference to deflect in T rather than N, although there was no distinction between -T or +T. The authors refer to the clock angle in an attempt to quantify the direction of switchback deflection. This is defined as the ``angle of the vector projected onto a plane perpendicular to the Parker spiral that also contains N", where 0$^{\circ}$, 90$^{\circ}$, 180$^{\circ}$ and 270$^{\circ}$ refer to +N, +T, -N, -T directions respectively. Unlike the entire switchback population, the longest switchbacks did show a preference for deflection direction, that often displayed clustering about a certain direction that was not correlated to the solar wind flow direction. Crucially, \citet{2020ApJS..246...45H} demonstrated a correlation between the duration of a switchback and the direction of deflection. They then asserted that the duration of a switchback was related to the way in which {\emph{PSP}} cut through the true physical shape. Since switchbacks are Alfv\'enic, the direction of the magnetic field deflection also creates a flow deflection. This, when combined with the S/C velocity (which had a maximum tangential component of +90~km~s$^{-1}$ during the first Enc.), sets the direction at which {\emph{PSP}} travels through a switchback. As a first attempt, they assumed the switchbacks were aligned with the radial direction or dHT, allowing for the angle of {\emph{PSP}} w.r.t. the flow to be calculated. The authors then demonstrated that as the angle to the flow decreased, the switchback duration increased, implying that these structures were long and thin along the flow direction, with transverse scales around $10^{4}$~km.

This idea was extended by \citet{2021AA...650A...1L} to more solar wind streams across the first two Encs. Instead of assuming a flow direction, they instead started with the idea that the structures were long and thin, and attempted to measure their orientation and dimensions. Allowing the average switchback width and aspect ratio to be free parameters they fit an expected model to the distribution of switchback durations, w.r.t. the S/C cutting angle. They applied this method while varying the switchback orientation, finding the orientation that was most consistent with the long, thin model. Switchbacks were found to be aligned away from the radial direction, towards to the Parker spiral. The statistical average switchback width was around $50,000$~km, with an aspect ratio of the order of 10, although there was a large variation. \citet{2021AA...650A...1L} again emphazised that the duration of a switchback is a function of how the S/C cut through the structure, which is in turn related to the switchback deflection, dimensions, orientation and S/C velocity. A similar conclusion was also reached by \citet{2020MNRAS.494.3642M} who argued that the direction of {\emph{Helios}} w.r.t. switchbacks could influence the statistics seen in the data.

Unlike the previous studies that relied on large statistics, \citet{2020ApJ...893...93K} analyzed several case study switchbacks during the first Enc., finding currents at the boundaries. They argued that these currents flowed along the switchback surface, and also imagined switchbacks to be cylindrical. Analysing the flow deflections relative to the S/C for three switchbacks, they found a transverse scale of $7,000$~km and $50,000$~km for a compressive and Alfv\'enic switchback, respectively. A similar method was applied to a larger set of switchbacks by \citet{2021AA...650A...3L}, who used minimum variance analysis (MVA) to find the normal directions of the leading and trailing edge. After calculating the width of the edges, an average normal velocity was multiplied by the switchback duration to give a final width. They found that the transverse switchback scale varied from several thousand km to a solar radius ($695,000$~km), with the mode value lying between $10^{4}$~km and $10^{5}$~km.

A novel approach to probe the internal structure of switchbacks was provided by \citet{2021AA...650L...4B}, who studied the behavior of energetic particles during switchback periods in the first five {\emph{PSP}} Encs. Energetic particles (80-200 MeV/nucleus) continued to stream anti-sunward during a switchback in 86\% of cases, implying that the radius of magnetic field curvature inside switchbacks was smaller or comparable to the ion gyroradius. Using typical solar wind parameters ($\Bm\sim50$ nT, ion energy $100$~eV) this sets an upper limit of $\sim4000$~km for the radius of curvature inside a switchback. Assuming a typical S-shaped curve envisaged by \citet{2019Natur.576..228K}, this would constrain the switchback width to be less than $\sim16,000$~km. 

A summary of the results is displayed in Table \ref{tab:shape_size}, which exhibits a large variation but a general consensus that the switchback transverse scale ranges from $10^{3}$~km to $10^{5}$~km. Future areas of study should be focused on how the switchback shape and size varies with distance from the Sun. However, a robust method for determining how {\emph{PSP}} cut through the switchback must be found for progress to be made in this area. For example, an increased current density or wave activity at the boundary may be used a signature of when {\emph{PSP}} is clipping the edge of a switchback. Estimates of switchback transverse scale, like \citet{2021AA...650A...3L}, could be constrained with the use of energetic particle data \citep{2021AA...650L...4B} on a case-by-case basis, improving the link between the duration measured by a S/C and the true physical size of the switchback.

\begin{table}[t]
\centering
\resizebox{\textwidth}{!}{
\begin{tabular}{l|l|l|l}
{\bf{Study}}                              & {\bf{Enc.}} & {\bf{Transverse Scale (km)}}      & {\bf{Aspect}}   \\ \hline
\citet{2020ApJS..246...45H}    & 1               & $10^{4}$ km                                & -         \\ \hline
\citet{2021AA...650A...1L}       & 1,\,2          & $50,000$ km                               & $\sim 10$ \\ \hline
\citet{2020ApJ...893...93K}     & 1               & $7000$ km for compressive         & -         \\ \hline
                                               &                   & $50,000$ km for Alfv\'enic            &           \\
\citet{2021AA...650A...3L}      & 1                & $10^{3}$ km - $10^{5}$ km          & -         \\ \hline
\citet{2021AA...650L...4B}      & 1-5             & $< 16,000$ km*                           & -         \\ \hline
\end{tabular}
}
\caption{Summary of the results regarding switchback shape and size, including which {\emph{PSP}} Encs. were used in the analysis. *assuming an S-shape structure. }\label{tab:shape_size}
\end{table}

\subsubsection{Occurrence and radial evolution in the solar wind}\label{sub:obs: occurrence}

\begin{figure}
\begin{center}
\includegraphics[width=.49\columnwidth]{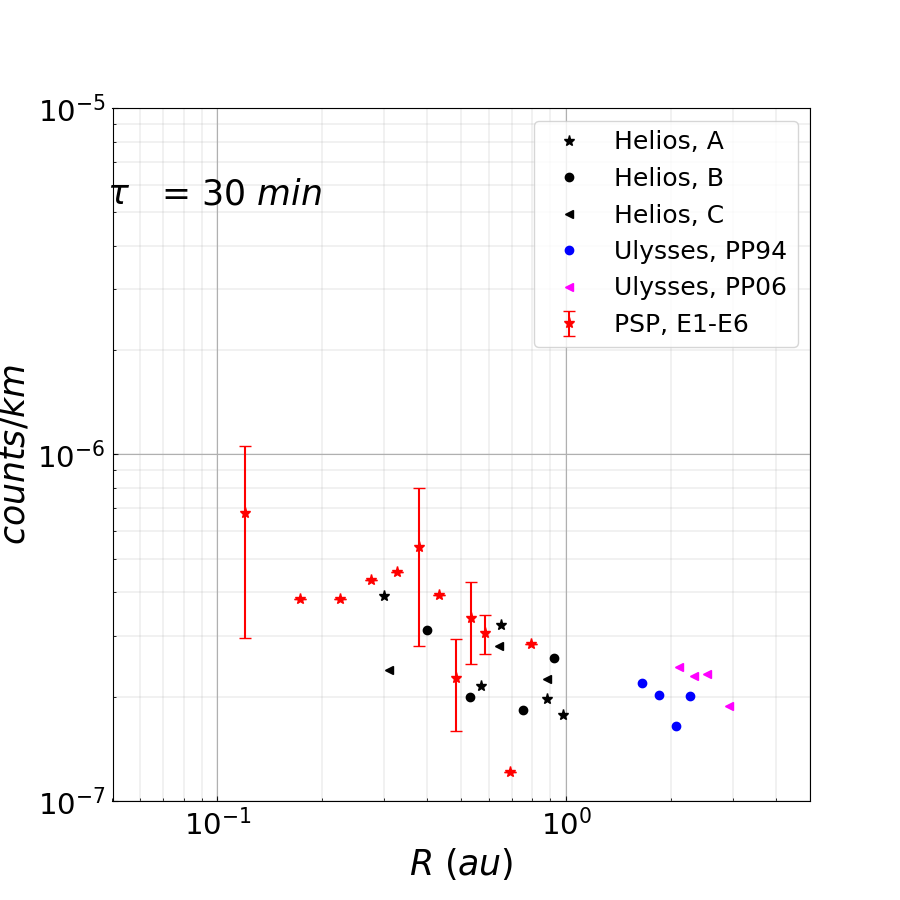}
\includegraphics[width=.49\columnwidth]{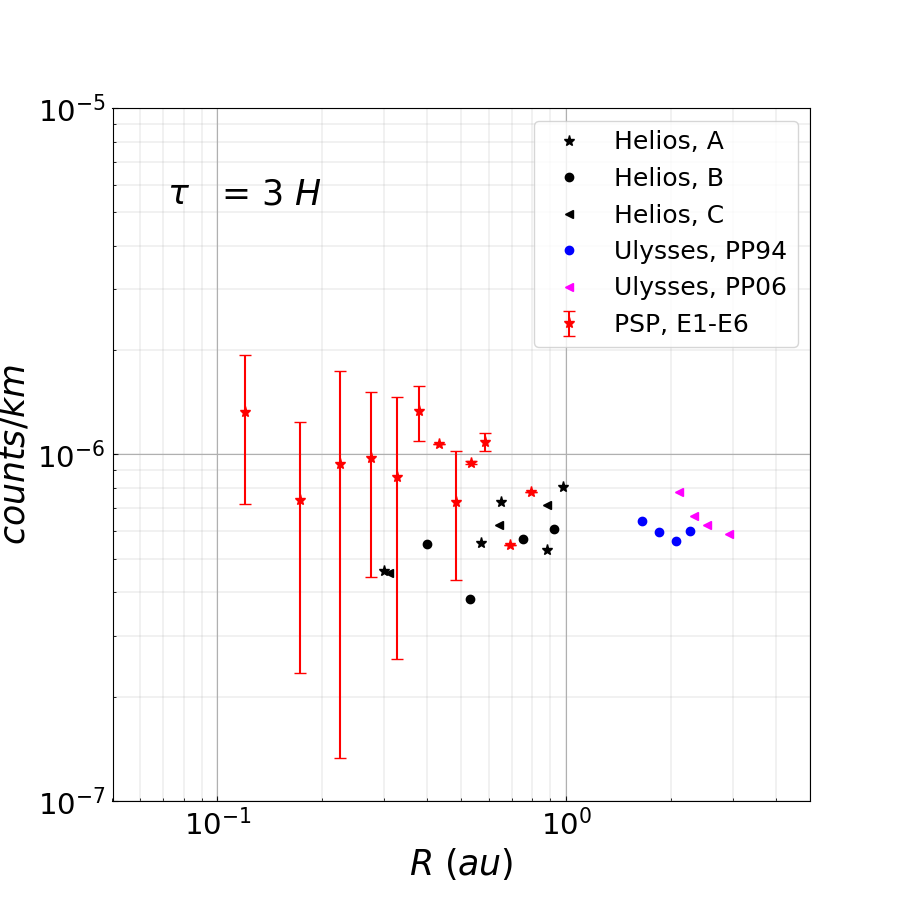}
\caption{Cumulative counts of switchbacks as a function of radial distance from {\emph{PSP}}, {\emph{Helios}} and two polar passes of {\emph{Ulysses}} (in 1994 and 2006). The left plot shows counts per km of switchbacks of duration up to 30 minute, while the right plot shows the same quantity but for switchbacks of duration up to 3 hours. {\emph{PSP}} data (43 in total) were binned in intervals of width $\Delta R=0.05$~AU. The error bars denote the range of data points in each bin \citep{2021ApJ...919L..31T}.}  
\label{fig_rates}
\end{center}
\end{figure}

Understanding how switchbacks evolve with radial distance is one of the key elements not only to determine their origin, but also to understand if switchbacks may contribute to the evolution of the turbulent cascade in the solar wind and to solar wind energy budget. Simulations (\S\ref{sub: theory SB propagation}) and observations (\S\ref{sec: obs: boundaries}) suggest that switchbacks may decay and disrupt as they propagate in the inner heliosphere. As a consequence, it is expected that the occurrence of switchbacks decreases with radial distance in the absence of an ongoing driver capable of reforming switchbacks {\emph{in~situ}}. On the contrary, the presence of an efficient driving mechanism is expected to lead to an increase, or to a steady state, of the occurrence of switchbacks with heliocentric distance. Based on this idea, \citet{2021ApJ...919...60M} analyzed the occurrence rate (counts per hour) of switchbacks with radial distance using data from  Encs.~3 through 7 of {\emph{PSP}}. The authors conclude that the occurrence rate depends on the wind speed, with higher count rates for higher wind speed, and that it and does not depend on the radial distance. Based on this result, \citet{2021ApJ...919...60M} exclude {\emph{in~situ}} generation mechanisms. However, it is interesting to note that counts of switchbacks observed by {\emph{PSP}} are highly scattered with radial distance, likely due to the mixing of different streams \citep{2021ApJ...919...60M}.   \citet{2021ApJ...919L..31T} also report highly scattered counts of switchbacks with radial distance, although they argue that the presence of decaying and reforming switchbacks might also contribute to such an effect. \citet{2021ApJ...919L..31T} analyzed the count rates  (counts per km) of switchbacks by complementing {\emph{PSP}} data with {\emph{Helios}} and {\emph{Ulysses}}. Their analysis shows that the occurrence of switchbacks is scale-dependent, a trend that is particularly clear in {\emph{Helios}} and {\emph{Ulysses}} data. In particular, they found that the fraction of switchbacks of duration of a few tens of seconds and longer increases with radial distance and that the fraction of those of duration below a few tens of seconds instead decreases. The overall cumulative counts per km, two examples of which are shown in Fig.~\ref{fig_rates}, show such a trend. Results from this analysis led \citet{2021ApJ...919L..31T} to conclude that  switchbacks in the solar wind can decay and reform in the expanding solar wind, with {\emph{in~situ}} generation being more efficient at the larger scales. They also found that the mean radial amplitude of switchbacks decays faster than the overall turbulent fluctuations, in a way that is consistent with the radial decrease of the mean radial field. They argued that this could be the result of a saturation of amplitudes and may be a signature of decay processes of switchbacks as they evolve and propagate in the inner Heliosphere.

\subsubsection{Thermodynamics and energetics}\label{sub: obs: thermodynamics}

An important question about switchbacks is whether the plasma inside these structures is different compared to the background surrounding plasma. We have seen already that switchbacks exhibit a bulk speed enhancement in the main core proton population. As this increase in speed corresponds to a net acceleration in the center of mass frame, the plasma kinetic energy is therefore larger in switchbacks than in the background solar wind. This result suggests these structures carry a significant amount of energy with them as the solar wind flows out into the inner heliosphere. A question that directly follows is whether the plasma is also hotter inside w.r.t. outside.

Attempting to answer this important question with SPC is non-trivial, since the measurements are restricted to a radial cut of the full 3D ion VDF \citep{2016SSRv..204..131K, 2020ApJS..246...43C}. While the magnetic field rotation in switchbacks enables the sampling of many different angular cuts as the S/C Encs. these structures, the cuts are not directly comparable as they represent different combinations of $T_\perp$ and $T_\|$ \cite[See for example,][]{2020ApJS..246...70H}:

\begin{equation} \label{SPCtemp}
w_{r}=\sqrt{w^{2}_{\parallel }\left( \hat{r} \cdot \hat{b} \right)^{2}  +w^{2}_{\perp }\left[ 1-\left( \hat{r} \cdot \hat{b} \right)^{2}  \right]},
\end{equation}

\noindent where $w_r$ is the measured thermal speed of the ions, related to temperature by $w=\sqrt{2 k_B T/m}$, and $\hat{\bm{b}}=\B/\Bm$. Therefore, SPC measurements of temperature outside switchbacks, where the magnetic field is typically radial, sample the proton parallel temperature, $T_{p\|}$. In contrast, as $\B$ rotates towards $90^\circ$ within a switchback, the SPC cut typically provides a better estimate of $T_{p\perp}$. To overcome this, \citet{2020ApJS..246...70H} investigated the proton temperature anisotropy statistically.  They assumed that the proton VDF does not vary significantly over the SPC sampling time as $\B$ deflects away from the radial direction, and then solved Eq.~\ref{SPCtemp} for both $w_\parallel$ and $w_\perp$. While this method does reveal some information about the the underlying temperature anisotropy, this approach is not suitable for the comparison of anisotropy within a single switchback since it assumes, $\textit{a priori}$, that the anisotropy is fixed compared to the background plasma.

\begin{figure}
\begin{center}
\includegraphics[width=\columnwidth]{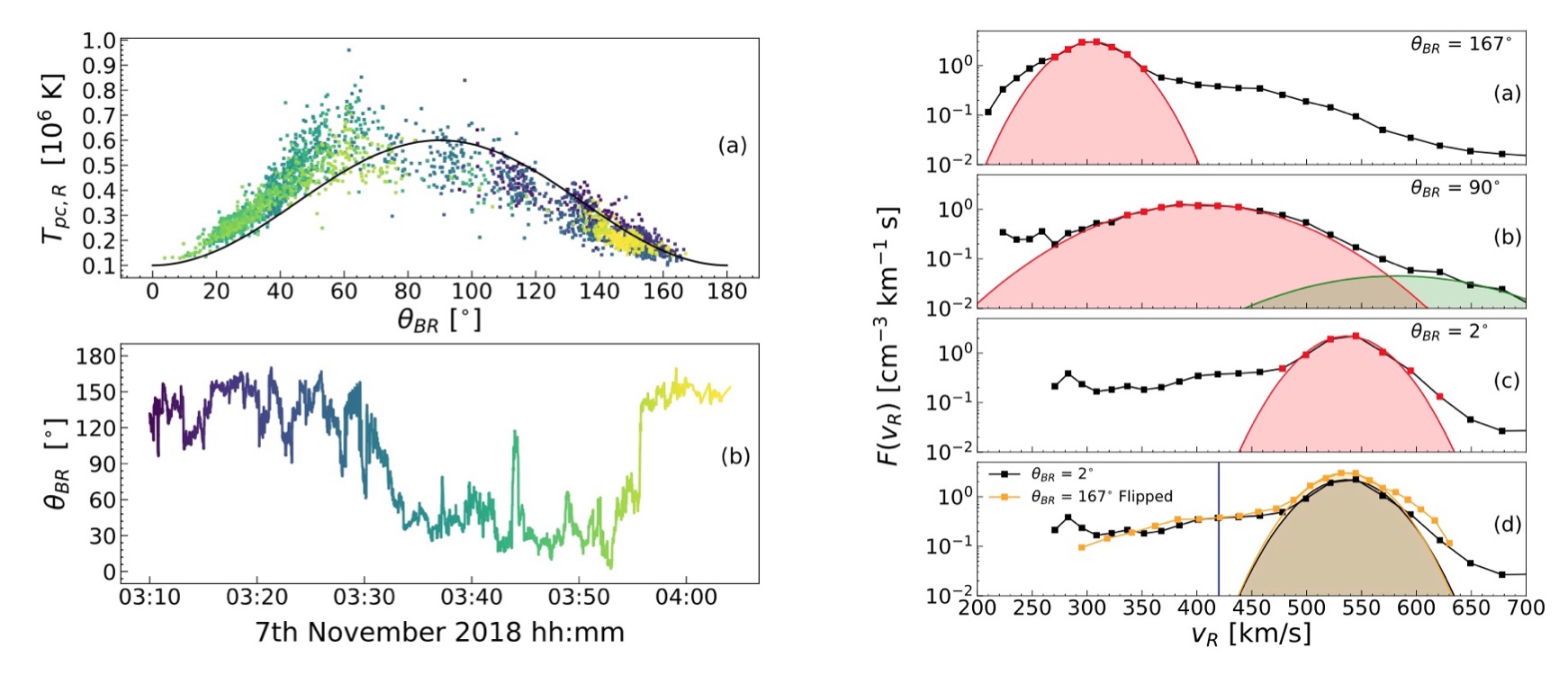}
\caption{{\it Left}: SPC measurements of the core proton radial temperature during a large amplitude switchback shown in the bottom panel. The measured core proton temperature (upper panel) is modulated by the B angle and it's maximum when measuring $T_{p\perp}$ at roughly $90^\circ$, consistent with a dominant $T_{p\perp}>T_{p\|}$ anisotropy in the background plasma. {\it Right}: cuts of the ion VDF made by SPC at different angles:  antiparallel (anti-radial), orthogonal and parallel (radial) to $\B$. The fit of the proton core is shown in pink. The bottom panel compares the radial and anti-radial VDFs, where the latter has been flipped to account for the field reversal inside the switchback. Figure adapted from \cite{2020MNRAS.498.5524W}.}
\label{fig_sb_spc}
\end{center}
\end{figure}

Another possibility is to investigate switchbacks that exhibit a  reversal in the  sign of $B_R$, in other words, $\theta_{BR}\simeq180^\circ$ inside the switchback for a radial background field.  This technique provides two estimates of $T_{p\|}$: outside the switchback, when the field is close to (anti-)radial, and inside, when $B_R$ is reversed. This is the only way to compare the same radial SPC cut of the VDF inside and outside switchbacks, leading then to a direct comparison between the two resulting $T_{p\|}$ values. \citet{2020MNRAS.498.5524W} first attempted this approach, and a summary of their results are presented in Fig.~\ref{fig_sb_spc}. A switchback with an almost complete reversal in the field direction is tracked in the left panels; the bottom panel shows the angle of the magnetic field, from almost anti-radial to radial and back again. The measured core proton temperature, $T_{cp\|}$ (upper left panel), increases with angle, $\theta_{BR}$, and reaches a maximum at $\theta_{BR}\simeq 90^\circ$, consistent with a dominant $T_{p\perp}>T_{p\|}$ anisotropy in the background plasma. On the other hand, when the SPC sampling direction is (anti-)parallel to $\Bm$ (approximately $0^\circ$ and $180^\circ$), \citet{2020MNRAS.498.5524W} find the same value for $T_{cp\|}$. Therefore, they concluded that the plasma inside switchbacks is not significantly hotter than the background plasma.

The right panels show radial cuts of the ion VDF made by SPC at different angles: anti-parallel (anti-radial), orthogonal and parallel (radial) to $\B$. The fit of the proton core is shown in pink. The bottom panel compares the measurement in the radial and anti-radial direction, once the latter has been flipped to account for the field reversal inside the switchback; the two distributions fall on top of each other, suggesting that core protons undergo a rigid rotation in velocity space inside the switchback, without a significant deformation of the VDF. The comparison in the panels also shows that the core temperature is larger for oblique angles $^\circ$ (large $T_{cp,\perp}$) and that the proton beam switches sides during the reversal, as discussed in \cite{2013AIPC.1539...46N}. They conclude that plasma inside switchbacks, at least those with the largest angular deflections, exhibits  a negligible  difference in the parallel temperature compared to the background, and therefore, the speed enhancement of the proton core inside these structures does not follow  the expected $T$-$V$ relation \citep[{\emph{e.g.}}, see][]{2019MNRAS.488.2380P}. This scenario is consistent with studies about turbulent properties and associated heating inside and outside switchbacks \citep{2020ApJ...904L..30B, 2021ApJ...912...28M}.

\begin{figure}
\begin{center}
\includegraphics[width=\columnwidth]{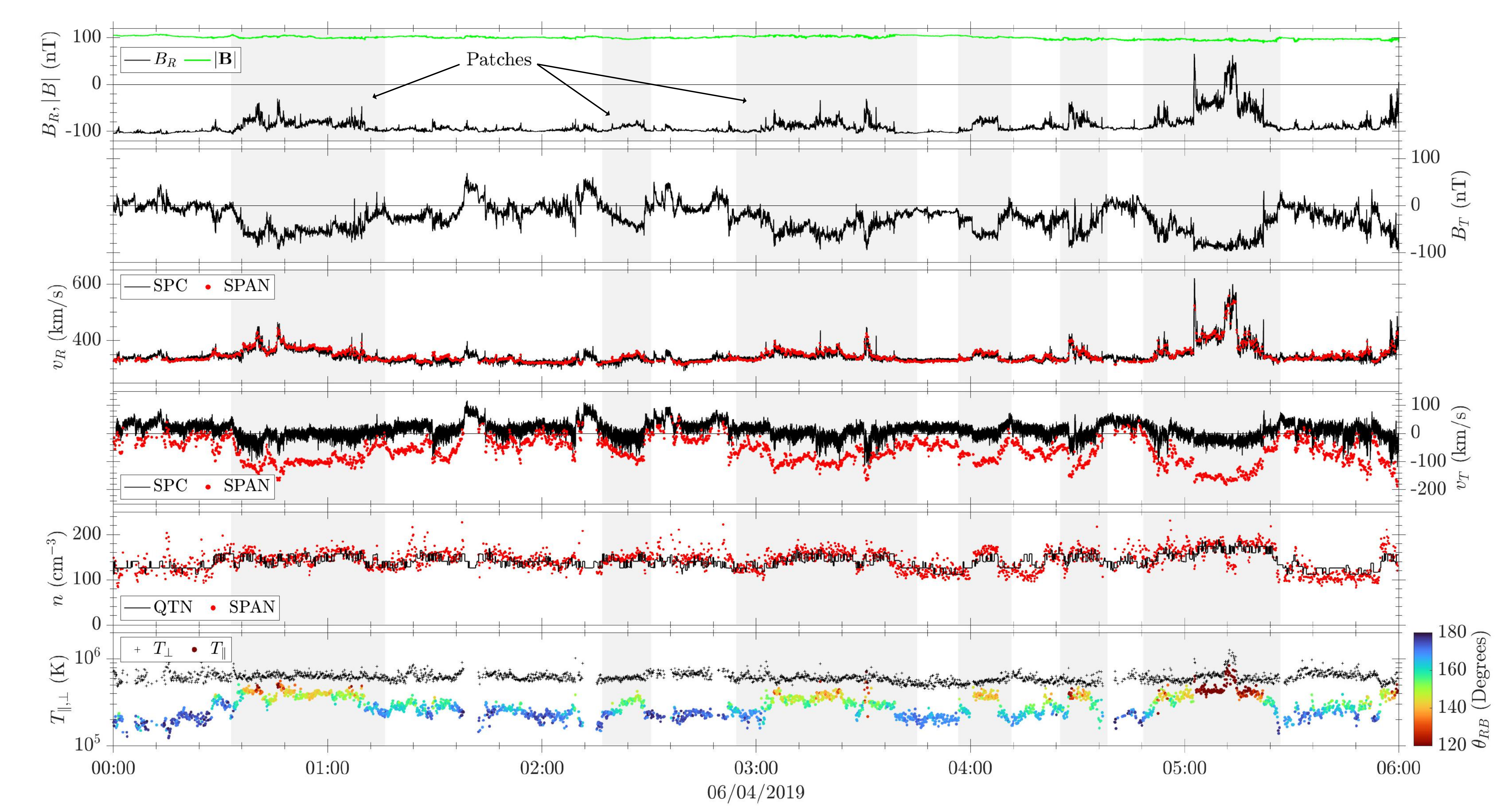}
\caption{Overview of plasma properties inside a group of switchback patches. The bottom panel shows the core proton parallel and perpendicular temperatures measured by SPAN. The colours in $T_{\|}$ encode the deflection of $\B$ from the radial direction. Patches (grey sectors exhibit systematically higher $T_{\|}$ than in quiet periods, while $T_{\perp}$ is mostly uniform throughout the interval. Figure adapted from \citet{2021A&A...650L...1W}.}
\label{fig_sb_span2}
\end{center}
\end{figure}

On the other hand, SPAN measurements of the core proton parallel and perpendicular temperatures show a large-scale modulation by patches of switchbacks \citep{2021A&A...650L...1W}. Fig.~\ref{fig_sb_span2} shows an overview of magnetic field and plasma properties through an interval that contains a series of switchback patches and quiet radial periods during Enc.~2. The bottom panel highlights the behavior of $T_{\perp}$  and $T_{\|}$ through the structures. The former is approximately constant throughout the interval, consistent with an equally roughly constant solar wind speed explained by the well-known speed-temperature relationship in the solar wind \citep[for example, see][and references therein]{2006JGRA..11110103M}. In contrast, the latter shows large variations, especially during patches when a systematic larger $T_{\|}$ is observed. As a consequence, increases in $T_{\|}$ are also correlated with deflections in the magnetic field directions (colors refer to the instantaneous angle $\theta_{BR}$). The origin of such a correlation between $\theta_{BR}$ and $T_{\|}$ is not fully understood yet, although the large-scale enhancement of the parallel temperature within patches could be a signature of some preferential heating of the plasma closer to the Sun  ({\emph{e.g.}}, by interchange reconnection), supporting a coronal origin for these structures.

\subsubsection{Switchback boundaries and small-scale waves}\label{sec: obs: boundaries}

\begin{figure}
\begin{center}
\includegraphics[width=\columnwidth]{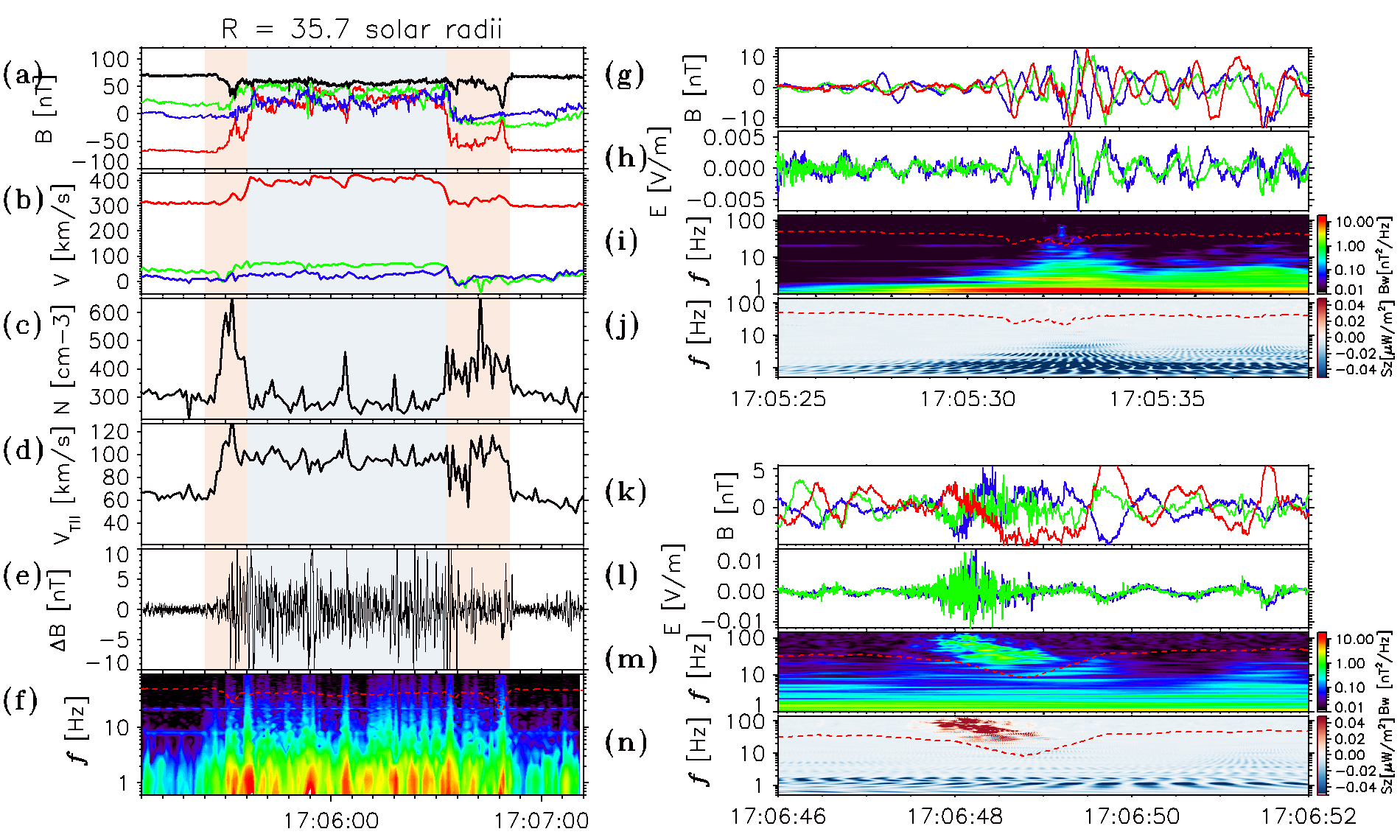}
\caption{The magnetic field dynamics for a typical deflection (switchback) of the magnetic field observed at heliocentric distance of $35.6~R_\odot$ during {\emph{PSP}}’s first solar Enc., on 4 Nov. 2018 (left) and at heliocentric distance of $\sim50~R_\odot$ on 10 Nov. 2018 (right). The radial component of the magnetic field (red curve in panel (a)) exhibits an almost complete inversion at the switchback boundary and becomes positive (anti-sunward). The transverse components are shown in blue (T, in the ecliptic plane) and in green (N –normal component, transverse to the ecliptic plane). The magnetic field magnitude is shown in black. Panel (b) represents plasma bulk velocity components (with a separate scale for the radial component $V_z$ shown in red) with the same color scheme as in panel (a). Panels (c) and (d) represent the proton density and temperature. Panel (e) presents the magnetic field waveforms from SCM (with the instrumental power cut-off below 3 Hz). The dynamic spectrum of these waveforms are shown in Panel (f), in which the red-dashed curve indicates the local lower hybrid ($f_{LH}$) frequency. Panels (g-j) represent the magnetic and eclectic field perturbations around the switchback leading boundary, the wavelet spectrum of the magnetic field perturbation, and radial component of the Poynting flux (blue color indicates propagation from the Sun and red sunward propagation). The same parameters for the trailing boundary are presented in panels (k-n).}
\label{fig:icx1}
\end{center}
\end{figure}

Switchback boundaries are plasma discontinuities, which separate two plasmas inside and outside the structure moving with different velocities that may have different temperatures and densities. Fig.~\ref{fig:icx1} shows a “typical” switchback, highlighting: (1) the sharp rotation of magnetic field as well as the dropouts in field intensity on the boundaries (Fig.~\ref{fig:icx1}a), in agreement with \citet{2020ApJS..249...28F}; (2) the increase of radial velocity showing the Alfv\'enicity  (Fig.~\ref{fig:icx1}b); (3) the plasma density enhancements at the boundaries of the switchback (Fig.~\ref{fig:icx1}c), from 300~cm$^{-3}$ to $\sim500$ and 400~cm$^{-3}$ at the leading and trailing edges respectively with some decrease of plasma density inside the structure \citep{2020ApJS..249...28F} down to $250-280$~cm$^{-3}$; and (4) enhanced wave activity inside the switchback and at the boundaries (Fig.~\ref{fig:icx1}d) predominantly below $f_{LH}$ with the higher amplitude wave bursts at the boundaries. The detailed superimposed epoch analysis of plasma and magnetic field parameters presented in \citet{2020ApJS..249...28F} showed that magnetic field magnitude dips and plasma density enhancement are the characteristic features associated with switchbacks boundaries.

It is further shown that wave activity decays with heliocentric distances. Together with the activity inside switchbacks, the boundaries also relax during propagation \citep{2020ApJS..246...68M, 2021ApJ...915...68F, 2021A&A...650A...4A} suggesting that the switchback boundary formation process is dynamic and evolving, even occurring near the {\emph{PSP}} observation point inside of $40~R_\odot$ \citep{2021ApJ...915...68F}.

\begin{figure}
\begin{center}
\includegraphics[width=1\columnwidth]{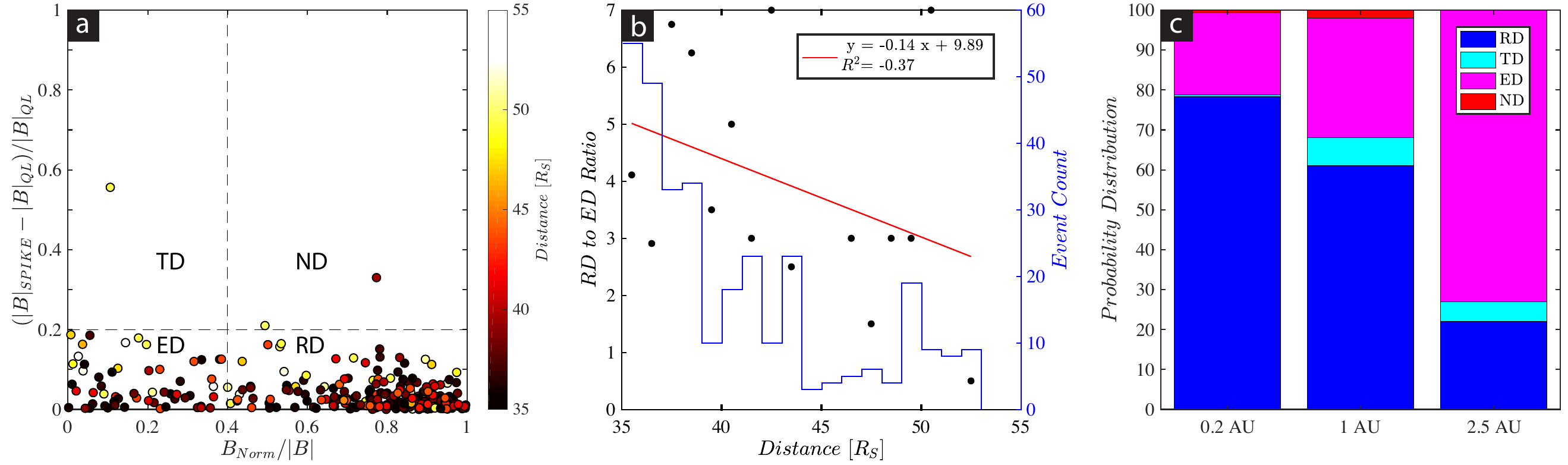}
\caption{(a) Discontinuity classification of 273 magnetic switchbacks. Scatter plot of relative normal component of magnetic field of upstream, pristine solar wind and relative variation in magnetic field intensity across switchbacks’ leading (QL-to-SPIKE) transition regions. The color shading indicates the switchbacks’ distance from the Sun. (b) Scatter plot of the ratio of number of RD events to that of ED as a function of distance from the Sun. The histogram of event count per radial distance (bin width = $1~R_\odot$) is provided on the right y-axis in blue for reference. (c) Stacked bar plots of the relative ratios of RD:TD:ED:ND discontinuities at 0.2~AU \citep[{\emph{PSP}};][]{2021A&A...650A...4A}, 1.0~AU \citep[{\emph{ISEE}};][]{1984JGR....89.5395N}, and $1.63-3.73$~AU \citep[{\emph{Ulysses}};][]{2002GeoRL..29.1383Y}.}
\label{fig:icx2}
\end{center}
\end{figure}

The analysis of MHD discontinuity types was performed by \citet{2021AA...650A...3L} who found that $32\%$ of switchbacks may be attributed to rotational discontinuities (RD), $17\%$ to tangential discontinuities (TD), about $42\%$ to the group of discontinuities that are difficult to unambiguously define (ED), and $9\%$ that do not belong to any of these groups (ND). Similarly, as shown in {\bf{Fig.~\ref{fig:icx2}}}, a recent study by \citet{2021A&A...650A...4A} reported that the relative occurrence rate of RD-type switchbacks goes down with heliocentric distance (Fig.~\ref{fig:icx2}b), suggesting that RD-type switchbacks may fully disappear past 0.3~AU. However, RD-type switchbacks have been observed at both Earth \citep[1~AU;][]{1984JGR....89.5395N} and near Jupiter \citep[2.5~AU;][]{2002GeoRL..29.1383Y}, though at smaller rates of occurrence (Fig.~\ref{fig:icx2}c) than that measured by {\emph{PSP}}. Future investigations are needed to examine (1) the mechanisms via which switchbacks may evolve, and (2) whether the dominant switchback evolution mechanism changes with heliocentric distance.

Various studies have also investigated wave activity on switchback boundaries \citep{2020ApJS..246...68M, 2020ApJ...891L..20A,2021AA...650A...3L}: the boundary surface MHD wave (observed at the leading edge of the switchback in Fig.~\ref{fig:icx1} and highlighted in panels (g-h)) and the localized whistler bursts in the magnetic dip (observed at the trailing edge of the switchback in Fig.~\ref{fig:icx1} and highlighted in panels (k-n)). The whistler wave burst in Fig.~\ref{fig:icx1}(k-n) had Poynting flux directed to the Sun that leaded to significant Doppler downshift of wave frequency in the S/C frame \citep{2020ApJ...891L..20A}. Because of their sunward propagation these whistler waves can efficiently scatter strahl electron population. These waves are often observed in the magnetic field magnitude minima at the switchback boundaries, {\emph{i.e.}}, can be considered as the regular feature associated with switchbacks.

Lastly, features related to reconnection are occasionally observed at switchback boundaries, albeit only in  about $1\%$ of the observed events \citep{2021A&A...650A...5F,2020ApJS..246...34P}. If occurring, reconnection on the boundary of switchbacks with the solar wind magnetic field may lead to the disappearance of some switchbacks \citep{2020AGUFMSH034..06D}. Surprisingly, there has been no evidence of reconnection on switchback boundaries at distances greater than $50~R_\odot$. \citet{2020ApJS..246...34P} explained that the absence of reconnection at these boundaries may be due to (a) large, albeit sub-Alfv\'enic, velocity shears at switchback boundaries which can suppress reconnection \citep{2003JGRA..108.1218S}, or that (b) switchback boundaries, commonly characterized as Alfv\'enic current sheets, are isolated RD-type discontinuities that do not undergo local reconnection. \citet{2021A&A...650A...4A} similarly showed that switchback boundaries theoretically favor magnetic reconnection based on their plasma beta and magnetic shear angle characteristics \citep{2003JGRA..108.1218S}. However, the authors concluded that negligible magnetic curvature, that is highly stretched magnetic field lines \citep{2019JGRA..124.5376A, 2019GeoRL..4612654A}, at switchback boundaries may inhibit magnetic reconnection. Further investigations are needed to explore whether and how magnetic curvature evolves with heliocentric distance.

\subsection{Theoretical models}\label{sec: theory switchbacks}
In this section, we outline the collection of theoretical models that have been formulated to explain  observations of switchbacks. These are based on a variety of physical effects, and there is, as of yet, no consensus about the key ingredients needed to explain observations. In the following we discuss each model and related works in turn, organized  by the primary physical effect that is assumed to drive switchback formation. These are (i) Interchange reconnection (\S\ref{sub: theory interchange }), (ii) Other solar-surface processes (\S\ref{sub: theory coronal jets}), (iii) Interactions between solar-wind streams (\S\ref{sub: theory stream interactions}), and (iv) Expanding AWs and turbulence (\S\ref{sub: theory alfven waves }).  Within each of these broad categories, we discuss the various theories and models, some of which differ in important ways.  In addition, some models naturally involve multiple physical effects, which we try to note as appropriate.

The primary motivation for understanding the origin of switchbacks is to understand their relevance to the heating and acceleration of the solar-wind. 
As discussed in, {\emph{e.g.}}, \citet{2009LRSP....6....3C}, magnetically driven wind models fall into the two broad classes of wave/turbulence-driven (WTD) and reconnection/loop-opening (RLO) models. A natural question is how switchbacks relate to the heating mechanism and what clues they provide as to the importance of different forms of heating in different types of wind. With this in mind, it is helpful to further, more broadly, categorize the mechanisms discussed above into ``{\emph{ex situ}}'' mechanisms (covering interchange reconnection and other solar-surface processes) -- in which switchbacks result from transient, impulsive events near the surface of the sun -- and ``{\emph{in situ}}'' mechanisms (covering stream interactions and AWs), in which switchbacks result from processes within the solar wind as it propagates outwards. An {\emph{ex situ}} switchback formation model, with its focus on impulsive events,  naturally ties into an RLO heating scenario; an {\emph{in situ}} formation process, by focusing on local processes in the extended solar wind, naturally ties into a WTD scenario. This is particularly true given the significant energy content of switchbacks in some {\emph{PSP}} observations (see \S\ref{sub:obs: velocity increase}), although there are also important caveats in some of the models. Thus, understanding the origin of switchbacks is key to understanding the origin of the solar wind itself. How  predictions from different models hold up when compared to observations may provide us with important clues. This is discussed in more detail in the summary of the implications of different  models  and  how they compare to observations in \S\ref{sub: sb summary theory}.

\subsubsection{Interchange reconnection}\label{sub: theory interchange }

\begin{figure}
\begin{center}
\includegraphics[width=.90\columnwidth]{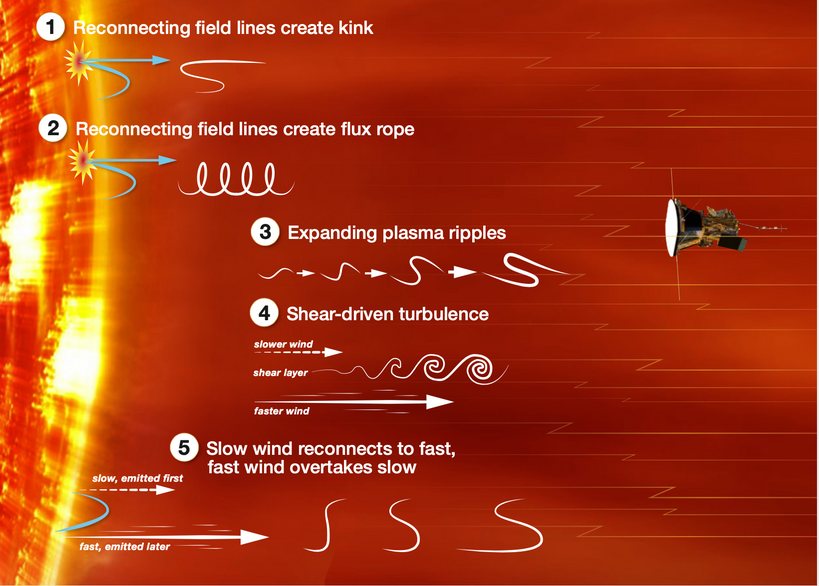}
\caption{Graphical overview covering most of  the various proposed switchback-generation mechanisms, reprinted from {\footnotesize \texttt{https://www.nasa.gov/feature/goddard/2021/} \texttt{switchbacks-science-explaining-parker-solar-probe-s-magnetic-puzzle}}. The mechanisms are classified into those that form switchbacks (1)  directly through interchange reconnection ({\emph{e.g.}}, \citealt{2020ApJ...894L...4F,2021ApJ...913L..14H,2020ApJ...896L..18S}); (2) through ejection of flux ropes by interchange reconnection \citep{2021A&A...650A...2D,2022ApJ...925..213A}; (3)  from expanding/growing AWs and/or Alfv\'enic turbulence \citep{2020ApJ...891L...2S,2021ApJ...918...62M,2021ApJ...915...52S}; (4)  due to roll up from nonlinear Kelvin-Helmholtz instabilities \citep{2020ApJ...902...94R}; and (5) through magnetic field lines that stretch between sources of slower and faster wind (\citealp{2021ApJ...909...95S}; see also \citealt{2006GeoRL..3314101L}).}
\label{fig:icx}
\end{center}
\end{figure}

Interchange reconnection refers to  the process whereby a region of open magnetic-field lines reconnect with a closed magnetic loop \citep{2005ApJ...626..563F}.  Since this process is expected to be explosive and suddenly change the shape and topology of the field, it is a good candidate for the origin of switchbacks and has been considered by several authors.  The basic scenario is shown in Fig.~\ref{fig:icx}. 

\citet{2020ApJ...894L...4F} first pointed out the general applicability of interchange reconnection to the {\emph{PSP}} observations (the possible  relevance to earlier {\emph{Ulysses}} observations had also been discussed in \citealt{2004JGRA..109.3104Y}).  They focus on the large transverse flows measured by {\emph{PSP}} as evidence for the global circulation of open flux enabled by the interchange reconnection process \citep{2001ApJ...560..425F,2005ApJ...626..563F}. Given that switchbacks tend to deflect preferentially in the transverse direction (see \S\ref{sub:obs: shapes and sizes}; \citealp{2020ApJS..246...45H}), they argue that these two observations are suggestively compatible: an interchange reconnection event that enables the transverse transport of open flux would naturally create a transverse switchback. 

Other authors have focused more on the plasma-physics process of switchback formation, including the reconnection  itself and the type of perturbation it creates. \citet{2021A&A...650A...2D} used two-dimensional (2D) particle-in-cell (PIC) simulations to study the hypothesis that switchbacks are flux-rope structures that are ejected by bursty interchange reconnection. They present two 2D simulations, the first focusing on the interchange reconnection itself and the second on the structure and evolution of a flux rope in the solar wind. They find generally positive conclusions: flux ropes with radial-field reversals, nearly constant $\Bm$, and temperature enhancements  are naturally generated by  interchange reconnection; and, flux-rope initial conditions relax into structures that match {\emph{PSP}} observations reasonably well. Further discussion of the evolution of such structures, in particular how they evolve and merge with radius, is given in \citet{2022ApJ...925..213A} (see also \S\ref{sub: theory SB propagation}) who also argue that the complex internal structure of observed switchbacks is consistent with the merging process. A  challenge of the scenario is to reproduce the high Alfv\'enicity ($\delta \B\propto \delta \bm{v}$) of {\emph{PSP}} observations, although the merging process of \citet{2022ApJ...925..213A} naturally halts once Alfv\'enic structures develop, suggesting we may be observing this end result at {\emph{PSP}} altitudes.

A somewhat modified reconnection geometry has been explored with 2D MHD simulations by \citet{2021ApJ...913L..14H}. They introduce an interchange reconnection process between open and closed regions with discontinuous guide fields, which is enabled by footpoint shearing motions and favors the emission of AWs from the reconnection site. They find quasi-periodic, intermittent emission of MHD waves, classifying the open-flux regions as ``un-reconnected,'' ``newly reconnected,'' and ``post-reconnected.'' Impulsive AWs, which can resemble switchbacks,  robustly propagate outwards in both the newly and post-reconnected regions. While both regions have enhanced temperatures, the newly-reconnected regions have more slow-mode activity and the post-reconnected regions have lower densities, features of the model that may be observable at higher altitudes by {\emph{PSP}}. They also see that  flux ropes, which are ejected into the open field lines, rapidly disappear after the secondary magnetic reconnection between the impacting flux rope and the impacted open field lines; it is unclear whether this difference with \citet{2021A&A...650A...2D} is a consequence of the MHD model or the different geometry. 

Finally, \cite{2020ApJ...903....1Z} focus more on the the evolution of magnetic-field structures generated by the reconnection process, which would often be in clustered in time as numerous open and closed loops reconnect over a short period. They argue that the strong radial-magnetic-field perturbations associated with switchbacks imply that their complex structures should propagate at the  fast magnetosonic speed (but see also \S\ref{sub: theory alfven waves } below), deriving an  equation from WKB ({\emph{i.e.}}, the Wentzel, Kramers, and  Brillouin approximation) theory for how the structures evolve as they propagate outwards from a reconnection site to {\emph{PSP}} altitudes. The model is compared to  data in more detail in \citet{2021ApJ...917..110L}, who use a Markov Chain Monte Carlo technique to fit the six free parameters of the model ({\emph{e.g.}}, wave angles and the initial perturbation) to seven observed variables taken from {\emph{PSP}} time-series data for individual switchbacks. They find reasonable agreement, with around half of the observed switchbacks accepted as good fits to the model.  \cite{2020ApJ...903....1Z}'s WKB evolution equation  implies that $|\delta \B|/|\B_0|$  grows in amplitude out to $\sim50~R_\odot$  (whereupon it starts decaying again),  and the shape of the proposed structures implies that   switchbacks should often be observed as closely spaced double-humped structures. Their assumed fast-mode polarization implies that  switchbacks that are more elongated in the radial direction will also exhibit  larger variation in   $\Bm$, because radial elongation, combined with $\nabla\cdot \B=0$, implies a mostly perpendicular wavenumber. This could  be tested directly (see \S\ref{sub:obs: shapes and sizes}) and is a distinguishing feature between the fast-mode and AW based models  (which generically predict  $\Bm\sim{\rm const}$; \S\ref{sub: theory alfven waves }).

Overall, we see that the various flavors of interchange-reconnection based models have a number of attractive features, in particular their natural explanation of the likely preferred tangential deflections of large switchbacks (\S\ref{sub:obs: shapes and sizes}; \citealp{2020ApJS..246...45H,2022MNRAS.517.1001L}), along with the bulk tangential flow \citep{2019Natur.576..228K}, and  of the possible observed temperature enhancements (\S\ref{sub: obs: thermodynamics}; although to our knowledge, there are not yet clear predictions for separate $T_\perp$ and $T_\|$ dynamics). However, a number of features remain unclear, including (depending on the model in question)  the Alfv\'enicity of the structures that are produced and how they survive and evolve as they propagate to {\emph{PSP}} altitudes (see \S\ref{sub: theory SB propagation}).

\subsubsection{Other solar-surface processes}\label{sub: theory coronal jets}

\cite{2020ApJ...896L..18S} present a phenomenological model for how switchbacks might form from the same process that creates coronal jets, which are small-scale filament eruptions observed in X-ray and extreme ultraviolet (EUV).  Their jet model (proposed in \citealt{2015Natur.523..437S}) involves jets originating as erupting-flux-rope ejections through a combination of internal and interchange reconnection (thus this model would also naturally belong to \S\ref{sub: theory interchange } above). Observations that suggest jets originate around regions of magnetic-flux cancellation ({\emph{e.g.}}, \citealt{2016ApJ...832L...7P}) support this concept.  \cite{2020ApJ...896L..18S} propose that the process can also produce a magnetic-field twist that propagates outwards as an AW packet that eventually evolves into a switchback. Although there is good evidence for equatorial jets reaching the outer corona (thus allowing the switchback propagation into the solar wind) their relation to switchbacks is somewhat circumstantial at the present time; further studies of this mechanism could, for instance, attempt to correlate switchback and jet occurrences by field-line mapping.  

Using 3D MHD simulations, \cite{2021ApJ...911...75M} examined how  photospheric motions at the base of a  magnetic flux tube might excite motions that resemble switchbacks. They introduced perturbations at the lower boundary of  a pressure-balanced magnetic-field solution, considering either a field-aligned, jet-like flow, or a transverse, vortical flows. Switchback-like fluctuations evolve in both cases: from the jet, a Rayleigh-Taylor-like instability that causes the field to from rolls; from the vortical perturbations, large-amplitude AWs that steepen nonlinearly. However, they also conclude that such perturbations are unlikely to enter the corona: the roll-ups fall back downwards due to gravity and the torsional waves unwind as the background field straightens. They conclude that while such structures are likely to be present in the chromosphere, it is unclear whether they are related to switchbacks as observed by {\emph{PSP}}, since propagation effects will clearly play a dominant role (see \S\ref{sub: theory SB propagation}).

\subsubsection{Interactions between wind streams}\label{sub: theory stream interactions}
There exist several models that relate the formation of switchbacks in some way to the interaction between neighbouring solar-wind streams with different speeds. These could be either large-scale variations between separate slow- and fast-wind regions, or smaller-scale ``micro-streams,'' which seem to be observed ubiquitously in imaging studies of the low corona \citep{2018ApJ...862...18D} as well as in {\emph{in~situ}} data \citep{2021ApJ...923..174B,2021ApJ...919...96F}.\footnote{We also caution, however, that switchbacks themselves  create large radial velocity perturbations (see \S\ref{sub:obs: velocity increase}), which  clearly could not be a cause of switchbacks.} Because these models require the stream shear to overwhelm the magnetic tension, they generically predict that switchbacks start forming primarily outside the Alfv\'en critical zone, once $V_R\gtrsim B$, and/or once $\beta\gtrsim1$. However, the mechanism of switchback formation differs significantly between the models. 

 \citet{2006GeoRL..3314101L} presented an early proposal of this form to explain {\emph{Ulysses}} observations. Using 2D MHD simulations, they studied the evolution of a large-amplitude parallel (circularly polarized)  AW propagating in a region that also includes   strong flow shear from a central smaller-scale velocity stream. They find that large-magnitude field reversals  develop across the stream due to the stretching of the field. However, the reversals  are also associated with large compressive fluctuations in the thermal pressure, $\Bm$, and plasma $\beta$. Although these match various {\emph{Ulysses}} datasets quite well, they are much less  Alfv\'enic then most switchbacks observed by {\emph{PSP}}.

\cite{2020ApJ...902...94R} consider the scenario where nonlinear Kelvin-Helmholtz instabilities develop across  micro-stream boundaries, with the resulting strong turbulence producing switchbacks. This is motivated in part by the Solar TErrestrial RElations Observaory \citep[{\emph{STEREO}};][]{2008SSRv..136....5K} observations of the transition between ``striated'' (radially elongated) and ``floculated'' (more isotropic) structures \citep{2016ApJ...828...66D} around the surface where $\beta\approx1$, which is around the Alfv\'en critical zone. Since this region is where the velocity shear starts to be able to overwhelm the stabilizing effect of the magnetic field, it is natural to imagine that the instabilities that develop will contribute to the change in fluctuation structure and the generation of switchbacks.  Comparing {\emph{PSP}} and {\emph{ex~situ}} observations with theoretical arguments and numerical simulations, \cite{2020ApJ...902...94R} argue that this scenario can account for a range of solar-wind properties, and that the conditions -- {\emph{e.g.}}, the observed Alfv\'en speed and prevalence of small-scale velocity shears -- are conducive to causing shear-driven turbulence. Their 3D MHD simulations of shear-driven turbulence generate a significant reversed-field fraction that is comparable to {\emph{PSP}} observations, with the distributions of $\Bm$, radial field, and tangential flows having a promising general shape. However, it remains unclear whether turbulence generated in this way is sufficiently Alfv\'enic to explain observations, since they see somewhat larger variation in $\Bm$ than observed in many {\emph{PSP}} intervals (but see \citealt{2021ApJ...923..158R}). A key prediction of this model is that switchback activity should generally increase with distance from the Sun, since the turbulence that creates the switchbacks should continue to be driven so long as there remains sufficient velocity shear between streams. This feature is a marked contrast to models that invoke switchback generation through interchange reconnection or other Solar-surface processes.

\cite{2021ApJ...909...95S} consider a  simpler geometric explanation -- that switchbacks result from global magnetic-field lines that stretch across streams with different speeds, rather than due to waves or turbulence generation. This situation is argued to naturally result from the global transport of magnetic flux as magnetic-field footpoints move between  sources of wind with different speeds, with the footpoint motions sustained by interchange reconnection to conserve magnetic flux \citep{2001ApJ...560..425F}. A field line that moves from a source of slower wind into faster wind (thus traversing faster to slower wind as it moves radially outwards) will naturally reverse its radial field across the boundary due to the stretching by velocity shear. This explanation focuses on  the observed asymmetry of the switchbacks -- as discussed in \S\ref{SB_obs}, the larger  switchback deflections seem to show a preference to be tangential and particularly in the +T (Parker-spiral) direction, which is indeed the direction expected from the global transport of flux through interchange reconnection.\footnote{Note, however, that \citealt{2020ApJS..246...45H} argue that the global tangential flow asymmetry is not a consequence of the switchbacks themselves.} Field reversals are argued to develop their Alfv\'enic characteristics beyond the Alfv\'en point, since the field kink produced by a coherent velocity shear does not directly produce $\delta \B\propto \delta \bm{v}$ or $\Bm\sim{\rm const}$ (as also seen in the simulations of \citealt{2006GeoRL..3314101L}).

\subsubsection{Expanding Alfv\'en waves and turbulence}\label{sub: theory alfven waves }

The final class of models relate to perhaps the simplest explanation: that switchbacks are  spherically polarized ($\Bm={\rm const}$) AWs (or Alfv\'enic turbulence) that have reached  amplitudes $|\delta \B|/|\B_0|\gtrsim 1$ (where $\B_0$ is the background field). The idea follows from the realisation \citep{1974sowi.conf..385G,1974JGR....79.2302B} that an Alfv\'enic perturbation -- one with $\delta \bm{v}=\B/\sqrt{4\pi\rho}$ and $\Bm$, $\rho$, and the thermal pressure all constant -- is an exact nonlinear solution to the MHD equations that propagates at the Alfv\'en velocity. This is true no matter the amplitude of the perturbation compared to $\B_0$, a property that seems unique among the zoo of hydrodynamic and hydromagnetic waves (other waves generally form into shocks at large amplitudes). Once $|\delta \B|\gtrsim |\B_0|$ such states will often reverse the magnetic field in order to maintain their spherical polarization (they  involve a perturbation $\delta \B$ parallel to $\B_0$). Moreover, as they propagate in an inhomogeneous medium, nonlinear AWs behave just like  small-amplitude waves \citep{1974JGR....79.1539H,1974JGR....79.2302B}; this implies that in the expanding solar wind, where the decreasing Alfv\'en speed causes $|\delta \B|/ |\B_0|$ to increase, waves propagating outwards from the inner heliosphere can grow to $|\delta \B|\gtrsim |\B_0|$, feasibly forming switchbacks from initially small-amplitude waves. In the process, they may develop the sharp discontinuities characteristic of {\emph{PSP}} observations if, as they grow, the constraint of constant $\Bm$ becomes incompatible with smooth $\delta \B$ perturbations. However, in the more realistic scenario where there exists a spectrum of waves, this wave growth  competes with the dissipation of the large-scale fluctuations due to turbulence induced by wave reflection \citep[see, {\emph{e.g.}},][]{1989PhRvL..63.1807V,2007ApJ...662..669V,2009ApJ...707.1659C,2022PhPl...29g2902J} or other effects \citep[{\emph{e.g.}},][]{1992JGR....9717115R}. If dissipation is too fast, it will stop the formation of switchbacks; so, in this formation scenario turbulence and switchbacks are inextricably linked (as is also the case in the scenario of \citealt{2020ApJ...902...94R}). Thus, understanding switchbacks will require understanding and accurately modelling the turbulence properties, evolution, and amplitude \citep{2018ApJ...865...25U,2019ApJS..241...11C,2013ApJ...776..124P}.

Several recent papers have explored the general scenario from different standpoints, finding broadly consistent results. \citet{2020ApJ...891L...2S} and \citet{2022PhPl...29g2902J} used expanding-box MHD simulations to understand how  AWs grow in amplitude and develop turbulence. The basic setup involves starting from a purely outwards propagating (fully imbalanced) population of moderate-amplitude randomly phased waves and expanding the box to grow the waves to larger amplitudes. Switchbacks form organically as the waves grow, with their strength ({\emph{e.g.}}, the strength and proportion of field reversals) and properties ({\emph{e.g.}}, the extent to which $\Bm$ is constant across switchbacks) depending on the expansion rate and the  wave spectrum. While promising, these simulations are highly idealized -- {\emph{e.g.}}, the expanding box applies only outside the Alfv\'en point, the equation of state was taken as isothermal. While this  has hindered the comparison to some observational properties, there are also some promising agreements \citep{2022PhPl...29g2902J}. Similar results were found by \cite{2021ApJ...915...52S} using more comprehensive and realistic simulations that capture 
the full evolution of the solar wind from coronal base to {\emph{PSP}} altitudes. Their simulation matches well the bulk properties of the slow-Alfv\'enic wind seen by {\emph{PSP}} and develops strong switchbacks beyond $\sim10-20~R_\odot$ (where the amplitude of the turbulence becomes comparable to the mean field). They find  switchbacks that are radially elongated, as observed, although the proportion of field reversals is significantly lower than observed (this was also the case in \citealt{2020ApJ...891L...2S}). It is unclear whether this discrepancy is simply due to insufficient numerical resolution or a more fundamental issue with the AW scenario. \cite{2021ApJ...915...52S}  do not see a significant correlation between switchbacks and density perturbations (see \S\ref{sub: obs: thermodynamics} for discussion), while  more complex correlations with kinetic thermal properties \citep{2020MNRAS.498.5524W} cannot be in addressed either this model or the simpler local ones \citep{2020ApJ...891L...2S,2022PhPl...29g2902J}. 

\citet{2021ApJ...918...62M} consider a complementary, analytic approach to the problem, studying how one-dimensional, large-amplitude AWs grow and change shape in an expanding plasma.  This shows that expansion necessarily generates small compressive perturbations in order to maintain the wave's spherical polarization as it grows to large amplitudes, providing specific $\beta$-dependent predictions for magnetic and plasma compressibility. The model has been extended to include the Parker spiral by {\citet{2022PhPl...29k2903S}, who find that the interaction with a rotating background field causes the development of tangential asymmetries in the switchback deflection directions. These, and the compressive predictions of \citet{2021ApJ...918...62M}, seem to  explain various  aspects of simulations  \citep{2022PhPl...29g2902J}.  Overall, these analyses provide simple, geometric explanations for various switchback properties, most importantly  the preference for switchbacks to be radially elongated (\S\ref{sub:obs: shapes and sizes} and Table~\ref{tab:shape_size}); however, they are clearly highly idealised, particularly concerning the neglect of  turbulence. The models also struggle to reproduce the extremely sharp switchback boundaries seen in {\emph{PSP}} data, which is likely a consequence of considering one-dimensional (1D) waves, since much sharper structures evolve in similar calculations with 2D or 3D structure \citep{2022arXiv220607447S}. 

Overall, AW models naturally recover the Alfv\'enicity ($\delta \B\propto\bm{v}$ and nearly constant $\Bm$) and  radial elongation of switchbacks seen in {\emph{PSP}} observations, but may struggle with some other features.  It remains unclear whether detailed aspects of the preferred  tangential  deflections of large switchbacks can be recovered \citep{2022A&A...663A.109F,2022MNRAS.517.1001L}, although large-ampliude  AWs  do develop tangentially asymmetries as a consequence of expansion and the rotating Parker spiral \citep{2022PhPl...29g2902J,2022PhPl...29k2903S}. Similarly, further work is needed to understand the compressive properties, in particular in a kinetic plasma.\footnote{Note, however, that AW  models do not predict an \emph{absence} of compressive features in switchbacks. Indeed, compressive flows are necessary to maintain spherical polarization as they grow in amplitude due to expansion \citep{2021ApJ...918...62M}.} AW models naturally predict an increase in switchback occurrence  with radial distance out to some maximum (whereupon it may decrease again), although the details depend on low-coronal conditions and the influence of turbulence, which remain poorly understood \citep{2022PhPl...29g2902J}. Computational models have also struggled to reproduce the very high switchback fractions observed by {\emph{PSP}}; whether this is due to numerical resolution or more fundamental issues remains poorly understood.

\subsubsection{Propagation and evolution of switchbacks}\label{sub: theory SB propagation}

A final issue to consider, particularly for understanding the distinction between {\emph{ex~situ}} and {\emph{in~situ}} generation mechanisms, is how a hypothetical switchback evolves as it propagates and is advected outwards in the solar wind. In particular, if switchbacks are to be formed at the solar surface, they must be able to propagate a long way without disrupting or dissolving. Further, different formation scenarios predict different occurrence rates and size statistics as a function of heliocentric radius (\S\ref{sub:obs: occurrence}), and it is important to understand how they change shape and amplitude in order to understand what solar-wind observations could tell us about coronal conditions. 

Various studies have focused on large-amplitude Alfv\'enic initial conditions, thus probing the scenario where Alfv\'enic switchback progenitors are released in the low corona ({\emph{e.g.}}, due to reconnection), perhaps with subsequent evolution resulting from the AW/turbulence effects considered in \S\ref{sub: theory alfven waves }. Using 2D MHD simulations, they  start from an analytic initial  condition with a magnetic perturbation that is large enough to reverse the  mean field and an Alfv\'enic  velocity $\delta \bm{v} = \pm \delta \B/\sqrt{4\pi \rho}$. While \citet{2005ESASP.592..785L} showed that such structures rapidly dissolve if  $\Bm$ is not constant across the wave, \citet{2020ApJS..246...32T} reached the opposite conclusion for switchbacks with constant $\Bm$ (as relevant to observations), with their initial conditions propagating unchanged for hundreds of Alfv\'en times  before eventually decaying due to parametric instability. They concluded that even relatively short wavelength switchbacks can in principle survive propagating out to tens of solar radii. Using the same initial conditions, \citet{2021ApJ...914....8M} extended the analysis to include switchbacks propagating through a radially stratified environment. They considered a fixed, near-exponential density profile and a  background magnetic field with different degrees of expansion, which changes the radial profile of $\va$ in accordance with different possible conditions in the low corona. Their basic results are consistent with the expanding-AW theory discussed above (\S\ref{sub: theory alfven waves }), with switchbacks in super-radially expanding background  fields maintaining strong field deflections, while those in radially expanding or non-expanding backgrounds unfold as they propagate outwards. The study also reveals a number of non-WKB effects from stratification, such as a gravitational damping from plasma entrained in the switchback. More generally, they point out that after propagating any significant distance in a radially stratified environment, a switchback will have deformed significantly compared to the results from \citet{2020ApJS..246...32T}, either changing shape  or unfolding depending on the background profile. This blurs the line between {\emph{ex~situ}} and {\emph{in~situ}} formation scenarios.

The above studies, by fixing $\delta \bm{v} = \pm \delta \B/\sqrt{4\pi \rho}$ and $\Bm={\rm const}$, effectively assume that switchbacks are Alfv\'enic. \citet{2022ApJ...925..213A} have considered the evolution and merging of flux ropes, that are ejected from interchange reconnection sites in the scenario of \citet{2021A&A...650A...2D}. They show that while flux ropes are likely to form initially with an aspect ratio of near unity, merging of ropes through slow reconnection of the wrapping magnetic field is energetically favorable. This merging continues until the axial flows inside the flux ropes increase to near Alfv\'enic values, at which point the  process becomes energetically unfavorable. This process also causes flux ropes to increasingly radially elongated with distance from the sun, which is observationally testable (see \S\ref{sub:obs: shapes and sizes}) and may be the opposite prediction to AW based models (since the wave vector rotates towards the parallel direction with expansion). \citet{2021A&A...650A...2D} also argue that the complex, inner structure of observed switchbacks is consistent with this merging process. 

The WKB fast-mode-like calculation of \citet{2020ApJ...903....1Z} produces somewhat modified scalings (which nonetheless predict a switchback amplitude that increases with radius), but does not address the stability or robustness of the structures. Considerations relating to the long-time stability of switchbacks are less relevant to the shear-driven models of \citet{2020ApJ...902...94R,2021ApJ...909...95S}, in which switchbacks are generated in the Alfv\'en zone and beyond (where $V_R\gtrsim \va$), so will not have propagated a significant distance before reaching {\emph{PSP}} altitudes.  

Overall, we see that  Alfv\'enic switchbacks are expected to be relatively robust, as are flux-rope structures, although they evolve significantly through merging. This suggests that source characteristics  could be retained  (albeit with significant changes in shape) as they propagate through the solar wind. If indeed switchbacks are of low-coronal origin, this is encouraging for the general program of using switchbacks to learn about the important processes that heat and accelerate the solar wind. 

\begin{figure}
\begin{center}
\includegraphics[width=.87\columnwidth]{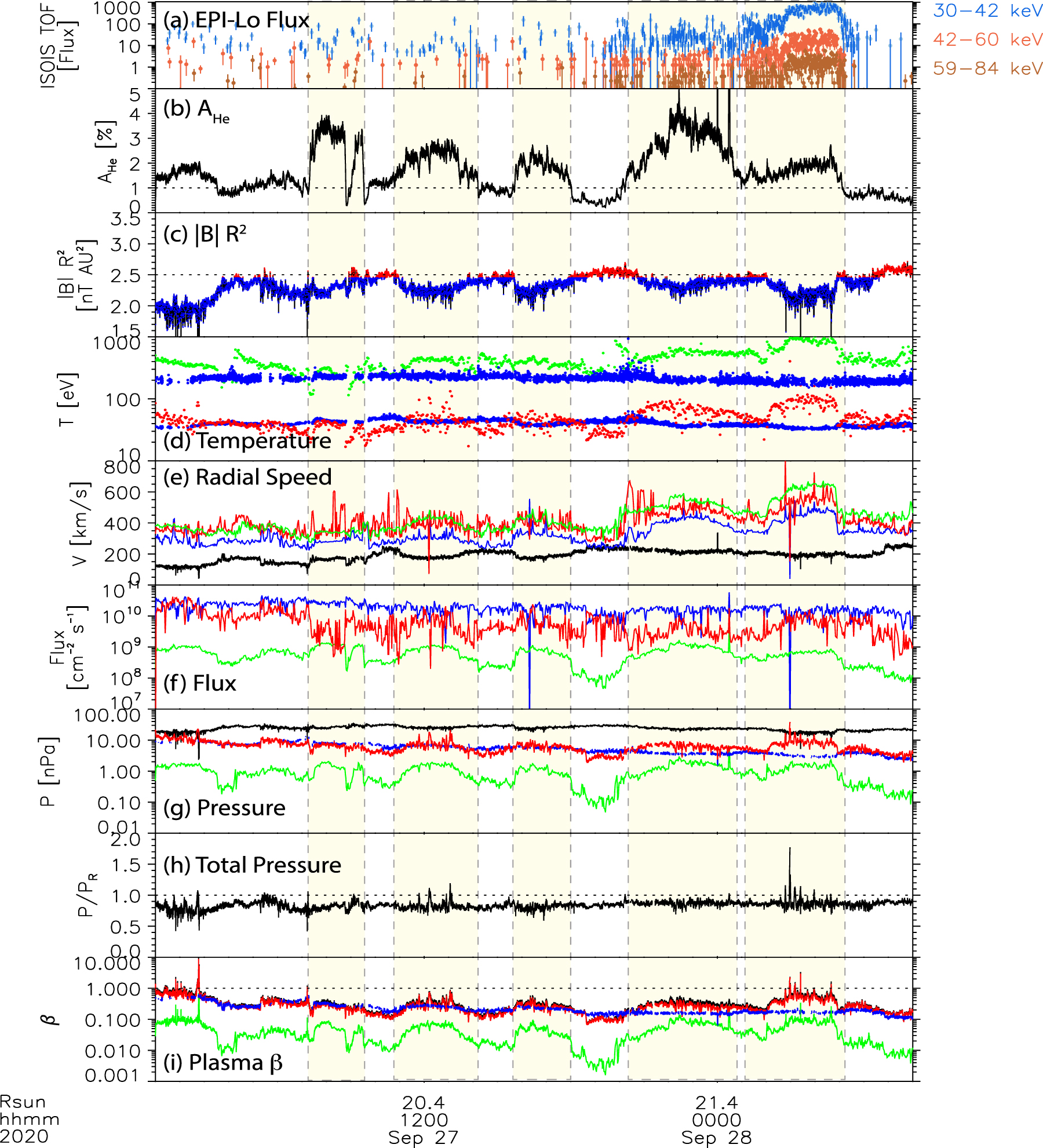}
\includegraphics[width=.87\columnwidth]{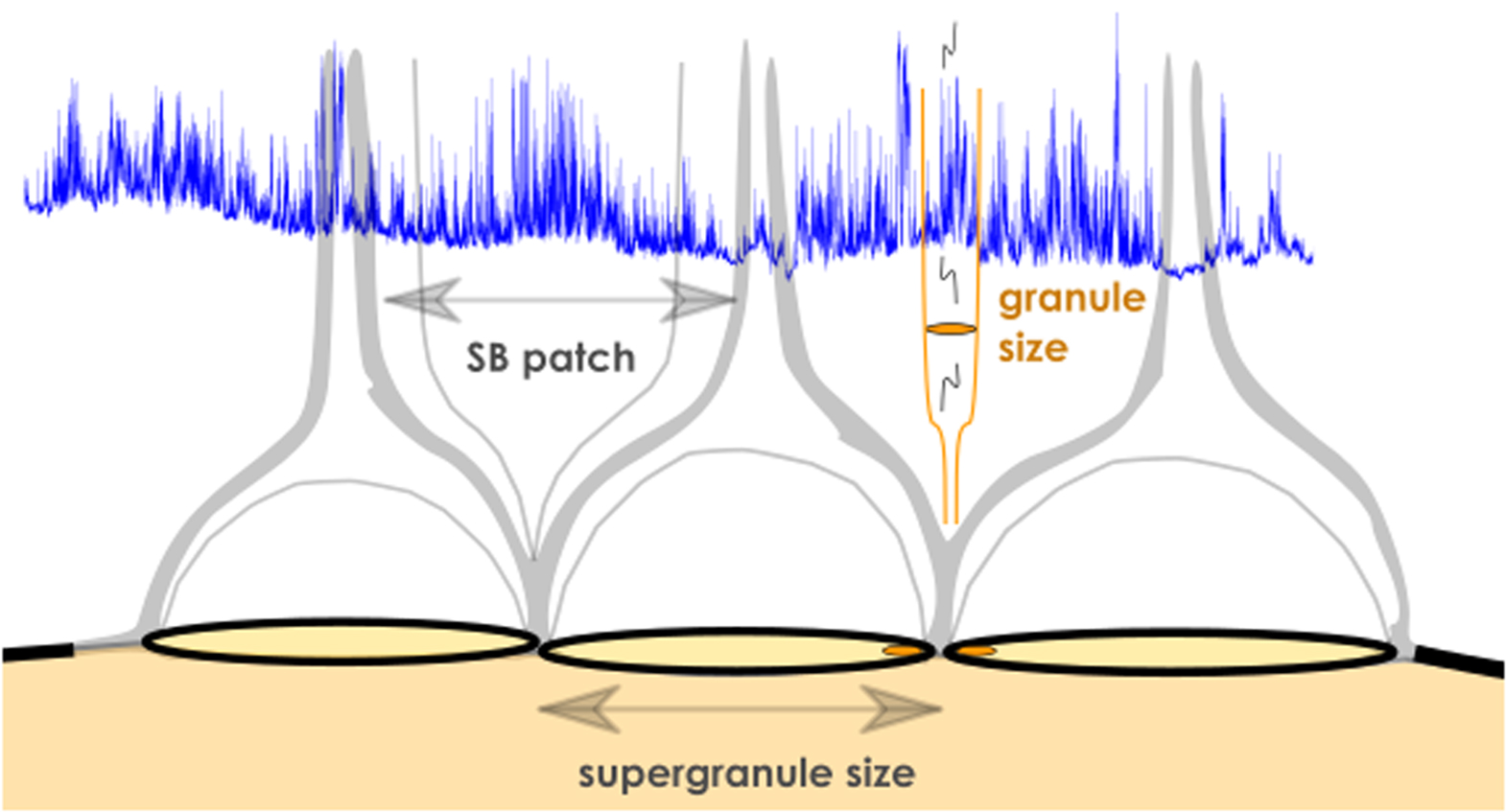}
\caption{{\it Top:} Variation of plasma properties observed during patches of switchbacks during Enc.~6. The second panel shows that Helium abundance is modulated with the patch profile too, with enhanced fractional density inside patches of switchbacks. The periodicity is consistent with the crossing of funnels emerging from the Solar atmosphere. Figure adapted from \cite{2021ApJ...923..174B} {\it Bottom:} Cartoon showing the association between switchback patches and their periodicity with supergranular and granular structure in Corona. Figure adapted from \cite{2021ApJ...919...96F}.}
\label{fig_sb_span}
\end{center}
\end{figure}

\subsection{Outlook and open questions}\label{SB_discussion}

\subsubsection{Connection to Solar sources}\label{sub:solar sources}
Because of their persistence in the solar wind, switchbacks can also be considered as tracers of processes occurring in the Solar atmosphere and therefore can be used to identify wind sources at the Sun. Recent work by \citet{2021MNRAS.508..236W} compared the properties of two switchback patches during different Encs. and suggested that patches could be linked to coronal hole boundary regions at the solar surface. \citet{2021MNRAS.508..236W} also showed that these periods, which had bulk velocities of $\sim300$~km~s$^{-1}$, could sometimes be characterized by a particularly low alpha abundance. The cause of this low alpha abundance is not known, but it could be related to the processes and mechanisms governing the release of the solar wind at the surface. Moreover, local modulation in the alpha fraction observed {\emph{in situ}} and crucially during switchback patches could be a direct signature of spatial modulation in solar sources; \cite{2021ApJ...923..174B} have identified funnels as a possible source for these structures. Such an interpretation is consistent with the finding presented by \cite{2021ApJ...919...96F} who have identified some periodicity in patches that is consistent with that expected from Solar supergranulation, and also some smaller scale signatures potentially related to granular structures inside funnels (see Fig.~\ref{fig_sb_span}).

Another interesting interpretation has been proposed by \citep{2021ApJ...920L..31N} who suggest that patches of switchbacks observed in the inner heliosphere by {\emph{PSP}} could then evolve with radial distance into structures with an overall higher bulk velocity, like the microstreams observed by {\emph{Ulysses}} in the polar wind \citep{1995JGR...10023389N}. According to the authors, then microstreams might be the result of accumulated and persistent velocity enhancements resulting from a series of switchbacks or patches. At the same time, \cite{2021ApJ...920L..31N} also propose that the individual switchbacks inside the patches could be generated by minifilament/flux rope eruptions that cause coronal jets \citep{2020ApJ...896L..18S}, so that microstreams are a consequence of a series of such jet-driven switchbacks occurring in close succession.

\subsubsection{Implications for our understanding of the solar wind}\label{sub: sb summary theory}

Switchback observations hold promise as a way to better constrain our understanding of the solar-wind itself.  In particular, most of the theoretical models discussed in \S\ref{sec: theory switchbacks} also suggest broader implications for coronal and solar-wind heating, and thus the origin of the solar wind. Given this, although there is currently no consensus about the key ingredients that form switchbacks, if a particular model does gain further observational support, this may lead to more profound shifts in our understanding of the heliosphere. Here, we attempt to broadly characterize the implications of the different  formation scenarios in order to highlight the general importance of understanding switchbacks. Further understanding will require constant collaboration between theory and observations, whereby theorists attempt to provide the most constraining, and unique, possible tests of their proposed mechanisms in order to better narrow down the possibilities. Such a program is strongly supported by the recent observations of switchback modulation on solar supergranulation scales, which suggest a direct connection to solar-surface processes (see above \S\ref{sub:solar sources}).

Above (\S\ref{sec: theory switchbacks}), we broadly  categorized  models into ``{\emph{ex~situ}}'' and ``{\emph{in~situ}},'' which involved switchbacks being formed on or near the solar surface, or in the bulk solar wind, respectively. These classes then naturally tied into RLO models of coronal heating for {\emph{ex~situ}} models, or to WTD coronal-heating theories for {\emph{in~situ}} switchback formation (with some modifications for specific models). But, the significant differences between different models narrow down the correspondence further than this. Let us consider some of the main proposals and what they will imply, if correct. In the discussion below, we consider some of the same models as discussed in  \S\ref{sec: theory switchbacks}, but grouped together by their consequence to the solar wind as opposed to the switchback formation  mechanism.
 
\citet{2020ApJ...894L...4F} and \citet{2021ApJ...909...95S} propose that switchbacks are intimately related to the global transport of open magnetic flux caused by interchange reconnection, either through the ejection of waves or due to the interaction between streams. This global circulation would have profound consequences more generally, {\emph{e.g.}},  for  coronal and solar-wind heating \citep{2003JGRA..108.1157F} or the magnetic-field structure of the solar wind structure (the sub- and super-Parker spiral; \citealp{2021ApJ...909...95S}). Some other interchange reconnection or impulsive jet mechanisms -- in particular, the ejection of flux ropes \citet{2021A&A...650A...2D,2022ApJ...925..213A}, or MHD waves \citet{2020ApJ...903....1Z,2021ApJ...913L..14H,2020ApJ...896L..18S} from the reconnection site -- do not necessarily involve the same open-flux-transport mechanism, but clearly favor an RLO-based coronal heating scenario, and have more specific consequences for each individual models; for example, the importance flux-rope structures to the energetics of the  inner heliosphere for \citet{2021A&A...650A...2D}, magnetosonic perturbations in \citet{2020ApJ...903....1Z}, or specific photospheric/chromospheric motions for \citet{2021ApJ...913L..14H,2020ApJ...896L..18S}. The  model of \citet{2020ApJ...902...94R} suggests a very different scenario, whereby shear-driven instabilities play  a crucial role  outside of the Alfv\'en critical zone (where $V_R\gtrsim\va$), where they would set the properties of the turbulent  cascade and change the energy budget by boosting heating and acceleration of slower regions. Finally,  in the Alfv\'enic turbulence/waves scenario,  switchbacks result from the evolution of turbulence to very large amplitudes $|\delta \B|\gtrsim |\B_0|$ \citep{2020ApJ...891L...2S,2021ApJ...915...52S,2021ApJ...918...62M}. In turn, given the low plasma $\beta$, this implies that the  energy contained in Alfv\'enic fluctuations is significant even at {\emph{PSP}} altitudes (at least  in switchback-filled regions), by which point they should already have contributed significantly to heating. Combined with low-coronal  observations \citep[{\emph{e.g.}},][]{2007Sci...318.1574D}, this is an important constraint on wave-heating models \citep[{\emph{e.g.}},][]{2007ApJS..171..520C}.
 
 Finally, despite the significant differences between models highlighted above, it  is also worth noting some similarities. In particular, features of various models are likely to coexist and/or feed into one another. For example, some of the explosive, {\emph{ex~situ}} scenarios (\S\ref{sub: theory interchange }--\ref{sub: theory coronal jets}) propose that such events  release AWs, which then clearly ties into the AW/expansion scenario of \ref{sub: theory alfven waves }. Indeed, as pointed out by \citet{2021ApJ...914....8M}, the subsequent evolution of switchbacks as they propagate in the solar wind cannot be \emph{avoided}, which muddies  the {\emph{in~situ}}/{\emph{ex~situ}} distinction. In this case, the distinction  between {\emph{in~situ}} and {\emph{ex~situ}} scenarios would be more related to the relevance  of distinct, impulsive events to wave launching, as opposed to slower, quasi-continuous stirring of motions.  Similarly, \citet{2020ApJ...902...94R} discuss how waves propagating upwards from the low corona could intermix and contribute to the shear-driven dynamics that form the basis for their model. These interrelationships, and  the coexistence of different mechanisms, should be kept in mind moving forward as we attempt to distinguish observationally between different mechanisms.

\subsubsection{Open Questions and Future Encs.}
Given their predominance in the solar wind plasma close to the Sun and because each switchback is associated with a positive increase in the bulk kinetic energy of the flow -- as they imply a net motion of the centre of mass of the plasma -- it is legitimate to consider whether switchbacks play any dynamical role in the acceleration of the flow and its evolution in interplanetary space. Moreover, it is  an open question if the kinetic energy and Poynting flux carried by these structures have an impact on the overall energy budget of the wind. To summarize, these are some of the main currently open questions about these structures and their role in the solar wind dynamics:
\begin{itemize}
    \item Do switchbacks play any role in solar wind acceleration?
    \item Does energy transported by switchbacks constitute a important possible source for plasma heating during expansion?
    \item Do switchbacks play an active role in driving and maintaining the turbulent cascade?
    \item Are switchbacks distinct plasma parcels with different properties than surrounding plasma?
    \item Do switchbacks continuously decay and reform during expansion?
    \item Are switchbacks signatures of key processes in the Solar atmosphere and tracers of specific types of solar wind sources? (Open field regions vs. streamer)
    \item Can switchback-like magnetic-field reversals exist close to the Sun, inside the Alfv\'en radius?
\end{itemize}
Answering these questions require taking measurements even closer to the Sun and accumulate more statistics for switchbacks in different types of streams.  These should include the fast solar wind, which has been so far seldomly encuntered by {\emph{PSP}} because of solar minimum and then a particularly flat heliospheric current sheet (HCS; which implies a very slow wind close to the ecliptic).

Crucially, future {\emph{PSP}} Encs. will provide the ideal conditions for answering these open questions, as the S/C will approach the solar atmosphere further, likely crossing the Alfv\'en radius. Further, this phase will  coincide with increasing solar activity, making possible  to sample coronal hole sources of fast wind directly.

\section{Solar Wind Sources and Associated Signatures}
\label{SWSAS}

A major outstanding research question in solar and heliophysics is establishing causal relationships between properties of the solar wind and the source of that same plasma in the solar atmosphere. Indeed, investigating these connections is one of the major science goals of {\emph{PSP}}, which aims to ``\textit{determine the structure and dynamics of the plasma and magnetic fields at the sources of the solar wind}" \cite[Science Goal \#2;][]{2016SSRv..204....7F} by making {\emph{in situ}} measurements closer to the solar corona than ever before. In this section, we outline the major outcomes, and methods used to relate {\emph{PSP}}’s measurements to specific origins on the Sun. In \S\ref{SWSAS:modeling} we review modeling efforts and their capability to identify source regions. In \S\ref{SWSAS:slowlafv} we outline how {\emph{PSP}} has identified significant contributions to the slow solar wind from coronal hole origins with high alfv\'enicity. In \S\ref{SWSAS:strmblt} we review  {\emph{PSP}}’s measurements of streams associated with the streamer belt and slow solar wind more similar to 1~AU measurements. In \S\ref{SWSAS:fstwnd} we examine {\emph{PSP}}’s limited exposure to fast solar wind, and the diagnostic information about its coronal origin carried by electron pitch angle distributions (PADs). In \S\ref{SWSAS:actvrgn} we recap {\emph{PSP}}’s measurements of energetic particles associated with solar activity and impulsive events, as well as how they can disentangle magnetic topology and identify pathways by which the Sun’s plasma can escape. Finally, in \S\ref{SWSAS:sbs} we briefly discuss clues to the solar origin of streams in which {\emph{PSP}} observes switchbacks and refer the reader to \S\ref{MagSBs} for much more detail.

\subsection{Modeling and Verifying Connectivity}
\label{SWSAS:modeling}

Association of solar wind sources with specific streams of plasma measured {\emph{in situ}} in the inner heliosphere requires establishing a connection between the Sun and S/C along which solar wind flows and evolves. One of the primary tools for this analysis is combined coronal and heliospheric modeling, particularly of the solar and interplanetary magnetic field which typically governs the large scale flow streamlines for the solar wind. 

In support of {\emph{PSP}}, there has been a broad range of such modeling efforts with goals including making advance predictions to test and improve the models, as well as making detailed connectivity estimates informed by {\emph{in situ}} measurements after the fact. The physics contained in coronal/heliospheric models lies on a continuum mediated by computational difficulty ranging from high computational tractability and minimal physics (Potential Field Source Surface [PFSS] models, \cite{1969SoPh....9..131A,1969SoPh....6..442S,1992ApJ...392..310W} to comprehensive (but usually still time-independent) fluid plasma physics \citep[MHD, {\emph{e.g.}},][]{1996AIPC..382..104M,2012JASTP..83....1R} but requiring longer computational run times. Despite these very different overall model properties, in terms of coronal magnetic structure they often agree very well with each other, especially at solar minimum \cite{2006ApJ...653.1510R}.

In advance of {\emph{PSP}}’s first solar Enc. in Nov. 2018, \cite{2019ApJ...874L..15R} and \cite{2019ApJ...872L..18V} both ran MHD simulations. They utilized photospheric observations from the Carrington rotation (CR) prior to the Enc. happening and model parameters which were not informed by any {\emph{in situ}} measurements. Both studies successfully predicted {\emph{PSP}} would lie in negative polarity solar wind during perihelion and cross the HCS shortly after (see Fig.~\ref{FIG:Badman2020}). Via field line tracing, \cite{2019ApJ...874L..15R} in particular pointed to a series of equatorial coronal holes as the likely sources of this negative polarity and subsequent HCS crossing.

This first source prediction was subsequently confirmed with careful comparison of {\emph{in~situ}} measurements of the heliospheric magnetic field and tuning of model parameters. This was done both with PFSS modeling \citep{2019Natur.576..237B,2020ApJS..246...23B,2020ApJS..246...54P}, Wang-Sheeley-Arge (WSA) PFSS $+$ Current Sheet Modeling \citep{2020ApJS..246...47S} and MHD modeling \citep{2020ApJS..246...24R,2020ApJS..246...40K,2021A&A...650A..19R}, all pointing to a distinct equatorial coronal hole at perihelion as the dominant solar wind source. As discussed in \S\ref{SWSAS:slowlafv} the predominant solar wind at this time was slow but with high alfv\'enicity. This first Enc. has proved quite unique in how distinctive its source was, with subsequent Encs. typically connecting to polar coronal hole boundaries and a flatter HCS such that {\emph{PSP}} trajectory skirts along it much more closely \citep{2020ApJS..246...40K, 2021A&A...650L...3C, 2021A&A...650A..19R}.

It is worth discussing how these different modeling efforts made comparisons with {\emph{in situ}} data in order to determine their connectivity. The most common is the timing and heliocentric location of current sheet crossings measured {\emph{in situ}} which can be compared to the advected polarity inversion line (PIL) predicted by the various models \citep[{\emph{e.g.}},][and Fig.~\ref{FIG:Badman2020}]{2020ApJS..246...47S}. By ensuring a given model predicts when these crossings occur as accurately as possible, the coronal magnetic geometry can be well constrained and provide good evidence that field line tracing through the model is reliable. Further, models can be tuned in order to produce the best agreement possible. This tuning process was used to constrain models using {\emph{PSP}} data to more accurately find sources. For example,  \citet{2020ApJS..246...54P} found evidence that different source surface heights (the primary free parameter of PFSS models) were more appropriate at different times during {\emph{PSP}}’s first two Encs. implying a non-spherical surface where the coronal field becomes approximately radial, and that a higher source surface height was more appropriate for {\emph{PSP}}’s second Enc. compared to its first. This procedure can also be used to distinguish between choices of photospheric magnetic field data \citep[{\emph{e.g.}},][]{2020ApJS..246...23B}. 

\begin{figure*}
    \centering
    \includegraphics[width=\textwidth]{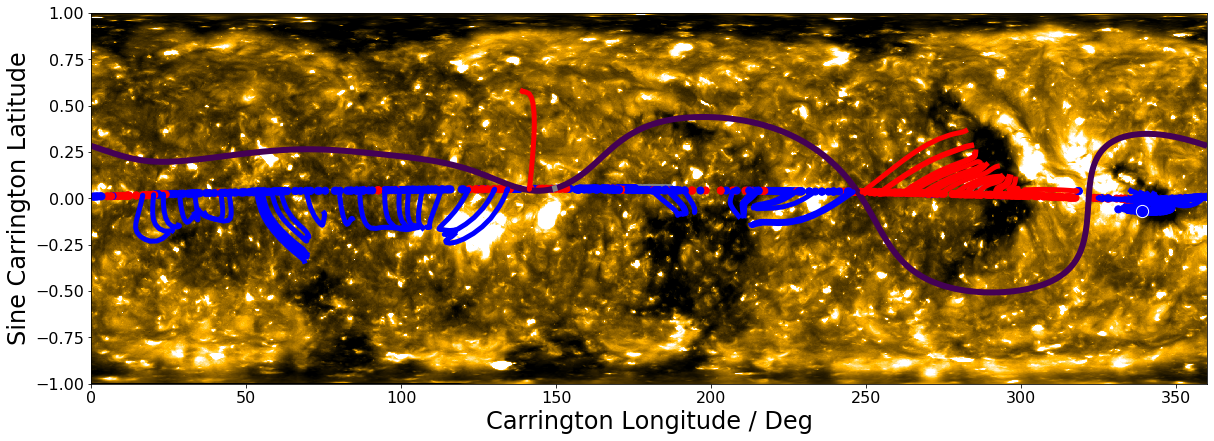}
    \caption{Illustration of mapping procedure and model validation. A PFSS model is run using a source surface height of 2.0~$R_\odot$ and a Global Oscillation Network Group \citep[{\emph{GONG}};][]{1988AdSpR...8k.117H} ZQS magnetogram from 6 Nov. 2018 (the date of the first {\emph{PSP}} perihelion). A black solid line shows the resulting PIL. Running across the plot is {\emph{PSP}}’s first solar Enc. trajectory ballistically mapped to the model outer boundary. The color (red or blue) indicated the magnetic polarity measured {\emph{in situ}} which compares well to the predicted crossings of the PIL. The resulting mapped field lines are shown as curves connecting {\emph{PSP}}’s trajectory to the Sun. A background map combining EUV images from the  {\emph{STEREO}}-A Extreme Ultraviolet Imager \citep[EUVI -- 195~{\AA};][]{2004SPIE.5171..111W} and the Advanced Imaging Assembly \citep[AIA -- 193~{\AA};][]{2012SoPh..275...17L} on the Solar Dynamic Observatory \citep[{\emph{SDO}};][]{2012SoPh..275....3P} shows the mapped locations correspond to coronal holes (dark regions) implying the locations of open magnetic field in the model are physical. Figure adapted from \citet{2020ApJS..246...23B}
    }
    \label{FIG:Badman2020}
\end{figure*}

Further validation of source estimates for {\emph{PSP}} have been made in different ways. For example, if a given source estimate indicates a specific coronal hole, or polar coronal hole extension or boundary, the model can be used to produce contours of the open field and compared with EUV observations of coronal holes to detect if the modeled source is empirically present, and if so if its size and shape are accurately captured \citep[{\emph{e.g.}},][]{2011SoPh..269..367L,2020ApJS..246...23B}. Other {\emph{in situ}} properties besides magnetic polarity have been compared in novel ways: \cite{2020ApJS..246...37R} showed for {\emph{PSP}}’s second Enc. a distinct trend in {\emph{in situ}} density depending on whether the S/C mapped to the streamer belt or outside (see \S\ref{SWSAS:strmblt} for more details). MHD models \citep[{\emph{e.g.}},][]{2020ApJS..246...24R,2020ApJS..246...40K,2021A&A...650A..19R} have also allowed direct timeseries predictions of other {\emph{in situ}} quantities at the location of {\emph{PSP}} which can be compared directly to the measured timeseries.  Kinetic physics such as plasma distributions have provided additional clues: \cite{2020ApJ...892...88B} showed cooler electron strahl temperatures for solar wind mapping to a larger coronal hole during a fast wind stream, and hotter solar wind mapping to the boundaries of a smaller coronal hole during a slow solar wind stream.  This suggests a connection between the strahl temperature and coronal origin (see \S\ref{SWSAS:fstwnd} for more details). \cite{2021A&A...650L...2S} showed further in {\emph{in situ}} connections with source type, observing an increase in mass flux (with {\emph{PSP}} and {\emph{Wind}} data) associated with increasing temperature of the coronal source, including variation across coronal holes and active regions (ARs).

Mapping sources with coronal modeling for {\emph{PSP}}’s early Encs. has nonetheless highlighted challenges yet to be addressed. The total amount of open flux predicted by modeling to escape the corona has been observed to underestimate that measured {\emph{in situ}} in the solar wind at 1~AU \citep{2017ApJ...848...70L}, and this disconnect persists at least down 0.13~AU \citep{2021A&A...650A..18B} suggesting there exist solar wind sources which are not yet captured accurately by modeling.  Additionally, due to its orbital plane lying very close to the solar equator, the solar minimum configuration of a near-flat HCS and streamer belt means {\emph{PSP}} spends a lot of time skirting the HCS \citep{2021A&A...650L...3C} and limits the types of sources that can be sampled. For example, {\emph{PSP}} has yet to take a significant sample of fast solar wind from deep inside a coronal hole, instead sampling primarily streamer belt wind and coronal hole boundaries. Finally, connectivity modeling is typically time-static either due to physical model constraints or computational tractability. However, the coronal magnetic field is far from static with processes such as interchange reconnection potentially allowing previously closed structures to contribute to the solar wind \citep{2020ApJ...894L...4F,2020JGRA..12526005V}, as well as disconnection at the tips of the streamer belt \citep{2020ApJ...894L..19L,2021A&A...650A..30N}. Such transient processes are not captured by the static modeling techniques discussed in this section, but have still been probed with {\emph{PSP}} particularly in the context of streamer blowouts (SBOs; \S\ref{SWSAS:strmblt}) via combining remote observations and {\emph{in situ}} observations, typically requiring multi-S/C collaboration and minimizing modeling uncertainty. Such collaborations are rapidly becoming more and more possible especially with the recent launch of Solar Orbiter \citep[{\emph{SolO}}; ][]{2020AA...642A...1M,2020A&A...642A...4V}, recently yielding for the first time detailed imaging of the outflow of a plasma parcel in the mid corona by {\emph{SolO}} followed by near immediate {\emph{in situ}} measurements by {\emph{PSP}} \citep{2021ApJ...920L..14T}.  

\subsection{Sources of Slow Alfv\'enic Solar Wind}
\label{SWSAS:slowlafv}
{\emph{PSP}}'s orbit has a very small inclination angle w.r.t. the ecliptic plane. It was therefore not surprising to find that over the first Encs. the solar wind streams were observed, with few exceptions, to have slower velocities than that expected for the high-speed streams (HSSs) typically originating in polar coronal holes around solar minimum. What however has been a surprise is that the slow solar wind streams were seen to have turbulence and fluctuation properties, including the presence of large amplitude folds in the magnetic field, i.re. switchbacks, typical of Alfv\'enic fluctuations usually associated with HSSs.

Further out from the Sun, the general dichotomy between fast Alfv\'enic solar wind and slow, non-Alfv\'enic solar wind \citep{1991AnGeo...9..416G} is broken by the so-called slow Alfv\'enic streams, first noticed in the {\emph{Helios}} data acquired at 0.3~AU \citep{1981JGR....86.9199M}. That slow wind interval appeared to have much of the same characteristics of the fast wind, including the presence of predominantly outwards Alfv\'enic fluctuations, except for the overall speed. These were observed at solar maximum, while Parker’s observations over the first four years have covered solar minimum and initial appearance of activity of the new solar cycle.

Alfv\'enic slow wind streams have also been observed at 1~AU \citep{2011JASTP..73..653D}, and been extensively studied in their composition, thermodynamic, and turbulent characteristics \citep{2015ApJ...805...84D, 2019MNRAS.483.4665D}. The results of these investigations point to a  similar origin \citep{2015ApJ...805...84D} for fast and Alfv\'enic slow wind streams. Instances of slow Alfv\'enic wind at solar minimum were found re-examining the {\emph{Helios}} data collected in the inner heliosphere \citep{2019MNRAS.482.1706S, 2020MNRAS.492...39S, 2020A&A...633A.166P}, again supporting a similar origin - coronal holes - for fast and slow Alfv\'enic wind streams. 

Reconstruction of the magnetic sources of the wind seen by Parker for the first perihelion clearly showed the wind origin to be associated with a small isolated coronal hole. Both ballistic backwards projection in conjunction with the PFSS method \citep{2020ApJS..246...54P, 2020ApJS..246...23B} and global MHD models showed {\emph{PSP}} connected to a negative-polarity equatorial coronal hole, within which it remained for the entire Enc. \citep{2019ApJ...874L..15R, 2020ApJS..246...24R}. The {\emph{in situ}} plasma associated with the small equatorial coronal hole was a highly Alfv\'enic slow wind stream, parts of which were also seen near Earth at L1 \citep{2019Natur.576..237B, 2020ApJS..246...54P}. The relatively high intermittency, low compressibility \citep{2020A&A...633A.166P}, increased turbulent energy level \citep{2020ApJS..246...53C}, and spectral break radial dependence are similar to fast wind \citep{2020ApJS..246...55D}, while particle distribution functions are also more anisotropic than in non-Alfv\'enic slow wind \citep{2020ApJS..246...70H}.

Whether Alfv\'enic slow wind always originates from small isolated or quasi-isolated coronal holes ({\emph{e.g.}}, narrow equatorward extensions of polar coronal holes), with large expansion factors within the subsonic/supersonic critical point, or whether the boundaries of large, polar coronal holes might also produce Alfv\'enic slow streams, is still unclear. 

There is however one possible implication of the overall high Alfv\'enicity observed by {\emph{PSP}} in the deep inner heliosphere: that all of the solar wind might be born Alfv\'enic, or rather, that Alfv\'enic fluctuations be a universal initial condition of solar wind outflow. Whether this is borne out by {\emph{PSP}} measurements closer to the Sun remains to be seen.

\subsection{Near Streamer Belt Wind}
\label{SWSAS:strmblt}

As already discussed in part in \S\ref{SWSAS:slowlafv}, the slow solar wind exhibits at least two states. One state has similar properties to the fast wind, it is highly Alfv\'enic, has low densities and low source temperature (low charge state) and appears to originate from inside coronal holes \citep[see for instance the review by][]{2021JGRA..12628996D}. The other state of the slow wind displays greater variability, higher densities and more elevated source temperatures. The proximity of {\emph{PSP}} to the Sun and the extensive range of longitudes scanned by the probe during an Enc. makes it inevitable that at least one of the many S/C (the Solar and Heliospheric Observatory [{\emph{SOHO}}; \citealt{1995SSRv...72...81D}], {\emph{STEREO}}, {\emph{SDO}}, \& {\emph{SolO}}) currently orbiting the Sun will image the solar wind measured {\emph{in situ}} by {\emph{PSP}}. Since coronagraph and heliospheric imagers tend to image plasma located preferably (but not only) in the vicinity of the so-called Thomson sphere (very close to the sky plane for a coronagraph), the connection between a feature observed in an image with its counterpart measured {\emph{in situ}} is most likely to happen when {\emph{PSP}} crosses the Thomson sphere of the imaging instrument. A first study exploited such orbital configurations that occurred during Enc.~2 when {\emph{PSP}} crossed the Thompson spheres of the {\emph{SOHO}} and {\emph{STEREO}}-A imagers \citep{2020ApJS..246...37R}. In this study, the proton speed measured by SWEAP was used to trace back ballistically the source locations of the solar wind to identify the source in coronagraphic observations.

\begin{figure*}
    \centering
    \includegraphics[width=\textwidth]{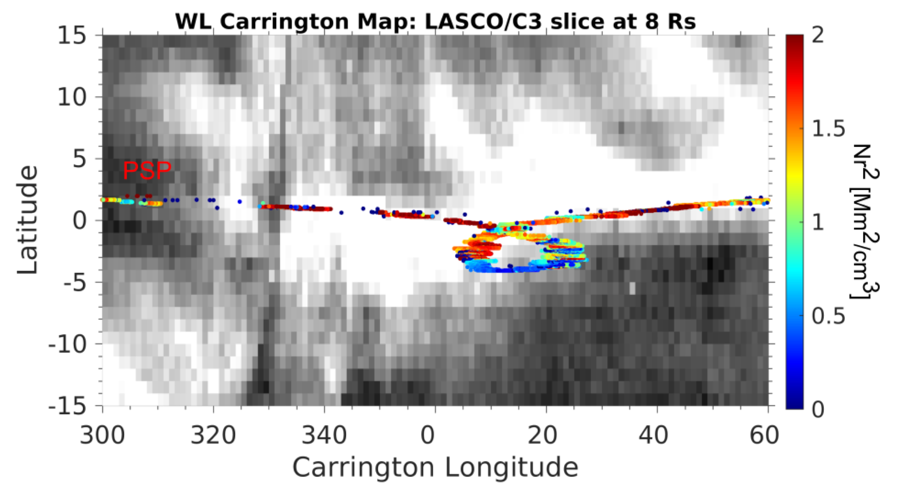}
    \caption{A zoomed-in view of a Carrington map built from LASCO-C3 bands of image pixels extracted at 8~R$_{\odot}$. The {\emph{PSP}} path corresponds to the points of magnetic connectivity traced back to the radial distance of the map (8~\textit{R}$_\odot$). The connectivity is estimated by assuming the magnetic field follows a Parker spiral calculated from the speed of the solar wind measured {\emph{in situ}} at {\emph{PSP}}. The color coding is defined by the density ($N\times r^2$) measured {\emph{in situ}} by {\emph{PSP}} with red corresponding to high densities and blue to low densities. Figure adapted from \cite{2020ApJS..246...37R}}.
    \label{FIG:Rouillard2020}
\end{figure*}

Fig.~\ref{FIG:Rouillard2020} presents, in a latitude versus longitude format, a comparison between the brightness of the solar corona observed by the Large Angle and Spectrometric COronagraph \citep[LASCO;][]{1995SoPh..162..357B} on {\emph{SOHO}} and the density of the solar wind measured {\emph{in~situ}} by {\emph{PSP}} and color-coded along its trajectory. The Figure shows that as long as the probe remained inside the bright streamers, the density of the solar wind was high but as soon as it exited the streamers (due to the orbital trajectory of {\emph{PSP}}), the solar wind density suddenly dropped by a factor of four but the solar wind speed remained the same around 300~km~s$^{-1}$ \citep{2020ApJS..246...37R}. \cite{2021ApJ...910...63G} exploited numerical models and ultraviolet imaging to show that as {\emph{PSP}} exited streamer flows it sampled slow solar wind released from deeper inside an isolated coronal hole. These measurements illustrate nicely the transitions that can occur between two slow solar wind types over a short time period. While switchbacks were observed in both flows, the patches of switchbacks were also different between the two types of slow wind with more intense switchbacks patches measured in the streamer flows \citep{2020ApJS..246...37R}, this is also seen in the spectral power of switchbacks \citep{2021A&A...650A..11F}. \cite{2021ApJ...910...63G} show that both types of solar winds can exhibit very quiet solar wind conditions (with no switchback occurrence). These quiet periods are at odds with theories that relate the formation of the slow wind with a continual reconfiguration of the coronal magnetic field lines due to footpoint exchange, since this should drive strong wind variability continually \citep[{\emph{e.g.}},][]{1996JGR...10115547F}.

\cite{2021A&A...650L...3C} also measured a distinct transition between streamer belt and non-streamer belt wind by looking at turbulence properties during the fourth {\emph{PSP}} perihelion when the HCS was flat and {\emph{PSP}} skirted the boundary for an extended period of time. They associated lower turbulence amplitude, higher magnetic compressibility, a steeper turbulence spectrum, lower Alfv\'enicity and a lower frequency spectral break with proximity to the HCS, showing that at {\emph{PSP}}'s perihelia distances {\emph{in situ}} data allows indirect distinction between solar wind sources.

Finally, in addition to steady state streamer belt and HCS connectivity, remote sensing and {\emph{in situ}} data on board {\emph{PSP}} has also been used to track transient solar phenomena erupting or breaking off from the streamer belt. \cite{2020ApJS..246...69K} detected the passage of a SBO coronal mass ejection (SBO-CME) during {\emph{PSP}} Enc.~1 and via imaging from {\emph{STEREO}}-A at 1~AU and coronal modeling with WSA associated it with a specific helmet streamer. Similarly, \cite{2020ApJ...897..134L} associated a SBO with a CME measured at {\emph{PSP}} during the second Enc. and via stereographic observations from 1~AU and arrival time analysis modeled the flux rope structure underlying the structure.

\subsection{Fast Solar Wind Sources}
\label{SWSAS:fstwnd}

During the first eight Encs. of {\emph{PSP}}, there are only very few observations of fast solar wind, {\emph{e.g.}}, 9 Nov. 2018 and 11 Jan. 2021. Most of the time, {\emph{PSP}} was inside the slow wind streams. Thus, exploration of the source of fast wind remains as a future work.

The first observed fast wind interval was included in the study by \cite{2020ApJ...892...88B}, who investigated the relation between the suprathermal solar wind electron population called the strahl and the shape of the electron VDF in the solar corona. Combining {\emph{PSP}} and {\emph{Helios}} observations they found that the strahl parallel temperature ($T_{s\parallel}$) does not vary with radial distance and is anticorrelated with the solar wind velocity, which indicates that $T_{s\parallel}$ is a good proxy for the electron coronal temperature. Fig.~\ref{FIG:Bercic2020} shows the evolution of $T_{s\parallel}$ along a part of the first {\emph{PSP}} orbit trajectory ballistically projected down to the corona to produce sub-S/C points. PFSS model was used to predict the magnetic connectivity of the sub-S/C points to the solar surface. The observed fast solar wind originates from the equatorial coronal hole \citep{2020ApJS..246...23B}, and is marked by low $T_{s\parallel}$ ($<$ 75 eV). These values are in excellent agreement with the coronal hole electron temperatures obtained via the spectroscopy technique  \citep{1998A&A...336L..90D,2002SSRv..101..229C}.

\begin{figure*}
    \centering
    \includegraphics[width=\textwidth]{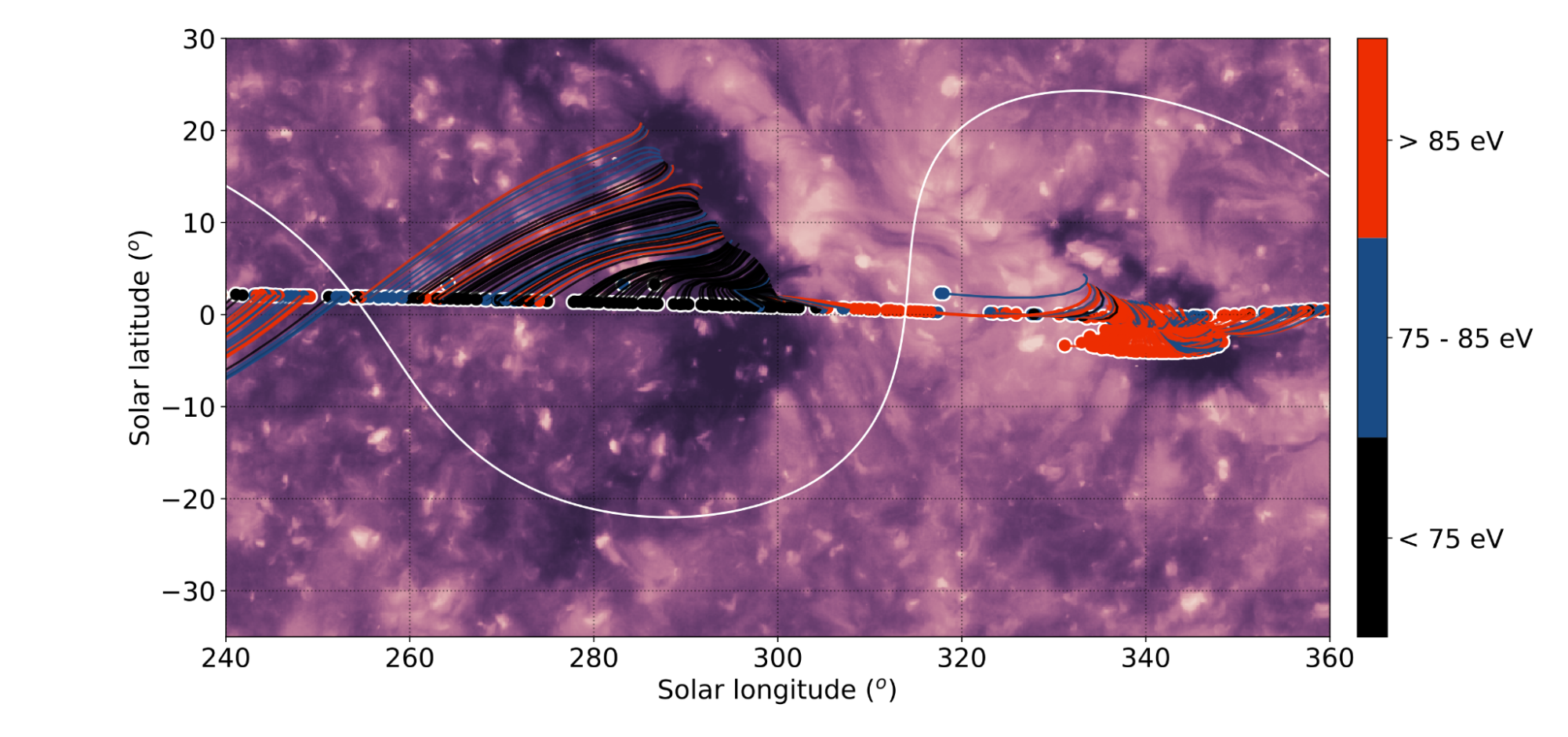}
    \caption{The evolution of $T_{s\parallel}$ with part of the {\emph{PSP}} orbit 1 between 30 Oct. 2018, 00:30~UT (Universal Time) and 23 Nov. 2018, 17:30~UT. The {\emph{PSP}} trajectory is ballistically projected down to the corona ($2~R_\odot$) to produce sub-S/C points.The colored lines denote the magnetic field lines mapped from the sub-S/C points to the solar surface as predicted by the PFSS model with source surface height $2~R_\odot$, the same as used in \cite{2019Natur.576..237B} and \cite{2020ApJS..246...23B}. The white line shows the PFSS neutral line. The points and magnetic field lines are colored w.r.t. hour-long averages of $T_{s\parallel}$. The corresponding image of the Sun is a synoptic map of the 193 Å emission synthesized from {\emph{STEREO}}/EUVI and {\emph{SDO}}/AIA for CR~2210, identical to the one used by \cite{2020ApJS..246...23B} in their Figs. 5 and 9. }
    \label{FIG:Bercic2020}
\end{figure*}

\begin{figure*}
    \centering
    \includegraphics[width=\textwidth]{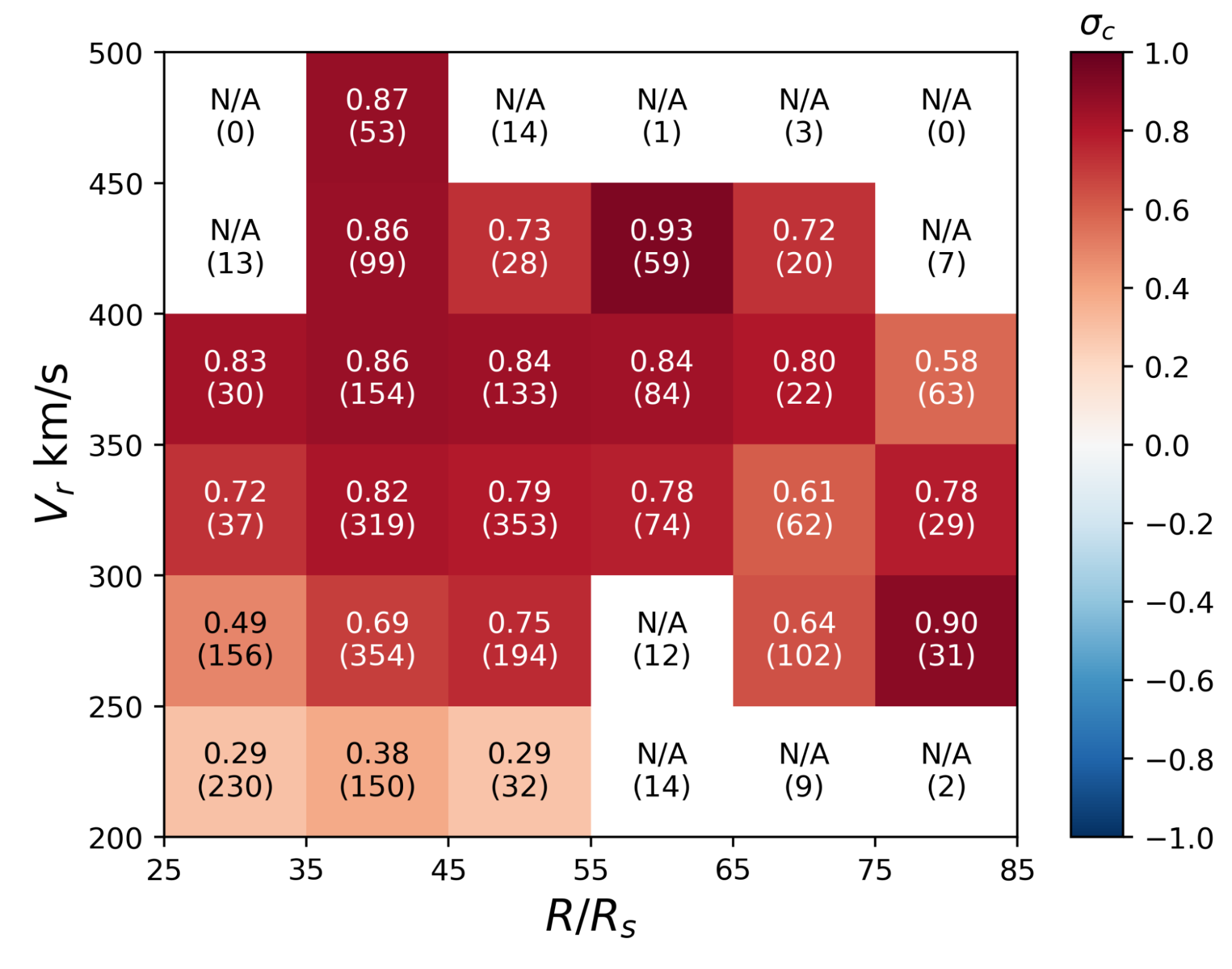}
    \caption{ Normalized cross helicity $\sigma_c$ of wave periods 112~s$-$56~s as a function of the radial distance to the Sun R (horizontal axis; $R_S\equiv R_\odot$) and radial speed of the solar wind $V_r$ (vertical axis). The colors of each block represent the median values of the binned data. Text on each block shows the value of the block and the number of data points (bracketed) in the block. Figure adapted from \citet{2021A&A...650A..21S}.}
    \label{FIG:Shi2020}
\end{figure*}

\cite{2021A&A...650A..21S} analyzed data from the first five Encs. of {\emph{PSP}} and showed how the Alfv\'enicity varies with the solar wind speed. Fig.~\ref{FIG:Shi2020} shows the statistical result of the normalized cross helicity $\sigma_c$ for waves with periods between 112 and 56 seconds, well inside the inertial range of the MHD turbulence. Although the result may be affected by certain individual wind streams due to the limited volume of data, overall, a positive $\sigma_c-V_r$ correlation is observed, indicating that the faster wind is generally more Alfv\'enic than the slower wind. We should emphasize that the result is acquired mostly from measurements of the slow solar wind as the fast wind was rarely observed by {\emph{PSP}}. Thus, the result implies that even for the slow solar wind, the faster stream is more Alfv\'enic than the slow stream. This could be a result of the shorter nonlinear evolution for the turbulence, which leads to a decrease of the Alfv\'enicity, in the faster stream. \cite{2020ApJS..246...54P} and \cite{2021A&A...650A..21S} showed that the slow wind originating from the equatorial pseudostreamers is Alfv\'enic while that originating from the boundary of the polar coronal holes is low-Alfv\'enic. Thus, this speed-dependence of Alfv\'enicity could also be related to the different sources of the slow wind streams. 

\subsection{Active Region Sources}
\label{SWSAS:actvrgn}

The magnetic structure of the corona is important to determining solar wind outflow in at least two different ways, closed coronal field lines providing a geometrical backbone determining the expansion rate of neighboring field lines, and the dynamics of the boundaries between closed and open fields provides time-dependent heating and acceleration mechanisms, as occurs with emerging ARs. 

Emerging ARs reconfigure the local coronal field, leading to the formation of new coronal hole or coronal hole corridors often at there periphery, and depending on the latitude emergence may lead to the formation of large pseudostreamer or reconfiguration of helmet streamers, therefore changing solar wind distributions. Such reconfiguration is also accompanied by radio bursts and energetic particle acceleration, with at least one energetic particle event seen by IS$\odot$IS, on 4 Apr. 2019, attributed to this type of process \citep{2020ApJS..246...35L,2020ApJ...899..107K}. 

The event seen by {\emph{PSP}} was very small, with peak 1~MeV proton intensities of $\sim 0.3$ particles~cm$^{-2}$~sr$^{-1}$~s$^{-1}$~MeV$^{-1}$. Temporal association between particle increases and small brightness surges in the EUV observed by {\emph{STEREO}}, which were also accompanied by type III radio emission seen by the Electromagnetic Fields Investigation on {\emph{PSP}}, provided evidence that the source of this event was an AR nearly $80^\circ$ east of the nominal {\emph{PSP}} magnetic footpoint, suggesting field lines expanding over a wide longitudinal range between the AR in the photosphere and the corona. Studies by \citep{2021A&A...650A...6C,2021A&A...650A...7H} further studied the ARs from these times with remote sensing and {\emph{in situ}} data, including the type III bursts, further associating these escaping electron beams with AR dynamics and open field lines indicated by the type III radiation.

The fractional contribution of ARs to the solar wind is negligible at solar minimum, and typically around $40\% \mbox{--} 60\%$ at solar maximum, scaling with sunspot number \citep{2021SoPh..296..116S}. The latitudinal extent of AR solar wind is highly variable between different solar cycles and varying from a band of about  $\pm30^\circ$ to $\pm60^\circ$ around the equator. As the solar cycle activity increases, {\emph{PSP}} is expected to measure more wind associated with ARs.  Contemporaneous measurements by multiple instruments and opportunities for quadratures and conjunctions with {\emph{SolO}} and {\emph{STEREO}} abound, and should shed light into the detailed wind types originating from AR sources.

\subsection{Switchback Stream Sources}
\label{SWSAS:sbs}
As discussed in previous sections, large amplitude fluctuations with characteristics of large amplitude AWs propagating away from the Sun, are ubiquitous in many solar wind streams. Though such features are most frequently found within fast solar wind streams at solar minimum, there are also episodes of Alfv\'enic slow wind visible both at solar minimum and maximum at 0.3~AU and beyond \citep[in the {\emph{Helios}} and {\emph{Wind}} data;][]{2020SoPh..295...46D,2021JGRA..12628996D}. A remarkable aspect of {\emph{PSP}} measurements has been the fact that Alfv\'enic fluctuations also tend to dominate the slow solar wind in the inner heliosphere. Part and parcel of this turbulent regimes are the switchback patches seen throughout the solar Encs. by {\emph{PSP}}, with the possible exception of Enc.~3. As the Probe perihelia get closer to the Sun, there are indications that the clustering of switchbacks into patches remains a prominent feature, though their amplitude decreases w.r.t. the underlying average magnetic field. The sources of such switchback patches appear to be open field coronal hole regions, of which at least a few have been identified as isolated coronal hole or coronal hole equatorial coronal holes (this was the case of the {\emph{PSP}} connection to the Sun throughout the first perihelion), while streams originating at boundaries of polar coronal holes, although also permeated by switchbacks, appear to be globally less Alfv\'enic.

The absence of well-defined patches of switchbacks in measurements at 1~AU or other S/C data, together with the association of patches to scales similar to supergranulation, when projected backwards onto the Sun, are indications that switchback patches are a signature of solar wind source structure. {\emph{PSP}} measurements near the Sun provide compelling evidence for the switchback patches being the remnants of magnetic funnels and supergranules \citep{2021ApJ...923..174B,2021ApJ...919...96F}.

\section{Kinetic Physics and Instabilities in the Young Solar Wind}
\label{KPIYSW}

In addition to the observation of switchbacks, the ubiquity of ion- and electron-scale waves, the deformation of the particle VDF from a isotropic Maxwellian, and the kinetic processes connecting the waves and VDFs has been a topic of focused study. 
The presence of these waves and departures from thermodynamic equilibrium was not wholly unexpected, given previous inner heliospheric observations by {\emph{Helios}} \citep{2012SSRv..172...23M}, but the observations by {\emph{PSP}} at previously unrealized distances has helped to clarify the role they play in the thermodynamics of the young solar wind.  In addition, the intensity and large variety of plasma waves in the near-Sun solar wind has offered new insight into the kinetic physics of plasma wave growth.

\subsection{Ion-Scales Waves \& Structures}
\label{KPIYSW.ion}

The prevalence of electromagnetic ion-scale waves in the inner heliosphere was first revealed by {\emph{PSP}} during Enc.~1 at $36-54~R_\odot$ by \cite{2019Natur.576..237B} and studied in more detail in \cite{2020ApJ...899...74B}; they implicated that kinetic plasma instabilities may be playing a role in ion-scale wave generation. A statistical study by \cite{2020ApJS..246...66B} showed that a radial magnetic field was a favorable condition for these waves, namely that $30\%-50\%$ of the circularly polarized waves were present in a quiet, radial magnetic field configuration. However, single-point S/C measurements obscure the ability to answer definitively whether or not the ion-scale waves still exist in non-radial fields, only hidden by turbulent fluctuations perpendicular to the magnetic field. Large-amplitude electrostatic ion-acoustic waves are also frequently observed, and have been conjectured to be driven by ion-ion and ion-electron drift instabilities \cite{2020ApJ...901..107M,2021ApJ...911...89M}. These ubiquitously observed ion-scale waves strongly suggest that they play a role in the dynamics of the expanding young solar wind.

The direction of ion-scale wave propagation, however, is ambiguous. The procedure for Doppler-shifting the wave frequencies from the S/C to plasma frame is nontrivial. A complimentary analysis of the electric field measurements is required, \cite[see][for a discussion of initially calibrated DC and low frequency electric field measurements from FIELDS]{2020JGRA..12527980M}. These electric field measurements enabled \cite{2020ApJS..246...66B} to constrain permissible wave polarizations in the plasma frame by Doppler-shifting the cold plasma dispersion relation and comparing to the S/C frame measurements. They found that a majority of the observed ion-scale waves are propagating away from the Sun, suggesting that both left-handed and right-handed wave polarizations are plausible. 

The question of the origin of these waves and their role in cosmic energy flow remains a topic of fervent investigation; {\emph{c.f.}} reviews in \cite{2012SSRv..172..373M,2019LRSP...16....5V}. An inquiry of the plasma measurements during these wave storms is a natural one, given that ion VDFs are capable of driving ion-scale waves after sufficient deviation from non-local thermal equilibrium \citep[LTE;][]{1993tspm.book.....G}. Common examples of such non-LTE features are relatively drifting components, {\emph{e.g.}}, a secondary proton beam, temperature anisotropies along and transverse to the local magnetic field, and temperature disequilibrium between components.
Comprehensive statistical analysis of these VDFs have been performed using {\emph{in situ}} observations from {\emph{Helios}} at 0.3~AU \citep{2012SSRv..172...23M} and at 1~AU ({\emph{e.g.}}, see review of {\emph{Wind}} observations in \citep{2021RvGeo..5900714W}). Many studies employing linear Vlasov theory combined with the observed non-thermal VDFs have implied that the observed structure can drive instabilities leading to wave growth. The question of what modes may dominate, {\emph{e.g.}}, right-handed magnetosonic waves or left-handed ion-cyclotron waves, under what conditions remains open, but {\emph{PSP}} is making progress toward solving this mystery.

 \begin{figure}
\centering
{\includegraphics[trim = 0mm 0mm 0mm 0mm, clip, width=1.0\textwidth]{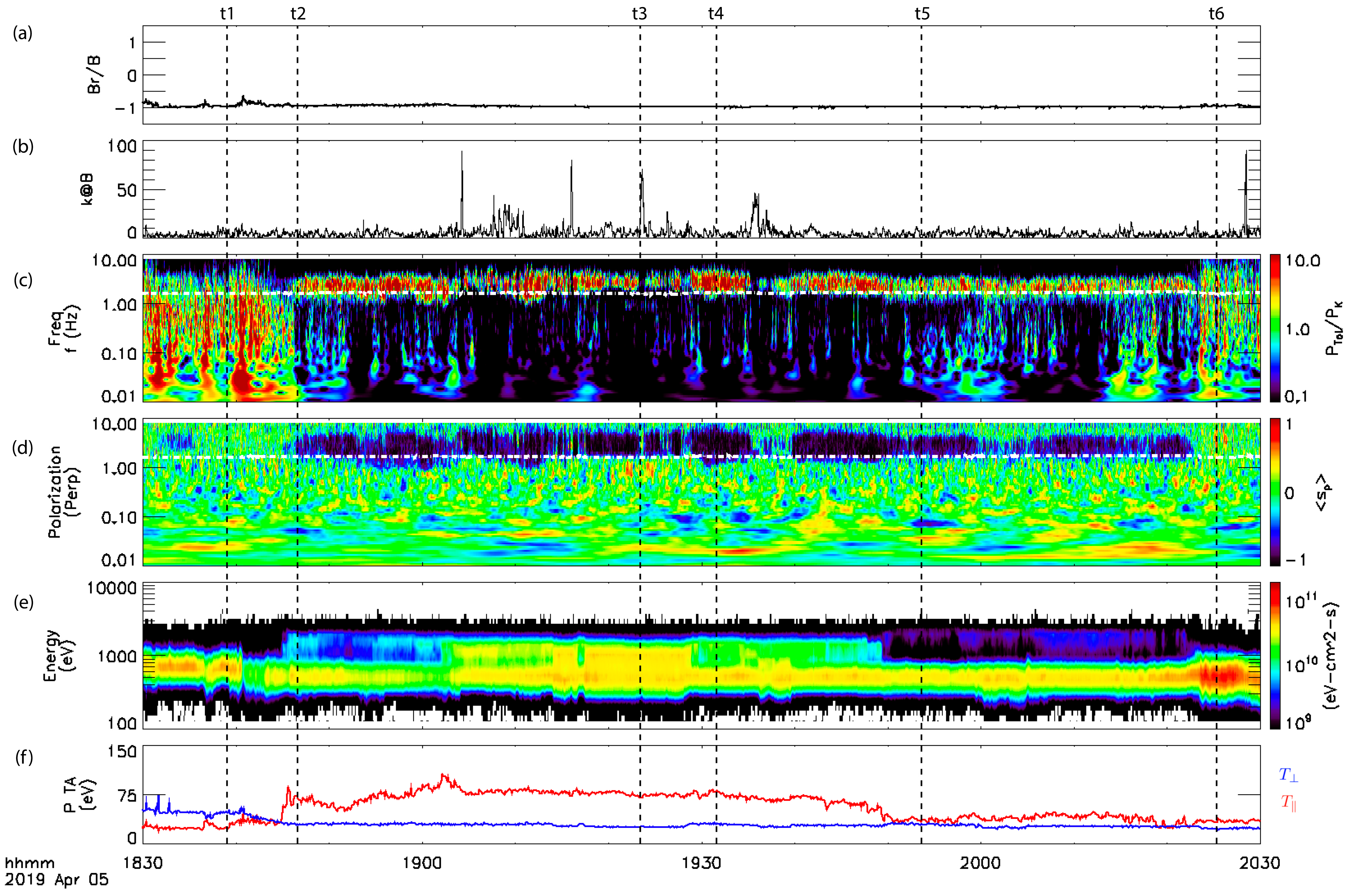}}
\caption{Example event on 5 Apr. 2019 (Event \#1) featuring a strong correlation between a proton beam and an ion-scale wave storm. Shown is the (a) radial magnetic field component, (b) angle of wave propagation w.r.t. B, (c) wavelet transform of B, (d) perpendicular polarization of B, (e) SPAN-i measured moment of differential energy flux, (f) SPAN-i measured moments of temperature anisotropy. In panels (c) and (d), the white dashed–dotted line represents the local $f_{cp}$. Figure adapted from \cite{2020ApJS..248....5V}. 
\label{fig:verniero2020f1}}
\end{figure}

 \begin{figure}
\centering
{\includegraphics[trim = 0mm 0mm 0mm 0mm, clip, width=0.95\textwidth]{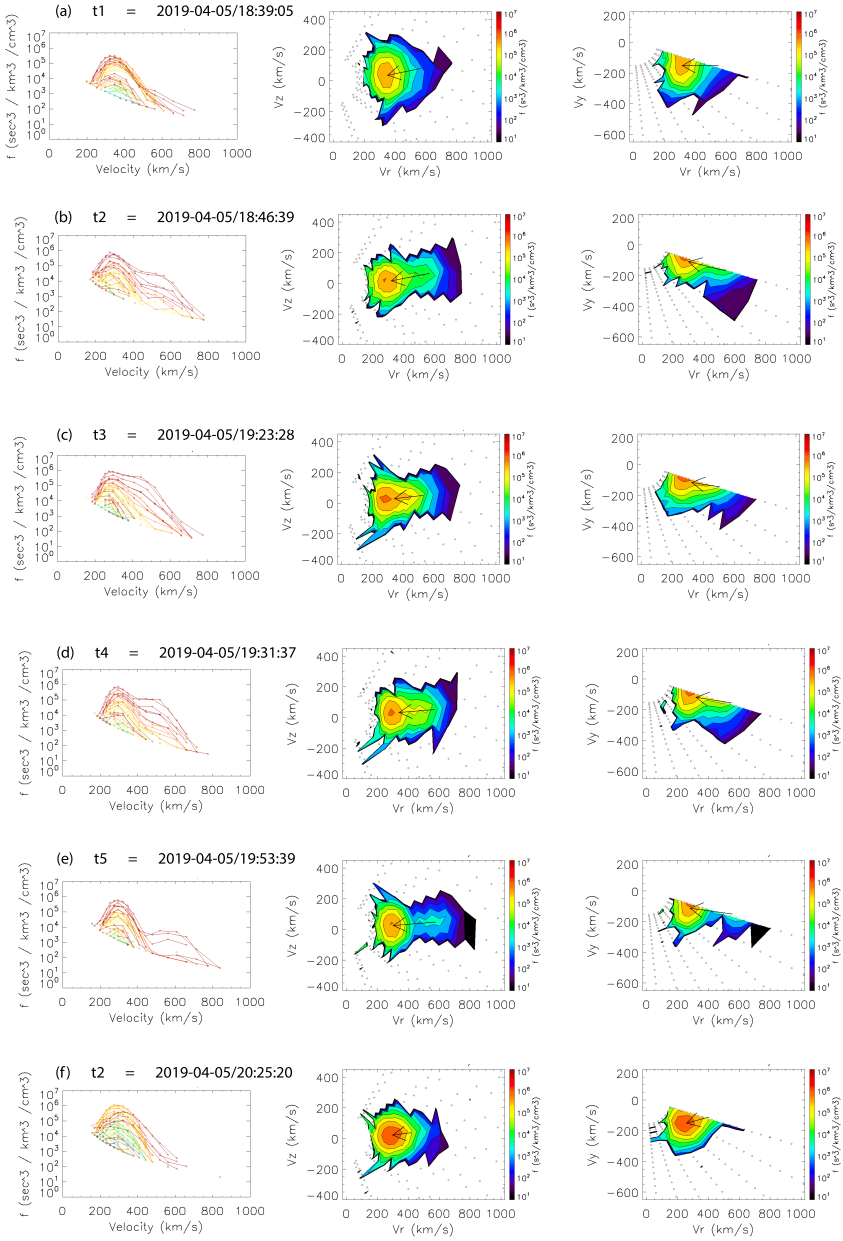}}
\caption{Beam evolution for times indicated by the dashed black lines in Fig.~\ref{fig:verniero2020f1}. Left: Proton VDFs, where each line refers to an energy sweep at different elevation angles. Middle: VDF contour elevations that are summed and collapsed onto the $\theta$-plane. Right: VDF contour elevations that are summed and collapsed onto the azimuthal plane. The black arrow represents the magnetic field direction in SPAN-i coordinates, where the head is at the solar wind velocity (measured by SPC) and the length is the Alfv\'en speed. Figure adapted from \cite{2020ApJS..248....5V}.
\label{fig:verniero2020f2}}
\end{figure}

During Enc.~2, {\emph{PSP}} witnessed intense secondary proton beams simultaneous with ion-scale waves at $\sim36~R_\odot$, using measurements from both SWEAP and FIELDS \citep{2020ApJS..248....5V}. 

The particle instrument suite, SWEAP is comprised of a Faraday Cup, called Solar Probe Cup \citep[SPC;][]{2020ApJS..246...43C} and top-hat electrostatic analyzers called Solar Probe ANalzers that measures electrons \citep[SPAN-e;][]{2020ApJS..246...74W} and ions \citep[SPAN-i;][]{10.1002/essoar.10508651.1}. The SPANs are partially obstructed by {\emph{PSP}}'s heat shield, leading to measurements of partial moments of the solar wind plasma. But, full sky coverage can be leveraged using SPC. The placement of SPAN-i on the S/C was optimal for detecting proton beams, both during initial and ongoing Encs. The time-of-flight capabilities on SPAN-i can separate protons from other minor ions, such as alpha particles. The instrument measures particle VDFs in 3D $(E,\theta,\phi)$ energy-angle phase-space.

These particle VDFs were showcased in \cite{2020ApJS..248....5V} where they displayed two events featuring the evolution of an intense proton beam simultaneous with ion-scale wave storms. The first of these, shown in Fig.~\ref{fig:verniero2020f1}, involved left-handed circularly polarized waves parallel propagating in a quiet, nearly radial magnetic field; the frequencies of these waves were near the proton gyrofrequency ($f_{cp}$). Analysis of the FIELDS magnetometer data shows in Fig.~\ref{fig:verniero2020f1}a the steady $B_r/|B|$; Fig.~\ref{fig:verniero2020f1}b shows from MVA the wave traveling nearly parallel to $\mathbf{B}$, and Fig.~\ref{fig:verniero2020f1}d shows the wavelet transform of $\mathbf{B}$ over a narrow frequency range about the $f_{cp}$, indicated by the white dashed horizontal line; Fig.~\ref{fig:verniero2020f1}d represents the wave polarization, where blue is left-handed in the S/C frame, and red is right-handed. The SPAN-i moments of differential energy flux is displayed in Fig.~\ref{fig:verniero2020f1}e, and the temperature anisotropy was extracted from the temperature tensor in Fig.~\ref{fig:verniero2020f1}f.

The evolution of proton VDFs reported in \cite{2020ApJS..248....5V} during this event (at the times indicated by the black dashed vertical lines in Fig.~\ref{fig:verniero2020f1}) are displayed in Fig.~\ref{fig:verniero2020f2}. The left column represents the proton VDF in 3D phase-space, where each line represents a different energy sweap at different elevation angles. The middle column represents contours of the VDF in SPAN-i instrument coordinates $v_r$-$v_z$, summed and collapsed onto the $\theta$-plane. The right column represents the VDF in the azimuthal plane, where one can notice the portion of the VDF that is obstructed by the heat shield. During this period of time, the proton core was over 50\% in the SPAN-i FOV, and therefore was determined as a suitable event to analyze. 

During both wave-particle interaction events described in \cite{2020ApJS..248....5V}, 1D fits were applied to the SPAN-i VDFs and inputted to a kinetic instability solver. Linear Vlasov analysis revealed many wave modes with positive growth rates, and that the proton beam was the main driver of the unstable plasma during these times.

\cite{2021ApJ...909....7K} further investigated the nature of proton-beam driven kinetic instabilities by using 3D fits of the proton beam and core populations during Enc.~4. Using the plasma instability solver, PLUMAGE \citep{2017JGRA..122.9815K}, they found significant differences in wave-particle energy transfer when comparing results from modeling the VDF as either one or two components. The differences between the waves predicted by the one- and two-component fits were not universal; in some instances, properly accounting for the beam simply enhanced the growth rate of the instabilities predicted by the one-component model while for other intervals, entirely different sets of waves were predicted to be generated.

\vspace{.5 in}
 \begin{figure}
\centering
{\includegraphics[trim = 0mm 0mm 0mm 0mm, clip, width=0.6\textwidth]{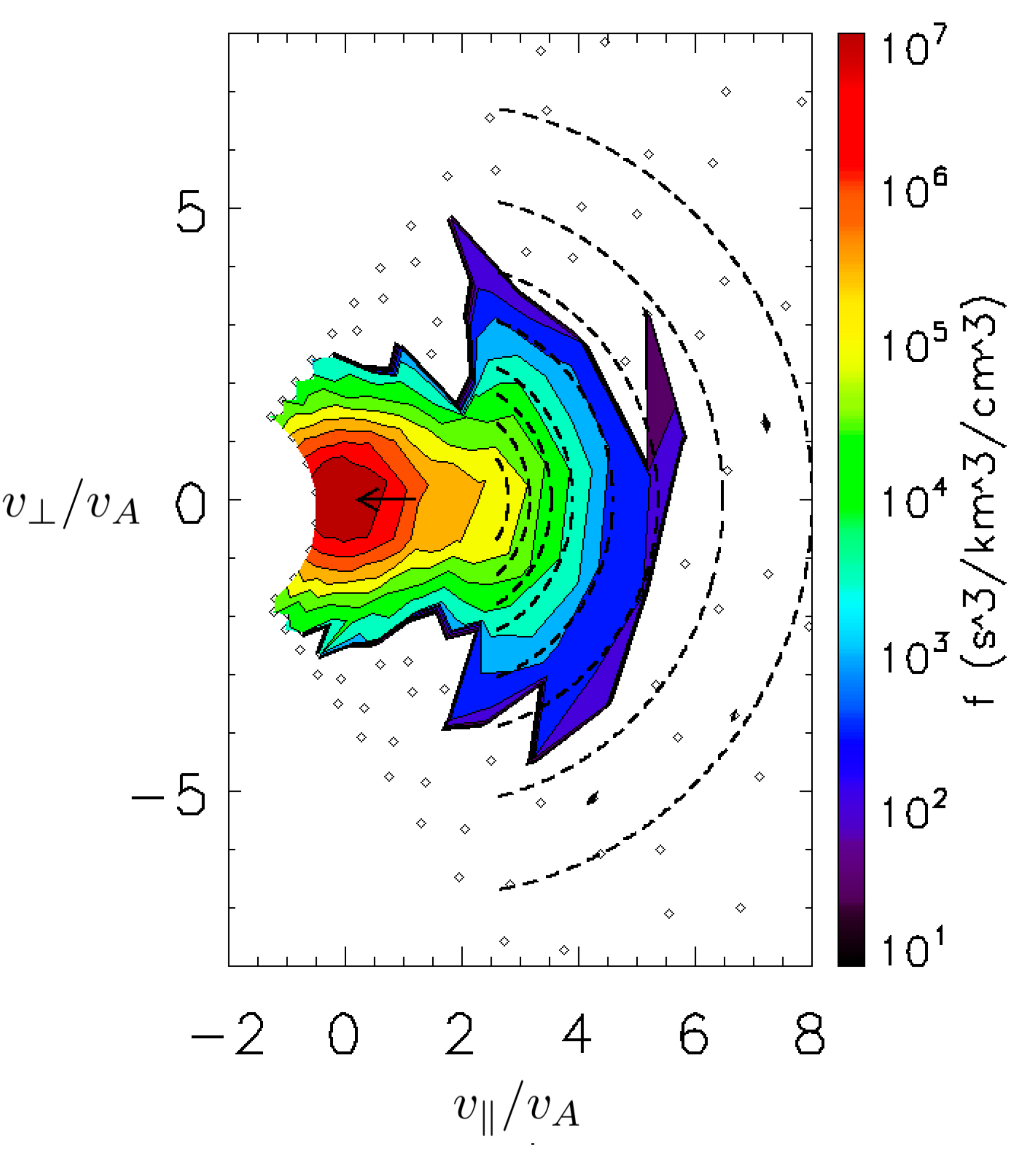}}
\caption{Example of proton VDF displaying a ``hammerhead" feature. The VDF was transformed from the SPAN-i $\theta$-plane to the plasma-frame in magnetic field-aligned coordinates. The black arrow represents the magnetic field, where the head is placed at the solar wind speed and the length is the Alfv\'en speed. The particles diffuse along predicted contours from quasilinear theory, seen by the dashed black curves. Figure adapted from \cite{2022ApJ...924..112V}.
\label{fig:verniero2021f1}}
\end{figure}

\vspace{.5 in}
 \begin{figure}
\centering
{\includegraphics[trim = 0mm 0mm 0mm 0mm, clip, width=1.0\textwidth]{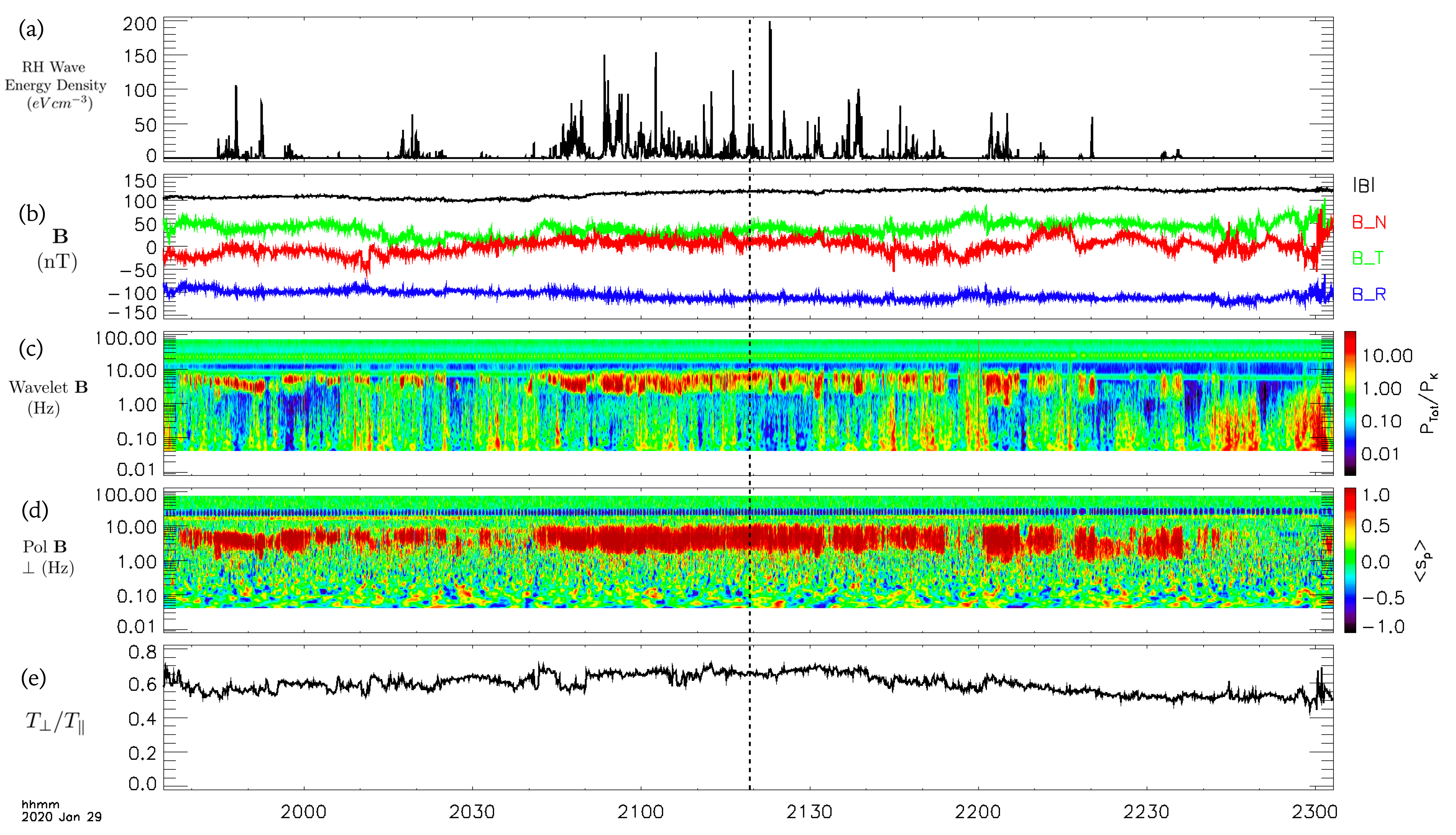}}
\caption{The proton VDF displays a `hammerhead' feature throughout this interval of enhanced Right-Handed wave power. The proton VDF at the time indicated by the vertical dashed black line is shown in Fig.~\ref{fig:verniero2021f1}. Figure adapted from \cite{2022ApJ...924..112V}.
\label{fig:verniero2021f2}}
\end{figure}

During Encs. 4 and 5, {\emph{PSP}} observed a series of proton beams in which the proton VDF was strongly broadened in directions perpendicular to the magnetic field, but only at $|v_\parallel| \gtrsim 2-3 v_{\rm A}$, where $v_\parallel$ is the proton velocity parallel to the magnetic field relative to the peak of the proton VDF. At $|v_\parallel | \lesssim 2-3 v_{\rm A}$, the beam protons' velocities were much more aligned with the magnetic-field direction. The resulting level contours of the proton VDF exhibited a  `hammerhead' shape \citep{2022ApJ...924..112V}. An example VDF is given in Fig.~\ref{fig:verniero2021f1}, at the time indicated by the vertical dashed line in Fig. \ref{fig:verniero2021f2}. These new complex distributions were quantified by modeling the proton VDF as a sum of three bi-Maxwellians, and using the temperature anisotropy of the third population as a proxy for the high energy asymmetries. In addition, the observations substantiate the need for multi-component proton VDF fitting to more accurately characterize the plasma conditions at kinetic scales, as discussed in \cite{2021ApJ...909....7K}.

\cite{2022ApJ...924..112V} found that these hammerhead distributions tended to occur at the same time as intense, right circularly polarized, plasma waves at wave lengths comparable to the proton inertial length. Moreover, the level contours of the VDF within the hammerhead region aligned with the velocity-space diffusion contours that arise in quasilinear theory when protons resonantly interact with parallel-propagating, right circularly polarized, fast-magnetosonic/whistler (FM/W) waves. These findings suggest that the hammerhead distributions occur when field-aligned proton beams excite FM/W waves and subsequently scatter off these waves. Resonant interactions between protons and parallel-propagating FM/W waves occur only when the protons' parallel velocities exceed $\simeq 2.6\ v_{\rm A}$, consistent with the observation that the hammerheads undergo substantial perpendicular velocity-space diffusion only at parallel velocities $\gtrsim 2.6\ v_{\rm A}$ \citep{2022ApJ...924..112V}.

Initial studies of the transfer of energy between the ion-scale waves and the proton distribution were performed in \cite{2021A&A...650A..10V}. During an Enc.~3 interval where an ion cyclotron wave (ICW) was observed in the FIELDS magnetometer data and SPC was operating in a mode where it rapidly measures a single energy bin, rather than scanning over the entire range of velocities, the work done by the perpendicular electric field on the measured region of the proton VDF was calculated. The energy transferred between the fields and particles was consistent with the damping of an ICW with a parallel $f$ wave-vector of order the thermal proton gyroradius.

\subsection{Electron-Scales Waves \& Structures}
\label{KPIYSW.electron}

Researchers in the 1970s recognized that the distributions should change dramatically as solar wind electrons propagated away from the Sun \citep{1971ApJ...163..383C,hollweg1974electron,feldman1975solar}. As satellites sampled regions from $\sim0.3$~AU to outside 5~AU, studies showed that the relative fractions of the field-aligned strahl and quasi-isotropic halo electrons change with radial distance in a manner that is inconsistent with adiabatic motion \citep{maksimovic2005radial,vstverak2009radial,2017JGRA..122.3858G}.  The changes in heat flux, which is carried predominantly by the strahl \citep{scime1994regulation, vstverak2015electron} are also not consistent with adiabatic expansion. Research assessing the relative roles of Coulomb collisions and wave-particle interactions in these changes has often concluded that wave-particle interactions are necessary \citep{phillips1990radial, scudder1979theory, 2019MNRAS.489.3412B, 2013ApJ...769L..22B}. The ambipolar electric field is another  mechanism that shapes the electron distributions \citep{lemaire1973kinetic,scudder2019steady}.

{\emph{PSP}} has provided new insights into electrons in the young solar wind, and the role of waves and the ambipolar electric field in their evolution. \cite{2020ApJS..246...22H}, in a study of the first two Encs., found that radial trends inside $\sim0.3$~AU were mostly consistent with earlier studies. The halo density, however, decreases close to the Sun, resulting in a large increase in the strahl to halo ratio. In addition, unlike what is seen at 1~AU, the core electron temperature is anti-correlated with solar wind speed \citep{2020ApJS..246...22H,2020ApJS..246...62M}. The core temperature may thus reflect the coronal source region, as also discussed for the strahl parallel temperature in \S4.4 \citep{2020ApJ...892...88B}. \cite{2022ApJ...931..118A} confirmed the small halo density, showing that it continued to decrease in measurements closer to the Sun, and also found that the suprathermal population (halo plus strahl) comprised only 1\% of the solar wind electrons at the closest distances sampled.

The electron heat flux carried primarily by strahl \citep{2020ApJ...892...88B, 2021A&A...650A..15H} is also anticorrelated with solar wind speed \citep{ 2020ApJS..246...22H}. Closer to the Sun (from 0.125 to 0.25~AU) the normalized electron heat flux is also anti-correlated with beta \citep{2021A&A...650A..15H}. This beta dependence is not consistent with a purely collisional scattering mechanism. 

The signature of the ambipolar electric field has also been clearly revealed in electron data \citep{2021ApJ...916...16H, 2021ApJ...921...83B} as a deficit of electrons moving back towards the Sun. This loss of electrons occurs more than 60\% of the time inside 0.2~AU. The angular dependence of the deficit is not consistent with Coulomb scattering, and the deficit disappears in the same radial distances as the increase in the halo. There is also a decrease in the normalized heat flux. Both observations provide additional support for the essential role of waves.

The role of whistler-mode waves in the evolution of solar wind electrons and regulation of heat flux has long been a topic of interest. Instability mechanisms including temperature anisotropy and several heat flux instabilities have been proposed \citep{1994JGR....9923391G,1996JGR...10110749G,2019ApJ...871L..29V}. Wave studies near 1~AU utilizing data from {\emph{Wind}}, {\emph{STEREO}}, {\emph{Cluster}} \citep{1997SSRv...79...11E} and the Acceleration, Reconnection, Turbulence and Electrodynamics of the Moon's Interaction \citep[{\emph{ARTEMIS}};][]{2011SSRv..165....3A} missions provided evidence for both low amplitude parallel propagating whistlers \citep{2014ApJ...796....5L,2019ApJ...878...41T} and large amplitude highly oblique  waves \citep{2010JGRA..115.8104B,2020ApJ...897..126C}.

The Fields instrument on {\emph{PSP}}, using waveform capture, spectral, and bandpass filter datasets using both electric fields and search coil magnetic fields, has enabled study of whistler-mode waves over a wide range of distances from the Sun and in association with different large-scale structures. This research, in concert with studies of solar wind electrons, has provided critical new evidence for the role of whistler-mode waves in regulation of electron heat flux and scattering of strahl electrons to produce the halo. Observational papers have motivated a number of theoretical studies to further elucidate the physics \citep{2020ApJ...903L..23M,2021ApJ...919...42M,2021ApJ...914L..33C,vo2022stochastic}.

Enc.~1 waveform data provided the first definitive evidence for the existence of sunward propagating whistler mode waves \citep{2020ApJ...891L..20A}, an important observation because, if the waves have wavevectors parallel to the background magnetic field,  only sunward-propagating waves can interact with the anti-sunward propagating strahl. The whistlers observed near the Sun occur with a range of wave angles from parallel to highly oblique \citep{2020ApJ...891L..20A,2021A&A...650A...8C,2022ApJ...924L..33C,Dudok2022_scm}. Because the oblique whistler waves are elliptically polarized (mixture of left and right hand polarized),  oblique waves moving anti-sunward can interact with electrons moving away from the Sun.
Particle tracing simulations\citep{2021ApJ...914L..33C,vo2022stochastic} and PIC simulations \citep{2020ApJ...903L..23M,2021ApJ...919...42M,2019ApJ...887..190R} have revealed details of wave-electron interactions.

Several studies have examined the association of whistlers with large-scale solar wind structures. There is a strong correlation between large amplitude waves and the boundaries of switchbacks \citep{2020ApJ...891L..20A}, and smaller waves can fill switchbacks \citep{2021A&A...650A...8C}. The whistlers are also seen primarily in the  slow solar wind  \citep{2021A&A...650A...9J,2022ApJ...924L..33C}, as initially observed near 1~AU \citep{2014ApJ...796....5L} and in the recent studies using the {\emph{Helios}} data between 0.3  to 1~AU \citep{2020ApJ...897..118J}. Several studies have found differences in the evolution of the non-thermal electron distributions between the slow and fast wind, suggesting that different scattering mechanisms are active in the fast and slow wind   \citep{pagel2005understanding,vstverak2009radial}.

\begin{figure}
	\centering
    \includegraphics[width=0.95\textwidth]{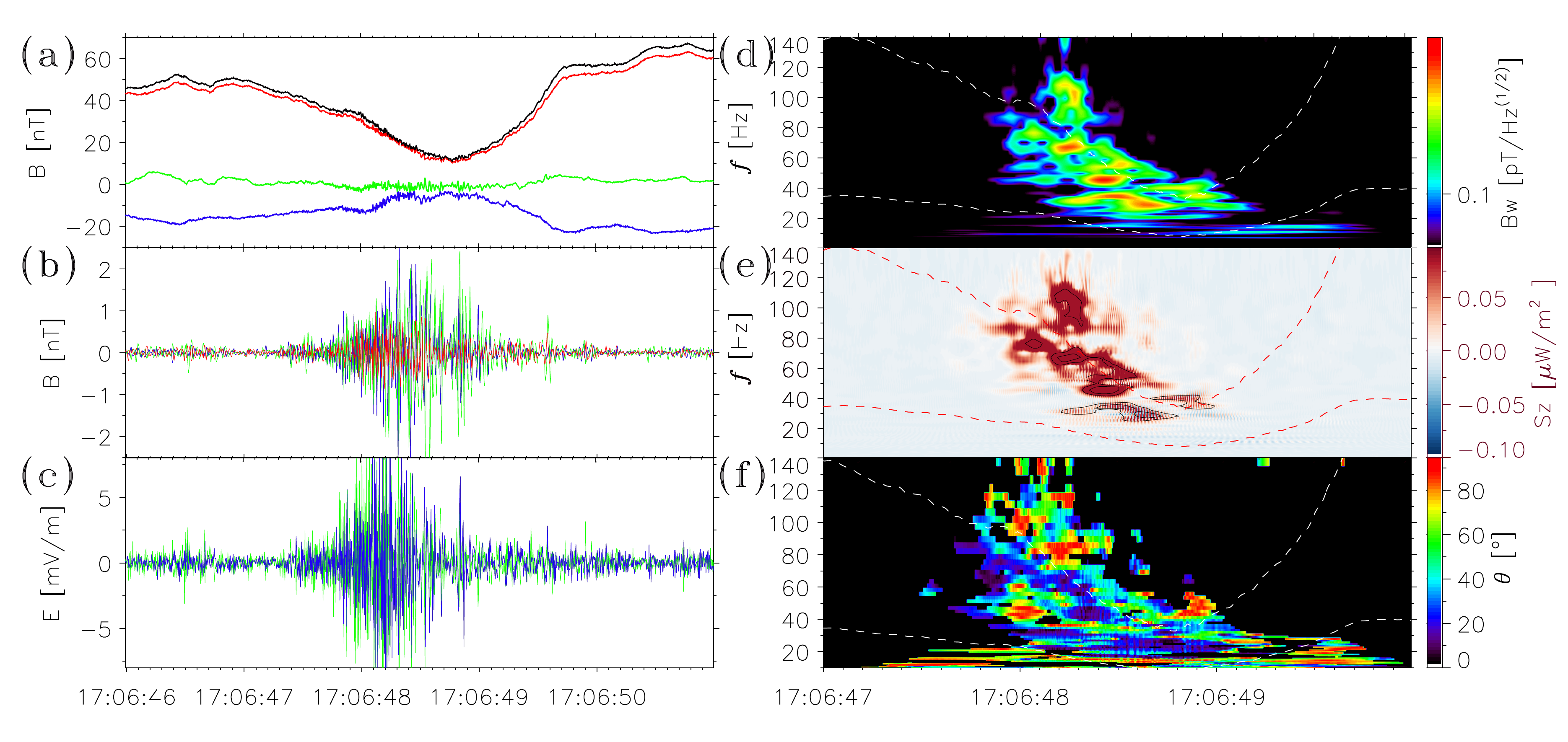}
\caption{ Enlargement of the trailing edge of the switchback of Fig.~\ref{fig:icx1}. Panel (a) shows the magnetic field from MAG with the same color code as in Fig \ref{fig:icx1}. Panels (b) and (c) show magnetic and electric field wave perturbations respectively. Panel (d) displays the dynamic spectrum of magnetic field perturbations $B_w$. The dashed curves in panels (d-f) represent the $f_{LH}$ frequency (bottom curve) and 0.1$f_{ce}$ (upper curve). Panel (e) displays the signed radial component of the Poynting flux. Red colors corresponds to a sunward propagation. Panel (f) shows the wave normal angle relative to the direction of the background magnetic field.}
    \label{fig:mag}
\end{figure}

Fig.~\ref{fig:mag} shows a zoom-in of the trailing edge of the switchback displayed in Fig.~\ref{fig:icx1}. Fig.~\ref{fig:mag}a emphasizes that the local dip in the magnetic field is essentially caused by a decrease of its radial component. This dip coincides with an increase of the ratio between electron plasma frequency ($f_{pe}$) and electron gyrofrequency $f_{ce}$ from 120 to $\approx500$. A polarization analysis reveals a right-handed circular polarization of the magnetic field and an elliptical polarization of the electric field perturbations with a $\pi/2$ phase shift. The dynamic spectrum in Fig.~\ref{fig:mag}d shows a complex inner structure of the wave packet, which consists of a series of bursts. The phase shift of the magnetic and electric field components transverse to the radial direction attest a sunward propagation of the observed waves. The sign of the radial component of the Poynting vector (Fig.~\ref{fig:mag}f) changes from positive (sunward) at high frequencies to negative (anti-sunward) at lower frequencies. The frequencies of these wave packets fall between $f_{LH}$ (lower dashed curve in Figs.~\ref{fig:mag}f and \ref{fig:mag}g) and one tenth of $f_{ce}$ (upper dashed curve). This suggests that the observed frequency range of the whistler waves is shifted down by the Doppler effect as the whistler phase velocity ($300-500$~km~s$^{-1}$) is comparable to that of the plasma bulk velocity. The observed whistlers are found to have a wide range of wave normal angle values from quasi-parallel propagation to quasi-electrostatic propagating close to the resonance cone corresponding to the complex structure of the dynamics spectrum (Fig.~\ref{fig:mag}b). Fig.~\ref{fig:mag}h thereby further supports the idea that our complex wave packet consists of a bunch of distinct and narrowband wave bursts. The wave frequency in the solar wind plasma frame, as reconstructed from the Doppler shift and the local parameters of plasma, are found to be in the range of \SIrange{100}{350}{Hz}, which corresponds to $0.2-0.5\ f_{ce}$ (Fig.~\ref{fig:mag}d). Incidentally, the reconstructed wave frequency can be used to accurately calibrate the electric field measurements, and determine the effective length of the electric field antennas, which was found to be approximately \SIrange{3.5}{4.5}{m} in the \SIrange{20}{100}{Hz} frequency range \citep{2020ApJ...891L..20A, 2020ApJ...901..107M}. More details can be found in \cite{2020ApJ...891L..20A}. 

The population of such sunward propagating whistlers can efficiently interact with the energetic particles of the solar wind and affect the strahl population, spreading their field-aligned PAD through pitch-angle scattering. For sunward propagating whistlers around \SIrange{100}{300}{Hz}, the resonance conditions occur for electrons with energies from approximately \SI{50}{eV} to \SI{1}{keV}. This energy range coincides with that of the observed strahl \citep{2020ApJS..246...22H} and potentially leads to efficient scattering of the strahl electrons. Such an interaction can be even more efficient taking into account that some of the observed waves are oblique. For these waves, the effective scattering is strongly enhanced at higher-order resonances \citep{2020ApJ...891L..20A,2021ApJ...911L..29C}, which leads to a regulation of the heat flux as shown by \cite{2019ApJ...887..190R}, and to an increase in the fraction of the halo distribution with the distance from the Sun. 

{\emph{PSP}} has provided direct evidence for scattering of strahl into halo by narrowband whistler-mode waves \citep{ 2020ApJ...891L..20A, 2021ApJ...911L..29C, 2021A&A...650A...9J}.  The increase in halo occurs in the same beta range and radial distance range as the whistlers, consistent with production of  halo by whistler scattering. Comparison of waveform capture data and electron distributions from Enc.~1 \citep{2021ApJ...911L..29C} showed that the narrowest strahl occurred when there were either no whistlers or very intermittent low amplitude waves, whereas the broadest distributions were associated with intense, long duration waves. Features consistent with an energy dependent scattering mechanism were observed in approximately half the broadened strahl distributions, as was also suggested by features in electrons displaying the signature of the ambipolar field \citep{2021ApJ...916...16H}.

In a study of bandpass filtered data from Encs. 1 through 9, \cite{ 2022ApJ...924L..33C} have shown that the narrowband whistler-mode waves that scatter strahl electrons and regulate heat flux are not observed inside approximately 30~$R_\odot$. Instead, large amplitude electrostatic (up to 200~mV/m) waves in the same frequency range (from the $f_{LH}$ frequency  up to a few tenths of $f_{ce}$) are ubiquitous, as shown in Fig.~\ref{fig:ESradial}. The peak amplitudes of the electrostatic (ES) waves ($\sim220$~mV/m) are an order of magnitude larger than those of the whistlers ($\sim40$~mV/m).  Within the same region where whistlers disappear, the deficit of sunward electrons is seen, and the density of halo relative to the total density decreases. The finding that, when the deficit was observed, the normalized heat flux-parallel electron beta relationship was not consistent with the whistler heat flux fan instability is consistent with loss of whistlers. The differences in the functional form of electron distributions due to this deficit very likely result in changes in the instabilities excited \citep{2021ApJ...916...16H, 2021A&A...656A..31B}. 

Theoretical studies have examined how changes in the distributions and background plasma properties might change which modes are destabilized. \cite{ lopez2020alternative} examined dependence  on beta and the ratio of the strahl speed to the electron Alfv\'en speed for different electromagnetic and electrostatic  instabilities.   This ratio decreases close to the Sun.  Other studies \citep{2021ApJ...919...42M, 2019ApJ...887..190R,schroeder2021stability} have concluded that multiple instabilities operate sequentially and/or at different distances from the Sun. 

\begin{figure}
\centering
{\includegraphics[width=0.8\textwidth]{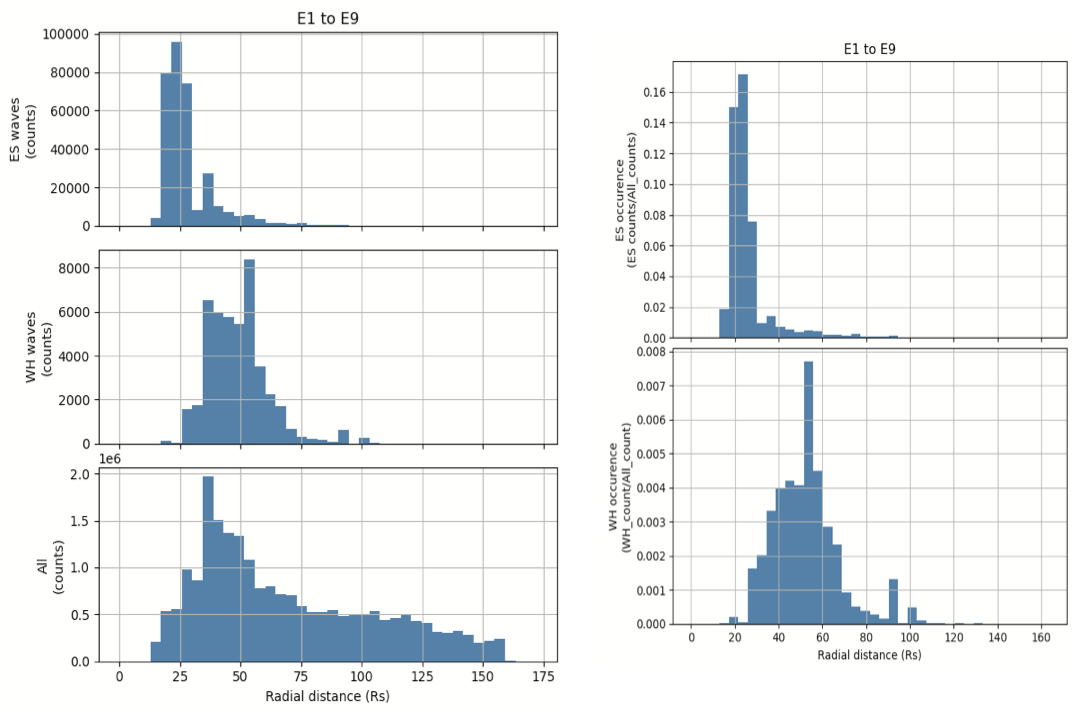}}
\caption{ Statistics of whistler-mode waves  and ES waves identified in bandpass filter data.  Left hand column, from top to bottom: the number of BBF samples identified as ES waves versus radial distance, the number of BBF samples identified as whistler-mode waves versus radial distance, and the total number of BBF samples in Enc.~1 through Enc.~9 versus radial distance. The right hand column: the electrostatic wave occurrence rate and the  whistler-mode wave occurrence rate , both versus radial distance. Note that the drop off outside 75~$R_\odot$ (0.3~AU) is associated with the impact of the decrease in frequency with radial distance on the algorithm used to identify waves. Figure adapted from \cite{2022ApJ...924L..33C}.
\label{fig:ESradial}}
\end{figure}

Closer to the Sun, other scattering mechanisms must operate, associated with the large narrowband ES waves and the nonlinear waves. Note that in some cases  these ES waves have been identified as ion acoustic waves \citep{2021ApJ...919L...2M}. \cite{ dum1983electrostatic} has discussed anomalous transport and heating associated with ES waves in the solar wind.  The lack of narrowband whistler-mode waves close to the sun and in regions of either low ($<1$) or high ($>10$) parallel electron beta may also be significant for the understanding and modeling the evolution of flare-accelerated electrons, other stellar winds, the interstellar medium, and intra-galaxy cluster medium.  

{\emph{PSP}} data has been instrumental in advancing the study of electron-resonant plasma waves other than whistler-mode waves.  \citet{Larosa2022} presented the first unambiguous observations of the Langmuir z-mode in the solar wind (Langmuir-slow extraordinary mode) using high frequency magnetic field data. This wave mode is thought to play a key role in the conversion of electron-beam driven Langmuir waves into type~III and type~II solar radio emission \citep[{\emph{e.g.}},][and references therein]{Cairns2018}.  However, progress understanding the detailed kinetic physics powering this interaction has been slowed by a lack of direct z-mode observations in the solar wind. Z-mode wave occurrence was found to be highly impacted by the presence of solar wind density fluctuations, confirming long-held theoretical assumptions.  

{\emph{PSP}} data also revealed the presence of unidentified electrostatic plasma waves near $f_{ce}$ in the solar wind (Fig.~\ref{fig:nearfce}).  \citet{Malaspina2020_waves} showed that these waves occur frequently during solar Encs., but only when fluctuations in the ambient solar wind magnetic field become exceptionally small.  \citet{2022ApJ...936....7T} identified that a necessary condition for the existence of these waves is the direction of the ambient solar wind magnetic field vector. They demonstrated that these waves occur for a preferential range of magnetic field orientation, and concluded their study by suggesting that these waves may be generated by S/C interaction with the solar wind.  \citet{Malaspina2021_fce} explored high-cadence observations of these waves, demonstrating that they are composed of many simultaneously present modes, one of which was identified as the electron Bernstein mode. The other wave modes remain inconclusively identified. \citet{Shi2022_waves} and \citet{Ma2021_waves} explored the possibility that these waves are created by nonlinear wave-wave interactions. Identifying the exact wave mode, the origin of these waves near $f_{ce}$, and their impact on the solar wind remain areas of active study. {\emph{PSP}} data from ever decreasing perihelion distances are expected to enable new progress.  

\begin{figure}
\centering
{\includegraphics[width=0.8\textwidth]{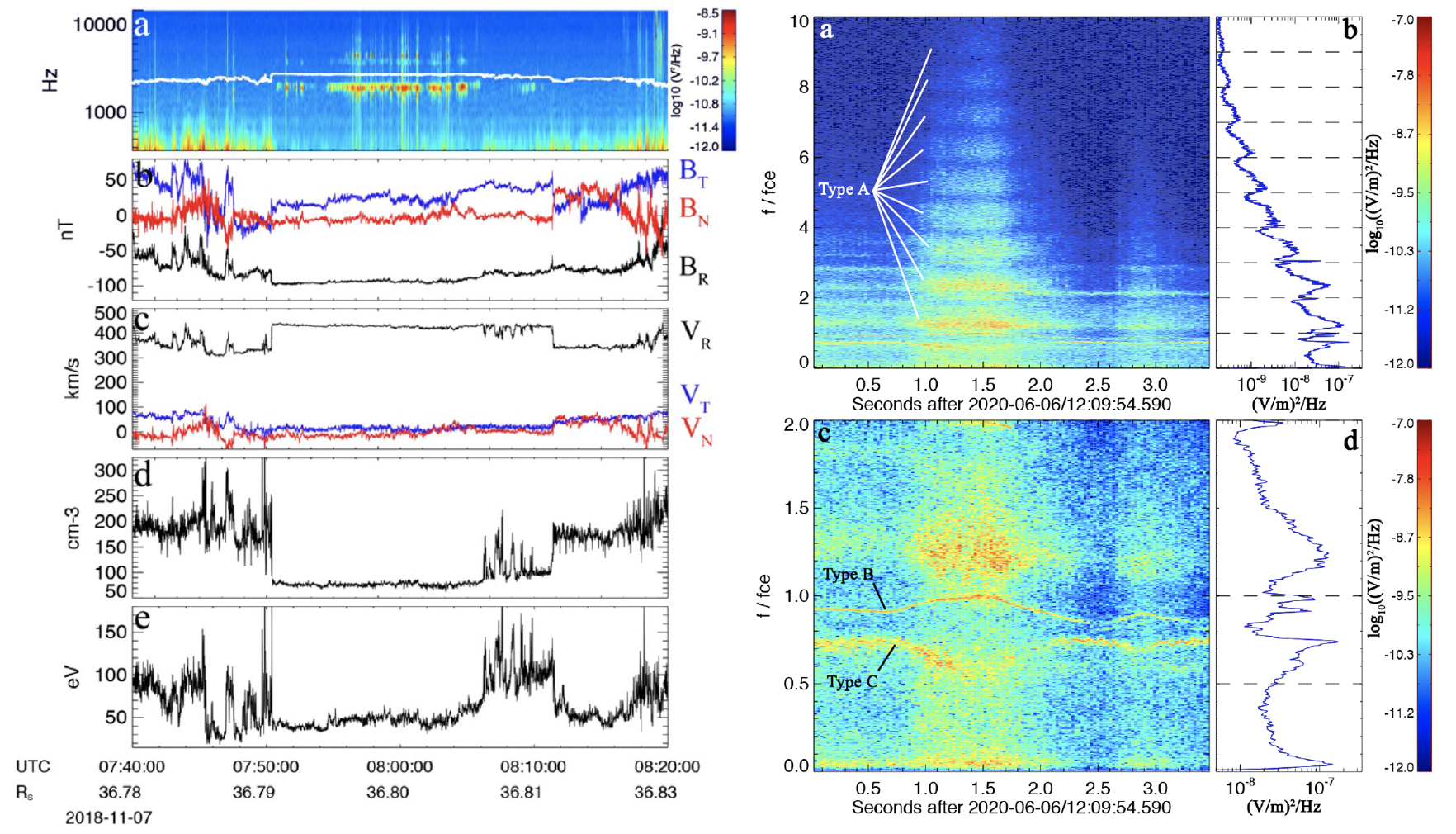}}
\caption{The left-column shows a spectrogram of electric field data, with $f_{ce}$ indicated by the white line.  The near-cyclotron waves are present at the center of the interval, where fluctuations in the ambient magnetic field (b), solar wind velocity (c), plasma density (d), and electron temperature (e) show decreased variability.  The right-column shows a high-cadence observation of near-cyclotron waves, where three distinct wave Types are identified (A,B,C). Type A was identified as an electron Bernstein wave. Figure adapted from \citet{Malaspina2021_waves} and \citet{Malaspina2020_waves}.
\label{fig:nearfce}}
\end{figure}

\section{Turbulence}
\label{TRBLCE}

Turbulence refers to a class of phenomena that characteristically occurs in fluids and plasmas when nonlinear effects are dominant. Nonlinearity creates complexity, involvement of many degrees of freedom and an effective lack of predictability. In contrast, linear effects such as viscous damping or waves in uniform media tend to operate more predictably on just a few coordinates or degrees of freedom. Statistical properties in both the space and time domains are crucial to fully understanding turbulence \citep{2004RvMP...76.1015Z, 2015RSPTA.37340154M}. The relative independence of spatial and temporal effects in turbulence presents particular  challenges to single S/C missions such as {\emph{PSP}}. Nevertheless various methods, ranging from Taylor's frozen-in hypothesis  \citep{2015ApJ...801L..18K,2019ApJS..242...12C,2021A&A...650A..22P} to more elaborate ensemble methods \citep{2019ApJ...879L..16B,2020PhRvR...2b3357P,2020ApJS..246...53C,2021ApJ...923...89C} have been brought to bear to reveal fundamental properties related to the turbulent dynamics of the newly explored regions of the solar atmosphere. This line of research is of specific importance to the {\emph{PSP}} mission in that turbulence is a likely contributor to heating of the corona and the origin of the solar wind -- two of the major goals of the {\emph{PSP}} mission. In this section we will review the literature related to {\emph{PSP}} observations of turbulence that has appeared in the first four years of the mission.

\subsection{Energy Range and Large-Scale (Correlation Scale) Properties}\label{sec:5_large_scale}

Turbulence is often described by a cascade process, which in simplest terms describes the transfer of energy across scales from largest to smallest scales. The largest relevant scales act as reservoirs and the smallest scales generally contribute to most of the dissipation of the turbulent  fluctuations. The drivers of turbulence are notionally the large ``energy-containing eddies'' which may be initially populated by a supply of fluctuations at the boundaries, or injection by ``stirring'' in the interior of the plasma. In the solar wind the dynamics at the photosphere is generally believed to inject a population of fluctuations, usually described as Alfv\'enic fluctuations. These propagate upwards into the corona triggering subsequent turbulent activity \citep{1999ApJ...523L..93M,2009ApJ...707.1659C,2011ApJ...736....3V,2013ApJ...776..124P}. However large scale organized shears in magnetic field and velocity field also exist in the solar atmosphere. While these are not initially in a turbulent state, they may become so at a later time, and eventually participate by enhancing the supply of cascading energy. Stream interaction regions (SIRs) are an example of the latter. Still further stirring mechanisms are possible, such as injection of waves due to scattering of reflected particles upstream of shocks, or by newly ionized particles associated with pickup of interstellar neutrals \citep{2020ApJ...900...94P}. We will not be concerned with this latter class of energy injection processes here. 

Among the earliest reports from {\emph{PSP}}, \citet{2020ApJS..246...53C} described a number of observations of relevance to the energy range. In particular the large-scale edge of the power-law inertial range (see \S5.2) indicates a transition, moving towards large scale, into the energy range. \citet{2020ApJS..246...53C} reports the presence of a shallower ``$1/f$'' range at these larger scales, a feature that is familiar from 1~AU observations \citep{1986PhRvL..57..495M}. It is important to recognize that in general turbulent theory provides no specific expectation for spectral forms at energy-containing scales, as these may be highly situation dependent. Indeed the implied range of scale at which $1/f$ is observed {\emph{in~situ}} corresponds rather closely to the scales and frequencies at which features are observed in the photospheric magnetic field \citep{2007ApJ...657L.121M} and in the deep corona \citep{2008ApJ...677L.137B}. This correspondence strongly hints that the $1/f$ signal is a vestige of dynamical processes very close to the sun, possibly in the dynamo. Still, \citet{2020ApJS..246...53C} point out that the measured nonlinear times at the {\it break scale} between inertial and energy ranges suggest that there is sufficient time in transit for the source region to develop nonlinear correlations at the that scale. This lends some credence to theories \citep[{\emph{e.g.}},][]{1989PhRvL..63.1807V,2018ApJ...869L..32M} offering explanation for local dynamical emergence of $1/f$ signals. This issue remains to be fully elucidated.

An important length scale that describes the energy range is the correlation scale, which is nominally the scale of the energy containing eddies \citep{2012JFM...697..296W}. The notion of a unique scale of this kind is elusive, given that a multiplicity of such scales exists, {\emph{e.g.}}, for MHD and plasmas. Even in incompressible MHD, one deals with one length for each of two Elsasser energies, as well as a separate correlation scale for magnetic field and velocity field fluctuations. Accounting for compressibility, the fluctuations of density \citep{2020ApJS..246...58P} also become relevant,
and when remote sensing ({\emph{e.g.}}, radio) techniques are used to probe density fluctuations observed by {\emph{PSP}} citep{2020ApJS..246...57K}, 
the notion of {\it{effective turbulence scale}} is introduced as a nominal characteristic scale.  

The correlation lengths themselves are formally defined as integrals computed from correlation functions, and are therefore sensitive to energy range properties. But this definition is notoriously sensitive to low frequency power \citep[or length of data intervals; see][]{2015JGRA..120..868I}. For this reason, simplified methods for estimating correlation lengths are often adopted. \cite{2020ApJS..246...53C} examined two of these in {\emph{PSP}} data -- the so called {\bf{``$1/e$''}} method \citep[see also][]{2020ApJS..246...58P} and the break point method alluded to above. As expected based on analytical spectral models \citep[{\emph{e.g.}},][]{2007ApJ...667..956M}, correlation  scales based on the break point and on $1/e$ are quite similar.

A number of researchers have suggested that measured correlation scales near {\emph{PSP}} perihelia are somewhat shorter than what may be expected based on extrapolations of 1~AU measurements \citep{2020AGUFMSH0160013C}. It is possible that this is partly explained as a geometric effect, noting that the standard Parker magnetic field is close to radial in the inner heliosphere, while it subtends increasing angles relative to radial at larger distances. One approach to explaining these observations is based on a 1D turbulence transport code \citep{2020ApJS..246...38A} that distinguishes ``slab'' and ``2D'' correlations scales, a parameterization of correlation anisotropy \citep{1994ApJ...420..294B}. Solutions of these equations \citep{2020ApJS..246...38A} were able to account for shorter correlation lengths seen in selected {\emph{PSP}} intervals with nearly radial mean magnetic fields. This represents a partial explanation but leaves open the  question of what sets the value of the slab (parallel) correlation scales closer to the sun.

The parallel and perpendicular correlation scales, measured relative to the ambient magnetic field direction, have obvious fundamental importance in parameterizing interplanetary turbulence. These scales also enter into expressions for the decay rates of energy and related quantities in von Karman similarity theory and its extensions \citep{1938RSPSA.164..192D,2012JFM...697..296W}. These length scales, or a subset of them, then enter into essentially all macroscopic theories of turbulence decay in the solar wind \citep{2008JGRA..113.8105B,2010ApJ...708L.116V,2014ApJ...782...81V,2014ApJ...796..111L,2017ApJ...851..117A,2018ApJ...865...25U,2019JPlPh..85d9009C}. While  the subject is complex and too lengthy for full exposition here, a few words are in order. First, the perpendicular scale may likely be set by the supergranulation scale in the photosphere. A reasonably well accepted number is 35,000~km, although smaller values are sometimes favored. The perpendicular scale is often adopted as a controlling parameter in the cascade, in that the cascade is known to be anisotropic relative to the ambient field direction, and favoring perpendicular spectral transfer \citep{1983JPlPh..29..525S,1995ApJ...438..763G,1999PhRvL..82.3444M}. The parallel correlation scale appears to be less well constrained in general, and may be regulated (at least initially in the photosphere) by the correlation {\it{time}} of magnetic field footpoint motions \citep[see, {\emph{e.g.}},][]{2006ApJ...641L..61G}. Its value at the innermost boundary remains not well determined, even as 
{\emph{PSP}} observations provide ever better determinations at lower perihelia, where the field direction is often radial. 

One interesting possibility is that shear driven nonlinear Kelvin Helmholtz-like rollups \citep{2006GeoRL..3314101L,2018MNRAS.478.1980H} drive solar wind fluctuations towards a state of isotropy as reported prior to {\emph{PSP}} based on remote sensing observations \citep{2016ApJ...828...66D,1929ApJ....69...49R}. Shear induced rollups of this type would tap energy in large scale shear flows, enhancing the energy range fluctuations, and supplementing pre-existing driving of the inertial range \citep{2020ApJ...902...94R}. Such interactions are likely candidates for inducing a mixing, or averaging, between the parallel and perpendicular turbulence length scales in regions of Kelvin-Helmholtz-like rollups \citep{2020ApJ...902...94R}.

In general, multi-orbit {\emph{PSP}} observations \citep{2020ApJS..246...53C,2021ApJ...923...89C,2020ApJS..246...38A} provide better determination of not only length scales but other parameters that characterize the energy containing scales of turbulence. Knowledge of energy range parameters impacts practical issues such as the selection of appropriate times for describing local bulk properties such as mean density, a quantity that enters into computations of cross helicity and other measures of the Alfv\'enicity in observed fluctuations \citep{2020ApJS..246...58P}.  

Possibly the most impactful consequence of energy range parameters is their potential control over the cascade rate, and therefore control over the plasma dissipation and heating, whatever the details of those processes may be. One approach is to estimate the energy supply into the smaller scales from the energy containing range by examining the evolution of the break point between the $1/f$ range and the inertial range \citep{2020ApJ...904L...8W}. This approach involves assumptions about the relationship of the $1/f$ range to the inertial range. Using three orbits of {\emph{PSP}} data, these authors evaluated the estimated energy supply rate from the radial break point evolution with the estimated perpendicular proton heating rate, finding a reasonable level of heating in fast and slow wind. Another approach to estimating heating rates in {\emph{PSP}} observations \cite{2020ApJS..246...30M} makes use of approximate connections between heating rate and radial gradient of temperature \citep{2007JGRA..112.7101V,2013ApJ...774...96B} along with theoretical estimates from a form of stochastic heating \citep{2010ApJ...720..503C}. Again, reasonable levels of correspondence are found. Both of these approaches \citep{2020ApJ...904L...8W,2020ApJS..246...30M} derive interesting conclusions based in part on assumptions about transport theory of temperature, or transport of turbulent fluctuations. An alternative approach is based entirely on turbulence theory extended to the solar wind plasma and applied locally to {\emph{PSP}} data \citep{2020ApJS..246...48B}. In this case two evaluations are made -- an energy range estimate adapted from von Karman decay theory \citep{2012JFM...697..296W} and a third order Yaglom-like law \citep{1998GeoRL..25..273P} applied to the inertial range. Cascade rates about 100 times that observed at 1~AU are deduced. The consistency of the estimates of cascade rates obtained from {\emph{PSP}} observations employing these diverse methods suggests a robust interpretation of interplanetary turbulence and the role of the energy containing eddies in exerting control over the cascade.

\subsection{Inertial Range}\label{sec:turbulence_inertial_range}

   During the solar minimum, fast solar wind streams originate near the poles from open magnetic flux tubes within coronal holes, while slow wind streams originate in the streamer belt within a few tens of degrees from the solar equator \citep{2008GeoRL..3518103M}. Because plasma can easily escape along open flux tubes, fast wind is typically observed to be relatively less dense, more homogeneous and characteristically more Alfv\'enic than slow streams, which are believed to arise from a number of sources, such as the tip helmet streamers prevalent in ARs \citep{1999JGR...104..521E,2005ApJ...624.1049L}, opening of coronal loops via interchange reconnection with adjacent open flux tubes \citep{2001ApJ...560..425F}, or from rapidly expanding coronal holes \citep{2009LRSP....6....3C}. 
   
   The first two {\emph{PSP}} close Encs. with the Sun, which occurred during Nov. 2018 (Enc.~1) and Apr. 2019 (Enc.~2), where not only much closer than any S/C before, but also remained at approximately the same longitude w.r.t. the rotating Sun, allowing {\emph{PSP}} to continuously sample a number of solar wind streams rooted in a small equatorial coronal hole \citep{2019Natur.576..237B,2019Natur.576..228K}. Measurements of velocity and magnetic field during these two Encs. reveal a highly structured and dynamic slow solar wind consisting of a quiet and highly Alfv\'enic background with frequent impulsive magnetic field polarity reversals, which are associated with so called switchbacks \citep[see also][]{2020ApJS..246...39D,2020ApJS..246...45H,2020ApJS..246...67M}. The 30~min averaged trace magnetic spectra for both quiet and switchbacks regions exhibit a dual power-law, with an inertial-range Kolmogorov spectral index of $-5/3$ at high heliocentric distances (as observed in previous observations near 1~AU) and Iroshnikov-Kraichnan index of $-3/2$ near 0.17~AU, consistent with theoretical predictions from MHD turbulence \citep{2016JPlPh..82f5302C}. These findings indicate that Alfv\'enic turbulence is already developed at $0.17$~AU. Moreover, the radial evolution of the turbulent dissipation rate between $0.17$~AU and $0.25$~AU, estimated using Politano-Pouquet third order law and the von Karman decay law, increases from $5\times 10^4~{\rm J~kg}^{-1}{\rm s}^{-1}$ at $0.25$~AU to $2\times 10^5~{\rm J~kg}^{-1}{\rm s}^{-1}$ at $0.17$~AU, which is up to 100 times larger at Perihelion 1 than its measured value at $1$~AU \citep{2020ApJS..246...48B} and in agreement with some MHD turbulent transport models \citep{2018ApJ...865...25U}. \citet{2021ApJ...916...49S} estimated the energy cascade rate at each scale in the inertial range, based on exact scaling laws derived for isentropic turbulent flows in three particular MHD closures corresponding to incompressible, isothermal and polytropic equations of state, and found the energy cascade rates to be nearly constant constant in the inertial range at approximately the same value of $2\times10^5~{\rm{J~kg}}^{-1}~{\rm{s}}^{-1}$  obtained by \citep{2018ApJ...865...25U} at $0.17$~AU, independent of the closure assumption.
    
    \citet{2020ApJS..246...71C} performed an analysis to decompose {\emph{PSP}} measurements from the first two orbits into MHD modes. The analysis used solar wind intervals between switchbacks to reveal the presence of a broad spectrum of shear Alfv\'en modes, an important fraction of slow modes and a small fraction of fast modes. The analysis of the Poynting flux reveals that while most of the energy is propagating outward from the sun, inversions in the Poynting flux are observed and are consistent with outward-propagating waves along kinked magnetic field lines. These inversions of the energy flux also correlate with the large rotations of the magnetic field. An observed increase of the spectral energy density of inward-propagating waves with increasing frequency suggests back-scattering of outward-propagating waves off of magnetic field reversals and associated inhomogeneities in the background plasma. Wave-mode composition, propagation and polarization within 0.3~AU was also investigated by~\citet{2020ApJ...901L...3Z} through the Probability Distribution Functions (PDFs) of wave-vectors within 0.3~AU with two main findings: (1) wave-vectors cluster nearly parallel and antiparallel to the local background magnetic field for $kd_i<0.02$ and shift to nearly perpendicular for $kd_i>0.02$. The authors also find that outward-propagating AW dominate over all scales and heliocentric distances, the energy fraction of inward and outward slow mode component increases with heliocentric distance, while the corresponding fraction of fast mode decreases.

    \cite{2020ApJS..246...53C} investigated the radial dependency of the trace magnetic field spectra, between 0.17~AU to about 0.6~AU, using MAG data from the FIELDS suite \citep{2016SSRv..204...49B} for each 24~h interval during the first two {\emph{PSP}} orbits. Their analysis shows that the trace spectra of magnetic fluctuations at each radii consists of a dual power-law, only this time involving the $1/f$ range followed by an MHD inertial-range with an spectral index varying from approximately $-5/3$ near 0.6~AU to about $-3/2$ at perihelion (0.17~AU). Velocity measurements obtained from SWEAP/SPC \citep{2016SSRv..204..131K} were used for the 24~h interval around Perihelion 1 to obtain the trace spectra of velocity and Elsasser field fluctuations, all of which show a power law with a spectral index closer to $-3/2$. The normalized cross-helicity and residual energy, which was also measured for each 24~h interval, shows that the turbulence becomes more imbalanced closer to the Sun, {\emph{i.e.}}, the dominance of outward-propagating increases with decreasing heliocentric distance. Additional measures of Alfv\'enicity of velocity and magnetic fluctuations as a function of their scale-size \citep{2020ApJS..246...58P} showed that both normalized cross-helicity and the scale-dependent angle between velocity and magnetic field decreases with the scale-size in the inertial-range, as suggested by some MHD turbulence models \citep{2006PhRvL..96k5002B,2009PhRvL.102b5003P,2010ApJ...718.1151P}, followed by a sharp decline in the kinetic range, consistent with observations at 1~AU \citep{2018PhRvL.121z5101P}. The transition from a spectral index of $-5/3$ to $-3/2$ with a concurrent increase in cross helicity is consistent with previous observations at 1~AU in which a spectral index of $-3/2$ for the magnetic field is prevalent in imbalanced streams \citep{2010PhPl...17k2905P,2013ApJ...770..125C,2013PhRvL.110b5003W}, as well as consistent with models and simulations of steadily-driven, homogeneous imbalanced Alfv\'enic turbulence \citep{2009PhRvL.102b5003P,2010ApJ...718.1151P}, and reflection-driven (inhomogeneous) Alfv\'en turbulence \citep{2013ApJ...776..124P,2019JPlPh..85d9009C}. A similar transition was also found by \citet{2020ApJ...902...84A}, where the Hilbert-Huang Transform (HHT) was used to investigate scaling properties of magnetic-field fluctuations as a function of the heliocentric distance, to show that magnetic fluctuations exhibit multifractal scaling properties far from the sun, with a power spectrum $f^{-5/3}$, while closer to the sun the corresponding scaling becomes monofractal with $f^{-3/2}$ power spectrum. Assuming ballistic propagation, \citet{2021ApJ...912L..21T} identified two 1.5~h intervals corresponding to the same plasma parcel traveling from 0.1 to 1~AU during the first {\emph{PSP}} and {\emph{SolO}} radial alignment, and also showed that the solar wind evolved from a highly Alfv\'enic, less developed turbulent state near the sun to a more fully developed and intermittent  state near 1~AU.

    \citet{2021A&A...650A..21S} performed a statistical analysis to investigate how the turbulence properties at MHD scales depend on the type of solar wind stream and distance from the sun. Their results show that the spectrum of magnetic field fluctuations steepens with the distance to the Sun while the velocity spectrum remains unchanged. Faster solar wind is found to be more Alfv\'enic and dominated by outward-propagating waves (imbalanced) and with low residual energy. Energy imbalance (cross helicity) and residual energy decrease with heliocentric distance. Turbulent properties can also vary among different streams with similar speeds, possibly indicating a different origin. For example, slow wind emanating from a small coronal hole has much higher Alfv\'enicity than a slow wind that originates from the streamer belt. \citet{2021A&A...650L...3C} investigated the turbulence and acceleration properties of the streamer-belt solar wind, near the HCS, using measurements from {\emph{PSP}}'s fourth orbit. During this close S/C, the properties of the solar wind from the inbound measurements are substantially different than from the outbound measurements. In the latter, the solar wind was observed to have smaller turbulent amplitudes, higher magnetic compressibility, a steeper magnetic spectrum (closer to $-5/3$ than to $-3/2$), lower Alfv\'enicity and a $1/f$ break at much smaller frequencies. The transition from a more Alfv\'enic (inbound) wind to a non-Alfv\'enic wind occurred at an angular distance of about $4^\circ$ from the HCS. As opposed to the inbound Alfv\'enic wind, in which the measured energy fluxes are consistent with reflection-driven turbulence models~\citep{2013ApJ...776..124P,2019JPlPh..85d9009C}, the streamer belt fluxes are significantly lower than those predicted by the same models, suggesting the streamer-belt wind may be subject to additional acceleration mechanisms. 

    Interpretation of the spectral analysis of temporal signals to investigate scaling laws in the inertial range thus far have relied on the validity of Taylor's Hypothesis (TH). \citet{2021A&A...650A..22P} investigated the applicability of TH in the first four orbits based on a recent model of the space-time correlation function of Alfv\'enic turbulence \citep[incompressible MHD;][]{2018ApJ...858L..20B,2020PhRvR...2b3357P}. In this model, the temporal decorrelation of the turbulence is dominated by hydrodynamic sweeping under the assumption that the turbulence is strongly anisotropic and Alfv\'enic. The only parameter in the model that controls the validity of TH is $\epsilon=\delta u_0/V_\perp$ where $\delta u_0$ is the rms velocity of the large-scale velocity field and $V_\perp$ is the velocity of the S/C in the plasma frame perpendicular to the local field. The spatial energy spectrum of turbulent fluctuations is recovered using conditional statistics based on nearly perpendicular sampling. The analysis is performed on four selected 24h intervals from {\emph{PSP}} during the first four perihelia. TH is observed to still be marginally applicable, and both the new analysis and the standard TH assumption lead to similar results. A general prescription to obtain the energy spectrum when TH is violated is summarized and expected to be relevant when {\emph{PSP}} get closer to the sun. 

    \citet{2021ApJ...915L...8D} investigated the anisotropy of slow Alfv\'enic solar wind in the kinetic range from {\emph{PSP}} measurements. Magnetic compressibility is consistent with kinetic Alfv\'en waves (KAW) turbulence at sub-ion scales. A steepened transition range is found between the (MHD) inertial and kinetic ranges in all directions relative to the background magnetic field. Strong power spectrum anisotropy is observed in the kinetic range and a smaller anisotropy in the transition range. Scaling exponents for the power spectrum in the transition range are found to be $\alpha_{t\|}=-5.7\pm 1.0$ and $\alpha_{t\perp}=-3.7\pm0.3$, while for the kinetic range the same exponent are $\alpha_{k\|}=-3.12\pm0.22$ and $\alpha_{k\perp}=-2.57\pm0.09$. The wavevector anisotropy in the transition and kinetic ranges follow the scaling $k_\|\sim k_\perp^{0.71\pm0.17}$ and $k_\|\sim k_\perp^{0.38\pm0.09}$, respectively.

\subsection{Kinetic Range, Dissipation, Heating and Implications} 

In-situ measurements have revealed that the solar wind is not adiabatically cooling, as both the ion and electron temperatures decay at considerably slower rates than the adiabatic cooling rates induced by the radial expansion effect of the solar wind \citep[{\emph{e.g.}},][]{2020ApJS..246...70H,2020ApJS..246...62M}. Thus, some heating mechanisms must exist in the solar wind. As the solar wind is nearly collisionless, viscosity, resistivity, and thermal conduction are unlikely to contribute much to the heating of the solar wind.  Hence, turbulence is believed to be the fundamental process that heats the solar wind plasma. In the MHD inertial range, the turbulence energy cascades from large scales to small scales without dissipation. Near or below the ion kinetic scale, various kinetic processes, such as the wave-particle interaction through cyclotron resonance or Landau damping, and the stochastic heating of the particles, become important. These kinetic processes effectively dissipate the turbulence energy and heat the plasma. As already discussed in \S\ref{sec:5_large_scale}, \citet{2020ApJ...904L...8W} show that the estimated energy supply rate at the large scales agrees well with the estimated perpendicular heating rate of the solar wind, implying that turbulence is the major heating source of the solar wind. However, to fully understand how the turbulence energy eventually converts to the internal energy of the plasma, we must analyze the magnetic field and plasma data at and below the ion scales.

\begin{figure}
    \centering
    \includegraphics[width=\hsize]{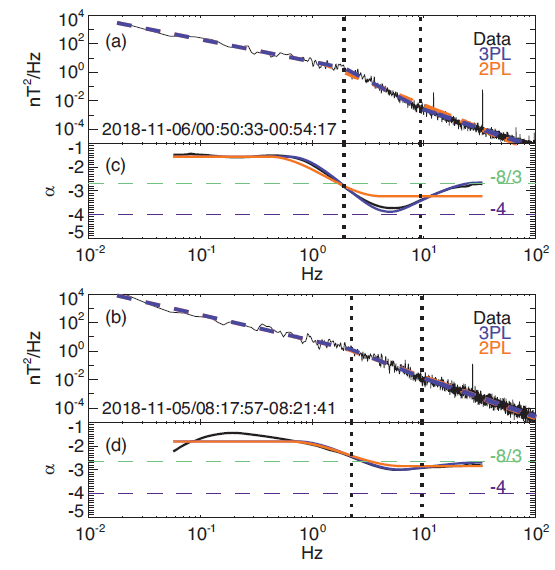}
    \caption{(a,b) Examples of {\emph{PSP}}/FIELDS magnetic field spectra
    with 3PL (three spectral range continuous power-law fit, blue) and 2PL (two spectral range continuous power-law fit, orange) fits. Vertical lines show 3PL spectral breaks. (c,d) spectral indices for data (black), 3PL (blue) and 2PL fits (orange). Horizontal lines are shown corresponding to spectral indices of $-8/3$ (teal) and $-4$ (purple). Top interval has statistically significant spectral steepening, while the bottom interval is sufficiently fit by 2PL. Figure adapted from \cite{2020PhRvL.125b5102B}.}.
    \label{fig:Bowen2020PRLFigure1}
\end{figure}

\citet{2020PhRvL.125b5102B} analyze data of the MAG and SCM onboard {\emph{PSP}} during Enc.~1 when a slow Alfv\'enic wind originating from an equatorial coronal hole was measured. They show that a very steep transition range in the magnetic field power spectrum with a spectral slope close to $-4$ is observed around the ion gyroscale ($k_\perp \rho_i \sim 1$ where $k_\perp$ is the perpendicular wave number and $\rho_i$ is the ion thermal gyroradius) (Fig.~ \ref{fig:Bowen2020PRLFigure1}). This transition range is steeper than both the inertial range ($k_\perp \rho_i \ll 1$) and the deep kinetic range ($k_\perp \rho_i \gg 1$). \citet{2020PhRvL.125b5102B} estimate that if the steep spectrum corresponds to some dissipation mechanism, then more than 50\% of the turbulence energy is dissipated at the ion scale transition range, which is a sufficient energy source for solar wind heating. \citet{2020ApJS..246...55D} conduct a statistical study of the magnetic field spectrum based on {\emph{PSP}} data from Enc.~2 and show that the break frequency between the inertial range and the transition range decreases with the radial distance to the Sun and is on the order of the ion cyclotron resonance frequency.

\citet{2020ApJ...897L...3H} find that, in a one-day interval during Enc.~1, the slow wind observed by {\emph{PSP}} contains both the right-handed polarized KAWs and the left-handed polarized Alfv\'en ion cyclotron waves (ACWs) at sub-ion scales. Many previous observations have shown that at 1~AU KAW dominates the turbulence at sub-ion scales \citep[{\emph{e.g.}},][]{2010PhRvL.105m1101S} and KAW can heat the ions through Landau damping of its parallel electric field. However, the results of \citep{2020ApJ...897L...3H} imply a possible role of ACWs, at least in the very young solar wind, in heating the ions through cyclotron resonance. \citet{2021ApJ...915L...8D}, by binning the Enc.~1 data with different $\mathbf{V-B}$ angles, show that the magnetic field spectrum is anisotropic in both the transition range and kinetic range with $k_\perp \gg k_\parallel$ and the anisotropy scaling relation between $k_\perp$ and $k_\parallel$ is consistent with the critical-balanced KAW model in the sub-ion scales (see \S\ref{sec:turbulence_inertial_range}). 

Another heating mechanism is the stochastic heating \citep{2010ApJ...720..503C}, which becomes important when the fluctuation of the magnetic field at the ion gyroscale is large enough such that the magnetic moment of the particles is no longer conserved. \citet{2020ApJS..246...30M} calculate the stochastic heating rate using data from the first two Encs. and show that the stochastic heating rate $Q_\perp$ decays with the radial distance as $Q_\perp \propto r^{-2.5}$. Their result reveals that the stochastic heating may be more important in the solar wind closer to the Sun. 

Last, it is known that development of turbulence leads to the formation of intermittency (see \S\ref{sec:5_intermit} for more detailed discussions). In plasma turbulence, intermittent structures such as current sheets are generated around the ion kinetic scale. \citet{2020ApJS..246...46Q} adopt the partial variance of increments (PVI) technique to identify intermittent magnetic structures using {\emph{PSP}} data from the first S/C. They show that statistically there is a positive correlation between the ion temperature and the PVI,  
indicating that the intermittent structures may contribute to the heating of the young solar wind through processes like the magnetic reconnection in the intermittent current sheets.

At the end of this subsection, it is worthwhile to make several comments. First, the Faraday Cup onboard {\emph{PSP}} (SPC) has a flux-angle operation mode, which allows measurements of the ion phase space density fluctuations at a cadence of 293~Hz \citep{2020ApJS..246...52V}. Thus, combination of the flux-angle measurements with the magnetic field data will greatly help us understand the kinetic turbulence. Second, more studies are necessary to understand behaviors of electrons in the turbulence, though direct measurement of the electron-scale fluctuations in the solar wind is difficult with current plasma instruments. \citet{2021A&A...650A..16A} show that by distributing the turbulence heating properly among ions and electrons in a turbulence-coupled solar wind model, differential radial evolutions of ion and electron temperatures can be reproduced. However, how and why the turbulence energy is distributed unevenly among ions and electrons are not fully understood yet and need future studies. Third, during the Venus flybys, {\emph{PSP}} traveled through Venus’s magnetosphere and made high-cadence measurements. Thus, analysis of the turbulence properties around Venus, {\emph{e.g.}}, inside its magnetosheath \citep{2021GeoRL..4890783B} will also be helpful in understanding the kinetic turbulence (see \S\ref{PSPVENUS}). Last, \citet{2021ApJ...912...28M} compare the turbulence properties inside and outside the magnetic switchbacks using the {\emph{PSP}} data from the first two Encs. They find that the stochastic heating rates and spectral slopes are similar but other properties such as the intermittency level are different inside and outside the switchbacks, indicating that some processes near the edges of switchbacks, such as the velocity shear, considerably affect the turbulence evolution inside the switchbacks (see \S\ref{SB_obs}).

\subsection{Intermittency and Small-scale Structure} \label{sec:5_intermit}

In the modern era to turbulence research, {\textit{intermittency}} has been established as a fundamental feature of turbulent flows \citep[{\emph{e.g.}},][]{1997AnRFM..29..435S}. Nonlinearly interacting fluctuations are expected to give rise to structure in space and time, which is characterized by sharp gradients, inhomogeneities, and departures from Gaussian statistics. In a plasma such as the solar wind, such ``bursty'' or ``patchy'' structures include current sheets, vortices, and flux tubes. These structures have been associated with enhanced plasma dissipation and heating \citep{2015RSPTA.37340154M}, and with acceleration of energetic particles \citep{2013ApJ...776L...8T}. Studies of intermittency may therefore provide insights into dissipation and heating mechanisms active in the weakly-collisional solar wind plasma. Intermittency also has implications for turbulence theory -- a familiar example is the evolution of hydrodynamic inertial range theory from the classical Kolmogorov \citeyearpar[K41;][]{1941DoSSR..30..301K} theory to the so-called refined similarity hypothesis \citep[][]{1962JFM....13...82K}; the former assumed a uniform energy dissipation rate while the latter allowed for fluctuations and inhomogeneities in this fundamental quantity.

Standard diagnostics of intermittency in a turbulent field include PDFs of increments, kurtosis (or flatness; fourth-order moment), and structure functions   \cite[{\emph{e.g.}},][]{2015RSPTA.37340154M}. Observational studies tend to focus on the magnetic field due to the availability of higher-quality measurements compared to plasma observations. In well-developed turbulence, one finds that the tails of PDFs of increments become wider (super-Gaussian) and large values of kurtosis are obtained at small scales within the inertial-range. A description in terms of \textit{fractality} is also employed -- monofractality is associated with structure that is non space-filling but lacking a preferred scale ({\emph{i.e.}}, scale-invariance). In contrast, multifractality also implies non space-filling structure but with at least one preferred scale \citep{1995tlan.book.....F}.\footnote{In hydrodynamic turbulence intermittency is often described in terms of the scaling of the structure functions $S^{(p)}_\ell \equiv \langle \delta u_\ell^p \rangle \propto \ell^{p/3 + \mu(p)}$, where \(\delta u_\ell = \bm{\hat{\ell}} \cdot [ \bm{u} (\bm{x} + \bm{\ell}) - \bm{u}(\bm{x}) ]\) is the longitudinal velocity increment, \(\bm{\ell}\) is a spatial lag, and \(\langle\dots\rangle\) is an appropriate averaging operator. In K41 (homogenous turbulence) the intermittency parameter \(\mu(p)\) vanishes. With intermittency one has non-zero \(\mu(p)\), and the scaling exponents \(\zeta (p) = p/3 + \mu(p) \) can be linear or nonlinear functions of \(p\), corresponding to monofractal or multifractal scaling, respectively \citep{1995tlan.book.....F}. The scale-dependent kurtosis \(\kappa\) can be defined in terms the structure functions: \(\kappa (\ell)\equiv S^{(4)}_\ell/ \left[S^{(2)}_\ell\right]^2\).}
 
Intermittency properties of solar wind turbulence have been extensively studied using observations from several S/C \citep[see, {\emph{e.g.}}, review by][]{2019E&SS....6..656B}. High-resolution measurements from the {\emph{Cluster}} and the Magnetospheric Multiscale \citep[{\emph{MMS;}}][]{2014cosp...40E.433B} missions have enabled such investigations within the terrestrial magnetosheath as well \citep[{\emph{e.g.}},][]{2009PhRvL.103g5006K,2018JGRA..123.9941C}, including kinetic-scale studies. {\emph{PSP}}'s trajectory allows us to extend these studies to the near-Sun environment and to examine the helioradial evolution of intermittency in the inner heliosphere. An advantage offered by {\emph{PSP}} is that the observations are not affected by foreshock wave activity that is often found near Earth's magnetosheath \citep[see, {\emph{e.g.}},][]{2012ApJ...744..171W}.

\begin{figure}
    \centering
    \includegraphics[width=0.7\textwidth]{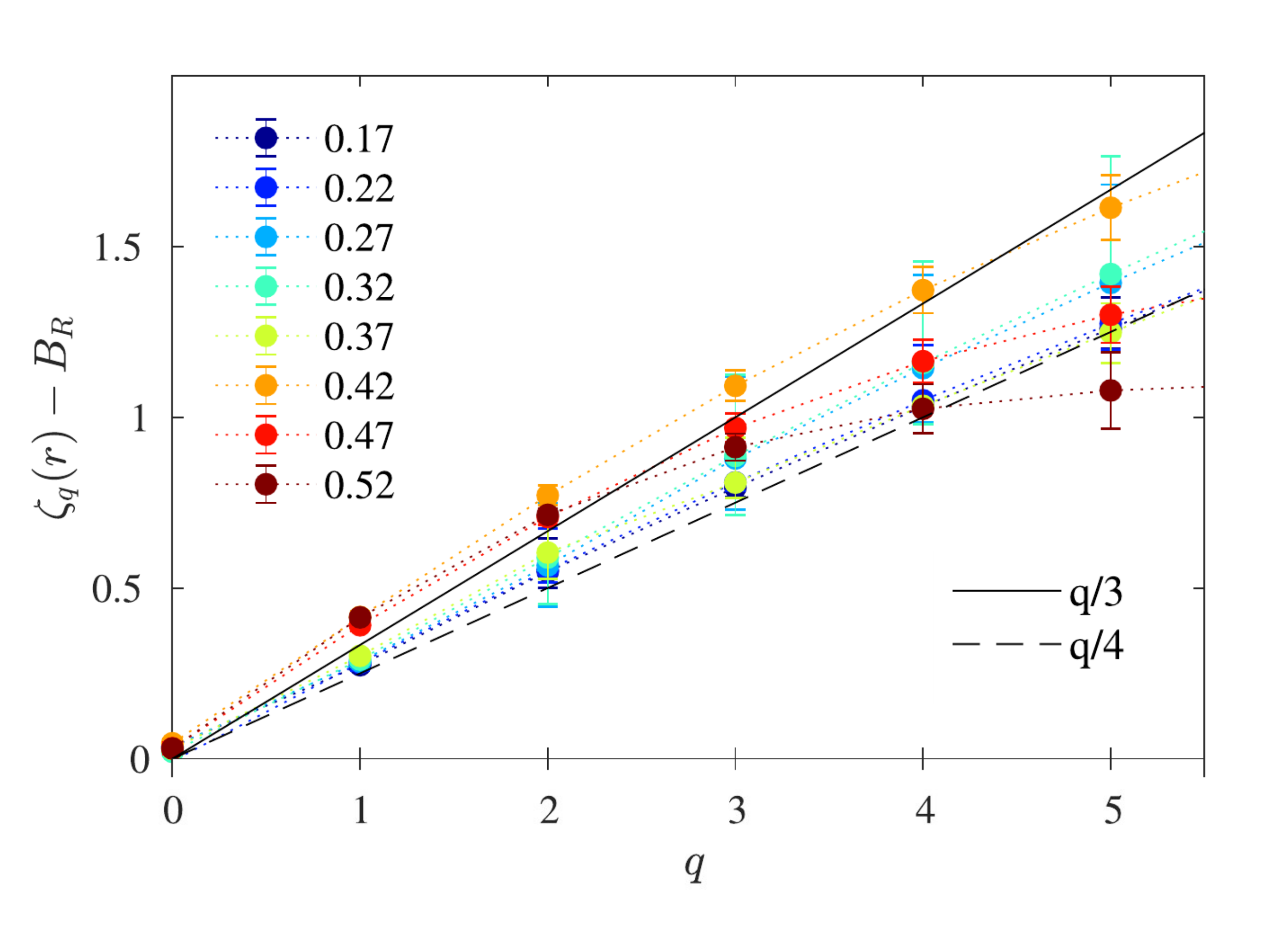}
    \includegraphics[width=0.68\textwidth]{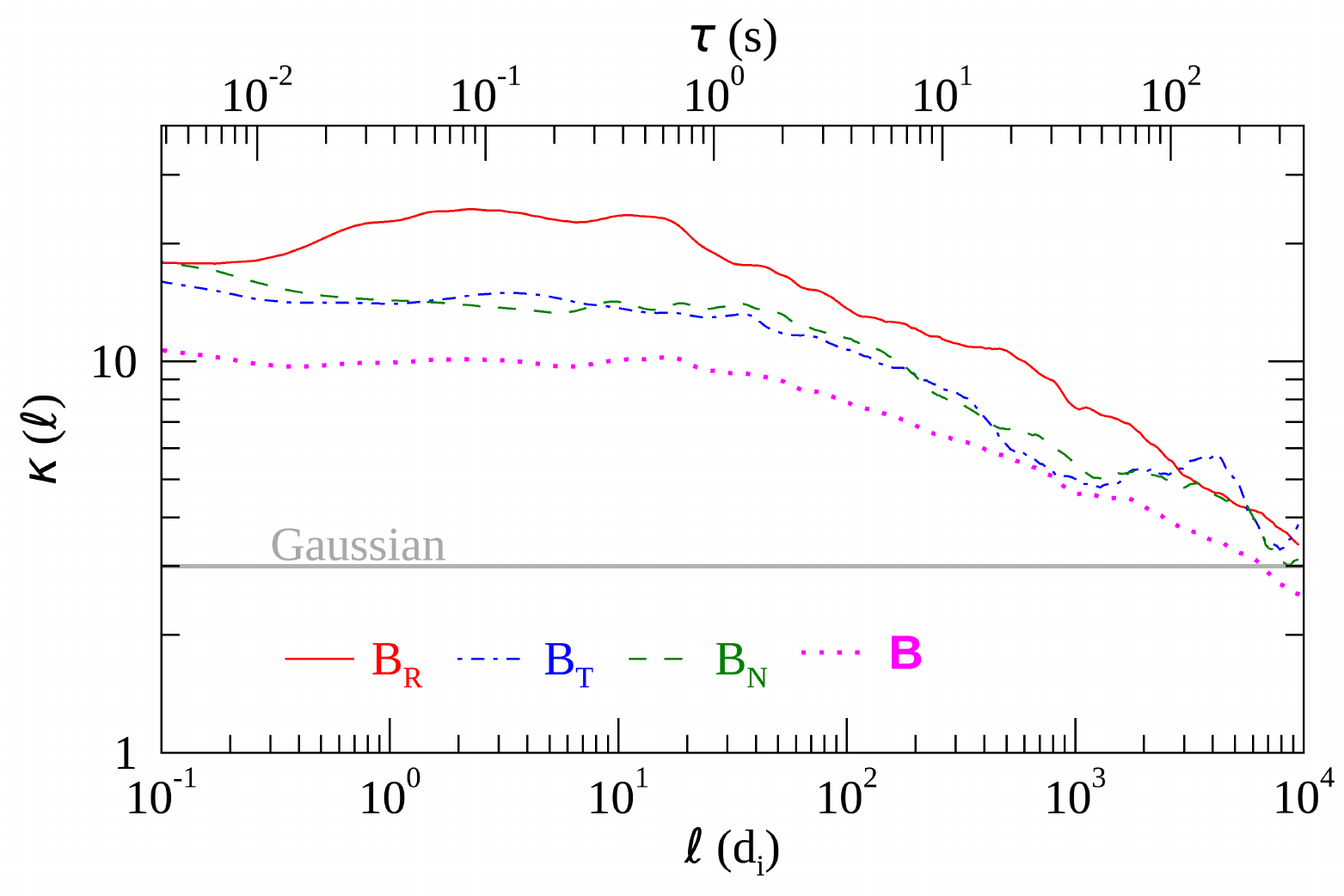}
    \caption{\textit{Top}: Scaling exponents \(\zeta_q\) (see text) of the radial magnetic field for different orders \(q\), observed by {\emph{PSP}} at different helioradii \(r\). A transition from monofractal linear scaling to multifractal scaling is observed for \(r>0.4\). Figure adapted from \cite{2020ApJ...902...84A}. \textit{Bottom}: Scale-dependent kurtosis of the magnetic field, as a function of lag. A transition from a multifractal inertial range to a monofractal kinetic range occurs near \(20\ d_\text{i}\) \citep{2021ApJ...911L...7C}.}
    \label{fig:Alberti_Chhib}
\end{figure}

The radial evolution of intermittency in inertial-range magnetic fluctuations measured by {\emph{PSP}} was investigated by \cite{2020ApJ...902...84A}, who found monofractal, self-similar scaling at distances below 0.4~AU. At larger distances, a transition to multifractal scaling characteristic of strongly intermittent turbulence was observed (see top panel of Fig.~\ref{fig:Alberti_Chhib}). A similar trend was observed by \citep{2021ApJ...912L..21T}, who used measurements during the first radial alignment of {\emph{PSP}} and {\emph{SolO}}. Strong intermittency was obtained in {\emph{SolO}} observations near 1~AU compared to {\emph{PSP}} observations near 0.1~AU, suggesting an evolution from highly Alfv\'enic and under-developed turbulence to fully-developed turbulence with increasing radial distance. Note that several prior studies have found that inertial-range magnetic turbulence at 1~AU is characterized by multifractal intermittency \citep[][]{2019E&SS....6..656B}. It is also worth noting (as in \S\ref{sec:5_large_scale}) that {\emph{PSP}} observations near the Sun may be statistically biased towards sampling variations in a lag direction that is (quasi-)parallel to the mean magnetic field \citep{2021PhPl...28h0501Z,2021ApJ...923...89C}. Future studies could separately examine parallel and perpendicular intervals \citep[{\emph{e.g.}},][]{2011JGRA..11610102R}, which would clarify the extent to which this geometrical sampling bias affects the measured radial evolution of intermittency.
 
A comparison of inertial range vs. kinetic-scale intermittency in near-Sun turbulence was carried out by \cite{2021ApJ...911L...7C} using the publicly available SCaM data product \citep{2020JGRA..12527813B}, which merges MAG and SCM measurements to obtain an optimal signal-to-noise ratio across a wide range of frequencies. They observed multifractal scaling in the inertial range, supported by a steadily increasing kurtosis with decreasing scale down to \(\sim 20\ d_\text{i}\). The level of intermittency was somewhat weaker in intervals without switchbacks compared to intervals with switchbacks, consistent with PVI-based analyses by \cite{2021ApJ...912...28M} (see also \S\ref{sec:5_switchback}). At scales below \(20\ d_\text{i}\), \cite{2021ApJ...911L...7C} found that the kurtosis flattened (bottom panel of Fig.~\ref{fig:Alberti_Chhib}), and their analysis suggested a monofractal and scale-invariant (but still intermittent and non-Gaussian) kinetic range down to the electron inertial scale, a finding consistent with near-Earth observations \citep{2009PhRvL.103g5006K} and some kinetic simulations \citep{2016PhPl...23d2307W}. From these results, they tentatively infer the existence of a scale-invariant distribution of current sheets between proton and electron scales. The SCaM data product was also used by \cite{2021GeoRL..4890783B} to observe strong intermittency at subproton scales in the Venusian magnetosheath. \cite{2020ApJ...905..142P} examined coherent structures at ion scales in intervals with varying switchback activity, observed during {\emph{PSP}}'s first S/C. Using a wavelet-based approach, they found that current sheets dominated intervals with prominent switchbacks, while wave packets were most common in quiet intervals without significant fluctuations. A mixture of vortex structures and wave packets was observed in periods characterized by strong fluctuations without magnetic reversals.

A series of studies used the PVI approach \citep{2018SSRv..214....1G} with {\emph{PSP}} data to identify intermittent structures and examine associated effects. \cite{2020ApJS..246...31C} found that the waiting-time distributions of intermittent magnetic and flow structures followed a power law\footnote{Waiting times between magnetic switchbacks, which may also be considered intermittent structures, followed power-law distributions as well \citep{2020ApJS..246...39D}.} for events separated by less than a correlation scale, suggesting a high degree of correlation that may originate in a clustering process. Waiting times longer than a correlation time exhibited an exponential distribution characteristic of a Poisson process. \cite{2020ApJS..246...61B} studied the association of SEP events with intermittent magnetic structures, finding a suggestion of positive correlation (see also \S\ref{sec:5_SEPs}). The association of intermittency with enhanced plasma heating (measured via ion temperature) was studied by \cite{2020ApJS..246...46Q}; their results support the notion that coherent non-homogeneous magnetic structures play a role in plasma heating mechanisms. These series of studies indicate that intermittent structures in the young solar wind observed by {\emph{PSP}} have certain properties that are similar to those observed in near-Earth turbulence.

In addition to the statistical properties described in the previous paragraphs, some attention has also been given to the identification of structures associated with intermittency, such as magnetic flux tubes and ropes. \cite{2020ApJS..246...26Z} used wavelet analysis to evaluate magnetic helicity, cross helicity, and residual energy in {\emph{PSP}} observations between 22 Oct. 2018 and 21 Nov. 2018. Based on these parameters they identified 40 flux ropes with durations ranging from 10 to 300 minutes. \cite{2020ApJ...903...76C} used a Grad-Shafranov approach to identify flux ropes during the first two {\emph{PSP}} Encs., and compared this method with the \cite{2020ApJS..246...26Z} approach, finding some consistency. \cite{2021A&A...650A..20P} developed a novel method for flux rope detection based on a real-space evaluation of magnetic helicity, and, focusing on the first {\emph{PSP}} orbit, found some consistency with the previously mentioned approaches. A subsequent work applied this method to orbit 5, presenting evidence that flux tubes act as transport boundaries for energetic particles \citep{2021MNRAS.508.2114P}. The characteristics of so-called interplanetary discontinuities (IDs) observed during {\emph{PSP}}'s $4^{\mathrm{th}}$ and $5^{\mathrm{th}}$ orbits were studied by \cite{2021ApJ...916...65L}, who found that the occurrence rate of IDs decreases from 200 events per day at 0.13~AU to 1 event per day at 0.9~AU, with a sharper decrease observed in RDs as compared with TDs. While the general decrease in occurrence rate may be attributed to radial wind expansion and discontinuity thickening, the authors infer that the sharper decrease in RDs implies a decay channel for these in the young solar wind.

We close this section by noting that the studies reviewed above employ a remarkable variety of intermittency diagnostics, including occurrence rate of structures, intensities of the associated gradients, and their fractal properties. The richness of the dataset that {\emph{PSP}} is expected to accumulate over its lifetime will offer unprecedented opportunities to probe the relationships between these various diagnostics and their evolution in the inner heliosphere.

\subsection{Interaction Between Turbulence and Other Processes}

\subsubsection{Turbulence Characteristics Within Switchbacks}\label{sec:5_switchback}
The precise definition of magnetic field reversal (switchbacks) is still ambiguous in the heliophysics community, but the common picture of switchbacks is that they incarnate the S-shaped folding of the magnetic field. Switchbacks are found to be followed by Alv\'enic fluctuations (and often by strahl electrons) and they are associated with an increase of the solar wind bulk velocity as observed recently by {\emph{PSP}} near 0.16~AU \citep{2019Natur.576..228K,2019Natur.576..237B,2020ApJS..246...39D,2020ApJS..246...67M,2020ApJ...904L..30B,2021ApJ...911...73W,2020ApJS..246...74W}. Switchbacks have been previously observed in the fast solar-wind streams near 0.3~AU \citep{ 2018MNRAS.478.1980H} and near or beyond 1~AU \citep{1999GeoRL..26..631B}. The switchback intervals are found to carry turbulence, and the characteristics of that turbulence have been investigated in a number of works using {\emph{PSP}} data. \citet{2020ApJS..246...39D} studied the the spectral properties of inertial range turbulence within intervals containing switchback structures in the first {\emph{PSP}} S/C. In their analysis they introduced the normalized parameter $z=\frac{1}{2}(1-\cos{\alpha})$ (where $\alpha$ is the angle between the instantaneous magnetic field, {\bf B}, and the prevalent field $\langle {\bf B} \rangle$) to determine the deflection of the field. Switchbacks were defined as a magnetic field deflection that exceeds the threshold value of $z=0.05$. They estimated the power spectral density of the radial component $B_R$ of the magnetic field for quiescent ($z < 0.05$) and active (all $z$) regimes. 

\begin{figure}
    \centering
    \includegraphics[width=0.7\textwidth]{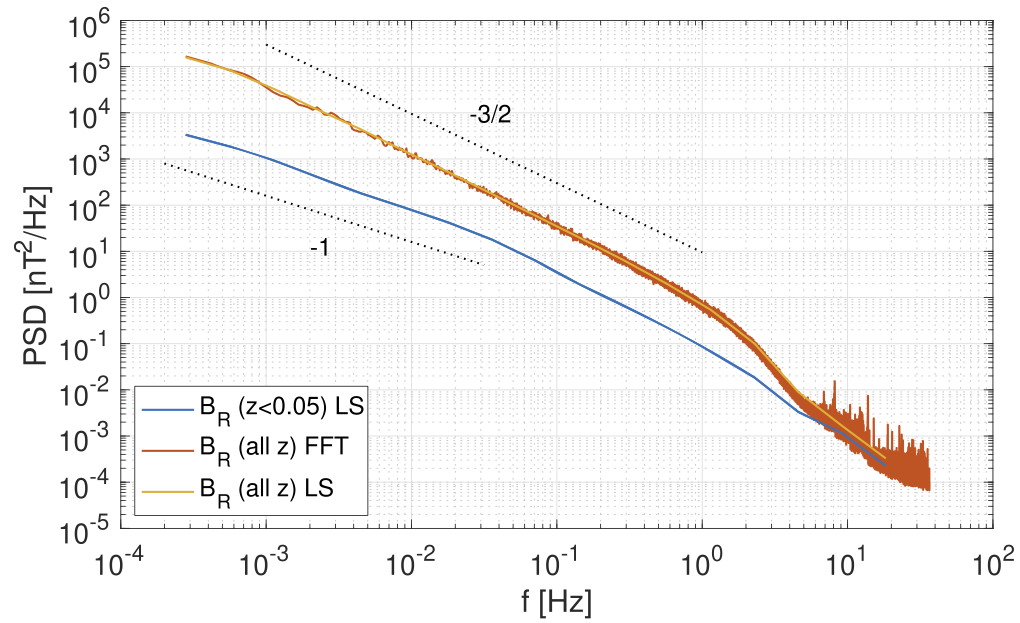}
    \caption{Power spectrum of radial magnetic field fluctuations for quiescent times ($z<0.05$) and for the entire interval (all $z$), during a period near perihelion of Enc.~1. The quiescent times show a lower overall amplitude, and a $1/f$ break at higher frequencies, suggestive of a less evolved turbulence. Figure adapted from \citet{2020ApJS..246...39D}.}
    \label{fig:Dudock20}
\end{figure}

Fig.~\ref{fig:Dudock20} shows the results of both power spectra. They found that the properties in active conditions are typical for MHD turbulence, with an inertial range whose spectral index is close to $-3/2$ and preceded by $1/f$ range (consistent with the observation by \citet{2020ApJS..246...53C}. Also, the break frequency (at the lower frequency part) was found to be located around 0.001~Hz. For the quiescent power spectrum, the break frequency moves up to 0.05~Hz showing a difference from the active power spectrum although both power spectra have similar spectral index (about $-3/2$) between 0.05~Hz and 1~Hz. The authors suggest that in the quiescent regime the turbulent cascade has only had time to develop a short inertial range, showing signature of a more pristine solar wind. 

 \citet{2020ApJS..246...67M} have studied the turbulent quantities such as the normalized residual energy, $\sigma_r(s, t)$, and cross helicity, $\sigma_c(s, t)$, during one day of {\emph{PSP}} first S/C as a function of wavelet scale $s$. Their study encompasses switchback field intervals. Overall, they found that the observed features of these turbulent quantities are similar to previous observations made by {\emph{Helios}} in slow solar wind \citep{2007PSS...55.2233B}, namely highly-correlated and Alfv\'enic fluctuations with ($\sigma_c\sim 0.9$ and $\sigma_r\sim -0.3$). However, a negative normalized cross helicity is found within switchback intervals, indicating that MHD fluctuations are following the local magnetic field inside switchbacks, {\emph{i.e.}}, the predominantly outward propagating Alfv\'enic fluctuations outside the switchback intervals become inward propagating during the field reversal. This signature is interpreted as that the field reversals are local kinks in the magnetic field and not due to small regions of opposite polarity of the field.
 
 \begin{figure}
    \centering
    \includegraphics[width=0.7\textwidth]{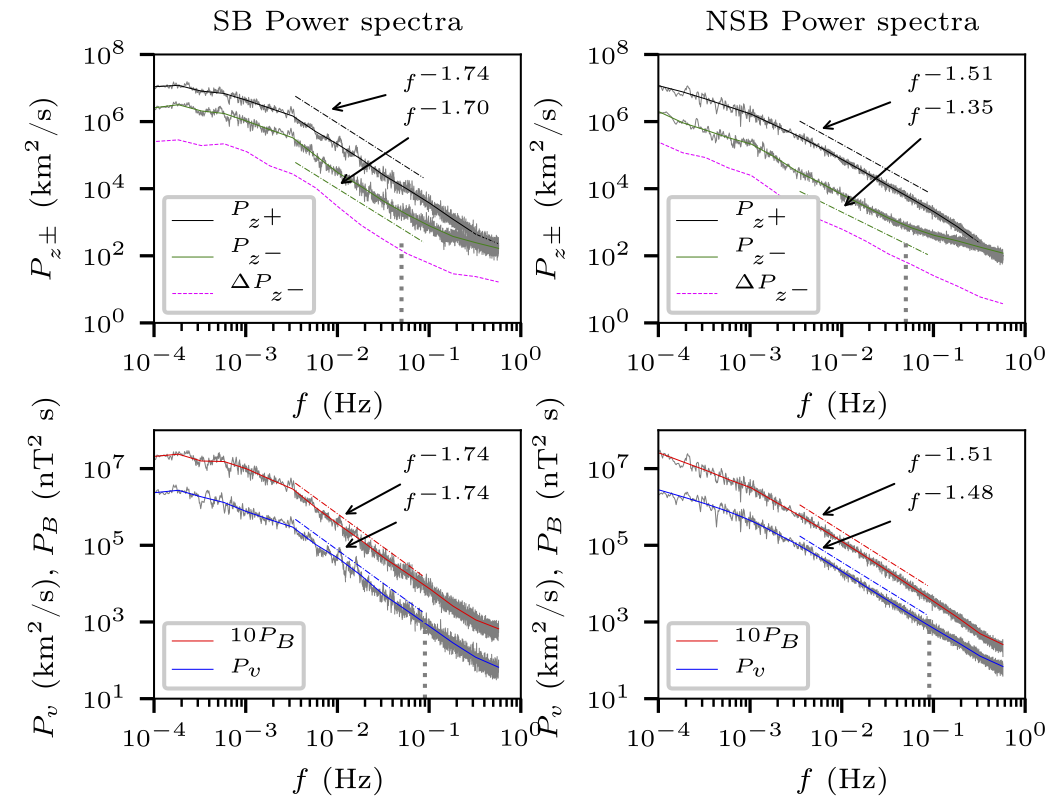}
    \caption{Power spectra of Elsasser variables (upper panels) and velocity/magnetic fluctuations (lower panels) for periods of switchbacks (SB; left panels) and periods not within switchbacks (NSB; right panels). Magnetic spectra are multiplied by a factor of 10. Power law fits are also indicated. There are notable differences in both the amplitudes and shape of the spectra between SB and NSB intervals. Figure adapted from \citet{2020ApJ...904L..30B}.}
    \label{fig:bourouaine20}
\end{figure}

 The turbulence characteristics associated with switchbacks have also been studied by \cite{2020ApJ...904L..30B} using 10 days of {\emph{PSP}} data during the first S/C. The authors used a technique that is based on the conditioned correlation functions to investigate the correlation times and the power spectra of the field $q$ that represents the magnetic field ${\bf B}$, the fluid velocity ${\bf V}$ and the Elsasser fields ${\bf z^\pm}$, inside and outside switchback intervals. In their study, the authors defined switchbacks as field reversals that are deflected by angles that are larger than $90^\circ$ w.r.t. the Parker spiral (the prevalent magnetic field). This work confirms that the dominant Alv\'enic fluctuations follow the field reversal. Moreover, the authors found that, in switchback intervals, the correlation time is about 2 minutes for all fields, but in non-switchback intervals, the correlation time of the sunward propagating Alfv\'enic fluctuations (the minor Elsasser field) is about 3 hr and longer than the ones of the other fields. This result seems to be consistent with previous 1~AU measurements \citep{2013ApJ...770..125C,2018ApJ...865...45B}. Furthermore, the authors estimated the power spectra of the corresponding fields (Fig.~\ref{fig:bourouaine20}), and found that the magnetic power spectrum in switchback intervals is steeper (with spectral index close to $-5/3$) than in switchback intervals, which have a spectral index close to $-3/2$. The analysis also shows that the turbulence is found to be less imbalanced with higher residual energy in switchback intervals. 
 
\citet{2021ApJ...911...73W} has investigated the turbulent quantities such as the normalized cross helicity and the residual energy in switchbacks using the first four Encs. of {\emph{PSP}} data. In their analysis they considered separate intervals of 100~s duration that satisfies the conditions of $B_R>0^\circ$ and $B_R>160^\circ$ for the switchbacks and non-switchbacks, respectively. Although, the analysis focuses on the time scale of 100~s, their findings seems to be consistent with the results of \cite{2020ApJ...904L..30B} (for that time scale), {\emph{i.e.}}, the switchback intervals and non-switchback intervals have distinct residual energy and similar normalized cross helicity suggesting that switchbacks have a different Alfv\'enic characteristics.

In another study, \citet{2021ApJ...912...28M} have investigated the spectral index and the stochastic heating rate at the inertial range inside and outside switchback intervals and found a fair similar behavior in both intervals. However, at the kinetic range, the kinetic properties, such as the characteristic break scale (frequency that separates the inertial and the dissipation ranges) and the level of intermittency differ inside and outside switchback intervals. The authors found that inside the switchbacks the level of intermittency is higher, which might be a signature of magnetic field and velocity shears observed at the edges.

\subsubsection{Impact of Turbulence and Switchbacks  on Energetic Particles and CMEs}\label{sec:5_SEPs}

Strahl electrons are observed to follow the reversed field within the switchbacks, however, it is not yet understood whether higher energy energetic particles reverse at the switchbacks as well. Recently, \cite{2021AA...650L...4B} examined the radial anisotropy of the energetic particles measured by the EPI-Lo (instrument of the IS$\odot$IS suite) in connection to magnetic switchbacks.  The authors investigated switchback intervals with $|\sigma_c|>0.5$ and $z\le 0.5$, respectively. The ratio $r= (F_{\mbox{away}}-F_{\mbox{toward}})/(F_{\mbox{away}}+F_{\mbox{outward}})$ has been used to determine the dominant flux direction of the energetic particles. Here "$\mbox{away}$" and "$\mbox{toward}$" refer to the direction of the measured radial particle fluxes ($F$) in the selected energy range. Fig.~3 of \citet{2021AA...650L...4B} displays the scattering points that correspond to the measurements of the first five Encs. plotted as a function of the $z$ parameter and the ratio $r$. The analysis shows that 80–200~keV energetic ions almost never reverse direction when the magnetic field polarity reverses in switchbacks. One of the reason is that particles with smaller gyroradii, such as strahl electrons, can reverse direction by following the magnetic field in switchbacks, but that larger gyroradii particles likely cannot. Therefore, from this analysis one can expect that particles with higher energies than those detectable by EPI-Lo will likely not get reversed in switchbacks.

\cite{2020ApJS..246...48B} studied the connection between the enhanced of the population of energetic particles (measured using IS$\odot$IS) and the intermittent structures (using FIELDS/MAG) near the sun using {\emph{PSP}} data. Intermittent structures are generated naturally by turbulence, and the PVI method was proposed previously to identify these structures 
\citep{2018SSRv..214....1G} 
For single S/C measurements, this method relies on the evaluation of the temporal increment in the magnetic field such as $|\Delta B(\tau,t)|=|B(t+\tau)-B(t)|$, and thus the so-called the $PVI(t)$ for a given $\tau$ index is defined as
 \begin{equation}
     PVI(t)=\sqrt{\frac{|\Delta(\tau,t)|^2}{|\langle \Delta(\tau,t)|^2\rangle}}
 \end{equation}
 where $\langle...\rangle$ denotes a time average computed along the time series. The analysis given in  \citet{2020ApJS..246...48B}  examined the conditionally averaged energetic-particle count rates and its connection to the intermittent structures using the PVI method. The results from the first two {\emph{PSP}} orbits seem to support the idea that SEPs are are likely correlated with coherent magnetic structures. The outcomes from this analysis may suggest that energetic particles are concentrated near magnetic flux tube boundaries.

Magnetic field line topology and flux tube structure may influence energetic particles and their transport in other ways. A consequence of the tendency of particles to follow field lines is that when turbulence is present the particle paths can be influenced by field line {\it random walk}. While this idea is familiar in the context of perpendicular diffusive transport, a recent study of SEP path lengths observed by {\emph{PSP}} \citep{2021A&A...650A..26C} suggested that random walking of field lines or flux tubes may account for apparently increased path of SEP path lengths. This is further discussed in \S\ref{SEPs}.

The inertial-range turbulent properties such as the normalized cross helicity, $\sigma_c$, and residual energy, $\sigma_r$  have been examined in magnetic clouds (MCs) using {\emph{PSP}} data by \citet{2020ApJ...900L..32G}. MCs are considered to be large-scale transient twisted magnetic structures that propagate in the solar wind having features of low plasma $\beta$ and low-amplitude magnetic field fluctuations. The analysis presented in \citet{2020ApJ...900L..32G} shows low $|\sigma_c|$ value in the cloud core while the cloud’s outer layers displays higher $|\sigma_c|$ and small residual energy. This study indicates that more balanced turbulence resides in the could core, and large-amplidude Alfv\'enic fluctuations characterize the cloud’s outer layers. These obtained properties suggest that low $|\sigma_c|$ is likely a common feature of magnetic clouds that have a have typical closed field structures.

\subsection{Implications for Large-Scale and Global Dynamics}

As well as providing information about the fundamental nature of turbulence, and its interaction with the various structures in the solar wind, {\emph{PSP}} has allowed us to study how turbulence contributes to the solar wind at the largest scales. Some of the main goals of {\emph{PSP}} are to understand how the solar wind is accelerated to the high speeds observed and how it is heated to the high temperatures seen, both close to the Sun and further out \citep{2016SSRv..204....7F}. {\emph{PSP}}'s orbits getting increasingly closer to the Sun are allowing us to measure the radial trends of turbulence properties, and directly test models of solar wind heating and acceleration.

To test the basic physics of a turbulence driven wind, \citet{2020ApJS..246...53C} compared {\emph{PSP}} measurements from the first two orbits to the 1D model of \citet{2011ApJ...743..197C}. In particular, they calculated the ratio of energy flux in the Alfv\'enic turbulence to the bulk kinetic energy flux of the solar wind (Fig.~\ref{FIG:Chen2020}). This ratio was found to increase towards the Sun, as expected, reaching about 10\% at 0.17~AU. The radial variation of this ratio was also found to be consistent with the model, leading to the conclusion that the wind during these first orbits could be explained by a scenario in which the Sun drives AWs that reflect from the Alfv\'en speed gradient, driving a turbulence cascade that heats and accelerates the wind. Consistent with this picture, \citet{2020ApJS..246...53C} also found that the inward Alfv\'enic fluctuation component grew at a rate consistent with the measured Alfv\'en speed gradient.

\begin{figure*}
    \centering
    \includegraphics[width=0.8\textwidth]{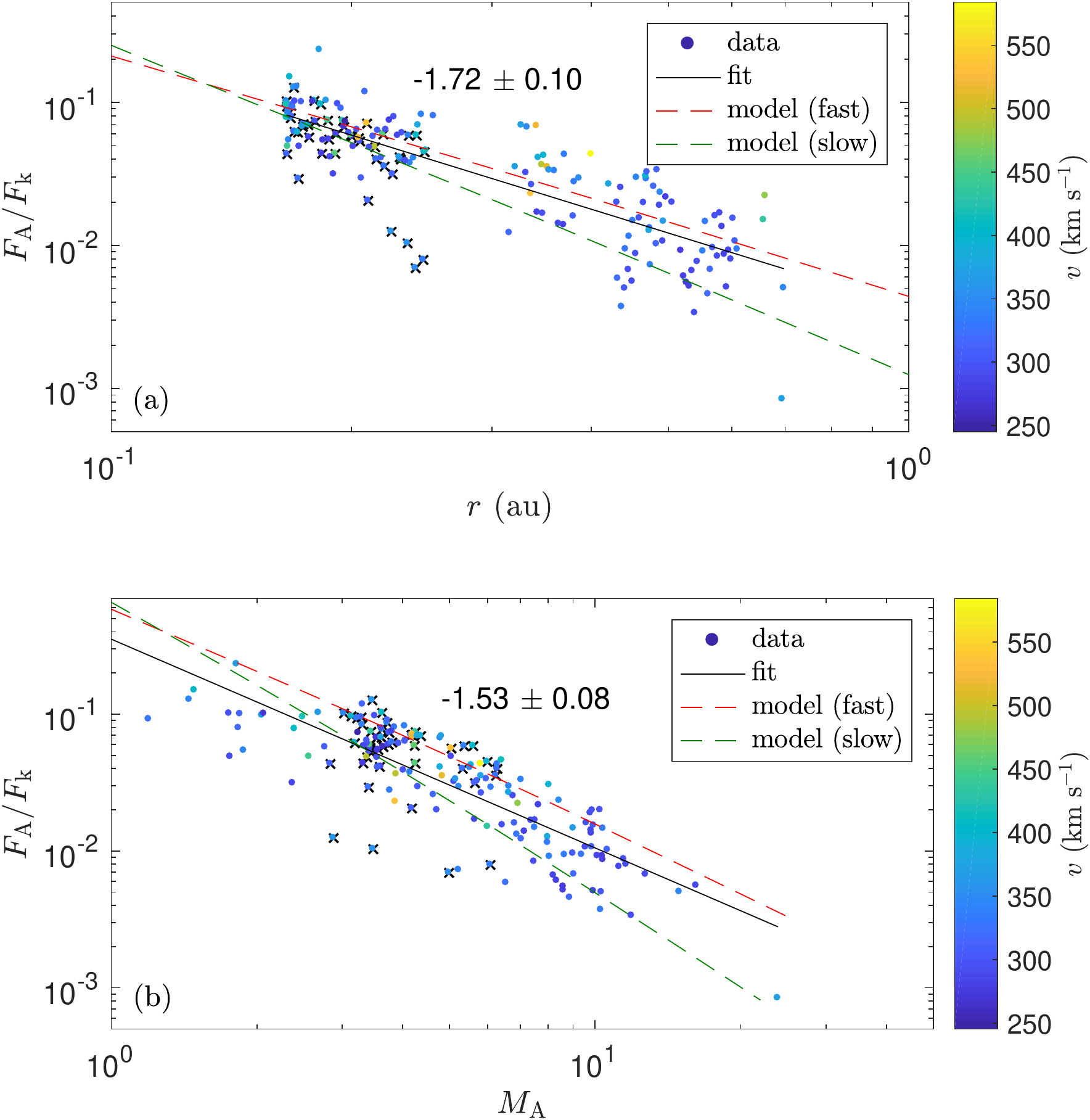}
    \caption{(a) Ratio of outward-propagating Alfv\'enic energy flux, $F_\mathrm{A}$, to solar wind bulk kinetic energy flux, $F_\mathrm{k}$, as a function of heliocentric distance, $r$. (b) The same ratio as a function of solar wind radial Alfv\'en Mach number, $M_\mathrm{A}$. In both plots, the black solid line is a power law fit, the red/green dashed lines are solutions to the \citet{2011ApJ...743..197C} model, the data points are colored by solar wind speed, $v$, and crosses mark times during connection to the coronal hole in Enc.~1. Figure adapted from \citet{2020ApJS..246...53C}.}
    \label{FIG:Chen2020}
\end{figure*}

To see if such physics can explain the 3D structure of the solar wind, the {\emph{PSP}} observations have also been compared to 3D turbulence-driven solar wind models. \citet{ 2020ApJS..246...48B} calculated the turbulent heating rates from {\emph{PSP}} using two methods: from the third-order laws for energy transfer through the MHD inertial range and from the von K\'arm\'an decay laws based on correlation scale quantities. These were both found to increase going closer to the Sun, taking values at 0.17~AU about 100 times higher than typical values at 1~AU. These were compared to those from the model of \citet{2018ApJ...865...25U}, under two different inner boundary conditions -- an untilted dipole and a magnetogram from the time of the S/C. The heating rates from both models were found to be in broad agreement with those determined from the {\emph{PSP}} measurements, although the magnetogram version provided a slightly better fit overall. \citet{2021ApJ...923...89C} later performed a comparison of the first five orbits to a similar 3D turbulence solar wind model (which captures the coupling between the solar wind flow and small-scale fluctuations), examining both mean-flow parameters such as density, temperature and wind speed, as well as turbulence properties such as fluctuation amplitudes, correlation lengths and cross-helicity. In general, the mean flow properties displayed better agreement with {\emph{PSP}} observations than the turbulence parameters, indicating that aspects of the turbulent heating were possibly being captured, even if some details of the turbulence were not fully present in the model. A comparison between the model and observations for orbit 1 is shown in Fig.~\ref{FIG:Chhiber2021}.

\begin{figure*}
    \centering
    \includegraphics[width=0.75\textwidth]{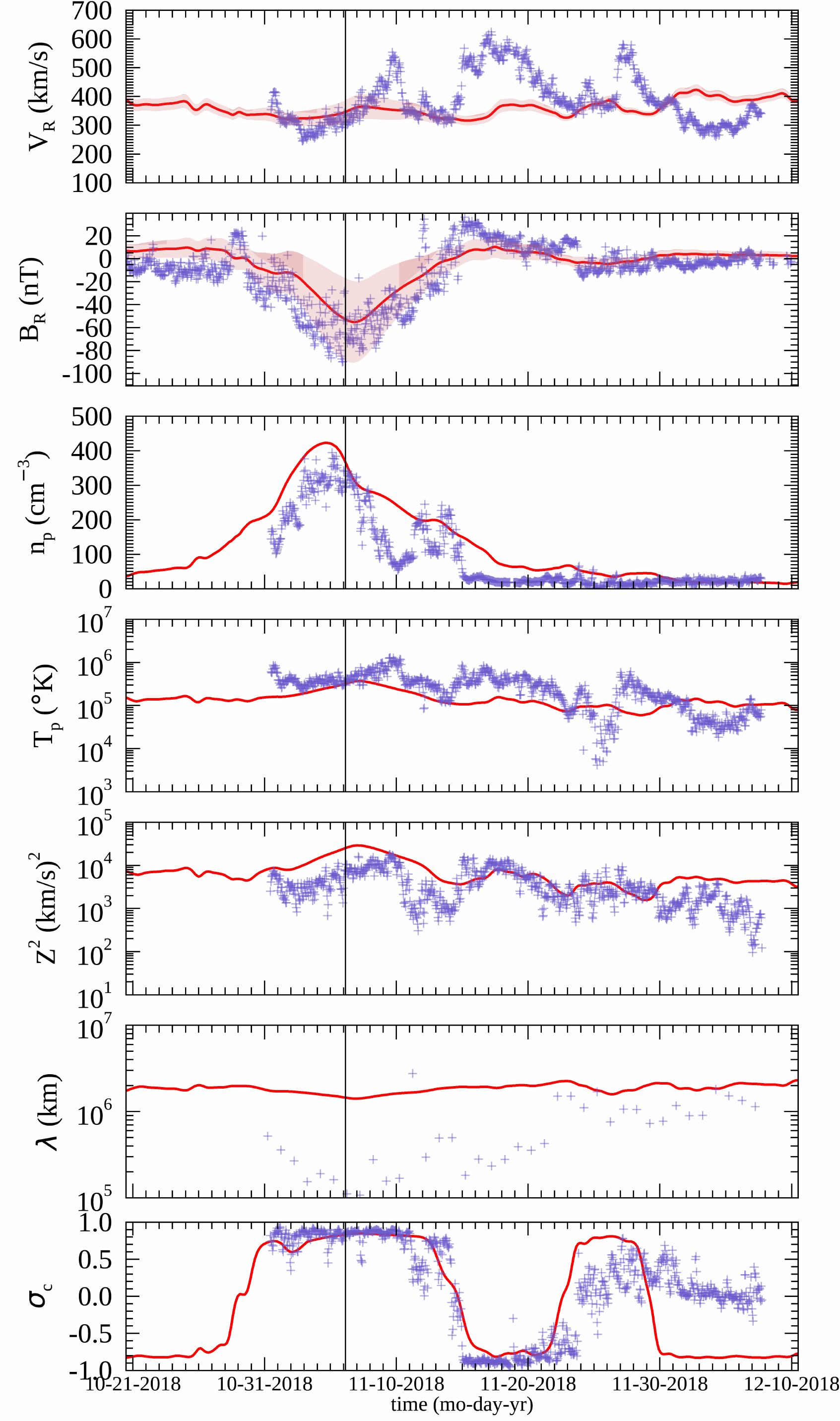}
    \caption{Blue `$+$' symbols show {\emph{PSP}} data from orbit 1, plotted at 1-hour cadence except \(\lambda\), for which daily values are shown. Red curve shows results from the model, sampled along a synthetic {\emph{PSP}} trajectory. Quantities shown are mean radial velocity of ions (\(V_R\)), mean radial magnetic field \(B_R\), mean ion density \(n_p\), mean ion temperature \(T_p\), mean turbulence energy \(Z^2\), correlation length of magnetic fluctuations \(\lambda\), and normalized cross helicity \(\sigma_c\). The shading in the top four panels marks an envelope obtained by adding and subtracting the local turbulence amplitude from the model to the mean value from the model (see the text for details). The vertical black line marks perihelion. The model uses ADAPT map with central meridian time 6 Nov. 2018, at 12:00 UT (Run I). Minor ticks on the time axis correspond to 1 day. Figure adapted from \citet{2021ApJ...923...89C}.}
    \label{FIG:Chhiber2021}
\end{figure*}

\citet{2020ApJS..246...38A} first compared the {\emph{PSP}} plasma observations to results from their turbulence transport model based on the nearly incompressible MHD description. They concluded that there was generally a good match for quantities such as fluctuating kinetic energy, correlation lengths, density fluctuations, and proton temperature. Later, \citet{2020ApJ...901..102A} developed a model that couples equations for the large scale dynamics to the turbulence transport equations to produce a turbulence-driven solar wind. Again, they concluded a generally good agreement, and additionally found the heating rate of the quasi-2D component of the turbulence to be dominant, and to be sufficient to provide the necessary heating at the coronal base.

Overall, these studies indicate a picture that is consistent with turbulence, driven ultimately by motions at the surface of the Sun, providing the energy necessary to heat the corona and accelerate the solar wind in a way that matches the {\emph{in situ}} measurements made by {\emph{PSP}}. Future work will involve adding even more realistic turbulence physics into these models, and testing them under a wider variety of solar wind conditions. One recent study in this direction is \citet{2021A&A...650L...3C}, which examined the turbulence energy fluxes as a function of distance to the HCS during Enc.~ 4. They found that the turbulence properties changed when {\emph{PSP}} was within 4$^\circ$ of the HCS, resembling more the standard slow solar wind seen at 1~AU, and suggesting this as the angular width of the streamer belt wind at these distances. Also, within this streamer belt wind, the turbulence fluxes were significantly lower, being on average 3 times smaller than required for consistency with the \citet{2011ApJ...743..197C} solar wind model. \citet{2021A&A...650L...3C} concluded, therefore, that additional mechanisms not in these models are required to explain the solar wind acceleration in the streamer belt wind near the HCS.

The coming years, with both {\emph{PSP}} moving even closer to the Sun and the the solar cycle coming into its maximum, will provide even better opportunities to further understand the role that turbulence plays in the heating and acceleration of the different types of solar wind, and how this shapes the large-scale structure of our heliosphere.

\section{Large-Scale Structures in the Solar Wind}
\label{LSSSW}

During these four years of mission (within the ascending phase of the solar cycle), {\emph{PSP}} crossed the HCS several times and also observed structures (both remotely and {\emph{in situ}}) with similar features to the internal magnetic flux ropes (MFRs) associated with the interplanetary coronal mass ejections (ICMEs). This section focuses on {\emph{PSP}} observations of large-scale structures, {\emph{i.e.}}, the HCS crossing, ICMEs, and CMEs.

Smaller heliospheric flux ropes are also included in this section because of their similarity to larger ICME flux ropes. The comparison of the internal structure of large- and small-scale flux ropes (LFRs and SFRs, respectively) is revealing. They can both store and transport magnetic energy. Their properties at different heliodistances provide insights into the energy transport in the inner heliosphere. {\emph{PSP}} brings a unique opportunity for understanding the role of the MFRs in the solar wind formation, evolution, and thus connecting scales in the heliosphere.

\subsection{The Heliospheric Current Sheet}
\label{LSSSWHCS}

The HCS separates the two heliospheric magnetic hemispheres: one with a magnetic polarity pointing away from the Sun and another toward the Sun. {\emph{PSP}} crossed the HCS multiple times in each orbit due to its low heliographic latitude orbit. Fig.~\ref{Orbit5_HCS} shows a comprehensive set of measurements with periods of HCS crossing identified as gray regions. These crossings are particularly evident in the magnetic field azimuth angle ($\phi_B$) and the PAD of suprathermal electrons. In the Radial-Tangential-Normal (RTN) coordinates used here, the outward and inward polarities have a near-zero degree and $180^{\circ}$ azimuth angle, respectively. Since the electron heat flux always streams away from the Sun, the streaming direction is parallel (antiparallel) to the magnetic field in the regions of the outward (inward) magnetic polarity, resulting in a magnetic pitch angle of $0^{\circ}$ ($180^{\circ}$).

\begin{figure*}
    \centering
    \includegraphics[width=0.95\textwidth]{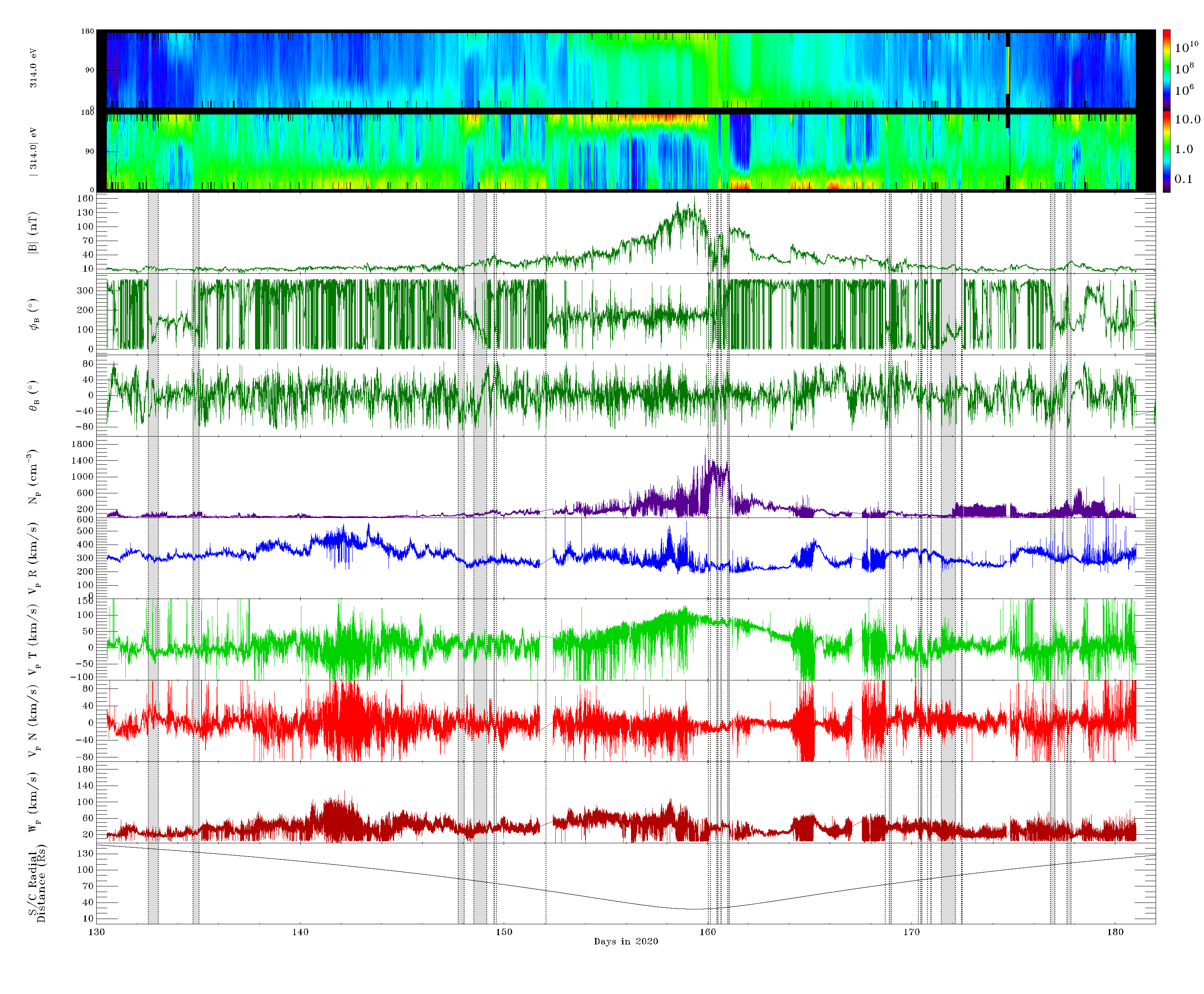}
    \caption{SP solar wind measurements during solar Enc.~5, 10 May 2020 (day of the year [DOY] 130) to 1 Jun. 2020 (DOY 182). The panels from top to bottom are the PAD of 314 eV suprathermal electrons; the normalized PAD of the 314 eV suprathermal electrons; the magnetic field magnitude; the azimuth angle of the magnetic field in the TN plane; the elevation angle of the magnetic field; the solar wind proton number density; the RTN components of the solar wind proton bulk speed; the thermal speed of the solar wind protons; and the S/C radial distance from the Sun. The gray bars mark periods of HCS crossings.}
    \label{Orbit5_HCS}
\end{figure*}

Comparing the observed locations of HCS crossings with PFSS model predictions yielded good agreement \citep{2020ApJS..246...47S, 2021A&A...652A.105L}. Lowering the source surface radius to 2~$R_\odot$ or even below would minimize the timing disagreements, though this would increase the amount of open magnetic flux to unreasonable values. The likely resolution is that the appropriate source surface radius is not a constant value but varies depending on the solar surface structures below. Other sources of disagreement between the model predictions and observations are the emergence of ARs not included in the photospheric magnetic maps used by the PFSS models and the presence of solar wind transients ({\emph{e.g.}}, ICMEs). \citet{2021A&A...652A.105L} also found that while the PFSS model predicted a relatively flat HCS, the observed current sheets had a much steeper orientation suggesting significant local corrugation. \citet{2020ApJS..246...47S} also compared the observed HCS crossing times to global MHD model predictions with similar results.

The internal structure of the HCS near the Sun is very complex \citep{2020ApJS..246...47S, 2020ApJ...894L..19L, 2021A&A...650A..13P}. \citet{2020ApJS..246...47S} has identified structures within the HCS region with magnetic field magnitude depressions, increased solar wind proton bulk speeds, and associated suprathermal electron strahl dropouts. These might evolve into the small density enhancements observed by \citet{2020ApJ...894L..19L} likely showing magnetic disconnections. In addition, small flux ropes were also identified inside or just outside the HCS, often associated with plasma jets indicating recent magnetic reconnection \citep{2020ApJS..246...47S, 2020ApJ...894L..19L, 2021A&A...650A..13P, 2021A&A...650A..12Z}. The near Sun HCS is much thicker than expected; thus, it is surprising that it is the site of frequent magnetic reconnection \citep{2021A&A...650A..13P}. Moreover, 1~AU observations of the HCS reveal significantly different magnetic and plasma signatures implying that the near-Sun HCS is the location of active evolution of the internal structures \citep{2020ApJS..246...47S}.

The HCS also appears to organize the nearby, low-latitude solar wind. \citet{2021A&A...650L...3C} observed lower amplitude turbulence, higher magnetic compressibility, steeper magnetic spectrum, lower Alfv\'enicity, and a $1/f$ break at much lower frequencies within $4^\circ$ of the HCS compared to the rest of the solar wind, possibly implying a different solar source of the HCS environment.

\subsection{Interplanetary Coronal Mass Ejections} \label{sec:icme}

The accurate identification and characterization of the physical processes associated with the evolution of the ICMEs require as many measurements of the magnetic field and plasma conditions as possible \citep[see][and references therein]{2003JGRA..108.1156C,2006SSRv..123...31Z}. 

Our knowledge of the transition from CME to ICME has been limited to the {\emph{in~situ}} data collected at 1~AU and remote-sensing observations from space-based observatories. {\emph{PSP}} provides a unique opportunity to link both views through valuable observations that will allow us to distinguish the evidence of the early transition from CME to ICME. Due to its highly elliptical orbit, {\emph{PSP}} measures the plasma conditions of the solar wind at different heliospheric distances. Synergies with other space missions and ground observatories allow building a complete picture of the phenomena from the genesis at the Sun to the inner heliosphere. 

In general, magnetic structures in the solar wind are MFRs, a subset of which are MCs \citep[][]{1988JGR....93.7217B}, and are characterized by enhanced magnetic fields where the field rotates slowly through a large angle. MCs are of great interest as their coherent magnetic field configuration and plasma properties drive space weather and are related to major geomagnetic storms. \citep{2000JGR...105.7491W}. Therefore, understanding their origin, evolution, propagation, and how they can interact with other transients traveling through space and planetary systems is of great interest. ICMEs are structures that travel throughout the heliosphere and transfer energy to the outer edge of the solar system and perhaps beyond.

\paragraph{{\textbf{Event of 11-12 Nov. 2018: Enc. 1 $-$ {\emph{PSP}} at (0.25~AU, -178$^{\circ}$)}}} $\\$
During the first orbit, {\emph{PSP}} collected {\emph{in~situ}} measurements of the thermal solar wind plasma as close as $35.6~R_\odot$ from the Sun. In this new environment, {\emph{PSP}} recorded the signatures of SBO-CMEs: the first on 31 Oct. 2018, at 03:36 UT as it entered the Enc. and the second on 11 Nov. 2018, at 23:50 UT as it exited the Enc.. The signature of the second SBO-CME crossing the S/C was a magnetic field enhancement ({\emph{i.e.}}, maximum strength 97~nT). The event was seen by {\emph{STEREO}}-A but was not visible from L1 or Earth-orbiting S/C as the event was directed away from Earth. The signature and characteristics of this event were the focus of several studies, \citep[see][]{2020ApJS..246...69K,2020ApJS..246...63N,2020ApJS..246...29G}. SBO-CMEs \citep{2008JGRA..113.9105C} are ICMEs that fulfill the following criteria in coronagraph data (1) slow speed ranging from 300 to 500~km~s$^{-1}$; (2) no identifiable surface or low coronal signatures (in this case from Earth point of view); (3) characterized by a gradual swelling of the overlying streamer (blowout type); and (4) follows the tilt of HCS. 

The source location was determined using remote sensing and {\emph{in situ}} observations, the WSA model \citep{2000JGR...10510465A}, and the Air Force Data Assimilative Photospheric Flux Transport (ADAPT) model \citep{2004JASTP..66.1295A}. Hydrodynamical analytical and numerical simulations were also utilized to predict the CME arrival time to {\emph{PSP}}. Using a CME propagation model, \cite{2020ApJS..246...69K} and \cite{2020ApJS..246...63N} explored the characteristics of the CME using {\emph{in situ}} data recorded closest to the Sun as well as the implications for CME propagation from the coronal source to {\emph{PSP}} and space weather. The CME was traveling at an average speed of $\sim391$~km~s$^{-1}$ embedded in an ambient solar wind flow of $\sim395$~km~s$^{-1}$ and a magnetic field of 37~nT. The difference in speed with the ambient solar wind suggests that drag forces drive the SBO-CME. 

The internal magnetic structure associated with the SBO displayed signatures of flux-rope but was characterized by changes that deviated from the expected smooth change in the magnetic field direction (flux rope-like configuration), low proton plasma beta, and a drop in the proton temperature. A detailed analysis of the internal magnetic properties suggested high complexity in deviations from an ideal flux rope 3D topology. Reconstructions of the magnetic field configuration revealed a highly distorted structure consistent with the highly elongated “bubble” observed remotely. A double-ring substructure observed in the FOV of COR2 coronagraph on the {\emph{STEREO}-A Sun-Earth Connection Coronal and Heliospheric Investigation \citep[SECCHI;][]{2008SSRv..136...67H}} may also indicate a double internal flux rope. Another possible scenario is described as a mixed topology of a closed flux rope combined with the magnetically open structure, justified by the flux dropout observed in the measurements of the electron PAD. In any case, the plethora of structures observed by the EUV imager (SECCHI-EUVI) in the hours preceding the SBO evacuation indicated that the complexity might be related to the formation processes \citep{2020ApJS..246...63N}.

Applying a wavelet analysis technique to the {\emph{in situ}} data from {\emph{PSP}}, \citet{2020ApJS..246...26Z} also identified the related flux rope. They inferred the reduced magnetic helicity, cross helicity, and residual energy. With the method, they also discussed that after crossing the ICME, both the plasma velocity and the magnetic field fluctuate rapidly and positively correlate with each other, indicating that Alfv\'enic fluctuations are generated in the region downstream of the ICME. 

Finally, \citet{2020ApJS..246...29G} also discussed the SBO-CME as the driver of a weak shock when the ICME was at 7.4 R$_{\odot}$ accelerating energetic particles. {\emph{PSP}}/IS$\odot$IS observed the SEP event (see Fig.~\ref{Giacalone_2020}). Five hours later, {\emph{PSP}}/FIELDS and {\emph{PSP}}/SWEAP detected the passage of the ICME (see \S\ref{EPsRad} for a detailed discussion). 

\paragraph{{\textbf{Event of 15 Mar. 2019: Enc. 2 $-$ {\emph{PSP}} at (0.547~AU, 161$^{\circ}$)}}} $\\$
An SBO-CME was observed by {\emph{STEREO}}-A and {\emph{SOHO}} coronagraphs and measured {\emph{in situ}} by {\emph{PSP}} at 0.547~AU on 15 Mar. 2019 from 12:14 UT to 17:45 UT. The event was studied in detail by \citet{2020ApJ...897..134L}. The ICME was preceded by two interplanetary shock waves, registered at 08:56:01 UT and 09:00:07 UT (see Fig.~\ref{Lario2020Fig1} in \S\ref{EPsRad}). This study's authors proposed that the shocks were associated with the interaction between the SBO-CME and a HSS. The analysis of the shocks' characteristics indicated that despite the weak strength, the successive structures caused the acceleration of energetic particles. This study aimed to demonstrate that although SBO-CMEs are usually slow close to the Sun, subsequent evolution in the interplanetary space might drive shocks that can accelerate particles in the inner heliosphere. 
The event is discussed in more detail in \S\ref{EPsRad}, showing that the time of arrival of energetic particles at {\emph{PSP}} (Fig.~\ref{Lario2020Fig2}) is consistent with the arrival of the ICME predicted by MHD simulations. With the simulations, \citet{2020ApJ...897..134L}  determined when the magnetic connection was established between {\emph{PSP}} and the shocks, potentially driven by the ICME.  

\paragraph{{\textbf{Event of 13 Oct. 2019: Enc. 3 $-$ {\emph{PSP}} at (0.81~AU, 75$^{\circ}$)}}} $\\$
The event observed during Enc.~3 was reported by \citet{2021ApJ...916...94W}. The ICME is associated with the stealth CME evacuation on 10 Oct. 2019, at 00:48 UT. It was characterized by an angular width of 19$^{\circ}$, position angle of 82$^{\circ}$, no signatures in {\emph{SDO}}/AIA and EUVI-A images, and reaching a speed of 282~km~s$^{-1}$ at 20 R$_{\odot}$. At the time of the eruption, two coronal holes were identified from EUV images and extrapolations of the coronal magnetic field topology computed using the PFSS model \citep{1992ApJ...392..310W}, suggesting that the stealth CME evolved between two HSSs originated at the coronal holes. The first HSS enabled the ICME to travel almost at a constant speed (minimum interaction time $\sim2.5$ days), while the second overtook the ICME in the later stages of evolution. 
 
The event was measured when {\emph{PSP}} was not taking plasma measurements due to its proximity to aphelion, and there are only reliable magnetic field measurements by {\emph{PSP}}/FIELDS instrument. Even with these observational limitations, this event is of particular interest as {\emph{STEREO}}-A was located, 0.15~AU in the radial distance, $<1^{\circ}$ in latitude and $-7.7^{\circ}$ in longitude, away of {\emph{PSP}}. 

The ICME arrival is characterized by a fast-forward shock observed by {\emph{PSP}} on 13 Oct. 2019, at 19:03 UT and by {\emph{STEREO}}-A on 14 Oct. 2019, at 07:44 UT. Both S/C observed the same main features in the magnetic field components (exhibiting flux rope-like signatures) except for the HSS that {\emph{STEREO}}-A observed as an increasing speed profile and shorter ICME duration. To show the similarity of the main magnetic field features and the effect of the ICME compression due to its interaction with the HSS, the magnetic field and plasma parameters of {\emph{STEREO}}-A were plotted (shown in Fig.~\ref{fig:Winslow2020}) and overlaid the {\emph{PSP}} magnetic field measurements scaled by a factor of $1.235$ and shifted to get the same ICME duration as observed by {\emph{STEREO}}-A.

\begin{figure}
    \centering
    \includegraphics[width=0.75\textwidth]{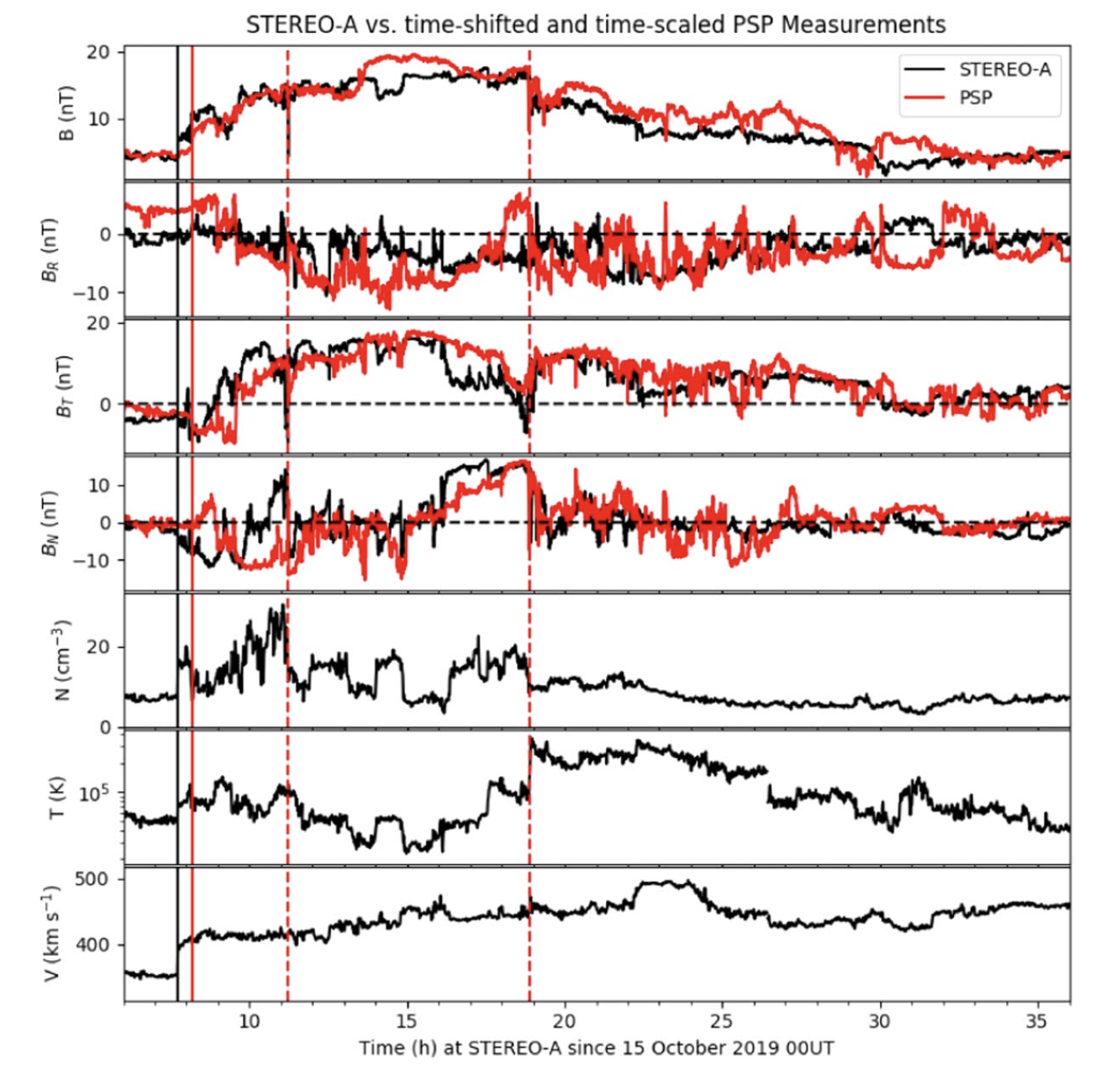}
    \caption{Overlay of the {\emph{in situ}} measurements by {\emph{STEREO}}-A (black) and {\emph{PSP}} (red). From top to bottom: the magnetic field strength and the radial (B$_R$), tangential (B$_T$) and normal (B$_N$) components, and for {\emph{STEREO}}-A only: the proton density (N), temperature (T), and velocity (V). The {\emph{PSP}} data are scaled (by a factor of 1.235) and time-shifted to obtain the same ICME duration delimit by the two dashed vertical red lines. The red vertical solid line marks the fast forward shock at {\emph{PSP}}, while at {\emph{STEREO}}-A with the black vertical solid line. Figure adapted from \citet{2021ApJ...916...94W}.}
    \label{fig:Winslow2020}
\end{figure}

\paragraph{{\textbf{Event of 20 Jan. 2020: Enc. 4 $-$ {\emph{PSP}} at (0.32~AU, 80$^{\circ}$)}}} $\\$
\citet{2021A&A...651A...2J} reported a CME event observed by {\emph{PSP}} on 18 Jan. 2020, at 05:30 UT. The event was classified as a stealth CME since the eruption signatures were identified on the Sun's surface. Coronal {\emph{SDO}}/AIA observations indicated the emission of a set of magnetic substructures or loops followed by the evacuation of the magnetic structure on 18 Jan. 2020, at 14:00 UT. The signatures of a few dispersed brightenings and dimmings observed in EUVI-A 195~{\AA} were identified as the source region \citep{2021A&A...651A...2J}.

The ICME arrived at {\emph{PSP}} on 20 Jan. 2020, at 19:00 UT, with a clear magnetic obstacle and rotation in the magnetic field direction but no sign of an IP shock wave. The event was also associated with a significant enhancement of SEPs. {\emph{PSP}} and {\emph{STEREO}}-A were almost aligned (separated by 5$^{\circ}$ in longitude). The ICME flew by both S/C, allowing for the examination of the evolution of the associated SEPs. Interestingly, this event established a scenario in which weaker structures can also accelerate SEPs. Thus, the presence of SEPs with the absence of the shock was interpreted as {\emph{PSP}} crossing the magnetic structure's flank, although no dense feature was observed in coronagraph images propagating in that direction \citep{2002ApJ...573..845G}. In \S\ref{EPsRad}, the event is discussed in detail, including the discussion of the associated {\emph{PSP}} observations of SEPs.

\paragraph{{\textbf{Event of 25 Jun. 2020: Enc. 5 $-$ {\emph{PSP}} at (0.5~AU, 20$^{\circ}$)}}} $\\$
\citet{2021ApJ...920...65P,2022SpWea..2002914K}, and \citet{2022ApJ...924L...6M} studied and modeled the event of 25 Jan. 2020, which occurred during Enc.~5. The lack of clear signatures on the solar surface and low corona led to the interpretation of this event as an SBO-CME and was the primary motivation for these studies.

The models were tested extensively to determine their capabilities to predict the coronal features and their counterparts in space. \citet{2021ApJ...920...65P} focused on predictions of the location of its source and the magnetic field configuration expected to be measured by {\emph{PSP}}. The {\emph{SDO}}/AIA and EUVI-A observations were used to determine the source location of the event. The increase in the solar activity around the source region was followed by a small eruption in the northern hemisphere on 21 Jun. 2020, at 02:00 UT (Fig.~\ref{fig:Palmerio2021}-left). This led to the outbreak of the SBO-CME on 23 Jun. 2020, at 00:54 UT. Using the PFSS model, the authors found that the SBO-CME was triggered by the interaction between the small eruption and the neighboring helmet streamer. The SBO-CME geometry and kinetic aspects were obtained by applying the graduated cylindrical shell \citep[GCS;][]{2011ApJS..194...33T} model to the series of coronagraph images resulting in the estimation of an average speed of 200~km~s$^{-1}$.

The magnetic field configuration from Sun to {\emph{PSP}} was obtained by modeling the event using OSPREI suite \citep{2022SpWea..2002914K}. The arrival at {\emph{PSP}} was also predicted to be on 25 Jun. 2020, at 15:50~UT (9 minutes before the actual arrival). {\emph{PSP}} was located at 0.5~AU and 20$^{\circ}$ west of the Sun-Earth line.

\begin{figure}
    \centering
    \includegraphics[width=0.45\textwidth]{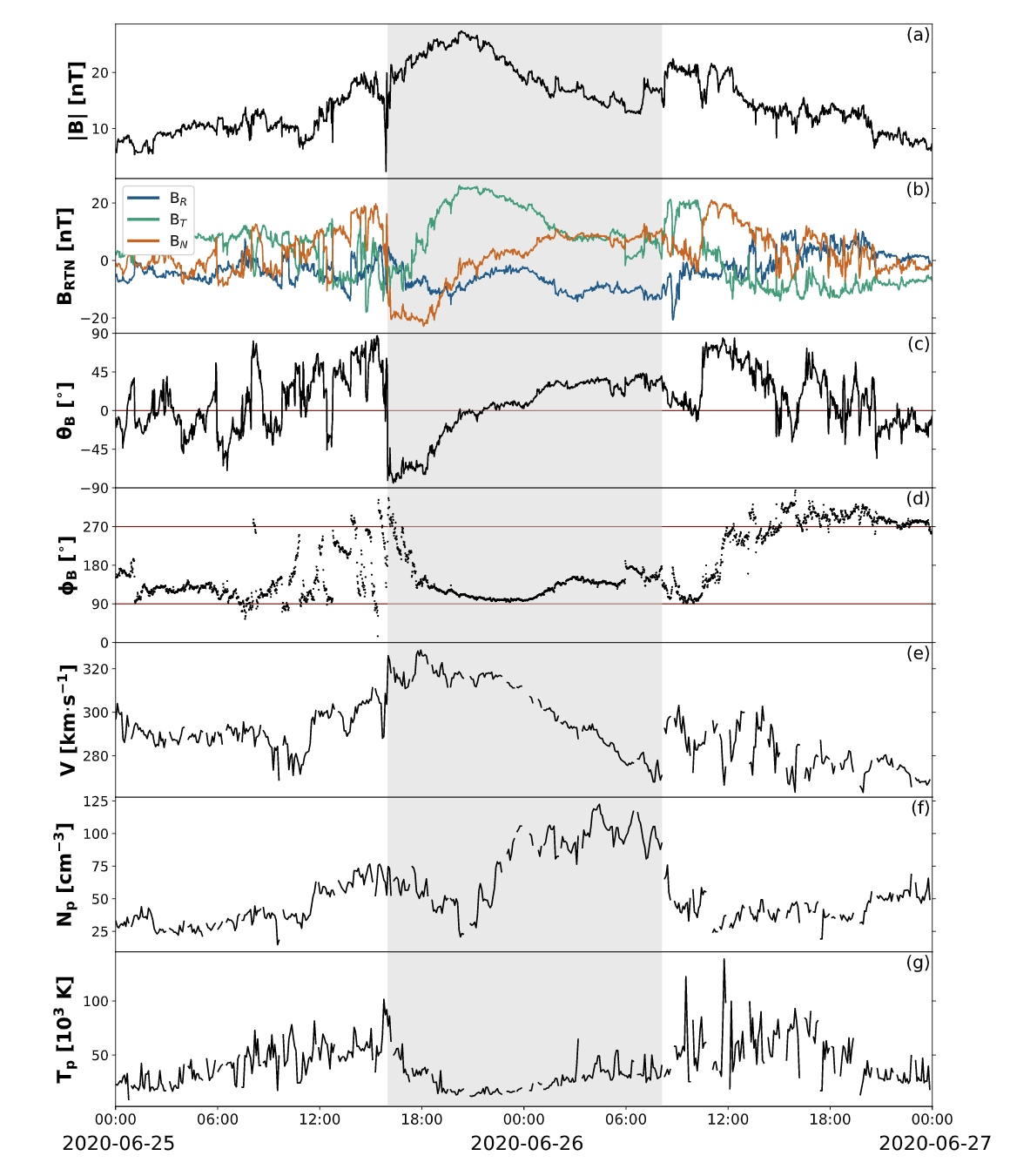}
    \includegraphics[width=0.45\textwidth]{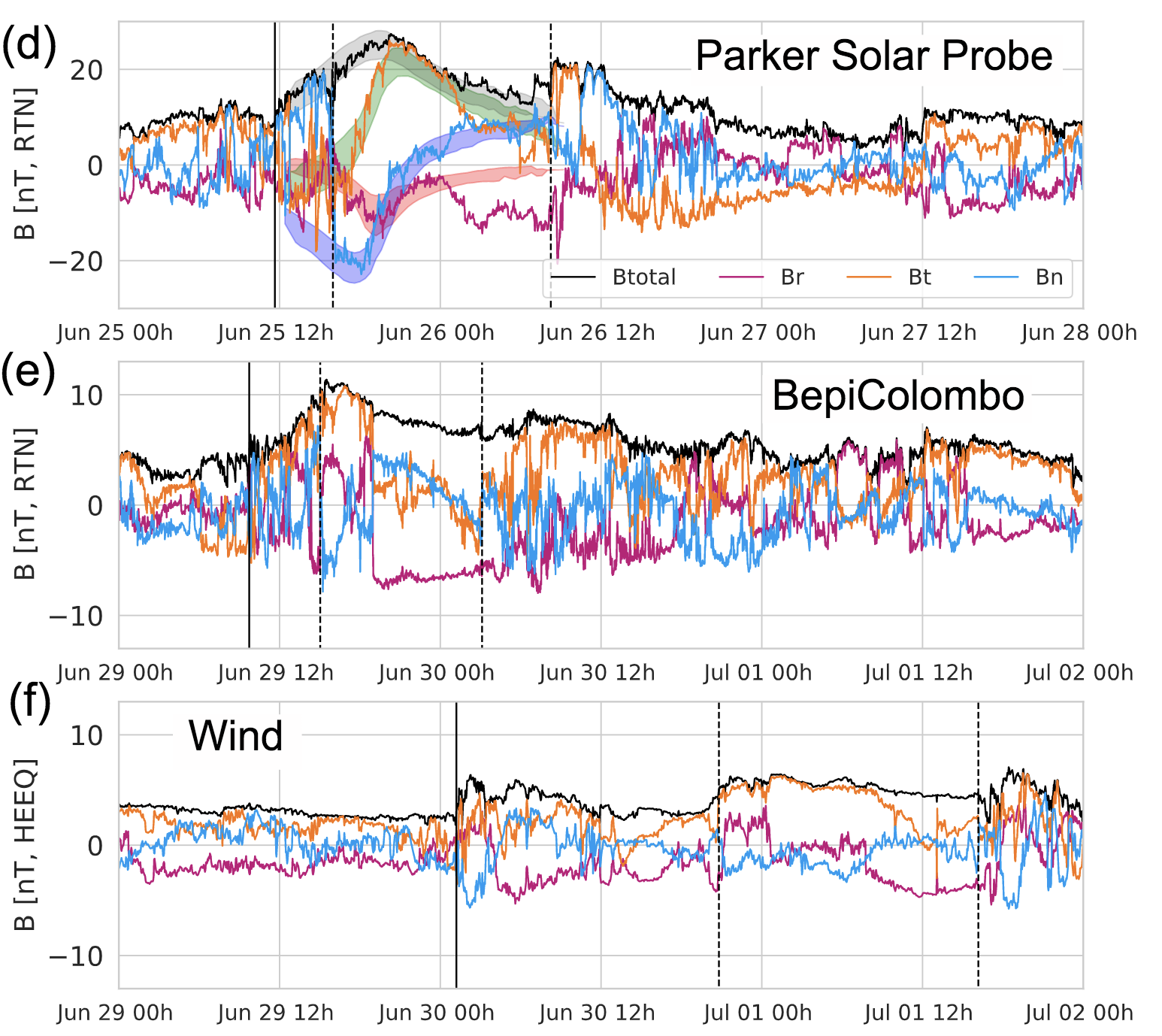}
\caption{Left: {\emph{PSP}} {\emph{in situ}} magnetic field and plasma measurements of the 21 Jun. 2020 stealth CME. From top to bottom:  magnetic field strength and components (B$_R$, B$_T$ and B$_N$), $\theta_B$ and (d)$\phi_B$ magnetic field angles, wind speed (V$_P$), proton number density (N$_P$), and proton temperature (T$_P$).  The flux rope interval is shaded in grey. Figure adapted from \citet{2021ApJ...920...65P}. Right: {\emph{in situ}} magnetic field data at {\emph{PSP}}, {\emph{BepiColombo}} and {\emph{Wind}}. Solid vertical lines indicate ICME start times, and dashed lines the boundaries of the magnetic obstacle. Figure adapted from \citet{2022ApJ...924L...6M}. }
    \label{fig:Palmerio2021}
\end{figure}

\citet[][; Fig.~\ref{fig:Palmerio2021}-right]{2022ApJ...924L...6M} studied the same event using multipoint measurements from {\emph{SolO}}, {\emph{BepiColombo}} \citep{2021SSRv..217...90B} (0.88~AU, -3.9$^{\circ}$), {\emph{PSP}}, {\emph{Wind}} (1.007~AU, -0.2$^{\circ}$), and {\emph{STEREO}}-A (0.96~AU, -69.6$^{\circ}$). The WSA/THUX \citep{2020ApJ...891..165R}, HELCATS, and 3DCORE \citep{2021ApJS..252....9W} models were used to infer the background solar wind in the solar equatorial plane, the height time plots, and the flux rope, respectively. With the multi-S/C observation, the authors attempted to explain the differences in the {\emph{in~situ}} signatures observed at different locations of a single CME. To accomplish this goal, they modeled the evolution of a circular front shape propagating at a constant speed. The {\emph{in situ}} arrival ICME speeds at {\emph{PSP}} and {\emph{Wind}} were 290~km~s$^{-1}$ and 326~km~s$^{-1}$, respectively. The arrival speed at {\emph{STEREO}}-A was computed using SSEF30 model described by \citet{2012ApJ...750...23D}. The discrepancies between the observed and predicted arrival times ranged from $-11$ to $+18$ hrs. The authors attributed this to a strong ICME deformation.

\paragraph{{\textbf{Event of 29 Nov. 2020: Enc. 6 $-$ {\emph{PSP}} at (0.81~AU, 104$^{\circ}$)}}} $\\$

The CME event of 2020 Nov. 29 has been widely studied and identified as the largest widespread SEP event of solar cycle 25 and the direct first {\emph{PSP}} observation of the interaction of two successive ICMEs \citep{2021A&A...656A..29C, 2021ApJ...919..119M, 2021A&A...656A..20K, 2021ApJ...920..123L, 2021A&A...656L..12M, 2022ApJ...924L...6M, 2022ApJ...930...88N}.

During this event, {\emph{PSP}}, {\emph{SolO}}, and {\emph{STEREO}}-A were located at respective radial distances of 0.81~AU, 0.87~AU,  and $\sim1$~AU. As seen from the Earth, they were at longitudinal angular positions of 104$^{\circ}$ east, 115$^{\circ}$ west, 65$^{\circ}$ east, respectively. The remote sensing observations show that at least four successive CMEs were observed during 24-29 Nov. 2020, although only two were directed toward {\emph{PSP}}.

During the SEP event, the particles spread over more than 230$^{\circ}$ in longitude close to 1~AU. \citet{2021A&A...656A..20K} compared the timing when the EUV wave intersects the estimated magnetic foot-points from different S/C with the particle release times from time shift and velocity dispersion analyses. They found that there was no EUV wave related to the event. The PAD and first-order anisotropies studies at {\emph{SolO}}, {\emph{STEREO}}-A, and {\emph{Wind}} S/C suggest that diffusive propagation processes were involved.

\citet{2022ApJ...930...88N} analyzed multi-S/C observations and included different models and techniques focusing on creating the heliospheric scenario of the CMEs' evolution and propagation and the impact on their internal structure at {\emph{PSP}}. The observations of {\emph{PSP}}, {\emph{STEREO}}-A, and {\emph{Wind}} of type II and III radio burst emissions indicate a significant left-handed polarization, which has never been detected in that frequency range. The authors identified the period when the interaction/collision between the CMEs took place using the results of reconstructing the event back at the Sun and simulating the event on the WSA-ENLIL+Cone and DBM models. They concluded that both ICMEs interacted elastically while flying by {\emph{PSP}}. The impact of such interaction on the internal magnetic structure of the ICMEs was also considered. Both ICMEs were fully characterized and 3D-reconstructed with the GCS, elliptical cylindrical (EC), and circular cylindrical (CC) models. The aging and expansion effects were implemented to evaluate the consequences of the interaction on the internal structure.

\citet{2021ApJ...919..119M} investigated key characteristics of the SEP event, such as the time profile and anisotropy distribution of near-relativistic electrons measured by IS${\odot}$IS/EPI-Lo. They observed the brief PAD with a peak between 40$^{\circ}$ and 90$^{\circ}$ supporting the idea of a shock-drift acceleration, noting that the electron count rate peaks at the time of the shock driven by the faster of the two ICMEs. They concluded that the ICME shock caused the acceleration of electrons and also discussed that the ICMEs show significant electron anisotropy indicating the ICME's topology and connectivity to the Sun. 

\citet{2021ApJ...920..123L} studied two characteristics of the shock and their impact on the SEP event intensity: (1) the influence of unrelated solar wind structures, and (2) the role of the sheath region behind the shock. The authors found that on arrival at {\emph{PSP}}, the SEP event was preceded by an intervening ICME that modified the low energy ion intensity-time profile and energy spectra. The low-energy ($\lesssim$220~keV) protons accelerated by the shock were excluded from the first ICME, resulting in the observation of inverted energy spectra during its passage.

\citet{2021A&A...656A..29C} analyzed the ion spectra during both the decay of the event (where the data are the most complete for H and He) and integrated over the entire event (for O and Fe). They found that the spectra follow a power law multiplied by an exponential with roll-over energies that decrease with the species' increasing rigidities. These signatures are typically found in SEP events where the dominant source is a CME-driven shock, supported by the He/H and Fe/O composition ratios. They also identified signatures in the electron spectrum that may suggest the presence of a population trapped between the ICMEs and pointed out the possibility of having the ICMEs interacting at the time of observation by noting a local ion population with energies up to $\sim1$~MeV. The SEP intensities dropped significantly during the passage of the MFR and returned to high values once {\emph{PSP}} crossed out of the magnetic structure. \citet{2021A&A...656L..12M} compared detailed measurements of heavy ion intensities, time dependence, fluences, and spectral slopes with 41 events surveyed by \citet{2017ApJ...843..132C} from previous solar cycles. They concluded that an interplanetary shock passage could explain the observed signatures. The observed Fe/O ratios dropped sharply above ~1 MeV nucleon$^{-1}$ to values much lower than the averaged SEP survey. They were a few MeV nucleon$^{-1}$ and $^3$He/$^4$He $<0.03$\% at {\emph{ACE}} and $<1$\% at {\emph{SolO}}. For further details on this SEP event, see the discussed in \S\ref{EPsCMENov}.

The second ICME hitting {\emph{PSP}} was also analyzed by \citet{2022ApJ...924L...6M}. The authors combined coronagraph images from {\emph{SOHO}} and {\emph{STEREO}} and applied the GCS model to obtain the ICME geometric and kinematic parameters, computing an average speed of 1637~km~s$^{-1}$ at a heliodistance ranging from 6 to 14~$R_{\odot}$. The ICME arrived at {\emph{PSP}} (0.80~AU and $-96.8^{\circ}$) on 1 Dec. 2020, at 02:22 UT and at {\emph{STEREO}}-A (0.95~AU and $-57.6^{\circ}$) on 1 Dec. 2020, at 07:28 UT. They also considered this event an excellent example of the background wind's influence on the possible deformation and evolution of a fast CME and the longitudinal extension of a high-inclination flux rope.

\subsection{Magnetic Flux Ropes}\label{7_mfr}

The {\emph{in situ}} solar wind measurements show coherent and clear rotations of the magnetic field components at different time scales. These magnetic structures are well known as MFRs. According to their durations and sizes, MFRs are categorized as LFRs \citep[few hours to few days;][]{2014SoPh..289.2633J} and SFRs \citep[tens of minutes to a few hours;][]{2000GeoRL..27...57M}.

At 1~AU, it has been found that 30\% to 60\% of the large-scale MFRs are related to CMEs \citep[][]{1990GMS....58..343G,2010SoPh..264..189R}. This subset of MFRs is known as MCs \citep{1988JGR....93.7217B}. On the other hand, the SFRs' origin is not well understood. Several studies proposed that SFRs are produced in the near vicinity of the Sun, while others can consider turbulence as a potential SFR source \citep[{\emph{i.e.}},][]{2019ApJ...881L..11P} or else that SFRs are related and originate from  SBO-CMEs. It is worth noticing that observations suggest that SBO-CMEs last a few hours, a time scale that falls in the SFR category.

To identify SFRs, \citet{2020ApJS..246...26Z} analyzed the magnetic field and plasma data from the {\emph{PSP}}'s first orbit from 22 Oct. to 21 Nov. 2018. They identified 40 SFRs by following the method described by \citet{2012ApJ...751...19T}. They applied a Morlet analysis technique to estimate an SFR duration ranging from 8 to 300 minutes. This statistical analysis suggests that the SFRs are primarily found in the slow solar wind, and their possible source is MHD turbulence. For the third and fourth orbits, they identified a total of 21 and 34 SFRs, respectively \citep{2021A&A...650A..12Z}, including their relation to the streamer belt and HCS crossing.

Alternatively, \citet{2020ApJ...903...76C} identified 44 SFRs by implementing an automated detection method based on the Grad-Shafranov reconstruction technique \citep{2001GeoRL..28..467H,2002JGRA..107.1142H} over the {\emph{in situ}} measurements in a 28-second cadence. They looked for the double-folding pattern in the relation between the transverse pressure and the magnetic vector potential axial component and removed highly Alfv\'enic structures with a threshold condition over a Wal\'en test. The SFRs were identified during the first two {\emph{PSP}} Encs. over the periods 31 Oct. $-$ Dec. 19 2018 ($\sim0.26-0.81$~AU), and 7 Mar. $-$ 15 May 2019 ($\sim0.66-0.78$~AU) with durations ranging from 5.6 to 276.3 min. They found that the monthly counts at {\emph{PSP}} (27 per month) are notably lower than the average monthly counts at {\emph{Wind}} (294 at 1~AU). The authors also noticed that some of the detected SFRs are related to magnetic reconnection processes \citep[two reported by][]{2020ApJS..246...34P} and HCS \citep[three reported by][]{2020ApJS..246...47S}. They argue that the SFR occurrence rate (being far less than at 1~AU) and a power-law tendency of the size-scales point towards an SFRs origin from MHD turbulence but note that the number of events analyzed is not sufficient to yield a statistically significant analysis result. 12 SFRs were also identified with the method proposed by \citet{2020ApJS..246...26Z} with similar duration and two cases with opposite helicity.

\subsection{Remote Sensing}\label{7_rs}

\subsubsection{Introduction} \label{InstIntro}

{\emph{PSP}}/WISPR is a suite of two white light telescopes akin to the heliospheric imagers \citep[HI-1 and HI-2;][]{2009SoPh..254..387E} of {\emph{STEREO}/SECCHI \citep{2008SSRv..136....5K}. The WISPR telescopes look off to the ram side of the S/C ({\emph{i.e.}}, in the direction of motion of {\emph{PSP}} in its counter-clockwise orbit about the Sun). When {\emph{PSP}} is in its nominal attitude ({\emph{i.e.}}, Sun-pointed and ``unrolled''), their combined FOV covers the interplanetary medium on the West side of the Sun, starting at a radial elongation of about $13.5^\circ$ from the Sun and extending up to about $108^\circ$. The FOV of WISPR-i extends up to $53.5^\circ$, while the FOV of WISPR-o starts at $50^\circ$ elongation, both telescopes covering  about $40^\circ$ in latitude. Since the angular size of the Sun increases as {\emph{PSP}} gets closer to the Sun, the radial offset from Sun center represents different distances in units of solar radii. For example, on 24 Dec. 2024, at the closest approach of {\emph{PSP}} of 0.046~AU ($9.86~R_\odot$), the offset of $13.5^\circ$ will correspond to $\sim2.3~R_\odot$.  

\subsubsection{Streamer Imaging with WISPR}
\citet{2019Natur.576..232H} reported on the first imaging of the solar corona taken by WISPR during {\emph{PSP}}’s first two solar Encs. ($0.16-0.25$~AU). The imaging revealed that both the large and small scale topology of streamers can be resolved by WISPR and that the temporal variability of the streamers can be clearly isolated from spatial effects when {\emph{PSP}} is corotating with the Sun \cite{ 2020ApJS..246...60P}, by exploiting synoptic maps based on sequential WISPR images, revealed the presence of multiple substructures (individual rays) inside streamers and pseudostreamers. This substructure of the streamers was noted in other studies  \citep{2006ApJ...642..523T, 2020ApJ...893...57M}. Noteworthy in the WISPR synoptic maps was the identification of a bright and narrow set of streamer rays located at the core of the streamer belt \citep{2020ApJS..246...60P,2020ApJS..246...25H}. The thickness of this bright region matches the thickness of the heliospheric plasma sheet (HPS) measured in the solar wind (up to 300Mm) at times of sector boundary crossings \citep{1994JGR....99.6667W}. Thus, WISPR may offer the first clear-cut connection between coronal imaging of streamers and the {\emph{in situ}} measurements of the rather narrow HPS.

Global PFSS and MHD models of the solar corona during the {\emph{PSP}} Encs. generally agree with the large-scale structure inferred from remote sensing observations \citep[{\emph{e.g.}},][]{2019ApJ...874L..15R,2020ApJS..246...60P}. As noted above, they have been used to interpret streamer sub-structure \citep{2020ApJS..246...60P} observed in WISPR observations, as well as during eclipses \citep{2018NatAs...2..913M}. Equally importantly, they have been used to connect remote solar observations with their {\emph{in~situ}} counterparts (\S\ref{LSSSWHCS}). Comparisons with white-light, but more importantly from emission images, provides crucial constraints for models that include realistic energy transport processes \citep{2019ApJ...872L..18V,2019ApJ...874L..15R}. They have already led to the improvement of coronal heating models \citep{2021A&A...650A..19R}, resulting in better matches with {\emph{in~situ}} measurements during multiple {\emph{PSP}} Encs. 

Images taken from a vantage point situated much closer to the Sun provide more detailed information on the population of transient structures released continually by helmet streamers. The fine-scale structure of streamer blobs is better resolved by WISPR than previous generations of heliospheric imagers. In addition the WISPR images have revealed that small-scale transients, with aspects that are reminiscent of magnetic islands and/or twisted 3D magnetic fields, are emitted at scales smaller than those of streamer blobs \citep{2019Natur.576..232H}. These very small flux ropes were identified {\emph{in situ}} as common structures during crossings of the HPS \citep{2019ApJ...882...51S} and more recently at {\emph{PSP}} \citep{2020ApJ...894L..19L}. They may also relate to the quasi-periodic structures detected by \citet{2015ApJ...807..176V} -- on-going research is evaluating this hypothesis. Recent MHD simulations have shown that the flux ropes observed in blobs and the magnetic islands between quasi-periodic increases in density could result from a single process known as the tearing-mode instability as the HCS is stretched by the adjacent out-flowing solar wind \citep{2020ApJ...895L..20R}.

\subsubsection{Coronal Mass Ejection Imaging with WISPR}
\begin{figure*}
    \centering
    \includegraphics[width=\textwidth]{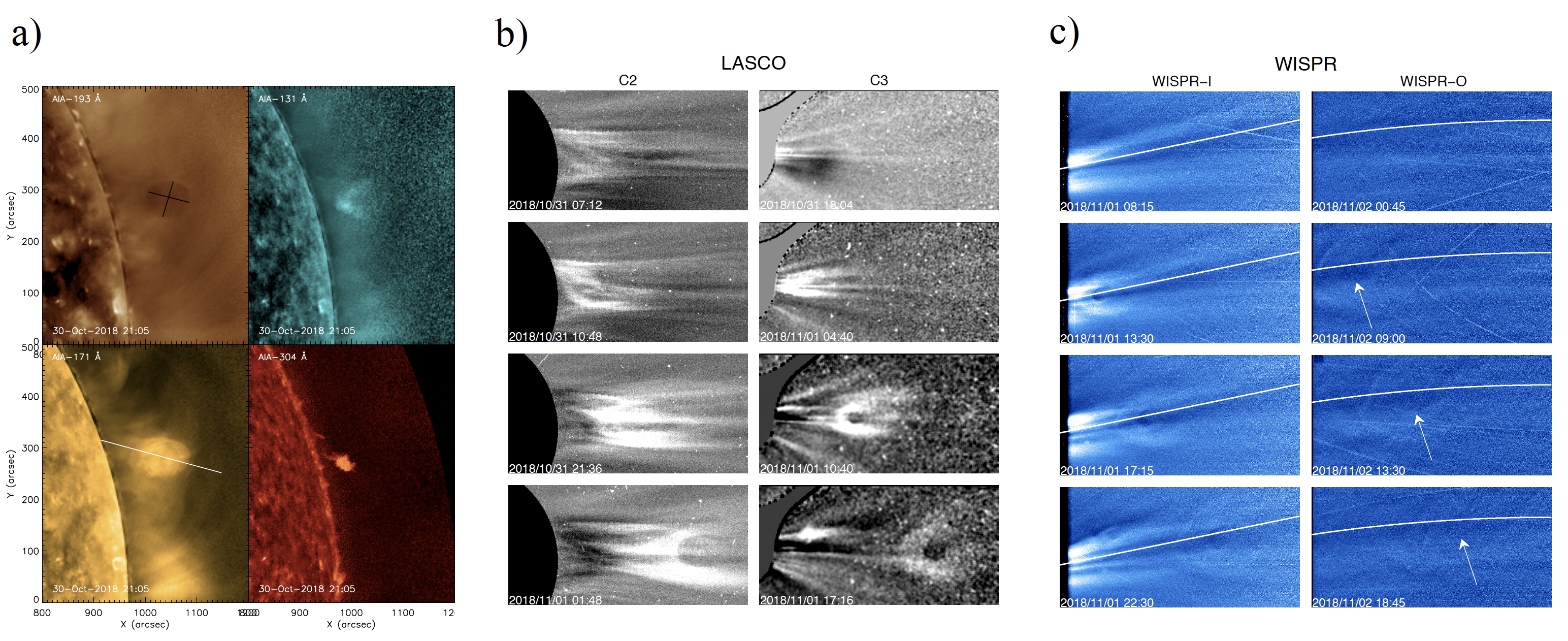}
    \caption{Multi-S/C observations of the first CME imaged by WISPR, on 1 Nov. 2018. (a) {\emph{SDO}}/AIA imaging of the CME. Different features are visible in each panel, including the dark, circular cavity (193 \AA; 1.6 MK and 20 MK), the bright trailing edge (131 \AA; 0.4 MK and 10 MK), a bright blob that is co-spatial with the cavity (171 \AA; 0.6 MK) and the prominence at the base of the eruption (304 \AA; 0.05 MK). The black line in the 193 \AA \ frame was used to calculate the size of the cavity. The white line in the 171 \AA \ frame is the approximate direction of motion and was used to measure the height and calculate the velocity of the cavity in AIA. (b) The CME as seen by the {\emph{SOHO}}/LASCO-C2 and -C3 coronagraphs. (c) The CME as seen by both {\emph{PSP}}/WISPR telescopes. The white line denotes the solar equatorial plane. The curvature of the line in WISPR-o is due to the distortion of the detector. Figure adapted from \citet{2020ApJS..246...25H}.}
    \label{FIG_NOV1}
\end{figure*}

Within a few hours of being turned on in preparation for the first {\emph{PSP}} perihelion passage, the WISPR imager observed its first CME. The WISPR cameras began taking science data at 00:00 UT on 1 Nov. 2018. By 11:00 UT a CME was visible in the inner telescope \citep{2020ApJS..246...25H}. Over the course of the next two days, the CME propagated along the solar equatorial plane throughout both WISPR telescopes, spanning $13.5^{\circ}-108.5^{\circ}$, with a speed of about 300~km~s$^{-1}$, consistent with SBO-CMEs \citep{2018ApJ...861..103V}. The WISPR observations are included in Fig.~\ref{FIG_NOV1}.

The CME was also observed from the Earth perspective by the {\emph{SDO}}/AIA EUV imager and the {\emph{SOHO}}/LASCO coronagraphs. In AIA, a small prominence was observed beneath a cavity, which slowly rose from the west limb in a non-radial direction. The cavity and prominence are both visible in the left panel Fig.~\ref{FIG_NOV1}. As this structure enters the LASCO-C2 FOV the cavity remains visible, as does a bright claw-like structure at its base, as seen throughout the middle panel of Fig.~\ref{FIG_NOV1}. The non-radial motion continues until the CME reaches the boundary of an overlying helmet streamer, at which point the CME is deflected out through the streamer along the solar equatorial plane. 

Because of the alignment of the S/C at the time of the eruption, WISPR was able to see the CME from a similar perspective as LASCO, but from a quarter of the distance. The inner FOV from WISPR was within the LASCO-C3 FOV, meaning that for a brief time WISPR and C3 observations were directly comparable. These direct comparisons demonstrate the improved resolution possible, even in a weaker event, from a closer observational position. This can be seen directly in Fig.~\ref{FIG_NOV1} in the LASCO frame at 17:16 UT and the WISPR-i frame at 17:15~UT.

The clarity of the observations of the CME cavity in WISPR allowed for tracking of the cavity out to $40~R{_\odot}$, as well as detailed modeling of the internal magnetic field of the CME \citep{2020ApJS..246...72R}. Both studies would have been impossible without the details provided by WISPR imaging.  

\begin{figure*}
   \centering
   \includegraphics[width=\textwidth]{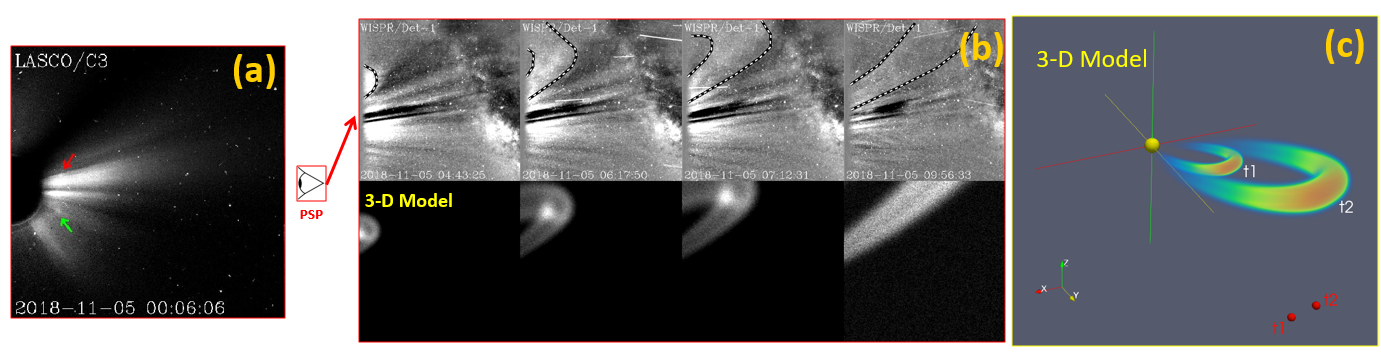}
   \caption{(a) A LASCO/C3 image from 5 Nov. 2018 showing two small streamer blobs marked by the two arrows.  The northern blob (red arrow) is observed by {\emph{PSP}}/WISPR during {\emph{PSP}}'s first perihelion passage.  (b) The upper panels are a sequence of four images from WISPR's inner detector of the northern streamer blob eruption from (a), with dotted lines outlining the transient.  Synthetic images are shown below the real images, based on the 3D reconstruction of the event.  (c) Reconstructed 3D flux rope structure of the streamer blob.  The flux rope is shown at two times, t1 and t2, corresponding to 03:48 UT and 09:40 UT, respectively.  The red circles indicate the location of {\emph{PSP}} at these two times, and the size of the Sun is to scale.  Figure adapted from \citet{2020ApJS..246...28W}.}
              \label{Wood2020Fig1}
\end{figure*} 

Another transient observed by WISPR during the first {\emph{PSP}} perihelion passage was a small eruption seen by the WISPR-i detector on 5 Nov. 2018, only a day before {\emph{PSP}}'s first close perihelion passage \citep{2020ApJS..246...28W}.  As shown in Fig.~\ref{Wood2020Fig1}(a), the LASCO/C3 coronagraph on board {\emph{SOHO}} observed two small jet-like eruptions on that day, with the northern of the two (red arrow) corresponding to the one observed by WISPR.  The appearance of the event from 1~AU is very consistent with the class of small transients called ``streamer blobs'' \citep{1997ApJ...484..472S,1998ApJ...498L.165W}, although it is also listed in catalogs of CMEs compiled from {\emph{SOHO}}/LASCO data, and so could also be described as a small CME.

At the time of the CME, {\emph{PSP}} was located just off the right side of the LASCO/C3 image in Fig.~\ref{Wood2020Fig1}a, lying almost perfectly in the C3 image plane.  The transient's appearance in WISPR images is very different than that provided by the LASCO/C3 perspective, being so much closer to both the Sun and the event itself.  This is the first close-up image of a streamer blob.  In the WISPR images in Fig.~\ref{Wood2020Fig1}b, the transient is not jet-like at all.  Instead, it looks very much like a flux rope, with two legs stretching back toward the Sun, although one of the legs of the flux rope mostly lies above the WISPR FOV. This leg basically passes over {\emph{PSP}} as the transient moves outward.

A 3D reconstruction of the flux rope morphology of the transient is shown in Fig.~\ref{Wood2020Fig1}c, based not only on the LASCO/C3 and {\emph{PSP}}/WISPR data, but also on images from the COR2 coronagraph on {\emph{STEREO}}-A, making this the first CME reconstruction performed based on images from three different perspectives that include one very close to the Sun.  Although typical of streamer blobs in appearance, a kinematic analysis of the 5 Nov. 2018 event reveals that it has a more impulsive acceleration than previously studied blobs.

\subsubsection{Analysis of WISPR Coronal Mass Ejections}

\label{intro}
The rapid, elliptical orbit of {\emph{PSP}} presents new challenges for the analysis of the white light images from WISPR due to the changing distance from the Sun.  While the FOV of WISPR’s two telescopes are fixed in angular size, the physical size of the coronal region imaged changes dramatically, as discussed in \citet{2019SoPh..294...93L}.  In addition, because of {\emph{PSP}}’s rapid motion in solar longitude, the projected latitude of a feature changes, as seen by WISPR, even if the feature has a constant heliocentric latitude. Because of these effects, techniques used in the past for studying the kinematics of solar ejecta may no longer be sufficient. The motion observed in the images is now a combination of the motion of the ejecta and of the S/C. On the other hand, the rapid motion gives multiple view points of coronal features and this can be exploited using triangulation. Prior to launch, synthetic white light WISPR images, created using the sophisticated ray-tracing software \citep{2009SoPh..256..111T}, were used to develop new techniques for analyzing observed motions of ejecta. \citet{2020SoPh..295...63N} performed extensive studies of the evolution of the brightness due to the motion of both the S/C and the feature. They concluded that the total brightness evolution could be exploited to obtain a more precise triangulation of the observed features than might be possible otherwise.

\begin{figure}
  \includegraphics[width=\textwidth]{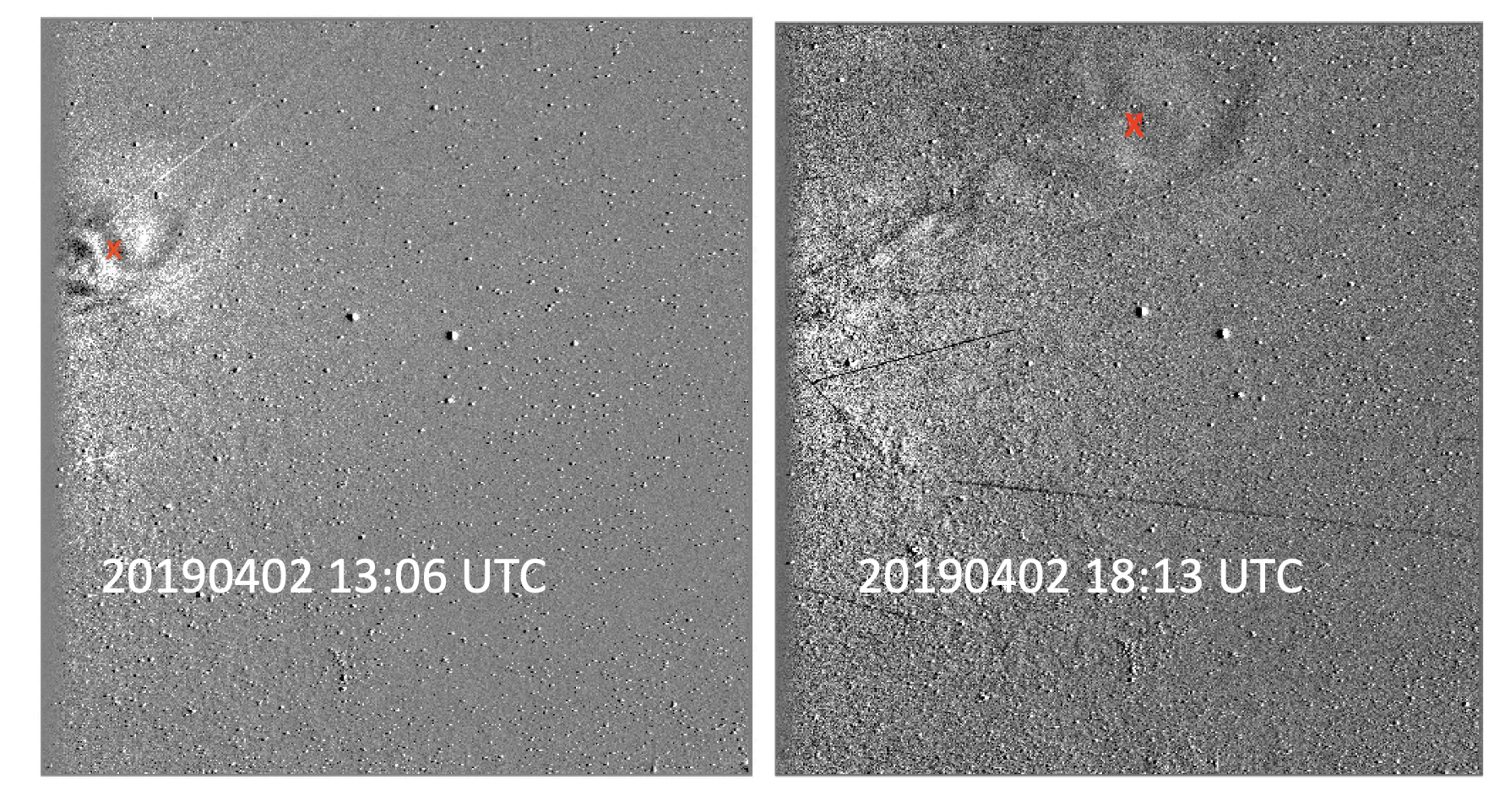}
 \caption{WISPR-i running-difference images at two times for the CME of 2 Apr. 2019 showing the tracked feature, the lower dark ``eye" (marked with red X's). The image covers approximately $13.5^{\circ} - 53.0^{\circ}$ elongation from the Sun center. The streaks seen in the images are due to reflections off debris created by dust impacts on the {\emph{PSP}} S/C.} 
\label{figPCL1}
\end{figure}

\smallskip
\paragraph{{\textbf{Tracking and Fitting Technique for Trajectory Determination}}} $\\$
\citet{2020SoPh..295..140L} developed a technique for determining the trajectories of CMEs and other ejecta that takes into account the rapid motion of {\emph{PSP}}. The technique assumes that the ejecta, treated as a point, moves radially at a constant velocity. This technique builds on techniques developed for the analysis of J-maps \citep{1999JGR...10424739S} created from LASCO and SECCHI white light images.  For ejecta moving radially at a constant velocity in a heliocentric frame, there are four trajectory parameters: longitude, latitude, velocity and radius (distance from the Sun) at some time $t_0$. Viewed from the S/C, the ejecta is seen to change position in a time sequence of images. The position in the image can be defined by two angles that specify the telescope’s LOS at that pixel location. We use a projective cartesian observer-centric frame of reference that is defined by the Sun-{\emph{PSP}} vector and the {\emph{PSP}} orbit plane. One angle ($\gamma$) measures the angle from the Sun parallel to the {\emph{PSP}} orbit plane and the second angle ($\beta$) measures the angle out of the orbit plane. We call this coordinate system the {\emph{PSP}} orbit frame. Using basic trigonometry, two equations were derived relating the coordinates in the heliocentric frame to those measured in the S/C frame ($\gamma$, $\beta$) as a function of time.  The geometry relating the ejecta's coordinates in the two frames is shown in Fig.~1 of \citet{2020SoPh..295..140L} for the case with the inclination of {\emph{PSP}}’s orbit plane w.r.t. the solar equatorial plane neglected. The coordinates of the S/C are  $[r_1, \phi_1, 0]$, and the coordinates of the ejecta are $[r_2, \phi_2, \delta_2]$.    The two equations are
\begin{equation}
\frac{\tan\beta (t)}{\sin\gamma (t)} = \frac{\tan\delta_2}{\sin[\phi_2 - \phi_1 (t)]},
\end{equation}
\begin{equation}
\cot\gamma(t) = \frac{r_1(t) - r_2(t)\cos\delta_2 \cos[\phi_2 - \phi_1(t)]}{r_2(t)\cos\delta_2 \sin[\phi_2 - \phi_1(t)]}.
\end{equation}

By tracking the point ejecta in a time sequence of WISPR images, we generate a set of angular coordinates $[\gamma(t_i), \beta(t_i)]$  for the ejecta in the {\emph{PSP}} orbit frame. In principle, ejecta coordinates in the S/C frame are only needed at two times to solve the above two equations for the four unknown trajectory parameters. However, we obtain more accurate results by tracking the ejecta in considerably more than two images. The ejecta trajectory parameters in the heliocentric frame are determined by fitting the above equations to the tracking data points $[\gamma(t_i), \beta(t_i)]$. Our fitting technique is described in \citet{2020SoPh..295..140L}, which also gives the equations with the corrections for the inclination of {\emph{PSP}}’s orbit w.r.t. the solar equatorial plane.

\begin{figure}
  \includegraphics[width=0.66\textwidth]{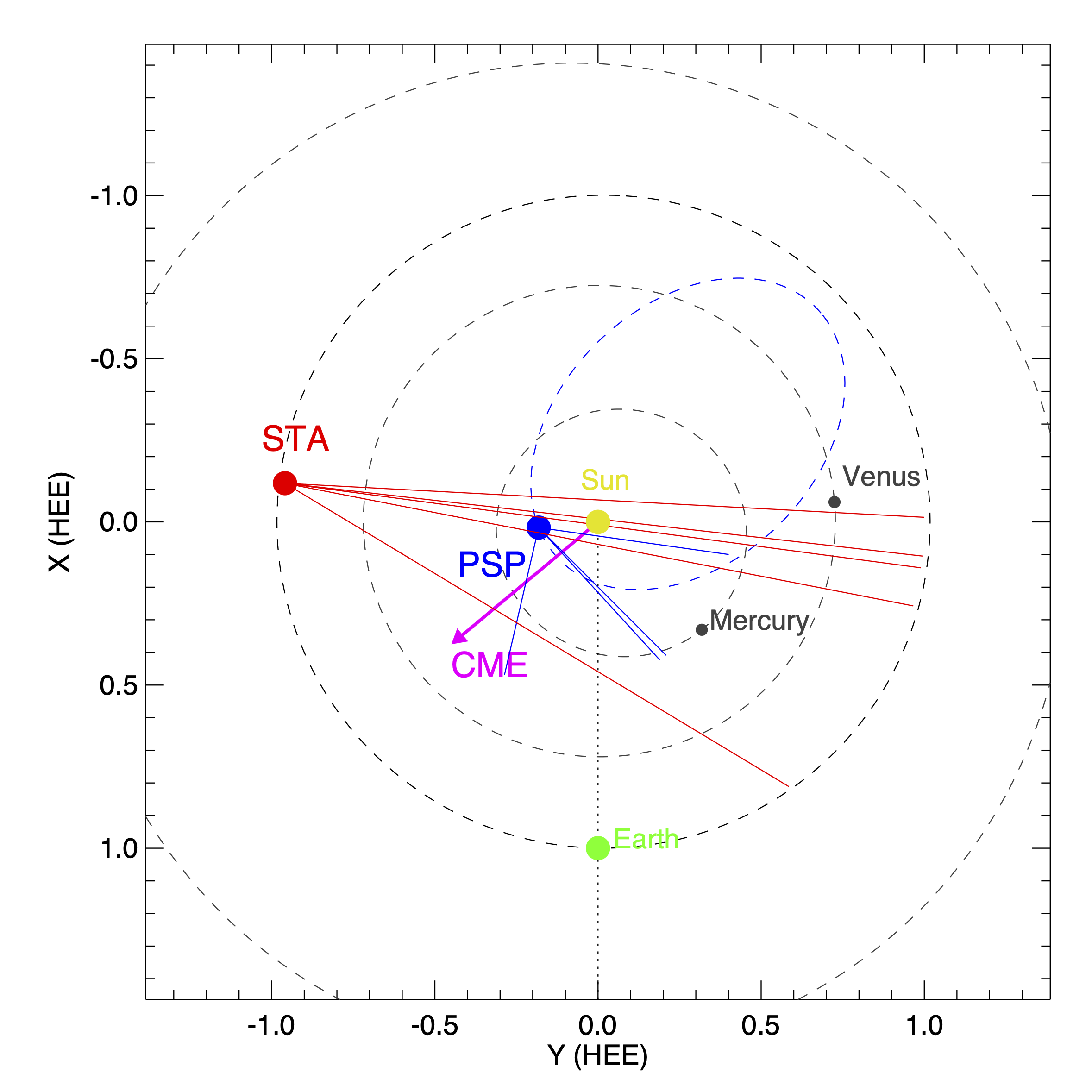}
 \caption{Trajectory solution for the 2 Apr. 2019 flux rope  (magenta arrow) shown in relation to {\emph{PSP}}, {\emph{STEREO}}-A, and Earth at 18:09 UT. The fine solid lines indicate the fields-of-view of the telescopes on {\emph{PSP}} and {\emph{STEREO-A}}. The CME direction was found to be  HCI longitude $= 67^{\circ} \pm 1^{\circ}$. Note that the arrow only indicates the direction of the CME, and it is not meant to indicate the distance from the Sun. The HCI longitudes of the Earth and {\emph{STEREO}}-A are 117$^{\circ}$ and 19$^{\circ}$, respectively. The blue dashed ellipse is {\emph{PSP}}'s orbit. The plot is in the Heliocentric Earth Ecliptic (HEE) reference frame and distances are in AU.}
\label{figPCL2}       
\end{figure}

\begin{figure}
  \includegraphics[width=.66\textwidth]{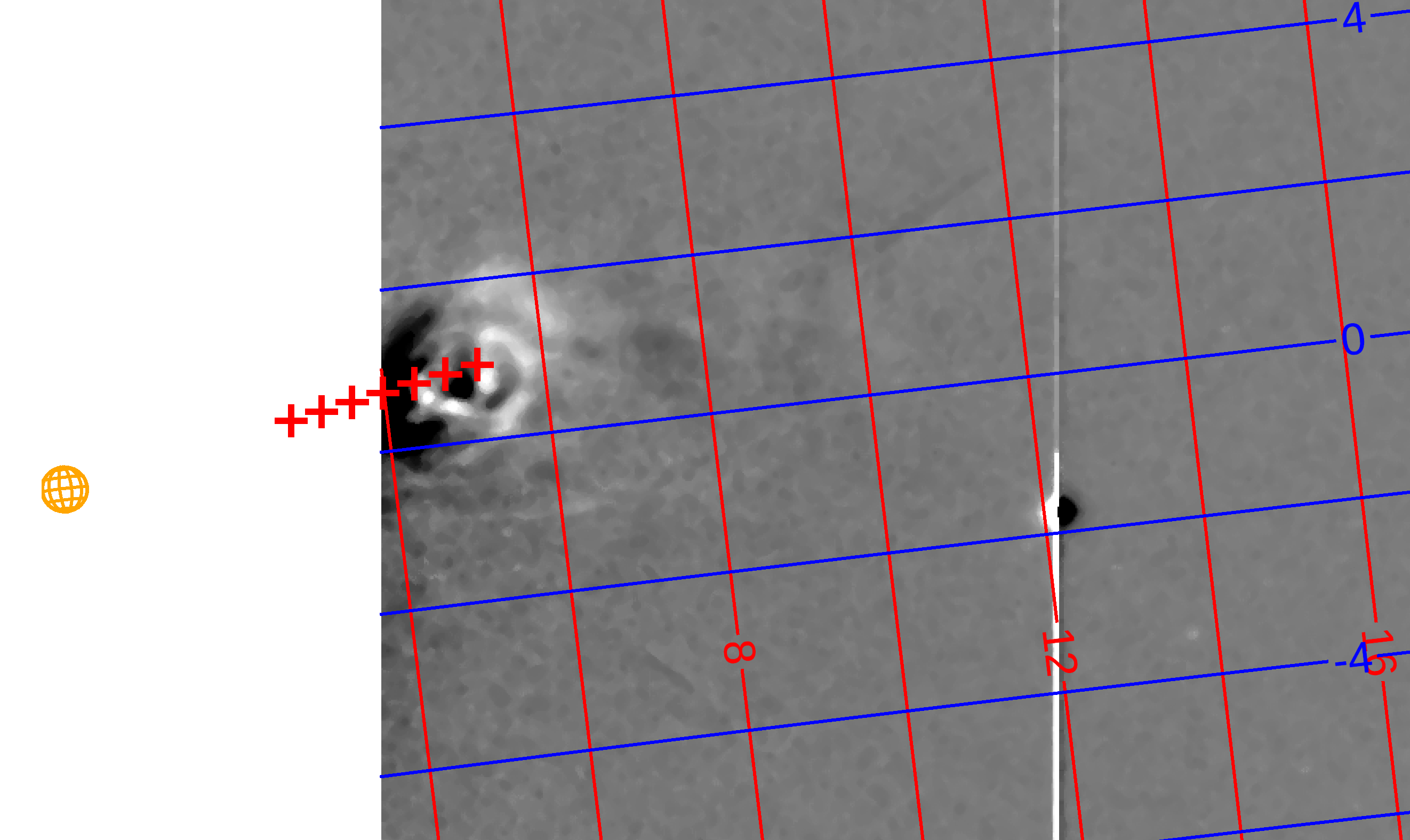}
 \caption{Trajectory of the flux rope of 2 Apr. 2019, found from the WISPR data using the tracking and fitting technique, projected to images from {\emph{STEREO}}-A/HI-1 at 18:09 UT. The trajectory from tracking and fitting  ({\textcolor{red}{\bf{+}}} signs)  is shown from 12:09 to 18:09 UT in hourly increments, as seen from {\emph{STEREO}}-A. The location of the prediction for the last time (18:09 UT) is in good agreement with the location of the tracked feature seen in the HI-1A image, thus verifying the trajectory. The grid lines are the coordinate lines of the WCS frame specified in the HI-1A FITS header. The size and location of the Sun (yellow globe) is shown to scale.}
\label{figPCL3}       
\end{figure}

The tracking and fitting technique was applied to a small CME seen by WISPR in the second solar Enc. on 1-2 Apr. 2019 \citep{2020SoPh..295..140L}. Fig.~\ref{figPCL1} shows two of the WISPR images used in the tracking; the feature tracked, an eye-like dark oval, is shown as the red X. The direction of the trajectory found for this CME is indicated with an arrow in Fig.~\ref{figPCL2}, which also shows the location and fields of view of {\emph{STEREO}}-A and {\emph{PSP}}. The trajectory solution in heliocentric inertial (HCI) coordinates was longitude $ = 67 \pm 1^{\circ}$, latitude $= 6.0\pm 0.3^{\circ}$, $V = 333 \pm 1$~km~s$^{-1}$, and $r_{2}(t_0) = 13.38 \pm 0.01$\,R$_{\odot}$,  where $t_0 $= 12:09 UT on 2 Apr. 2019. There were simultaneous observations of the CME from {\emph{STEREO}}-A and {\emph{PSP}}, which enabled us to use the second viewpoint observation of {\emph{STEREO}}-A/HI-1 to verify the technique and the results.  This was done by generating a set of 3D trajectory points using the fitting solution above that included the time of the {\emph{STEREO}}-A/HI-1 observations. These trajectory points were then projected  onto an image from HI-1A using the World Coordinate System (WCS) information in the Flexible Image Transport System (FITS) image header. This is illustrated in Fig.~\ref{figPCL3}, which shows the trajectory points generated from the solutions from 12:09 to 18:09 UT in hourly increments projected onto the HI-1A image at 18:09 UT on 2 Apr. 2019. Note that the last point, corresponding to the time of the HI-1A image, falls quite near a similar feature on the CME as was tracked in the WISPR images (Fig.~\ref{figPCL1}).  Thus, the trajectory determined from the WISPR data agrees with the {\emph{STEREO}}-A observations from a second view point. This technique was also applied to the first CME seen by WISPR on 2 Nov. 2018. Details of the tracking and the results are in \citet{2020SoPh..295..140L}, and independent analyses of the CME kinematics and trajectory for the 2 Apr. 2019 event were carried out by \citet{2021A&A...650A..31B} and \citet{2021ApJ...922..234W}, with similar results.

\begin{figure}
  \includegraphics[width=.66\textwidth]{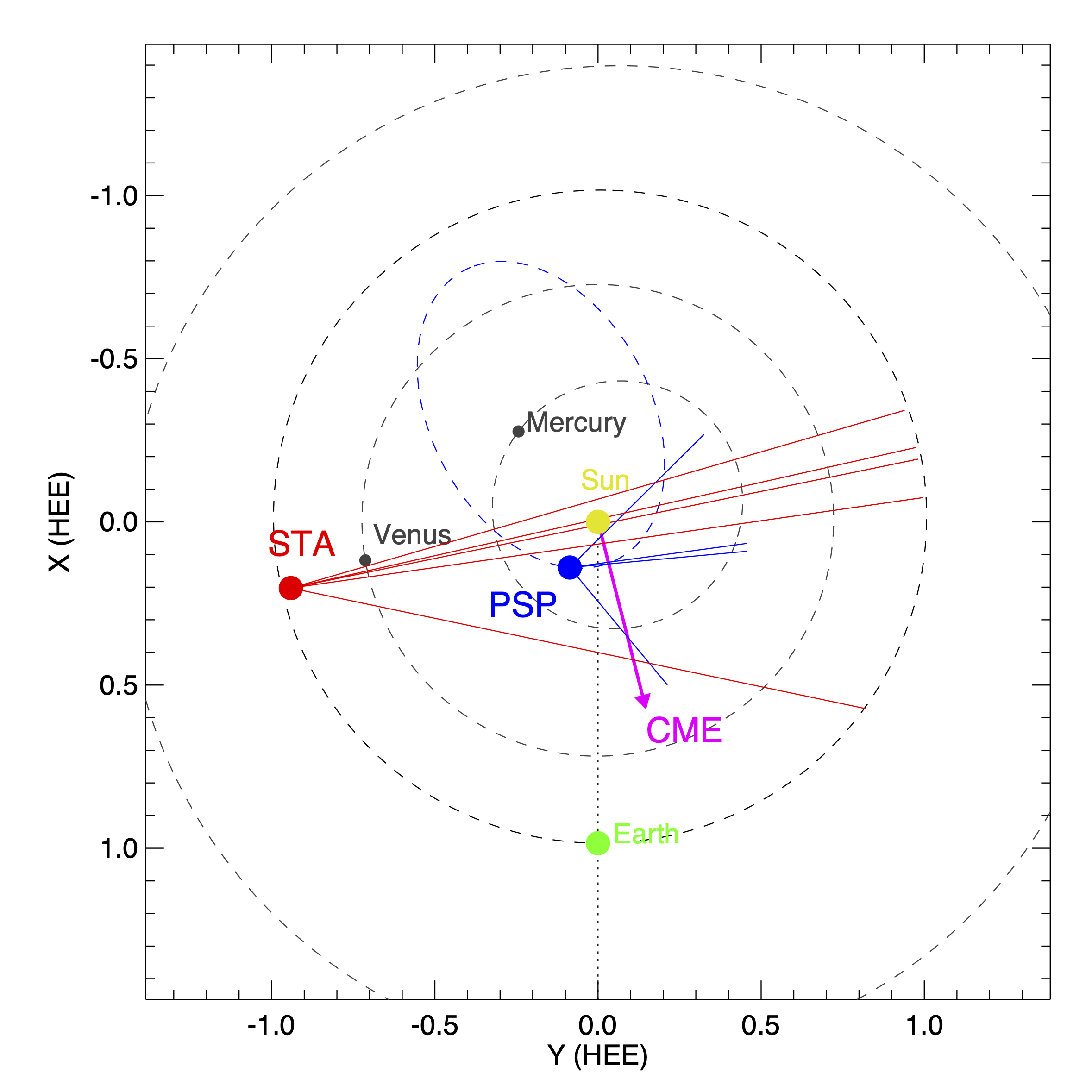}
 \caption{ Trajectory of the 26-27 Jan. 2020 CME  (magenta arrow) in relation to {\emph{PSP}}, {\emph{STEREO}}-A, and Earth, projected in the HEE reference frame on 26 Jan. 2020, at 20:49 UT. The fields-of-view of the WISPR-i and WISPR-o telescopes on {\emph{PSP}} and COR2 and HI-1 on {\emph{STEREO}}-A are indicated by solid lines. The plot is in the HEE coordinate frame and distances are in AU.}
\label{figPCL4}       
\end{figure}

\begin{figure}
  \includegraphics[width=\textwidth]{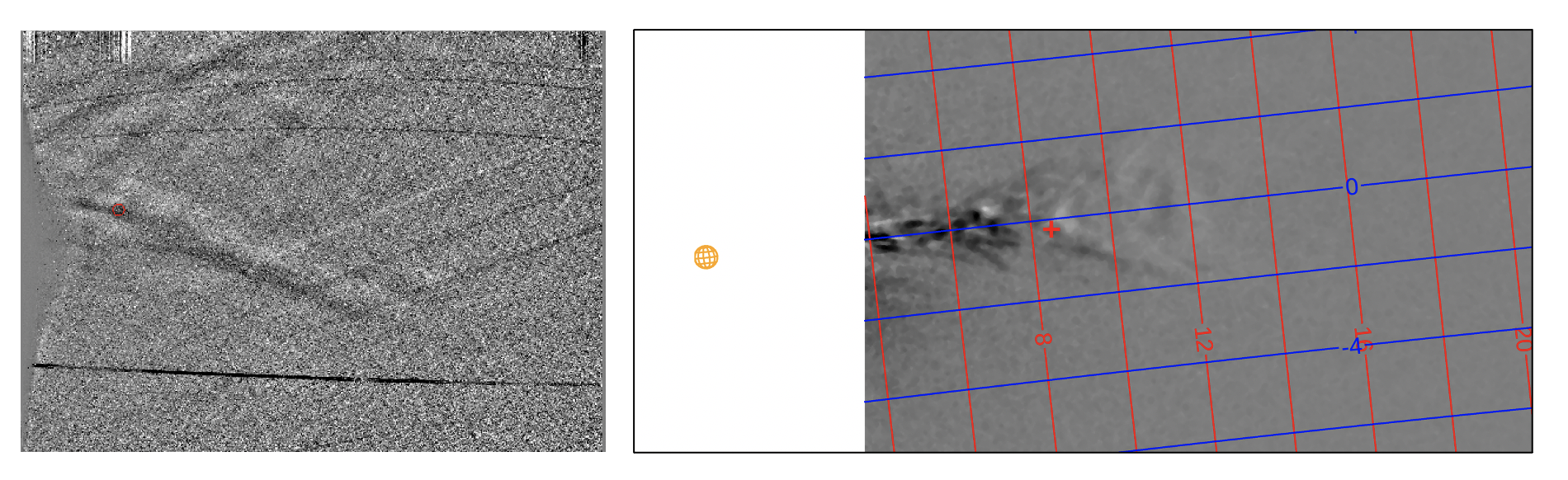}
\caption{Image pair used to determine the location of the 26-27 Jan. 2020 CME by triangulation. Left: WISPR-o image of 26 Jan. 2020, at 20:49 UT with the selected feature circled in red.Right: Simultaneous HI-1 image showing the location of the same feature identified in the WISPR-o image. The location of the CME found by triangulation of this feature was in excellent agreement with the trajectory found from tracking and fitting (see core text).  The Sun (yellow globe) in the panel on the right is shown to scale. The HI-1 image is projected in the Helioprojective Cartesian (HPC) system (red and blue grid lines) with the Sun (yellow globe) drawn to scale.}
\label{figPCL5}       
\end{figure}

\smallskip
\paragraph{{\textbf{WISPR and {\emph{STEREO}} Observations of the Evolution of a Streamer Blowout CME}}} $\\$
WISPR obtained detailed images of the flux rope structure of a CME on 26-27 Jan. 2020. The tracking and fitting procedure was also used here to determine the trajectory. The direction of the trajectory is shown in Fig.~\ref{figPCL4}, along with locations and FOVs of {\emph{STEREO}} A and WISPR. The trajectory solution parameters are HCI longitude and latitude = (65$^{\circ} \pm $2$^{\circ}$, 2$^{\circ} \pm$ 2$^{\circ}$), $v=248 \pm 16$~km~s$^{-1}$ and  $r_{2}(t_0$)/R$_{\odot}$= $30.3 \pm 0.3$ at $t_0$ = 20:04 UT on 26 Jan. 2020. The CME was also observed by {\emph{STEREO}}-A/COR2 starting on 25 Jan. 2020. The brightening of the streamer before the eruption, the lack of a bright leading edge, and the outflow following the ejecta led to its identification as a SBO-CME \citep{2021A&A...650A..32L}.  Data from {\emph{STEREO}}-A/EUVI suggested that this CME originated as a rising flux rope on 23 Jan. 2020, which was constrained in the corona until its eruption of 25 Jan. 2020. The detail of the observations and the supporting data from AIA and the Helioseismic and Magnetic Imager \citep[HMI;][]{2012SoPh..275..207S} on {\emph{SDO}} can be found in \citet{2021A&A...650A..32L}. The direction determined from the tracking and fitting was consistent with this interpretation; the direction was approximately $30^{\circ}$  west of a new AR, which was also a possible source. 

To verify the CME’s trajectory determined by tracking and fitting, we again made use of simultaneous observations from {\emph{STEREO}}-A, but in this case we used triangulation to determine the 3D location of the CME at the time of a simultaneous image pair. This was only possible because details of the structure of the CME were evident from both viewpoints so that the same feature could be located in both images. Fig.~\ref{figPCL5} shows the simultaneous WISPR-o and HI-1A images of the CME at time 26 Jan. 2020, at 20:49 UT when the S/C were separated by 46$^{\circ}$. The red X in each image marks the what we identify as the same feature in both images (a dark spot behind the bright V). Using a triangulation technique on this image pair  \citet{2021A&A...650A..32L} gave the result distance from the Sun $r/R_{\odot}$ = 31$ \pm 2$, HCI longitude and latitude = ( $66^{\circ}\pm  3^{\circ}$,  $-2^{\circ} \pm 2^{\circ}$). These angles are in excellent agreement with the longitude found by tracking and fitting given above. The distance from the Sun is also in excellent agreement with the predicted distance at this time of $r_2/R_{\odot}$= $31.2 \pm 0.3$, validating our trajectory solution. Thus the trajectory was confirmed, which further supported our interpretation of the evolution of this slowly evolving SBO-CME.

\section{Solar Radio Emission}
\label{SRadE}

At low frequencies, below $\sim10-20$ MHz, radio emission cannot be observed well from ground-based observatories due to the terrestrial ionosphere. Solar radio emission at these frequencies consists of radio bursts, which are signatures of the acceleration and propagation of non-thermal electrons. Type II and type III radio bursts are commonly observed, with the former resulting from electrons accelerated at shock fronts associated with CMEs, and the latter from electron beams accelerated by solar flares (see Fig.~\ref{Wiedenbeck2020Fig}C).

Solar radio bursts offer information on the kinematics of the propagating source, and are a remote probe of the properties of the local plasma through which the source is propagating. Radio observations on {\emph{PSP}} are made by the FIELDS RFS, which measures electric fields from 10 kHz to 19.2 MHz \citep{2017JGRA..122.2836P}. At frequencies below and near $f_{pe}$, the RFS measurements are dominated by the quasi-thermal noise (QTN).

{\emph{PSP}} launched at solar minimum, when the occurrence rate of solar radio bursts is relatively low. Several {\emph{PSP}} Encs. (Enc.~1, Enc.~3, Enc.~4) near the start of the mission were very quiet in radio, containing only a few small type III bursts. The second {\emph{PSP}} Enc. (Enc.~2), in Apr. 2019, was a notable exception, featuring multiple strong type III radio bursts and a type III storm \citep{2020ApJS..246...49P}. As solar activity began rising in late 2020 and 2021, with Encs.~5 and beyond, the occurrence of radio bursts is also increasing. Taking advantage of the quiet Encs. near the start of the mission, \citet{ChhabraThesis} applied a correlation technique developed by \citet{2013ApJ...771..115V} for imaging data to RFS light curves, searching for evidence of heating of the coronal by small-scale nanoflares which are too faint to appear to the eye in RFS spectrograms.

During {\emph{PSP}} Encs., the cadence of RFS spectra range is typically 3.5 or 7 seconds, which is higher than the typical cadence of radio spectra available from previous S/C such as {\emph{STEREO}} and {\emph{Wind}}. The relatively high cadence of RFS data is particularly useful in the study of type III radio bursts above 1 MHz (in the RFS High Frequency Receiver [HFR] range), which typically last $\lesssim1$~minute at these frequencies. Using the HFR data enabled \citet{2020ApJS..246...49P} to measure circular polarization near the start of several type III bursts in Enc.~2. \citet{2020ApJS..246...57K} characterized the decay times of type III radio bursts up to 10 MHz, observing increased decay times above 1 MHz compared to extrapolation using previous measurements from {\emph{STEREO}}. Modeling suggests that these decay times may correspond to increased density fluctuations near the Alfv\'en point.

Recent studies have used RFS data to investigate basic properties of type III bursts, in comparison to previous observations and theories. \cite{2021ApJ...915L..22C} examined a single Enc.~2 type IIIb burst featuring fine structure (striae) in detail. \cite{2021ApJ...915L..22C} found consistency between RFS observations and results of a model with emission generated via the electron cyclotron maser instability \citep{2004ApJ...605..503W}, over the several-MHz frequency range corresponding to solar distances where $f_{ce} > f_{pe}$.

\cite{2021ApJ...913L...1M} performed a statistical survey of the lower cutoff frequency of type III bursts using the first five {\emph{PSP}} Encs., finding a higher average cutoff frequency than previous observations from {\emph{Ulysses}} and {\emph{Wind}}. \cite{2021ApJ...913L...1M} proposed several explanations for this discrepancy, including solar cycle and event selection effects.

The launch of {\emph{SolO}} in Feb. 2020 marked the first time four S/C ({\emph{Wind}}, {\emph{STEREO}}-A, {\emph{PSP}}, and {\emph{SolO}}) with radio instrumentation were operational in the inner heliosphere. \cite{2021A&A...656A..34M} combined observations from these four S/C along with a model of the burst emission to determine the directivity of individual type III radio bursts, a measurement previously only possible using statistical analysis of large numbers of bursts.

\section{Energetic Particles}
\label{EPsRad}

The first four years of the {\emph{PSP}} mission have provided key insights into the acceleration and transport of energetic particles in the inner heliosphere and has enabled a comprehensive understanding into the variability of solar radio emission. {\emph{PSP}} observed a multitude of solar radio emissions, SEP events, CMEs, CIRs and SIRs, inner heliospheric anomalous cosmic rays (ACRs), and energetic electron events; all of which are critical to explore the fundamental physics of particle acceleration and transport in the near-Sun environment and throughout the universe.

\subsection{Solar Energetic Particles}
\label{SEPs}

On 2 and 4 Apr. 2019, {\emph{PSP}} observed two small SEP events \citep{2020ApJS..246...35L, 2020ApJ...899..107K, 2020ApJ...898...16Z} while at $\sim0.17$~AU (Fig.~\ref{Leske2020Fig}). The event on 4 Apr. 2019 was associated with both a type III radio emission seen by {\emph{PSP}} as well as surges in the EUV observed by {\emph{STEREO}}-A, all of which determined the source was an AR $\sim80^{\circ}$ east of the {\emph{PSP}} footpoint \citep{2020ApJS..246...35L}. To better understand the origin of these SEP events, \citet{2020ApJ...899..107K} conducted a series of simulations constrained by remote sensing observations from {\emph{SDO}}/AIA, {\emph{STEREO}}-A/EUVI and COR2, {\emph{SOHO}}/LASCO, and {\emph{PSP}}/WISPR to determine the magnetic connectivity of {\emph{PSP}}, model the 3D structure and evolution of the EUV waves, investigate possible shock formation, and connect these simulations to the SEP observations. This robust simulation work suggests that the SEP events were from multiple ejections from AR 12738. The 2 Apr. 2019 event likely originated from two ejections that formed a shock in the lower corona \citep{2020ApJ...899..107K}. Meanwhile, the 4 Apr. 2019 event was likely the result of a slow SBO, which reconfigured the global magnetic topology to be conducive for transport of solar particles away from the AR toward {\emph{PSP}}. Interestingly, however, \citet{2020ApJS..246...35L} did not observe \textsuperscript{3}He for this event, as would be expected from flare-related SEPs. \citet{2020ApJ...898...16Z} explained gradual rise of the 4 Apr. 2019 low-energy H\textsuperscript{+} event compared to the more energetic enhancement on 2 Apr. 2019 as being indicative of different diffusion conditions.

\begin{figure*}
   \centering
   \includegraphics[width=\textwidth]{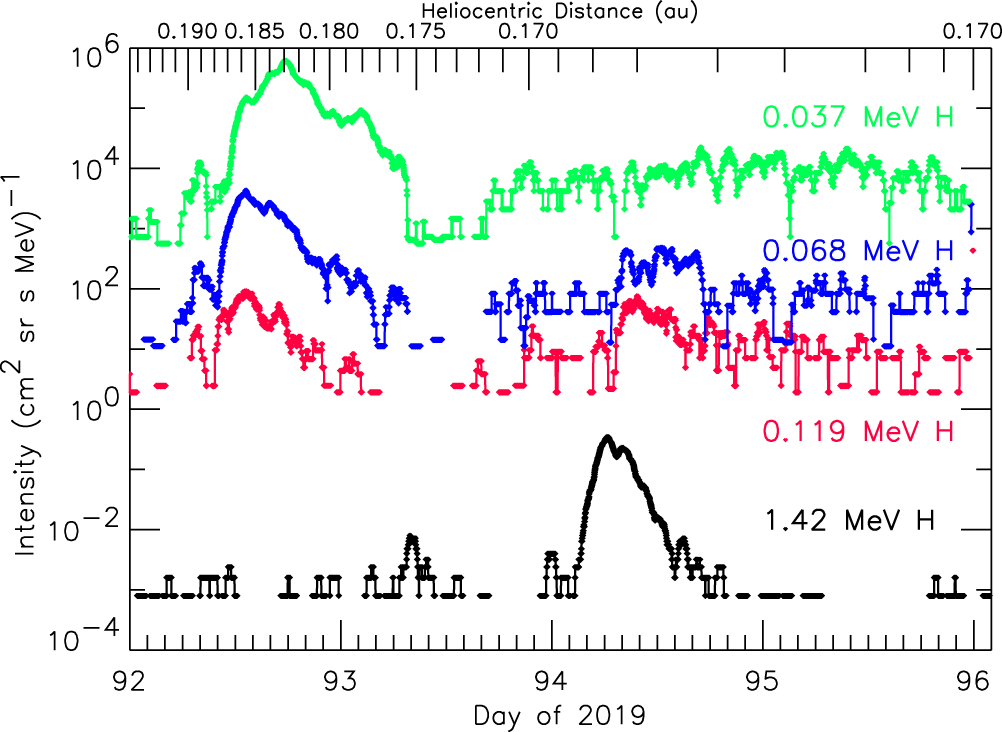}
   \caption{IS$\odot$IS/EPI-Lo time-of-flight measurements for the two SEP events on 2 Apr. 2019 (DOY 92) and 4 Apr. 2019 (DOY 94) are shown in green, blue, and red for the stated energies. IS$\odot$IS/EPI-Hi/LET1 observations are shown in black. Figure adapted from \citet{2020ApJS..246...35L}.}
              \label{Leske2020Fig}
\end{figure*} 

The same AR (AR 12738) was later responsible for a \textsuperscript{3}He-rich SEP event on 20-21 Apr. 2019 observed by {\emph{PSP}} at $\sim0.46$~AU that was also measured by {\emph{SOHO}} at $\sim1$~AU (shown in Fig.~\ref{Wiedenbeck2020Fig}) \citep{2020ApJS..246...42W}. This SEP event was observed along with type III radio bursts and helical jets. The \textsuperscript{3}He/\textsuperscript{4}He ratios at {\emph{PSP}} and {\emph{SOHO}} were $\sim250$ times the nominal solar wind ratio; such large enhancements are often seen in impulsive SEP events. This event demonstrated the utility of IS$\odot$IS/EPI-Hi to contribute to our understanding of the radial evolution of \textsuperscript{3}He-rich SEP events, which can help constrain studies of potential limits on the amount of \textsuperscript{3}He that can be accelerated by an AR \citep[{\emph{e.g.}},][]{2005ApJ...621L.141H}.

\begin{figure*}
   \centering
   \includegraphics[width=\textwidth]{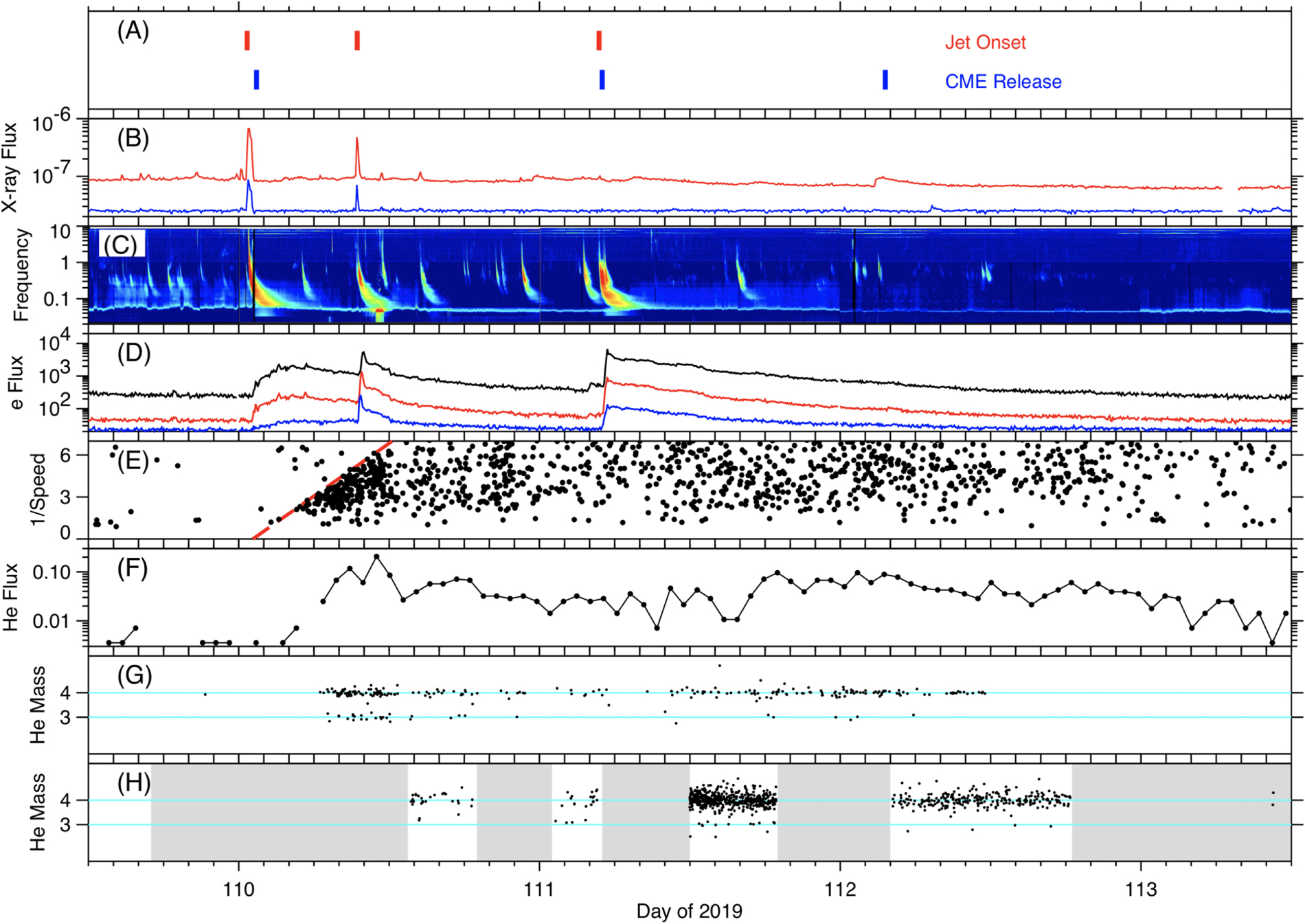}
   \caption{Remote and {\emph{in situ}} observations for the 20-21 Apr. 2019 \textsuperscript{3}He-rich SEP event. (a) Jet onset times and CME release times as reported by \citet{2020ApJS..246...33S}, (b) 5-min 0.05-0.4 nm (blue) and $0.1-0.8$ nm (red) X-ray flux from the Geostationary Operational Environmental Satellite ({\emph{GOES}}), (c) radio emissions from {\emph{Wind}}/WAVES \citep{1995SSRv...71..231B}, (d) electron fluxes for 53 (black), 79 (red), and 133 (blue) keV from the {\emph{ACE}} Electron Proton Alpha Monitor \citep[EPAM;][]{1998SSRv...86..541G}, (e) velocity dispersion with red line indicating the dispersion slope from the {\emph{ACE}} Ultra Low Energy Isotope Spectrometer \citep[ULEIS;][]{1998SSRv...86..409M}, (f) {\emph{ACE}}/ULEIS 1 MeV He flux, (g) {\emph{ACE}}/ULEIS He mass vs. time, and (h) {\emph{PSP}}/IS$\odot$IS/EPI-Hi mass vs. time. Grey boxes in panel (h) indicate times without IS$\odot$IS observations. Figure adapted from \citet{2020ApJS..246...42W}.}
              \label{Wiedenbeck2020Fig}
\end{figure*} 

\citet{2021A&A...650A..23C} later investigated the helium content of six SEP events from May to Jun. 2020 during the fifth orbit of {\emph{PSP}}. These events demonstrated that SEP events, even from the same AR, can have significantly different \textsuperscript{3}He/\textsuperscript{4}He and He/H ratios. Additionally, EUV and coronagraph observations of these events suggest that the SEPs were accelerated very low in the corona. Using velocity-dispersion analysis, \citet{2021A&A...650A..26C} concluded that the path length of these SEP events to the source was $\sim0.625$~AU, greatly exceeding that of a simple Parker spiral. To explain the large path length of these particles, \citet{2021A&A...650A..26C} developed an approach to estimate how the random-walk of magnetic field lines could affect particle path length, which well explained the computed path length from the velocity-dispersion analysis. 
        
During the first orbit of {\emph{PSP}}, shortly after the first perihelion pass, a CME was observed locally at {\emph{PSP}}, which was preceded by a significant enhancement in SEPs with energies below a few hundred keV/nuc \cite[][]{2020ApJS..246...29G,2020ApJS..246...59M}. The CME was observed to cross {\emph{PSP}} on 12 Nov. 2018 (DOY 316), and at this time, {\emph{PSP}} was approximately 0.23~AU from the Sun. The CME was observed remotely by {\emph{STEREO}}-A which was in a position to observe the CME erupting from the east limb of the Sun (w.r.t. {\emph{STEREO}}-A) and moving directly towards {\emph{PSP}}. {\emph{PSP}} was on the opposite side of the Sun relative to Earth. Through analysis of {\emph{STEREO}}-A coronagraph images, the speed of the CME was determined to be 360~km~s$^{-1}$, which is slower than typical SEP-producing CMEs seen by S/C near 1~AU. Moreover, in the few days that preceded the CME, there were very few energetic particles observed, representing a very quiet period. Thus, this represented a unique observation of energetic particles associated with a slow CME near the Sun.

Fig.~\ref{Giacalone_2020} shows a multi-instrument, multi-panel plot of data collected during this slow-CME/SEP event. Fig.~\ref{Giacalone_2020}a shows the position of the CME as a function of time based on {\emph{STEREO}}-A observations as well as {\emph{PSP}} (the cyan point), while the lower Fig.~\ref{Giacalone_2020}e-f shows $30-300$~keV energetic particles from the IS$\odot$IS/EPI-Lo instrument. The CME was observed to erupt and move away from the Sun well before the start of the SEP event, but the SEP intensities rose from the background, peaked, and then decayed before the CME crossed {\emph{PSP}}. There was no shock observed locally at {\emph{PSP}}, and there is no clear evidence of local acceleration of SEPs at the CME crossing. It was suggested by \citet{2020ApJS..246...29G} that the CME briefly drove either a weak shock or a plasma compression when it was closer to the Sun, and was capable of accelerating particles which then propagated ahead of the CME and observed by {\emph{PSP}}. In fact, modeling of the CME and local plasma parameters, also presented in this paper, suggested there may have been a weak shock over parts of the (modeled) CME-shock surface, but is not clear whether {\emph{PSP}} was magnetically connected to these locations. The energetic particle event was characterized by a clear velocity dispersion in which higher-energy particles arrived well before the lower energy particles. Moreover, the time-intensity profiles at specific energies, seen in Fig.~\ref{Giacalone_2020}e, show a relatively smooth rise from the background to the peak, and a gradual decay. The particles were observed to be initially anisotropic, moving radially away from the Sun, but at the peak of the event were observed to be more isotropic. \citet{2020ApJS..246...29G} interpreted this in terms of the diffusive transport of particles accelerated by the CME starting about the time it was 7.5~$R_\odot$ and continuing with time but with a decreasing intensity. They used a diffusive-transport model and fit the observed time-intensity profiles, which gave values for the scattering mean-free path, parallel to the magnetic field, of 30~keV to 100~keV protons to be from $0.04-0.09$~AU at the location of {\emph{PSP}}.

\begin{figure*}
   \centering
   \includegraphics[width=\textwidth]{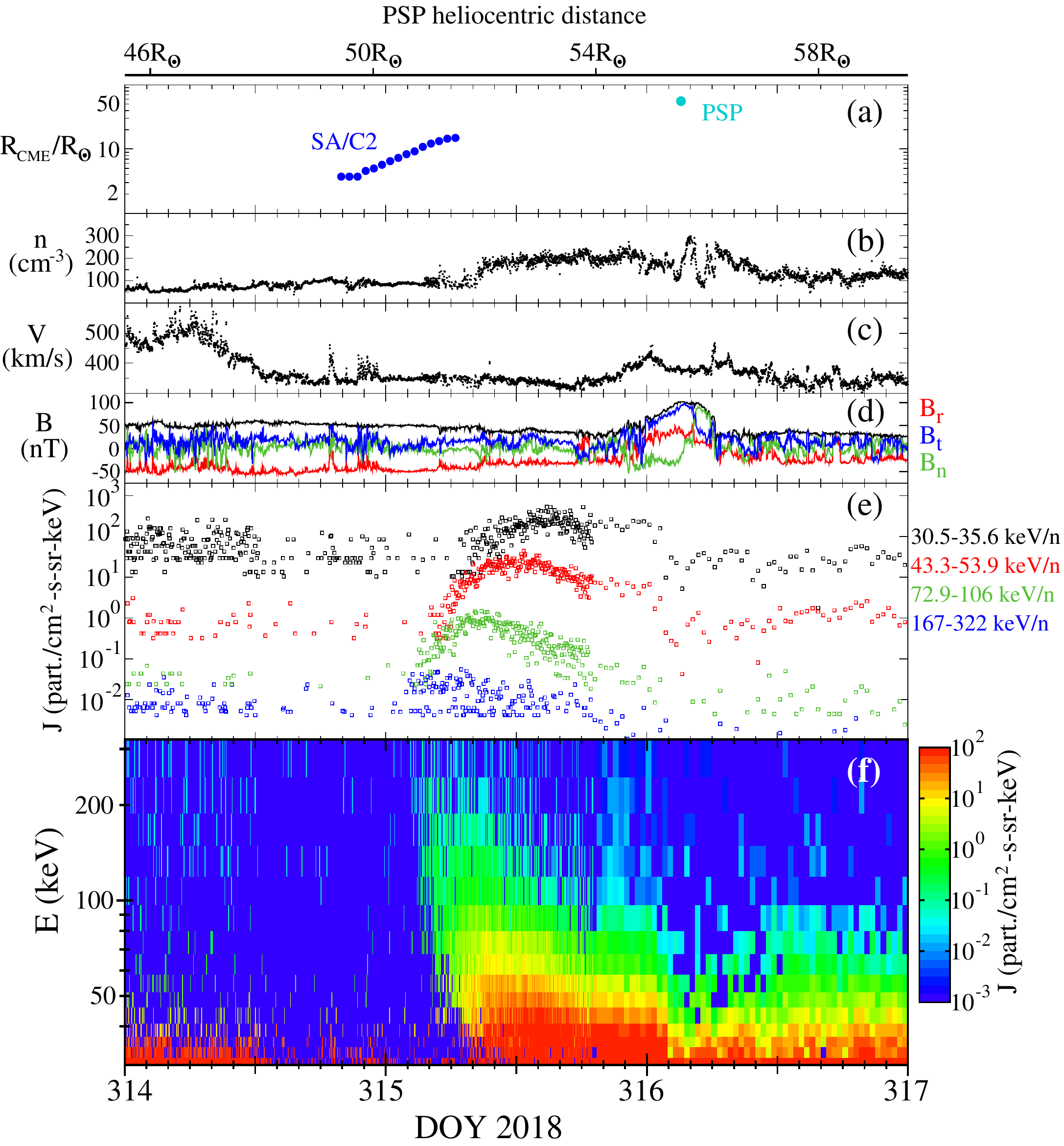}
   \caption{Multi-instrument data for a CME and SEP event observed by {\emph{PSP}} and {\emph{STEREO}}-A. (a) shows the heliocentric distance of the CME, (b-c) show solar wind density and solar wind speed from the SWEAP instrument, (d) shows the vector and magnitude of the magnetic field from the FIELDS instrument, and (e-f) show data from the IS$\odot$IS/EPI-Lo instrument. Figure adapted from \citet{2020ApJS..246...29G} where further details are provided.}
              \label{Giacalone_2020}
\end{figure*} 

Another important feature of this event was the generally steep energy spectrum of the low-energy ions. This suggested a very weak event. In the comparison between the model used by \citet{2020ApJS..246...29G} and the observations, it was found that a source spectrum, assumed to be at the CME when it was close to the Sun, had an approximately $E^{-5.5}$ power-law dependence. 

At the time, this was the closest to the Sun that a CME-related SEP event had been observed {\emph{in situ}}. \citet{2020ApJS..246...29G} used their diffusive transport model to estimate the total fluence that this event would have had at 1~AU, in order to compare with previous observations of CME-related SEP events. It was determined that this event would have been so weak to not even appear on a figure showing a wide range of values of the SEP fluence as a function of CME speed produced by \citet{2001JGR...10620947K}.

\citet[][]{2020ApJS..246...59M} presented a separate analysis of this same CME-SEP event suggested an alternative acceleration mechanism. They noted that since {\emph{PSP}} did not observe a shock locally, and modeling of the CME suggested it may not have ever driven a shock, the acceleration mechanism was not likely diffusive shock acceleration. Instead, they suggested it may be similar to that associated with aurora in planetary magnetospheres \cite[{\emph{e.g.}},][and references therein]{2009JGRA..114.2212M}. This study focused on two important observed aspects: the velocity dispersion profile and the composition of the SEP event. In the proposed mechanism, which was referred to as ``the pressure cooker'' \cite[{\emph{e.g.}},][]{1985JGR....90.4205G}, energetic particles are confined below the CME in the solar corona in a region bound by an electric potential above and strong magnetic fields below. The electric field is the result of strong field-aligned electric currents associated with distorted magnetic fields and plasma flow, perhaps associated with magnetic reconnection, between the CME and corona during its eruption. Particles are confined in this region until their energy is sufficient to overcome the electric potential. There are two key results from this process. One is that the highest energy particles will overcome this barrier earlier, and, hence, will arrive at {\emph{PSP}} earlier than low energy particles, which are presumably released much later when the CME has erupted from the Sun. The other is that the mechanism produces a maximum energy that depends on the charge of the species. Although the event was quite weak, there were sufficient counts of He, O, and Fe, that when combined with assumptions about the composition of these species in the corona, agreed with the observed high-energy cut-off as a function of particle species.

{\emph{PSP}} was in a fortunate location, during a fortuitously quiet period, and provided a unique opportunity to study energetic particles accelerated by a very slow and weak CME closer to the Sun that had been see {\emph{in situ}} previously. On the one hand, the observations suggests that very weak shocks, or even non-shock plasma compressions driven by a slow CME, are capable of accelerating particles. On the other hand, the pressure cooker method provides an interesting parallel with processes that occur in planetary ionospheres and magnetosphere. Moreover, the observation of the SEP event provided the opportunity to determine the parallel mean-free path of the particles, at 0.23~AU, as the particles were transported from their source to {\emph{PSP}}. 

In Mar. 2019, {\emph{PSP}} encountered a SBO-CME with unique properties which was analyzed by \citet{2020ApJ...897..134L}. SBO-CMEs are generally well-structured, slow CMEs that emerge from the streamer belt in extended PILs outside of ARs. Fig.~\ref{Lario2020Fig1} shows an overview of the plasma, magnetic field, electron and energetic particle conditions associated with the CME. Despite the relatively low speed of the SBO-CME close to the Sun determined by remote observation from {\emph{SOHO}} and {\emph{STEREO}}-A ($\sim311$~km~s$^{-1}$), the transit time to {\emph{PSP}} indicated a faster speed and two shocks were observed at {\emph{PSP}} prior to the arrival of the CME. The low initial speed of the SBO-CME makes it unlikely that it would have driven a shock in the corona and \citet{2020ApJ...897..134L} proposed that the formation of the shocks farther from the Sun were likely caused by compression effects of a HSS that followed the CME and that the formation of the two-shock structure may have been caused by distortions in the CME resulting from the HSS. This demonstrates the importance of the surrounding plasma conditions on the viability of energetic particle acceleration in CME events, though the associated energetic particle event in this case was limited to low energies $<$100~kev/nuc.

\begin{figure*}
   \centering
   \includegraphics[width=0.95\textwidth]{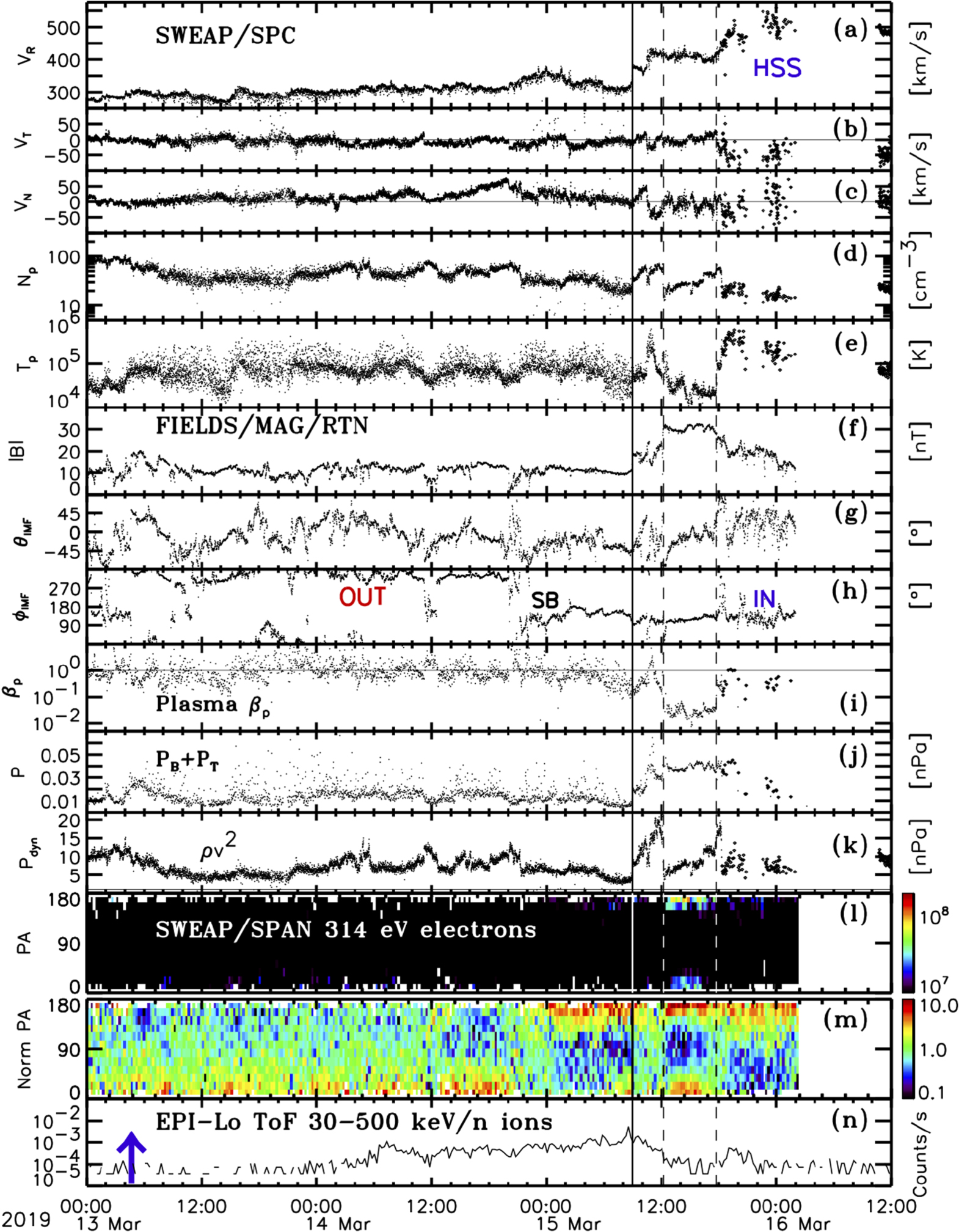}
   \caption{Overview of plasma (measured by SWEAP), magnetic field (measured by FIELDS), electron (SWEAP) and energetic particle (measured by IS$\odot$IS conditions during the Mar. 2019 SBO-CME event observed by {\emph{PSP}}. From top to bottom: (a) radial velocity, (b) tangential velocity component, and (c) normal velocity components of the solar wind proton velocity in RTN coordinates, (d) solar wind proton density, (e) solar wind proton temperature, magnetic field (f) magnitude, (g) elevation and (h) azimuth angles in RTN coordinates, (i) proton plasma beta, (j) the sum of the magnetic and thermal pressures, (k) ram pressure, (l) 314 eV electron PADs, (m) normalized 314 eV electron PADs, and (n) $\sim30-500$ keV TOF-only ion intensities measured by IS$\odot$IS/EPI-Lo. The vertical solid line indicates the passing of the two shocks associated with the CME which are too close in time to be separately resolved here, the vertical dashed lines indicate the boundaries of the CME, and the blue arrow indicates the eruption time of the SBO-CME at the Sun. Figure adapted from \citet{2020ApJ...897..134L}.}
              \label{Lario2020Fig1}
\end{figure*} 

In order to determine the point at which {\emph{PSP}} would have been magnetically connected to the CME, \citet{2020ApJ...897..134L} ran two ENLIL simulations, one with just ambient solar wind conditions and another including the CME. By evaluating the solar wind speed along the magnetic field line connecting {\emph{PSP}} to the Sun, they find the point at which the solar wind speed in the CME simulation exceeds that of the ambient simulation, which establishes the point at which {\emph{PSP}} is connected to the CME, termed the ``Connecting with the OBserving" point or "cobpoint" \citep{1995ApJ...445..497H}. Fig.~\ref{Lario2020Fig2} shows the coordinates of the cobpoint determined by this analysis alongside energetic particle anisotropy measurements. The energetic particle event is shown to be highly anisotropic, with enhanced particle intensities seen in the sunward-facing sensors of the instrument, and that the onset of energetic particles coincided with the establishment of the cobpoint, connecting the CME to {\emph{PSP}}. Also notable is that the increase in the speed jump at the cobpoint coincides with an increase in the measured energetic particle intensities prior to shock arrival. This analysis demonstrates the importance of energetic particle measurements made by IS$\odot$IS in constraining modelling of large scale magnetic field structures such as CMEs.

\begin{figure*}
   \centering
   \includegraphics[width=\textwidth]{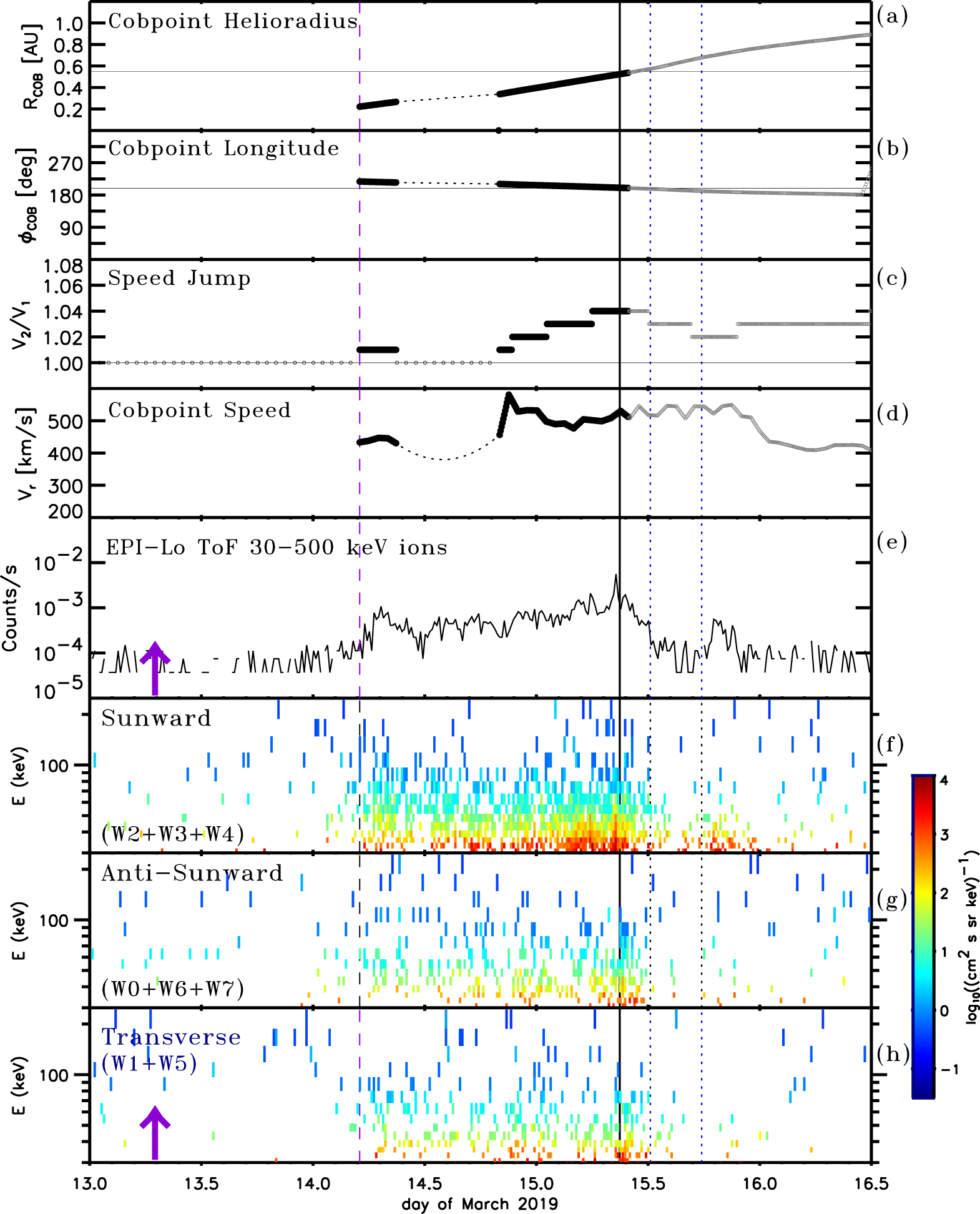}
   \caption{Cobpoint characteristics as determined from ENLIL simulations and TOF-only energetic ion data measured by IS$\odot$IS/EPI-Lo. From top to bottom: (a) heliocentric radial distance of {\emph{PSP}} cobpoint, (b) Heliocentric Earth Equatorial (HEEQ) longitude of {\emph{PSP}} cobpoint, (c) speed jump ratio measured at {\emph{PSP}} cobpoint by comparing the ENLIL background simulation to the simulation including the CME, (d) speed of the cobpoint, (e) ion intensities $\sim20-500$, (f) ion intensities measured in the Sun-facing wedges of EPI-Lo, (g) ion intensities measured in the EPI-Lo wedges facing away from the Sun, (h) ion intensities measured in the transverse wedges of EPI-Lo. The verticle solid line indicate the passage of the two shocks, the vertical dotted lines show the boundaries of the CME, the vertical purple dashed lie indicates the time when {\emph{PSP}} became connected to the compression region in front of the CME, and the purple vertical arrows indicate the time that the SBO-CME was accelerated at the Sun. Figure adapted from \citet{2020ApJ...897..134L}.}
              \label{Lario2020Fig2}
\end{figure*} 

\citet{2021A&A...651A...2J} analyzed a CME that was measured by {\emph{PSP}} on 20 Jan. 2020, when the S/C was 0.32~AU from the Sun. The eruption of the CME was well imaged by both {\emph{STEREO}}-A and {\emph{SOHO}} and was observed to have a speed of $\sim380$~km~s$^{-1}$, consistent with the transit time to {\emph{PSP}}, and possessed characteristics indicative of a stealth-type CME. Fig.~\ref{Joyce2021CMEFig1} shows a unique evolution of the energetic particle anisotropy during this event, with changes in the anisotropy seeming to coincide with changes of the normal component of the magnetic field ($B_N$). Of particular interest is a period where $B_N$ is close to zero and the dominant anisotropy is of energetic particles propagating toward the Sun (highlighted in yellow in Fig.~\ref{Joyce2021CMEFig1}), as well as two periods when $B_N$ spikes northward and there is a near complete dropout in energetic particle flux (highlighted in orange). The period dominated by particles propagating toward the Sun has the highest fluxes extending to the highest energies in the event and \citet{2021A&A...651A...2J} argued that this may be evidence that {\emph{PSP}}, located on the western flank of the CME throughout the event, may have briefly been connected to a region of stronger energetic particle acceleration, likely closer to the nose of the CME where the compression is likely strongest.

\begin{figure*}
   \centering
   \includegraphics[width=\textwidth]{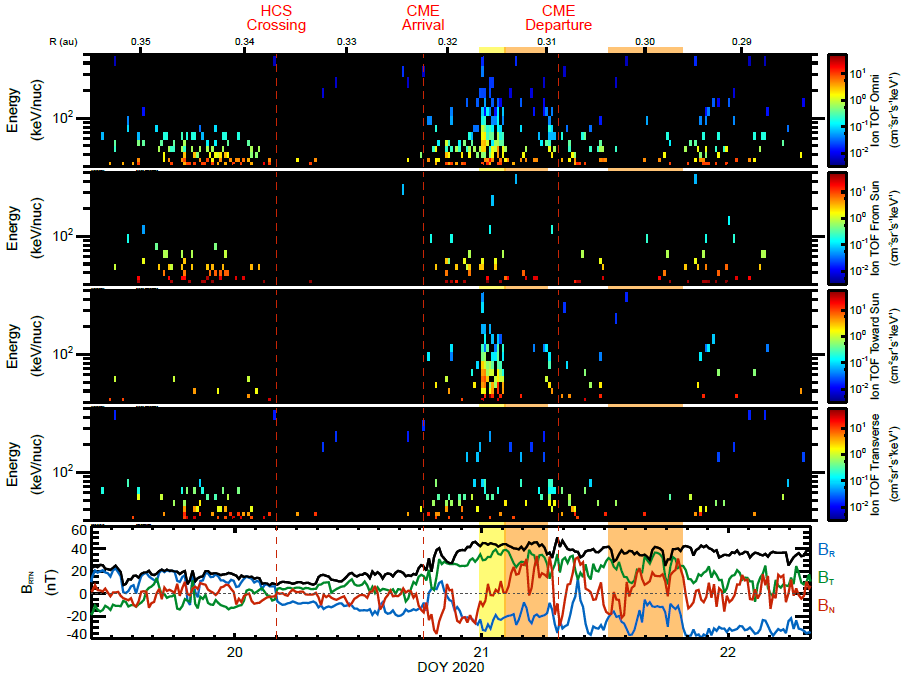}
   \caption{Overview of energetic particle anisotropy and magnetic field conditions during the Jan. 2020 CME. Energetic particle measurements are from the TOF-only channel of IS$\odot$IS/EPI-Lo and magnetic field data are from FIELDS. From top to bottom: omnidirectional ion spectrogram, ion spectrum from Sun ($0-60^\circ$ from nominal Parker spiral direction), ion spectrogram toward the Sun ($120-180^\circ$), ion spectrogram in the transverse direction ($60-120^\circ$), and the magnetic field vector in RTN coordinates, with the magnetic field magnitude in black. The period highlighted in yellow shows a strong influx of particles propagating toward the Sun, while periods of energetic particle dropouts are highlighted in orange. Figure adapted from \citet{2021A&A...651A...2J}.}
              \label{Joyce2021CMEFig1}
\end{figure*} 

{\emph{STEREO}}-A was well-aligned radially with {\emph{PSP}} during this time period and observed the same CME also from the western flank. Fig.~\ref{Joyce2021CMEFig2} shows the comparison between energetic particle spectrograms and magnetic field vectors measured by both instruments. Particularly striking is the remarkable similarity of the magnetic field vector measured by both S/C, suggesting that they both encountered a very similar region of the magnetic structure, contrasted with the dissimilarity of the energetic particle observations, with those at {\emph{STEREO}}-A lacking the fine detail and abrupt changes in anisotropy (not shown here) that are seen closer to the Sun.  This is likely due to transport effects such as scattering and diffusion which have created a much more uniform distribution of energetic particles by the time the CME has reached 1~AU. This demonstrates the importance of measurements of such events close to the Sun, made possible by {\emph{PSP}}/IS$\odot$IS, when it is still possible to distinguish between different acceleration mechanisms and source regions that contribute to energetic particle populations before these fine distinctions are washed out by transport effects. Such detailed measurements will be critical in determining which mechanisms play an important role in the acceleration of energetic particles close to the Sun.

\begin{figure*}
   \centering
   \includegraphics[width=\textwidth]{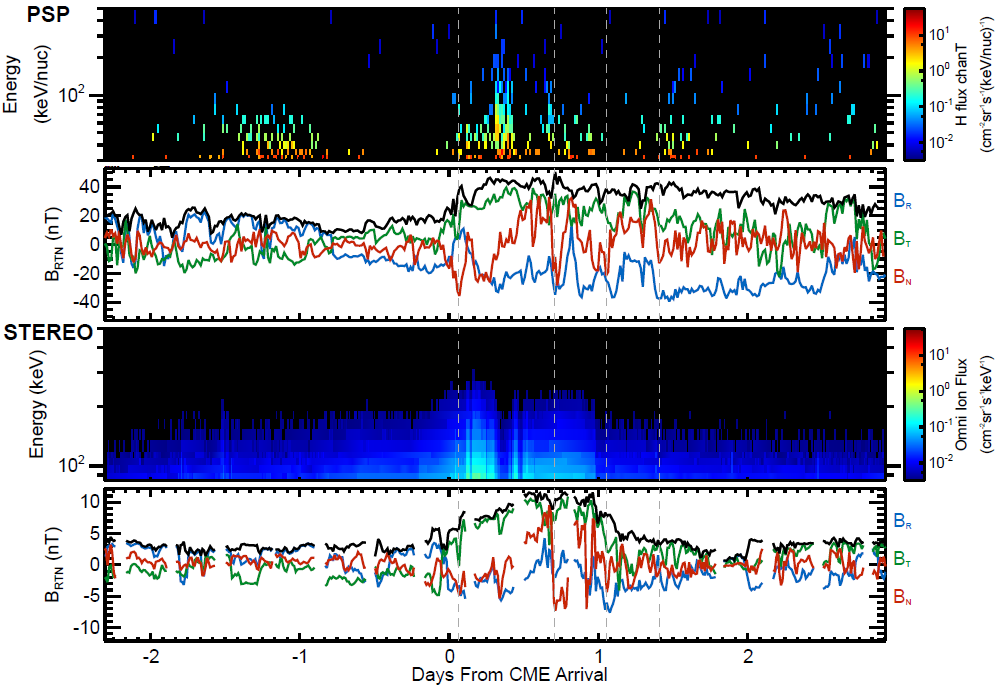}
   \caption{Comparison of energetic particle and magnetic field measurements of the same CME event observed at both {\emph{PSP}} and {\emph{STEREO}}-A. The data have been lined up by the arrival time of the CME arrival and the {\emph{PSP}} data have been stretched in time by a factor of 1.3 to match the magnetic field features seen by both S/C. Gray dotted lines indicate reference points used to line up the measurements from both S/C.}
              \label{Joyce2021CMEFig2}
\end{figure*} 

\subsubsection{The Widespread CME Event on 29 Nov. 2020}
\label{EPsCMENov}

The beginning of solar cycle 25 was marked by a significant SEP event in late Nov. 2020. The event has gained substantial attention and study, not only as one of the largest SEP events in several quiet years, but also because it was a circumsolar event spanning at least 230$^{\circ}$ in longitude and observed by four S/C positioned at or inside of 1~AU (see Fig.~\ref{Kollhoff2021Fig}). Among those S/C was {\emph{PSP}} and {\emph{SolO}}, providing a first glimpse of coordinated studies that will be possible between the two missions. The solar source was AR 12790 and the associated M4.4 class flare (as observed by {\emph{GOES}} at 12:34~UT on 29 Nov.) was at (E99$^{\circ}$,S23$^{\circ}$) (as viewed from Earth), 2$^{\circ}$ east of {\emph{PSP}}’s solar longitude. A CME traveling at $\sim1700$~km~s$^{-1}$ was well observed by {\emph{SOHO}}/LASCO and {\emph{STEREO}}-A/COR2, both positioned west of {\emph{PSP}} \citep{2021A&A...656A..29C}. {\emph{STEREO}}-A/EUVI also observed an EUV wave propagating away from the source at $\sim500$~km~s$^{-1}$, lasting about an hour and traversing much of the visible disk \citep{2021A&A...656A..20K}.

\begin{figure*}
   \centering
   \includegraphics[width=\textwidth]{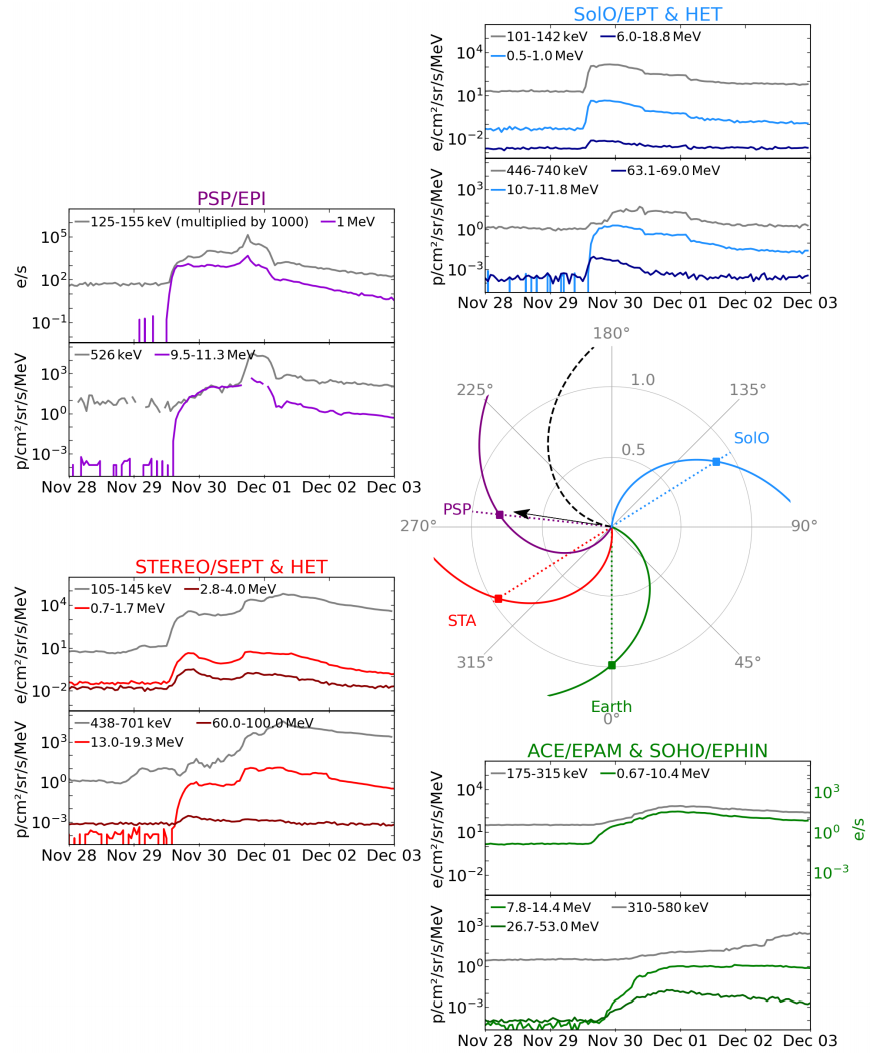}
   \caption{Overview of the widespread CME event on 19 Nov. 2020. Counterclockwise from the top right: {\emph{SolO}}, {\emph{PSP}}, {\emph{STEREO}}-A, and {\emph{ACE}} energetic particle observations are shown along with the relative location of all S/C. The direction of the CME is given by the black array, while curved lines in the orbit plot indicate nominal Parker Spiral magnetic field lines each S/C would be connected to. Figure adapted from \citet{2021A&A...656A..20K}.}
              \label{Kollhoff2021Fig}
\end{figure*} 

Protons at energies $>50$~MeV and $>1$~MeV electrons were observed by {\emph{PSP}}, {\emph{STEREO}}-A, {\emph{SOHO}}, and {\emph{SolO}} with onsets and time profiles that were generally organized by the S/C’s longitude relative to the source region, as has been seen in multi-S/C events from previous solar cycles \citep{2021A&A...656A..20K}. However, it was clear that intervening solar wind structures such as a slower preceding CME and SIRs affected the temporal evolution of the particle intensities. Analysis of the onset times of the protons and electrons observed at the four S/C yielded solar release times that were compared to the EUV wave propagation. The results were inconsistent with a simplistic scenario of particles being released when the EUV wave arrived at the various S/C magnetic footpoints, suggesting more complex particle transport and/or acceleration process.

Heavy ions, including He, O and Fe, were observed by {\emph{PSP}}, {\emph{STEREO}}-A, {\emph{ACE}} and {\emph{SolO}} and their event-integrated fluences had longitudinal spreads similar to those obtained from three-S/C events observed in cycle 24 \citep{2021A&A...656L..12M}. The spectra were all well described by power-laws at low energies followed by an exponential roll-over at higher energies (Fig.~\ref{Mason2021CMEFig}). The roll-over energy was element dependent such that Fe/O and He/H ratios decreased with increasing energy; a signature of shock-acceleration that is commonly seen in SEP events \citep{2021A&A...656A..29C, 2021A&A...656L..12M}. The overall composition (relative to O) at 0.32 – 0.45 MeV/nuc was fairly typical of events this size, with the exception of Fe/O at {\emph{PSP}} and {\emph{ACE}} where it was depleted by a factor of $\sim2$ \citep{2021A&A...656L..12M}.

\begin{figure*}
   \centering
   \includegraphics[width=\textwidth]{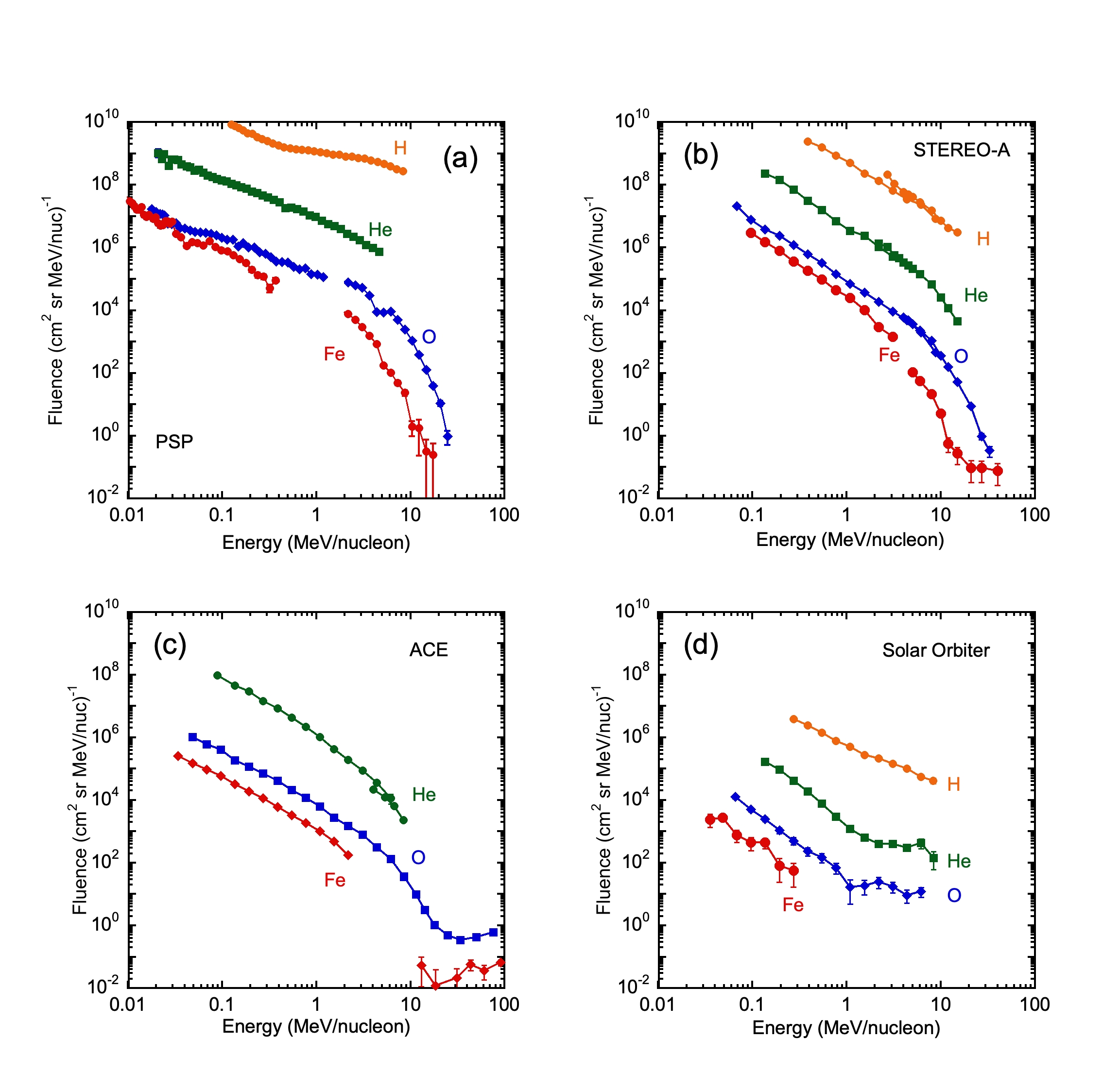}
   \caption{Multi-species fuence spectra of the 29 Nov. 2020 event from (a) {\emph{PSP}}, (b) {\emph{STEREO}}-A, (c) {\emph{ACE}}, and (d) {\emph{SolO}}. Figure adapted from \citet{2021A&A...656L..12M}.}
              \label{Mason2021CMEFig}
\end{figure*} 

Due to the relative positioning of the source region and {\emph{PSP}}, the CME passed directly over the S/C. It was traveling fast enough to overtake a preceding, slower CME in close proximity to {\emph{PSP}}, creating a dynamic and evolving shock measured by FIELDS \citep{2021ApJ...921..102G, 2022ApJ...930...88N}. Coincident with this shock IS$\odot$IS observed a substantial increase in protons up to at least 1 MeV, likely due to local acceleration. More surprisingly, an increase in energetic electrons was also measured at the shock passage (Fig.~\ref{Kollhoff2021Fig}). Acceleration of electrons by interplanetary shocks is rare \citep{2016A&A...588A..17D}, thus it is more likely this increase is a consequence of a trapped electron distribution, perhaps caused by the narrowing region between the two CMEs \citep{2021A&A...656A..29C}.

\begin{figure*}
   \centering
   \includegraphics[width=\textwidth]{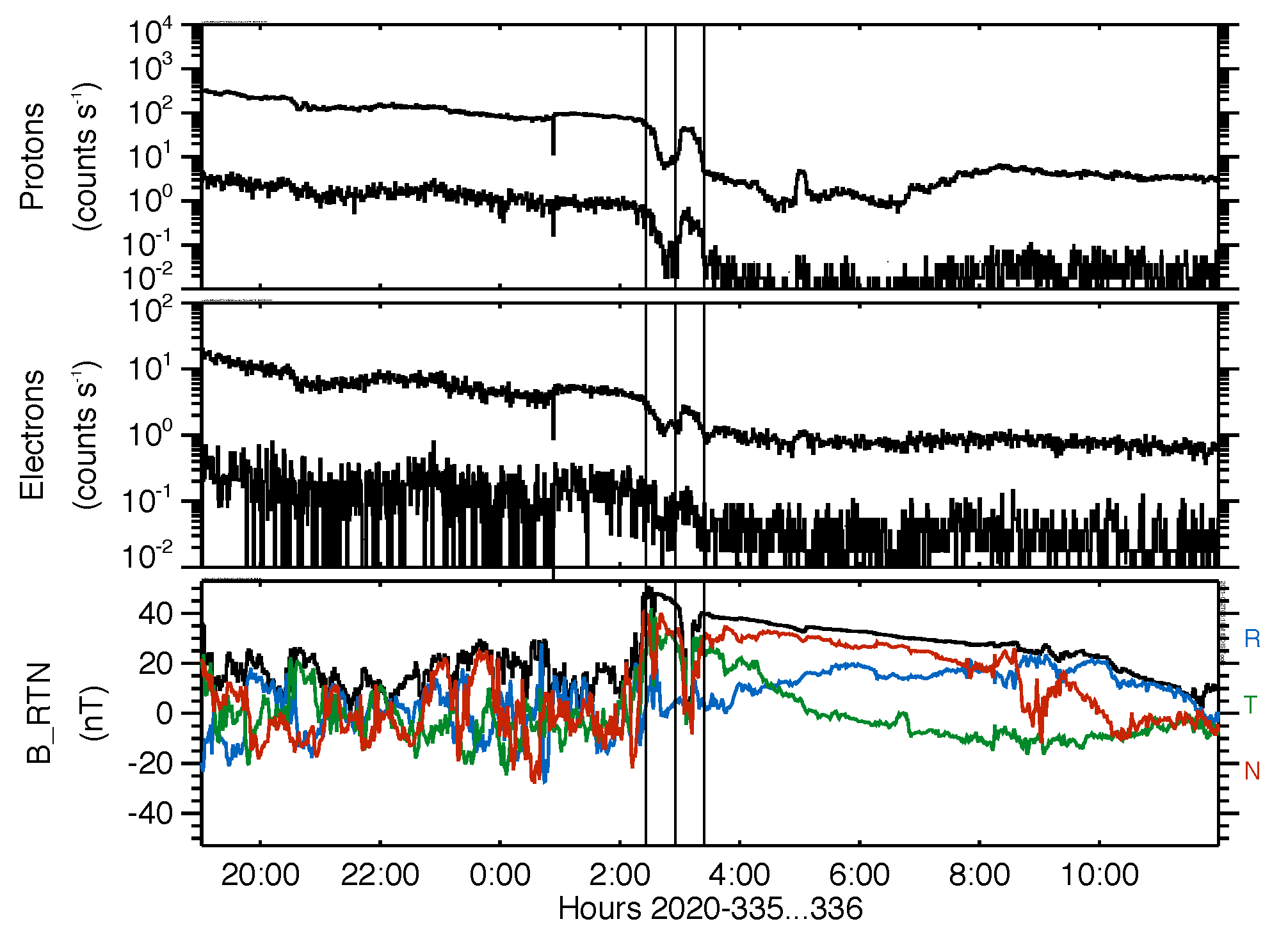}
   \caption{Time profile of energetic protons stopping in the third and fifth detector of HET (top panel, upper and lower traces, respectively) and electrons stopping in the third and fourth detector of HET (middle panel, upper and lower traces, respectively) with the magnetic field in the bottom panel for the 29 Nov. 2020 CME. The over plotted virticle lines illustrate that the variations in the particle count rates occur at the same time as changes in the magnetic field. See \citet{2021A&A...656A..29C} for more information.}
              \label{Cohen2021Fig}
\end{figure*} 

The MC of the fast CME followed the shock and sheath region, with a clear rotation seen in the magnetic field components measured by FIELDS (Fig.~\ref{Cohen2021Fig}). At the onset of the cloud, the particle intensities dropped as is often seen due to particles being unable to cross into the magnetic structure (Fig.~\ref{Cohen2021Fig}). During this period there was a 30 minute interval in which all the particle intensities increased briefly to approximately their pre-cloud levels. This was likely the result of {\emph{PSP}} exiting the MC, observing the surrounding environment populated with SEPs, and then returning to the interior of the MC.

Although several of the properties of the SEP event are consistent with those seen in previous studies, the 29 Nov. 2020 event is noteworthy as being observed by four S/C over 230$^{\circ}$ longitude and as the first significant cycle 25 event. The details of many aspects of the event (both from individual S/C and multi-S/C observations) remain to be studied more closely. In addition, modeling of the event has only just begun and will likely yield significant insights regarding the evolution of the CME-associated shock wave \citep{2022A&A...660A..84K} and the acceleration and transport of the SEPs throughout the inner heliosphere \citep{2021ApJ...915...44C}.

\subsection{Energetic Electrons}
\label{EE}

The first observations of energetic electrons by {\emph{PSP}}/IS$\odot$IS were reported by \citet{2020ApJ...902...20M}, who analyzed a series of energetic electron enhancements observed during {\emph{PSP}}’s second Enc. period, which reached a perihelion of 0.17~AU. Fig.~\ref{Mitchell2020Fig} shows a series of four electron events that were observed at approximately 03:00, 05:00, 09:00, and 15:40~UT on 2 Apr. 2019. The events are small compared with the background and are only observable due to the small heliocentric distance of {\emph{PSP}} during this time. Background subtraction is applied to the electron rates data to help resolve the electron enhancements as well as a second-degree Savitzky–Golay smoothing filter over 7 minutes is applied to reduce random statistical fluctuations \citep{1964AnaCh..36.1627S}. While the statistics for these events are very low, the fact that they are observed concurrently in both EPI-Hi and EPI-Lo and that they coincide with either abrupt changes in the magnetic field vector or that they can plausibly be connected to type III radio bursts observed by the FIELDS instrument which extend down to $f_{pe}$ suggests that these are real electron events. These are the first energetic electron events which have been observed within 0.2~AU of the Sun and suggest that such small and short-duration electron events may be a common feature close to the Sun that was not previously appreciated since it would not be possible to observe such events farther out from the Sun. This is consistent with previous observations by {\emph{Helios}} between 0.3 and 1~AU \citep{2006ApJ...650.1199W}. More observations and further analysis are needed to determine what physical acceleration mechanisms may be able to produce such events.

\begin{figure*}
   \centering
   \includegraphics[width=\textwidth]{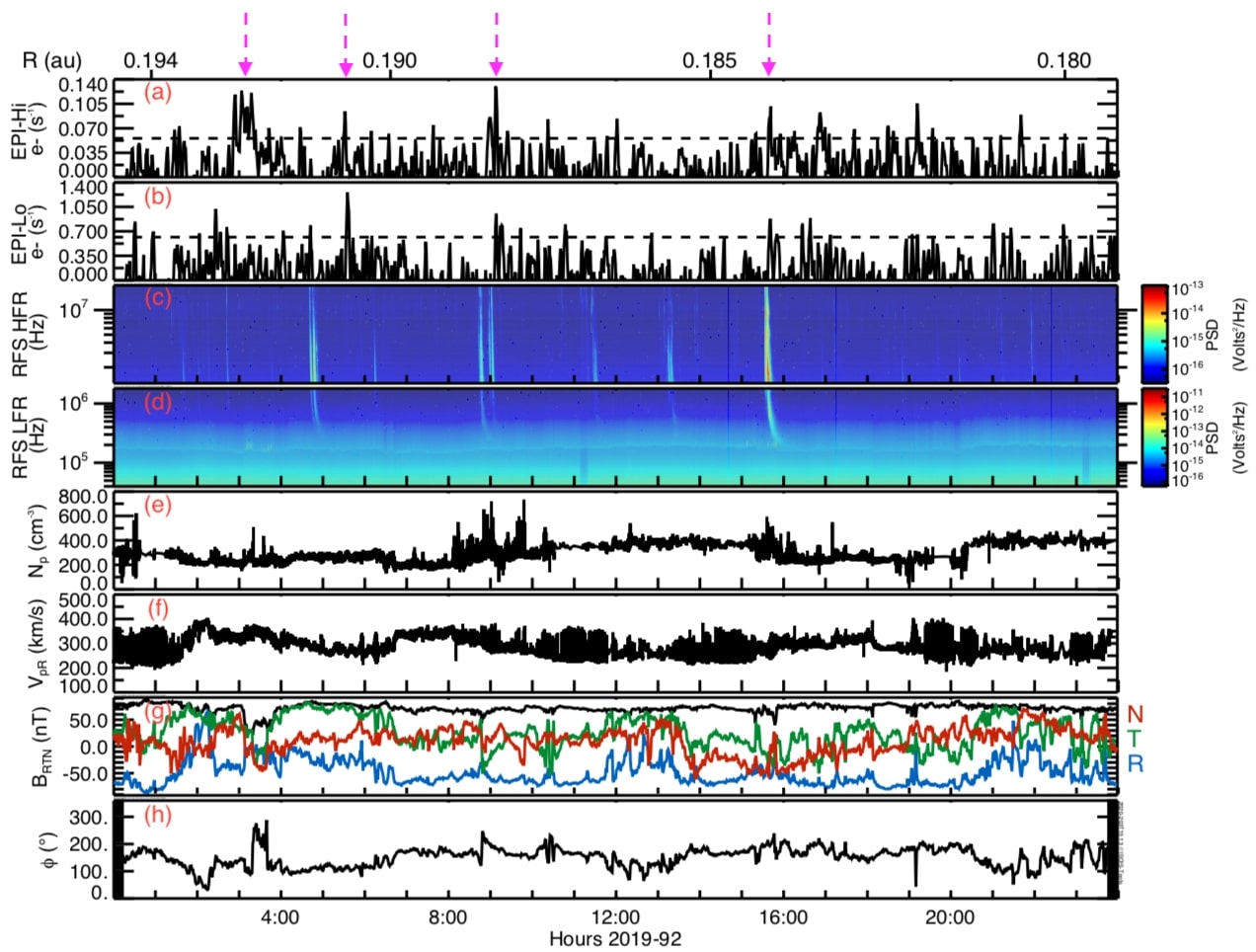}
   \caption{Overview of {\emph{PSP}} observations during 2 Apr. 2019. Panels show the following: (a) EPI-Hi electron count rate (0.5–6 MeV) with a background subtraction and 7 minutes Savitzky–Golay smoothing applied and with a dashed line to indicate 2$\sigma$ deviation from the mean, (b) EPI-Lo electron count rate (50–500 keV) with a background subtraction and 7 minutes Savitzky–Golay smoothing applied and with a dashed line to indicate 2$\sigma$ deviation from the mean, (c) FIELDS high-frequency radio measurements (1.3–19.2 MHz), (d) FIELDS low-frequency radio measurements (10.5 kHz–1.7 MHz), (e) SWEAP solar wind ion density ($\sim5$ measurements per second), (f) SWEAP radial solar wind speed ($\sim5$ measurements per second), (g) FIELDS 1 minute magnetic field vector in RTN coordinates (with magnetic field strength denoted by the black line). A series of electron events are observed (in the top two panels), occurring at approximately 03:00, 05:00, 09:00, and 15:40~UT, as well as a series of strong type III radio bursts (seen in panels c and d). Figure adapted from \citet{2020ApJ...902...20M}.}
              \label{Mitchell2020Fig}
\end{figure*} 

In late Nov. 2020, {\emph{PSP}} measured an SEP event associated with two CME eruptions, when the S/C was at approximately 0.8~AU.  This event is the largest SEP event observed during the first 8 orbits of {\emph{PSP}}, producing the highest ion fluxes yet observed by IS$\odot$IS \citep{2021A&A...656A..29C,2021ApJ...921..102G,2021A&A...656A..20K,2021A&A...656L..12M}, and also produced the first energetic electron events capable of producing statistics sufficient to register significant anisotropy measurements by IS$\odot$IS as reported by \citet{2021ApJ...919..119M}. Fig.~\ref{Mitchell2021Fig1} shows an overview of the electron observations during this period along with magnetic field data to provide context. Notable in these observations is the peaking of the EPI-Lo electron count rate at the passage of the second CME shock, which is quite rare, though not unheard of, due to the inefficiency of CME driven shocks in accelerating electrons. This may indicate the importance of quasi-perpendicular shock acceleration in this event, which has been shown to be a more efficient acceleration of electrons \citep{1984JGR....89.8857W,1989JGR....9415089K,1983ApJ...267..837H,2010ApJ...715..406G,2012ApJ...753...28G,2007ApJ...660..336J}. The notable dip in the EPI-Hi electron count rate at this time is an artifact associated with dynamic threshold mode changes of the EPI-Hi instrument during this time \citep[for details, see][]{2021A&A...656A..29C}.
 
 \begin{figure*}
   \centering
   \includegraphics[width=\textwidth]{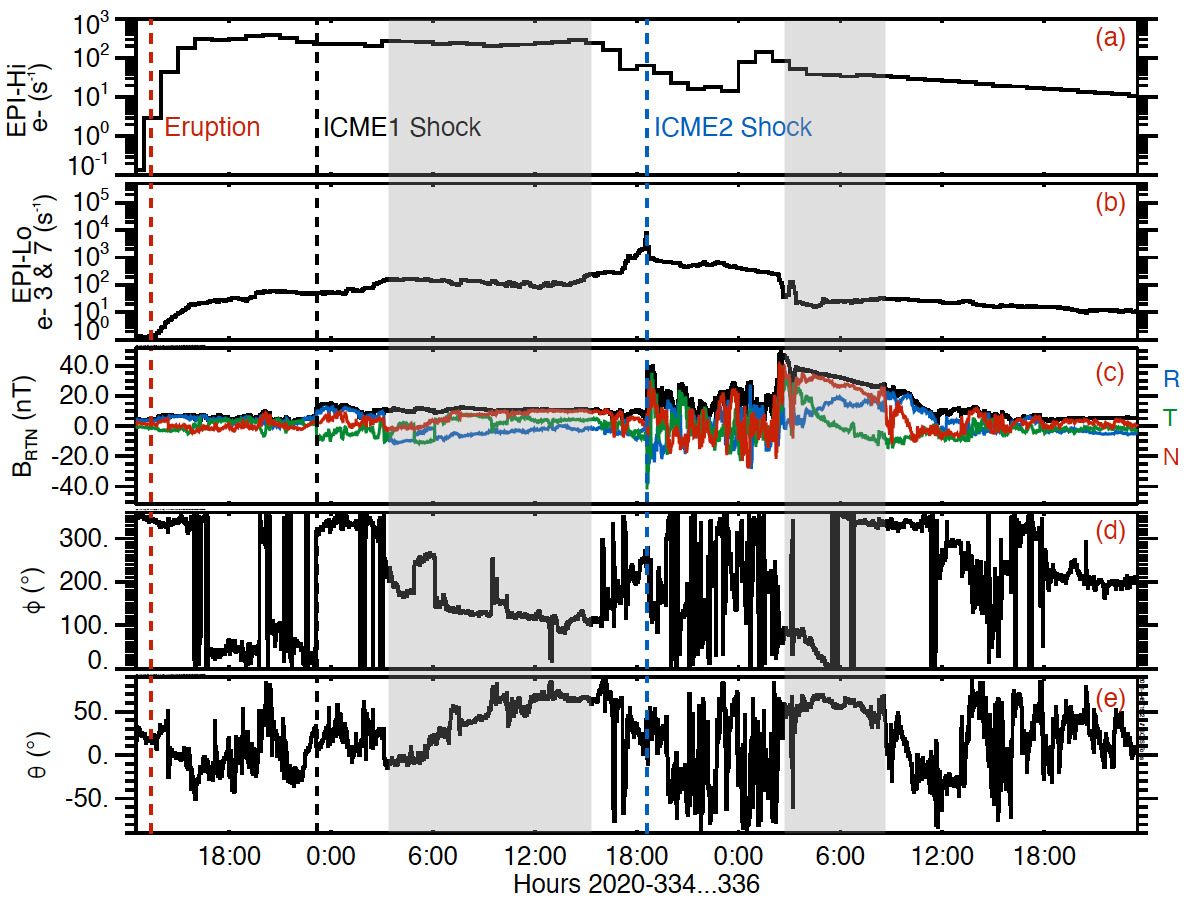}
   \caption{Overview of electron observations associated with the Nov. 29 CMEs Panels are as follows: (a) shows the EPI-Hi electron count rates summed in all 5 apertures ($\sim0.5-2$~MeV), (b) shows the EPI-Lo electron count rate from wedges 3 and 7 ($\sim57-870$~keV), (c) shows the FIELDS magnetic field vector in RTN coordinates, and (d) and (e) show the magnetic field vector angles. Vertical lines show the eruption of the second CME and the passage of the shocks associated with both CMEs. Flux-rope like structures are indicated by the shaded grey regions. The decrease in the EPI-Hi electron count rate seen at the passage of the second CME and the overall flat profile are artifacts caused by EPI-Hi dynamic threshold mode changes (explained in detail by Cohen et al. 2021). Figure adapted from \citet{2021ApJ...919..119M}.}
              \label{Mitchell2021Fig1}
\end{figure*} 

Fig.~\ref{Mitchell2021Fig2} shows the electron and magnetic field measurements during a three-hour period around the shock crossing associated with the second CME, including the electron PAD. Because of the off-nominal pointing of the S/C during this time, the pitch angle coverage is somewhat limited, however the available data shows that the highest intensities to be in the range of $\sim40-90^{\circ}$ at the time of the shock crossing. Distributions with peak intensities at pitch angles of around $90^{\circ}$ may be indicative of the shock-drift acceleration mechanism that occurs at quasi-perpendicular shocks \citep[{\emph{e.g.}},][]{2007ApJ...660..336J,1974JGR....79.4157S,2003AdSpR..32..525M}. This, along with the peak electron intensities seen at the shock crossing, further supports the proposition that electrons may be efficiently accelerated by quasi-perpendicular shocks associated with CMEs. Other possible explanations are that the peak intensities may be a result of an enhanced electron seed population produced by the preceding CME \citep[similar to observations by][]{2016A&A...588A..17D}, that energetic electrons may be accelerated as a result of being trapped between the shocks driven by the two CMEs \citep[a mechanism proposed by][]{2018A&A...613A..21D}, and that enhanced magnetic fluctuations and turbulence created upstream of the shock by the first CME may increase the efficiency of electron acceleration in the shock \citep[as proposed by][]{2015ApJ...802...97G}.
 
\begin{figure*}
   \centering
   \includegraphics[width=\textwidth]{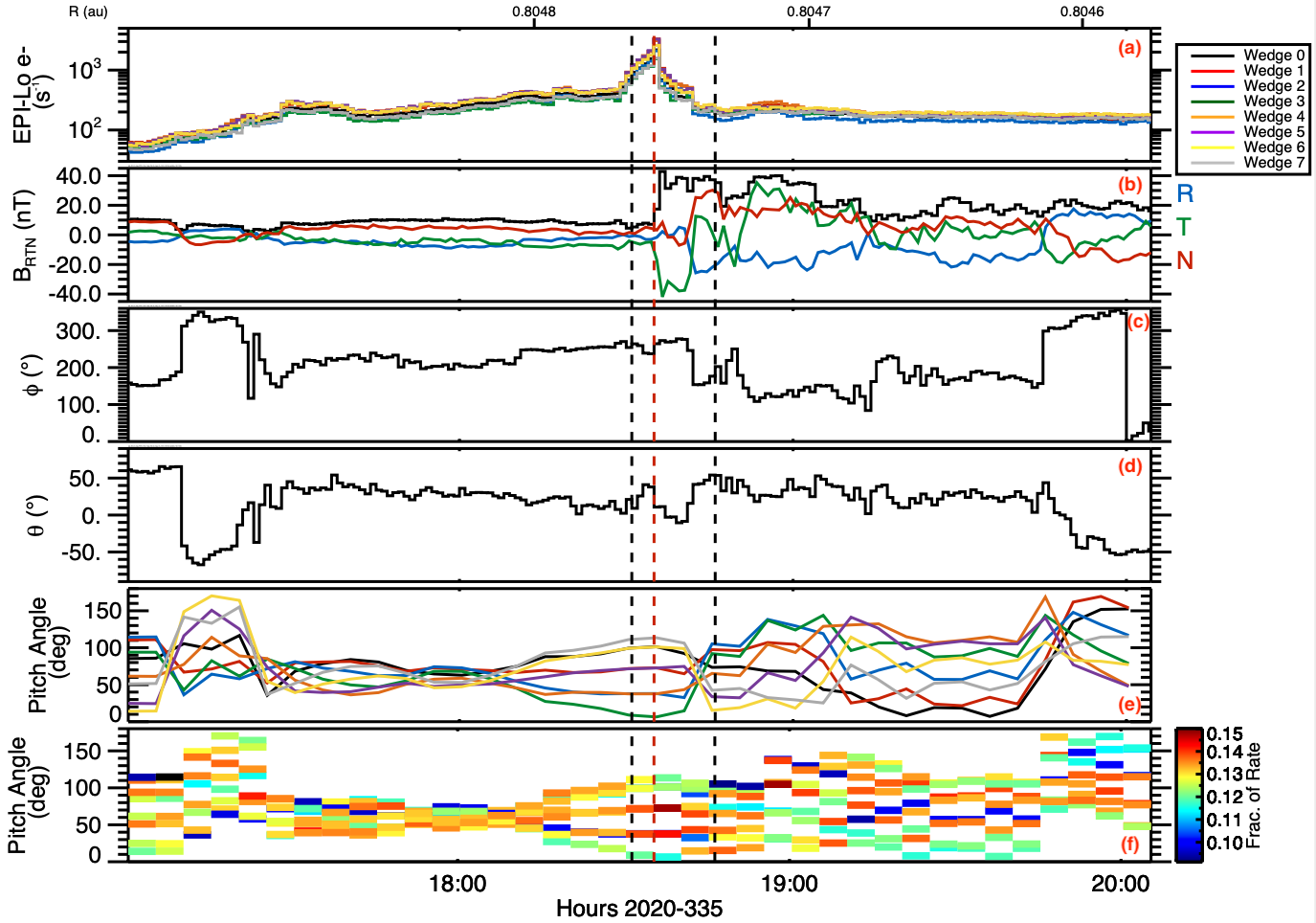}
   \caption{Electron and magnetic field measurements around the time of the second CME shock crossing (indicated by the vertical red dashed line). Panels show the following: (a) EPI-Lo electron measurements ($\sim130-870$ keV) in each of its 8 wedges, (b) FIELDS magnetic field vector in RTN coordinates, (c) azimuthal angle of the magnetic field, (d) polar angle of the magnetic field, (e) pitch angle of the geometric center of each EPI-Lo wedge (each with a width of $\sim30^{\circ}$), and (f) pitch angle time series for each EPI-Lo wedge ($80-870$~keV) showing the fraction at each time bin to the total count rate over the entire interval. Figure adapted from \citet{2021ApJ...919..119M}.}
              \label{Mitchell2021Fig2}
\end{figure*} 
 
While observations of energetic electron events thus far in the {\emph{PSP}} mission have been few, the measurements that have been made have shown IS$\odot$IS to be quite capable of characterizing energetic electron populations. Because of its close proximity to the Sun during {\emph{PSP}}’s Enc. phases, IS$\odot$IS has been shown to be able to measure small, subtle events which are not measurable farther from the Sun, but which may provide new insights into electron acceleration close to the Sun. The demonstrated ability to provide detailed electron anisotropy analyses is also critical for determining the acceleration mechanisms for electrons (particularly close to the Sun where transport effects have not yet influenced these populations) and for providing insight into the magnetic topology of magnetic structures associated with SEP events. As the current trend of increasing solar activity continues, we can expect many more unique observations and discoveries related to energetic electron events in the inner heliosphere from {\emph{PSP}}/IS$\odot$IS.

\subsection{Coronating/Stream Interaction Region-Associated Energetic Particles}
\label{CIREPs}

SIRs form where HSSs from coronal holes expand into slower solar wind \citep{1971JGR....76.3534B}. As the coronal hole structure corotates on the Sun, the SIR will corotate, as well, and becoming a CIR after one complete solar co-rotation. As the HSS flows radially outward, both forward and reverse shocks can form along the SIR/CIR, often at distances beyond 1~AU \citep[{\emph{e.g.}},][]{2006SoPh..239..337J, 2008SoPh..250..375J, 1978JGR....83.5563P, 1976GeoRL...3..137S}, which act as an important source of energetic ions, particularly during solar minimum. Once accelerated at an SIR/CIR-associated shock, energetic particles can propagate back towards the inner heliosphere along magnetic field lines and are subject to adiabatic deceleration and scattering \citep{1980ApJ...237..620F}. The expected result of these transport effects is a softening of the energetic particle spectra and a hardening of the lower-energy suprathermal spectra \citep[see][]{1999SSRv...89...77M}. This spectral variation, however, has not always been captured in observations, motivating the formulation of various other SIR/CIR-associated acceleration processes such as compressive, non-shock related acceleration \citep[{\emph{e.g.}},][]{2002ApJ...573..845G,2012ApJ...749...73E,2015JGRA..120.9269C} which can accelerate ions into the suprathermal range at lower heliospheric distances. Additionally, footpoint motion and interchange reconnection near the coronal hole boundary has been proposed to lead to more radial magnetic field lines on the HSS side of the SIR/CIR, resulting in more direct access, and less modulation, of energetic particles \citep{2002GeoRL..29.2066M, 2002GeoRL..29.1663S, 2005GeoRL..32.3112S}. Observations within 1~AU, by {\emph{PSP}}, are therefore particularly well suited to detangle these acceleration and transport effects as the SIR/CIR-associated suprathermal to energetic ion populations are further from shock-associated acceleration sites that are usually beyond 1~AU.

\begin{figure*}
   \centering
   \includegraphics[width=\textwidth]{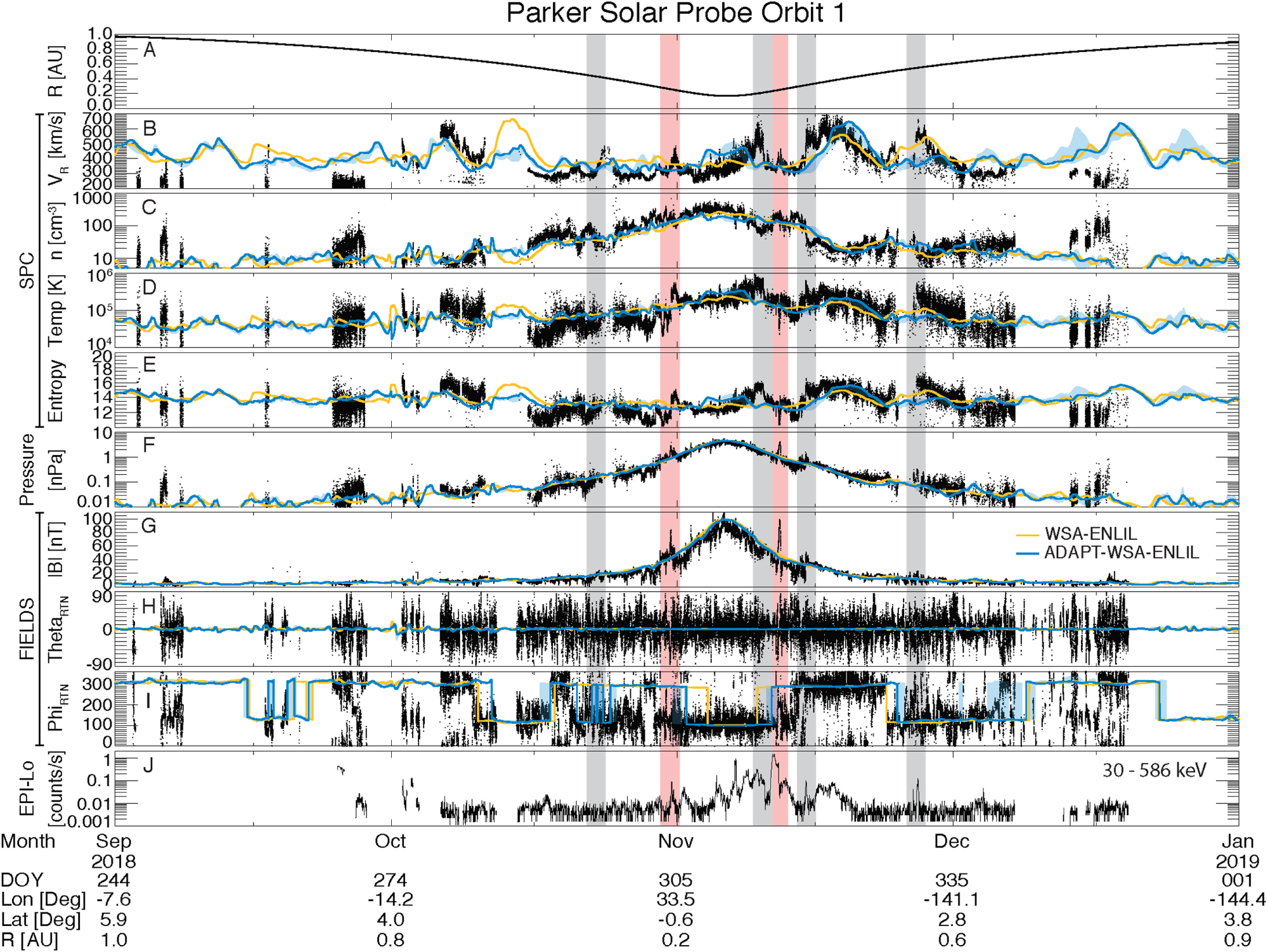}
   \caption{Overview of four months around the first perihelion (6 Nov. 2018). Panels show (a) the heliographic distance of {\emph{PSP}}; bulk proton (b) radial velocity, (c) density, (d) temperature, and (e) entropy; (f) summation of the magnetic and bulk proton thermal plasma pressure; (g) magnitude of the magnetic field, (h) $\Theta$, and (i) $\Phi$ angels of the magnetic field; and (j) EPI-Lo ion time-of-flight count rate for energies from 30 to 586 keV. Simulated quantities from two simulations are shown by the yellow and blue lines \citep[see][for more information]{2020ApJS..246...36A}. The four HSSs investigated in \citet{2020ApJS..246...36A} are indicated by the grey shaded regions, while pink shaded regions denote CMEs. Figure adapted from \citet{2020ApJS..246...36A}.}
              \label{Allen_2020}
\end{figure*} 

During the first orbit of {\emph{PSP}}, \citet{2020ApJS..246...36A} reported on four HSSs observed by {\emph{PSP}}, illustrated in Fig.~\ref{Allen_2020}, and compared these to observations of the streams at 1~AU using observations from L1 ({\emph{ACE}} and {\emph{Wind}}) and {\emph{STEREO}}-A. Many of these nascent SIR/CIRs were associated with energetic particle enhancements that were offset from the interface of the SIR/CIR. One of the events also had evidence of local compressive acceleration, which was previously noted by \citet{2019Natur.576..223M}. \citet{2020ApJS..246...20C} further analyzed energetic particle increases associated with SIR/CIRs during the first two orbits of {\emph{PSP}} (Fig.~\ref{Cohen_2020}). They found He/H abundance ratios similar to previous observations of SIR/CIRs at 1~AU with fast solar wind under 600~km~s$^{-1}$, however the proton spectra power laws, with indices ranging from $-4.3$ to 6.5, were softer than those often observed at 1~AU. Finally, \citet{2020ApJS..246...56D} investigated the suprathermal-to-energetic ($\sim0.03-3$ MeV/nuc) He ions associated with these SIR/CIRs from the first two orbits. They found that the higher energy He ions would arrive further in the rarefaction region than the lower-energy ions. The He spectra behaved as flat power laws modulated by exponential roll overs with an e-folding at energies of $\sim0.4$ MeV/nuc, suggesting acceleration at shocks further out in the heliosphere. \citet{2020ApJS..246...56D} interpreted the tendency for the suprathermal ion peak to be within the rarefaction regions with acceleration further out in the heliosphere as evidence that the rarefaction regions allowed easier access for particles than other regions in the SIR/CIR structure.

\begin{figure*}
   \centering
   \includegraphics[width=\textwidth]{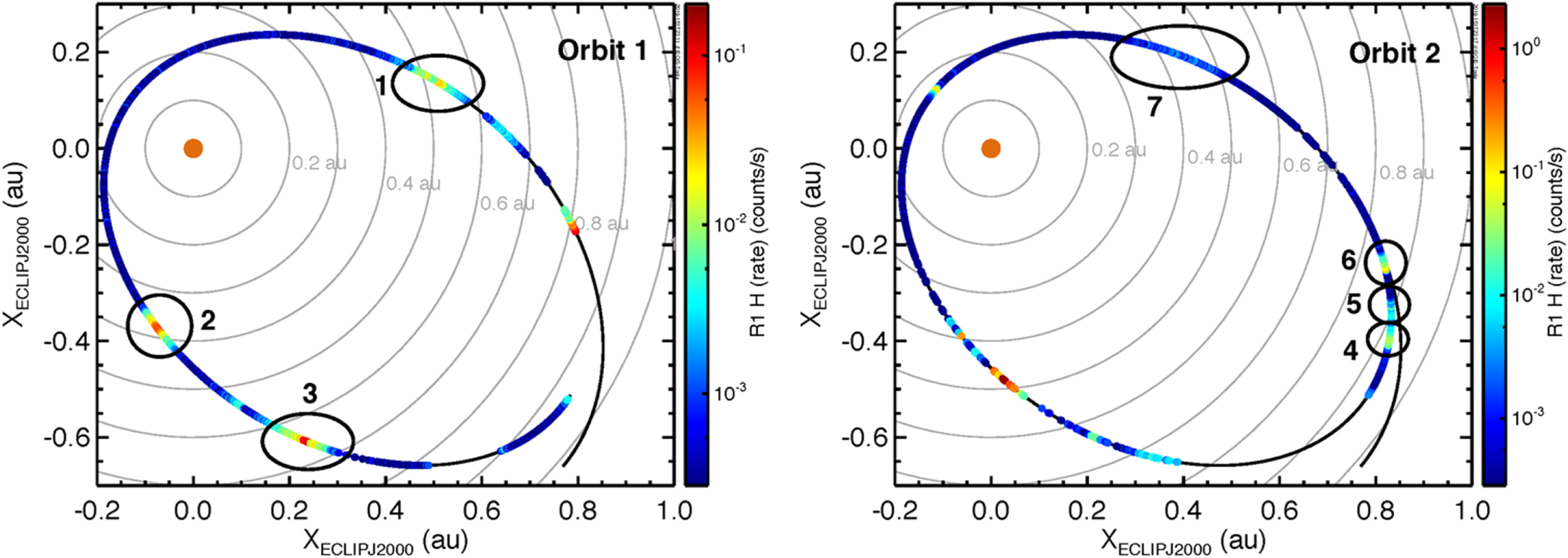}
   \caption{Summary of EPI-Hi LET $\sim1-2$ MeV proton observations from the first two orbiter of {\emph{PSP}}. SIR-associated energetic particle events studied by \citet{2020ApJS..246...20C} are denoted by the numbered circles. Figure adapted from \citet{2020ApJS..246...20C}.}
              \label{Cohen_2020}
\end{figure*} 

One fortuitus CIR passed {\emph{PSP}} on 19 Sep. 2019, during the third orbit of {\emph{PSP}}, when {\emph{PSP}} and {\emph{STEREO}}-A were nearly radially aligned and $\sim0.5$~AU apart \citep{2021A&A...650A..25A, 2021GeoRL..4891376A}. As shown in Fig.~\ref{Allen_2021}, while the bulk plasma and magnetic field observations between the two S/C followed expected radial dependencies, the CIR-associated suprathermal ion enhancements were observed at {\emph{PSP}} for a longer duration in time than at {\emph{STEREO}}-A \citep{2021GeoRL..4891376A}. Additionally, the suprathermal ion spectral slopes between {\emph{STEREO}}-A total ions and {\emph{PSP}} H\textsuperscript{+} were nearly identical, while the flux at {\emph{PSP}} was much smaller, suggesting little to no spectral modulation from transport. \citet{2021GeoRL..4891376A} concluded that the time difference in the CIR-associated suprathermal ion enhancement might be related to the magnetic topology between the slow speed stream ahead of the CIR interface, where the enhancement was first observed, and the HSS rarefaction region, where the suprathermal ions returned to background levels. \citet{2021ApJ...908L..26W} furthered this investigation by simulating the {\emph{PSP}} and {\emph{STEREO}}-A observations using the European Heliopheric FORecasting Information Asset \citep[EUHFORIA;][]{2018JSWSC...8A..35P} model and the Particle Radiation Asset Directed at Interplanetary Space Exploration \citep[PARADISE;][]{2019AA...622A..28W, 2020AA...634A..82W} model, suggesting that this event provides evidence that CIR-associated acceleration does not always require shock waves. 

\begin{figure*}
   \centering
   \includegraphics[width=\textwidth]{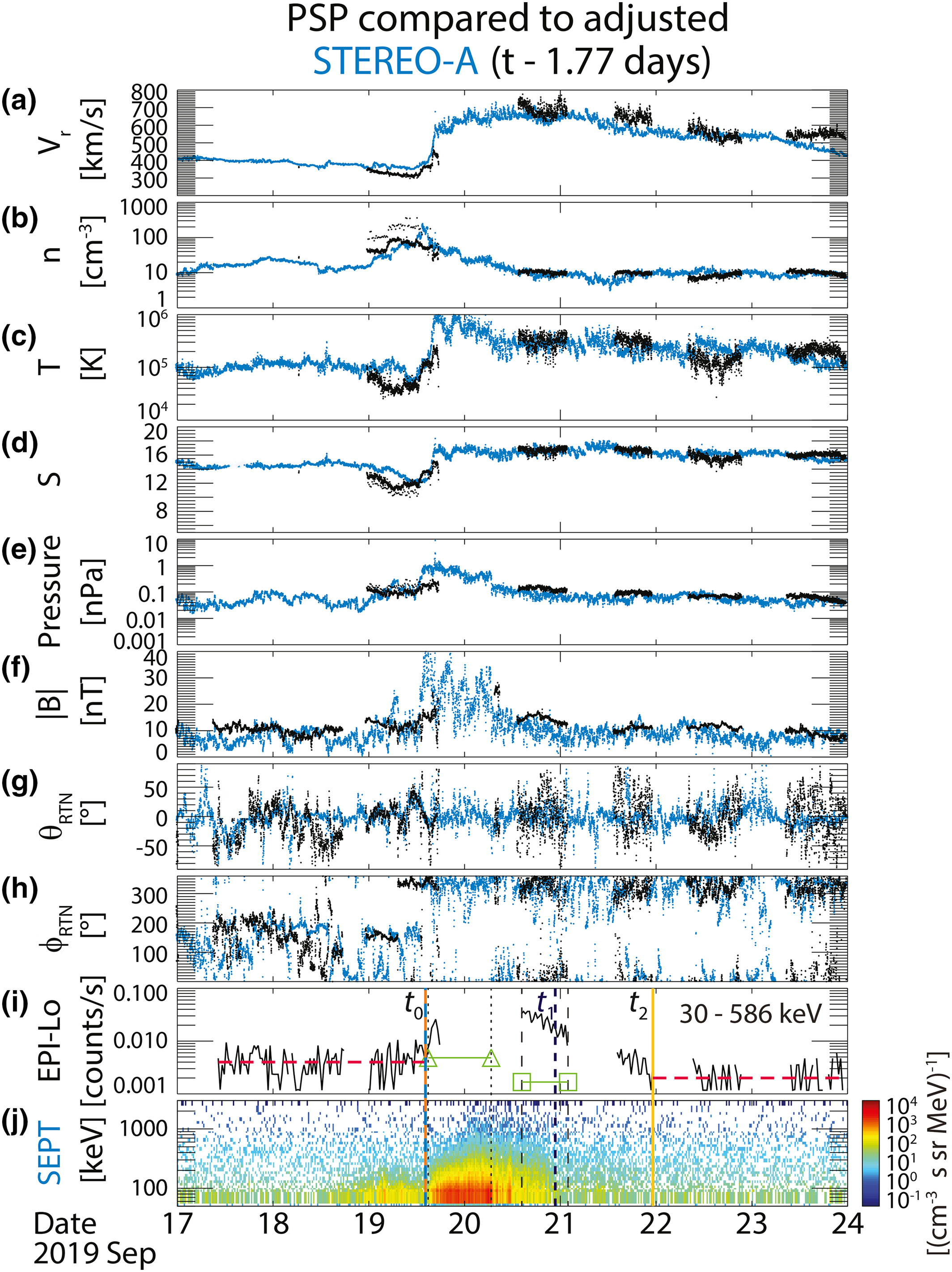}
   \caption{Comparison of {\emph{PSP}} observations (black) and time-shifted and radially corrected {\emph{STEREO}}-A observations (blue) for the CIR that passed over {\emph{PSP}} on 19 Sep. 2019. While the bulk solar wind and magnetic field observations match well after typical scaling factors are applied \citep[a-h, see][for more information]{2021GeoRL..4891376A}, the energetic particle are elevated at {\emph{PSP}} (i) for longer than at {\emph{STEREO}}-A (j). Figure adapted from \citet{2021GeoRL..4891376A}.}
              \label{Allen_2021}
\end{figure*} 

An SIR that passed over {\emph{PSP}} on 15 Nov. 2018 when the S/C was $\sim0.32$~AU from the Sun providing insight into energetic particle acceleration by SIRs in the inner heliosphere and the importance of the magnetic field structures connecting the observer to the acceleration region. Fig.~\ref{SSR_SIR_Fig} shows an overview of the energetic particle, plasma and magnetic field conditions during the passage of the SIR and the energetic particle event that followed it, which started about a day after the passage of the compression and lasted for about four days.The spectral analysis provided by \citet{2021A&A...650L...5J}, showed that for the first day of the event, the spectra resembled a simple power law, which is commonly associated with local acceleration, despite being well out of the compression region by that point. The spectrum for the remaining three days of the event was shown to be fairly constant, a finding that is inconsistent with the traditional model of SIR energetic particle acceleration provided by \citet{1980ApJ...237..620F}, which models energetic particle acceleration at distant regions where SIRs have steepened into shocks and predicts changes in the spectral shape with distance from the source region. Within this paradigm, we would expect that the the distance along the magnetic field connecting to the source region would increase during the event and that the observed spectrum would change. This combined with the simple power law spectrum observed on the first day, seems to indicate that the source region is much closer to the observer than is typically thought as we do not see the expected transport effects and that acceleration all along the compression, not only in the distant regions where the SIR may steepen into shocks, may play an important role in energetic particle acceleration associated with SIRs (consistent with previous studies by \citealt{2000JGR...10523107C} and \citealt{2015JGRA..120.9269C}).

\begin{figure*}
   \centering
   \includegraphics[width=\textwidth]{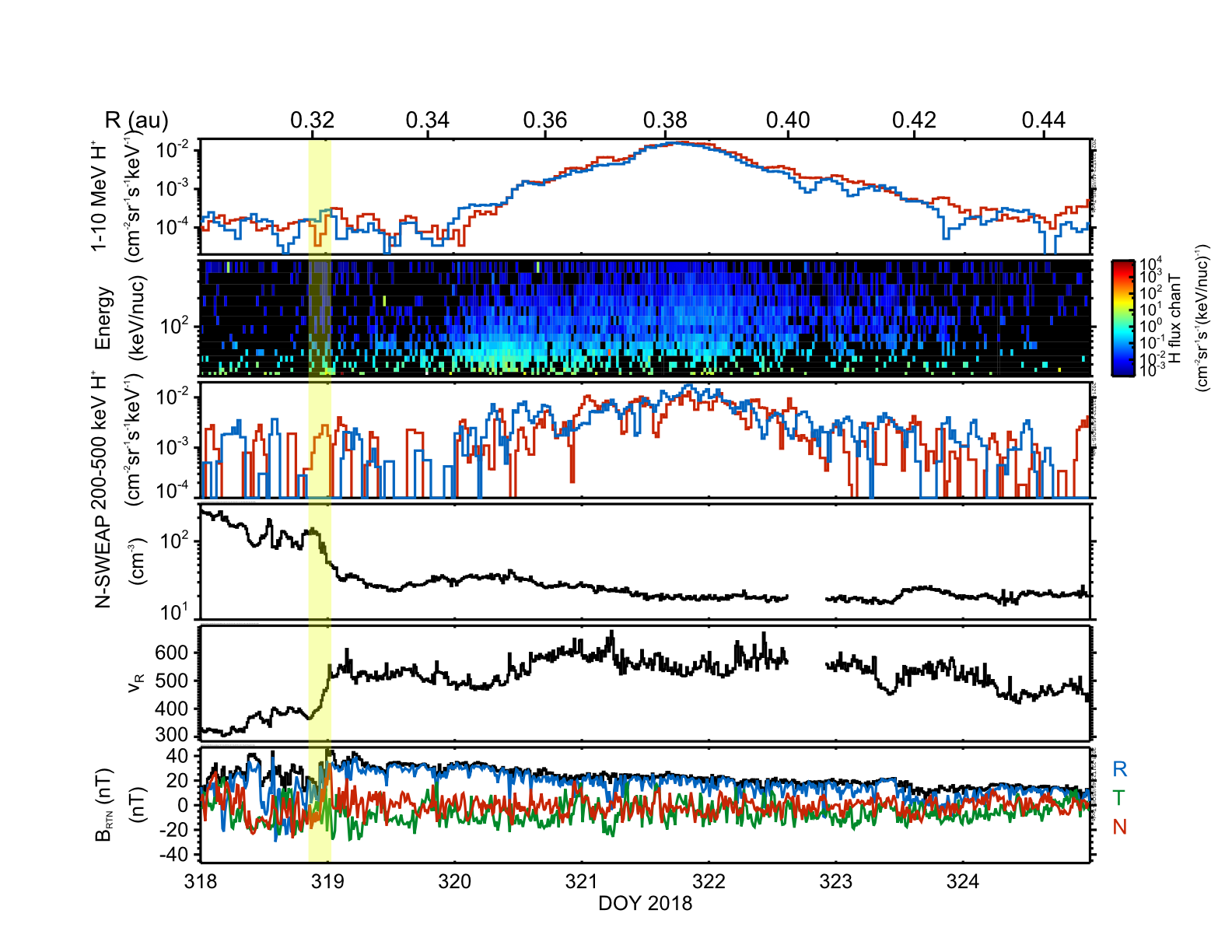}
   \caption{Overview of energetic particle observations associated with the SIR that passed over {\emph{PSP}} on 15 Nov. 2018. Plasma data is provided by the SWEAP instrument and magnetic field data by the FIELDS instrument. The compression associated with the passage of the SIR is highlighted in yellow. Figure is updated from figures shown in \citet{2021A&A...650A..24S} and \citet{2021A&A...650L...5J}.}
              \label{SSR_SIR_Fig}
\end{figure*} 

\citet{2021A&A...650A..24S} analyzed the same event, also noting that the long duration of the energetic particle event following the passage of the CME suggests a non-Parker spiral orientation of the magnetic field and proposed that the observations may be explained by a sub-Parker magnetic field structure~\citep{2002GeoRL..29.2066M,2002GeoRL..29.1663S,2005GeoRL..32.3112S,2005JGRA..110.4104S}. The sub-Parker spiral structure forms when magnetic footpoints on the Sun move across coronal hole boundaries, threading the magnetic field between the fast and slow solar wind streams that form the compression and creating a magnetic field structure that is significantly more radial than a nominal Parker spiral. Fig.~\ref{Schwadron2021SIRFig}a shows a diagram of the sub-Parker spiral and Fig.~\ref{Schwadron2021SIRFig}b shows a comparison between the energetic particle fluxes measured by IS$\odot$IS in two different energy regimes compared with modeled fluxes for both the Parker and sub-Parker spiral magnetic field orientations. The modeling includes an analytic solution of the distribution function at the SIR reverse compression/shock and numerical modeling of the propagation of the particles back to the S/C \citep[details in][]{2021A&A...650A..24S}. The modeled fluxes for the sub-Parker spiral match the observed fluxes much better than the nominal Parker spiral, demonstrating that the sub-Parker spiral structure is essential for explaining the extended duration of the energetic particle event associated with the SIR. The sub-Parker spiral is often seen in rarefaction regions, such as those that form behind SIRs, and thus is likely to play a significant role in the observed energetic particle profiles associated with such events. Both the \citet{2021A&A...650L...5J} and \citet{2021A&A...650A..24S} demonstrate the importance of IS$\odot$IS observations of SIRs in understanding the large scale structure of the magnetic field in the inner heliosphere, the motion of magnetic footpoints on the Sun and the propagation of energetic particles, helping us to understand the variability of energetic particles and providing insight into the source of the solar wind.

\begin{figure*}
   \centering
   \includegraphics[width=0.7\textwidth]{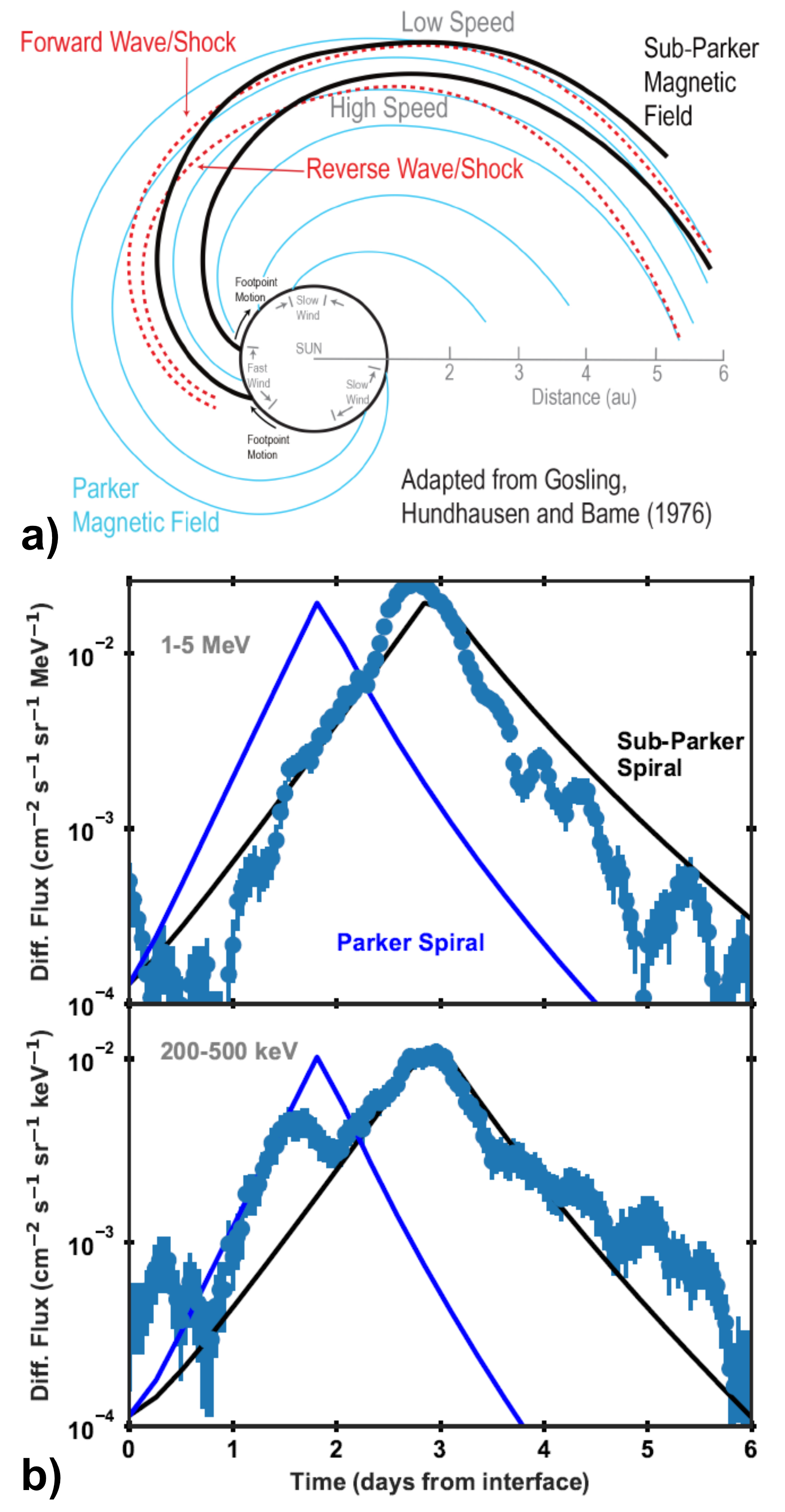}
   \caption{(a) shows the magnetic field structure associated with an SIR, with the red dashed lines representing the compression region where the fast solar wind overtakes the slow solar wind, the blue lines represent the nominal Parker spiral configuration, and the clack lines represent the sub-Parker spiral field lines that are threaded between the fast and slow solar wind streams as a result of footpoint motion across the coronal hole boundary. (b) shows a comparison between IS$\odot$IS energetic particle fluxes in two energy ranges (blue data points) and modeled energetic particle fluxes for both the Parker spiral (blue lines) and sub-Parker spiral (black lines) magnetic field configurations. Figure adapted from \citet{2021A&A...650A..24S}.}
\label{Schwadron2021SIRFig}
\end{figure*}
        
\subsection{Inner Heliospheric Anomalous Cosmic Rays}
\label{ACRs}

The ability of {\emph{PSP}} to measure the energetic particle content at unprecedentedly close radial distances during a deep solar minimum has also allowed for detailed investigations into the radial dependence of ACRs in the inner heliosphere. ACRs are mainly comprised of singly ionized hydrogen, helium, nitrogen, oxygen, neon, and argon, with energies of $\sim5-50$ MeV/nuc \citep[{\emph{e.g.}},][]{1973ApJ...182L..81G, 1973PhRvL..31..650H, 1974IAUS...57..415M, 1988ApJ...334L..77C, 1998SSRv...83..259K, 2002ApJ...578..194C, 2002ApJ...581.1413C, 2013SSRv..176..165P}. The source of these particles are neutral interstellar particles that are part of the interstellar wind \citep{2015ApJS..220...22M} before becoming ionized near the Sun \citep{1974ApJ...190L..35F}. Once the particles become ionized, they become picked-up by the solar wind convective electric field and are convected away from the Sun as pick-up ions. A small portion of these pick-up ions can become accelerated to high energies (tens to hundreds of MeV) further out in the heliosphere before returning into the inner heliosphere, thus becoming ACRs \citep{1996ApJ...466L..47J, 1996ApJ...466L..43M, 2000AIPC..528..337B, 2012SSRv..173..283G}.

While the acceleration of ACRs is primarily thought to occur at the termination shock \citep{1981ApJ...246L..85P, 1992ApJ...393L..41J}, neither {\emph{Voyager}}~1 or {\emph{Voyager}}~2 \citep{1977SSRv...21...77K} observed a peak in ACR intensity when crossing the termination shock \citep{2005Sci...309.2017S, 2008Natur.454...71S}. As a result, numerous theories have been proposed to explain this including a ``blunt” termination shock geometry in which the ACR acceleration occurs preferentially along the termination shock flanks and tail \citep{2006GeoRL..33.4102M, 2008ApJ...675.1584S} away from the region the {\emph{Voyager}} S/C crossed, magnetic reconnection at the heliopause \citep{2010ApJ...709..963D}, heliosheath compressive turbulence \citep{2009AdSpR..43.1471F}, and second-order Fermi processes \citep{2010JGRA..11512111S}. 

After being accelerated, ACR particles penetrate back into the heliosphere, where their intensities decrease due to solar modulation \citep[{\emph{e.g.}},][]{1999AdSpR..23..521K, 2002ApJ...578..194C, 2006GeoRL..33.4102M}. The radial gradients of ACRs in the heliosphere have primarily been studied from 1~AU outward through comparing observations at the {\emph{IMP-8}}\footnote{The Interplanetary Monitoring Platform-8} at 1~AU to observations from {\emph{Pioneer}}~10, {\emph{Pioneer}}~11, {\emph{Voyager}}~1, and {\emph{Voyager}}~2 in the outer heliosphere. These comparisons revealed that the helium ACR intensity varied as $r^{-0.67}$ from 1 to $\sim41$~AU \citep{1990ICRC....6..206C}. Understanding this modulation provides insight into the various processes that govern global cosmic ray drift paths throughout the heliosphere.
        
\begin{figure*}
   \centering
   \includegraphics[width=\textwidth]{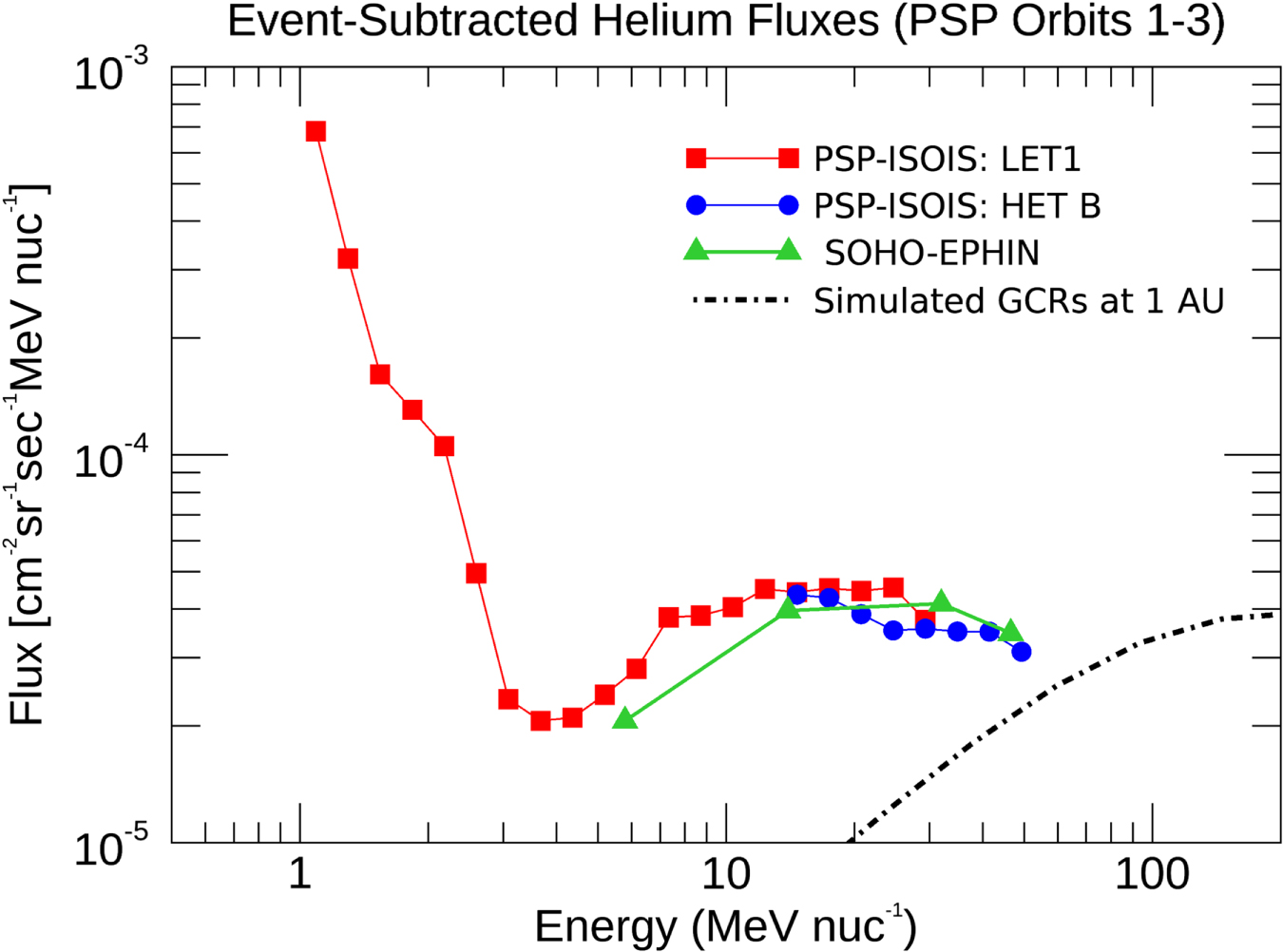}
   \caption{Helium spectra over the first three orbits of {\emph{PSP}} after removing transient events \citep[see][for more information]{2021ApJ...912..139R} at {\emph{PSP}} (red and blue) and at {\emph{SOHO}} (green). A simulated GCR spectrum at 1~AU is included (black) from HelMOD (version 4.0.1, 2021 January; www.helmod.org). Figure adapted from \citet{2021ApJ...912..139R}.}
              \label{Rankin_2021_1}
\end{figure*} 

The orbit of {\emph{PSP}} is well suited to investigate ACR radial variations due to its sampling of a large range of radial distances near the ecliptic. Additionally, {\emph{PSP}} enables investigations into the ACR populations at distances closer to the Sun than previously measured. \citet{2021ApJ...912..139R} utilized the {\emph{PSP}}/IS$\odot$IS/EPI-Hi instrument to study the radial variation of the helium ACR content down to $35.6~R_\odot$ (0.166~AU) and compare these observations to ACR observations at 1~AU measured the {\emph{SOHO}} mission. To ensure that the particles included in the comparisons were ACRs, rather than SEPs, only “quiet-time” periods were used \citep[see the Appendix in][]{2021ApJ...912..139R}. The resulting quiet-time EPI-Hi and {\emph{SOHO}} spectra over the first three orbits of {\emph{PSP}} is shown in Fig.~\ref{Rankin_2021_1}. The ACR intensity was observed to increase over energies from $\sim5$ to $\sim40$ MeV/nuc, a characteristic feature of ACR spectra.

Figs.~\ref{Rankin_2021_2}a and \ref{Rankin_2021_2}b show normalized ACR fluxes from the {\emph{SOHO}} Electron Proton Helium INstrument \citep[EPHIN;][]{1988sohi.rept...75K} and {\emph{PSP}}/IS$\odot$IS/EPI-Hi, respectively. The ratio of the ACR fluxes (Fig.~\ref{Rankin_2021_2}c) correlate well with the heliographic radial distance of {\emph{PSP}} (Fig.~\ref{Rankin_2021_2}d). This presents clear evidence of radial-dependent modulation, as expected. However, the observed radial gradient is stronger ($\sim25\pm5$\%~AU) than observed beyond 1~AU. Better understanding the radial gradients of ACRs in the inner heliosphere may provide needed constraints on drift transport and cross-field diffusion models, as cross-field diffusion will become more dominant in the inner heliosphere \citep{2010JGRA..11512111S}. Future studies will also be aided by the addition of ACR measurements by {\emph{SolO}}, such as those reported in \citet{2021A&A...656L...5M}.
       
\begin{figure*}
   \centering
   \includegraphics[width=\textwidth]{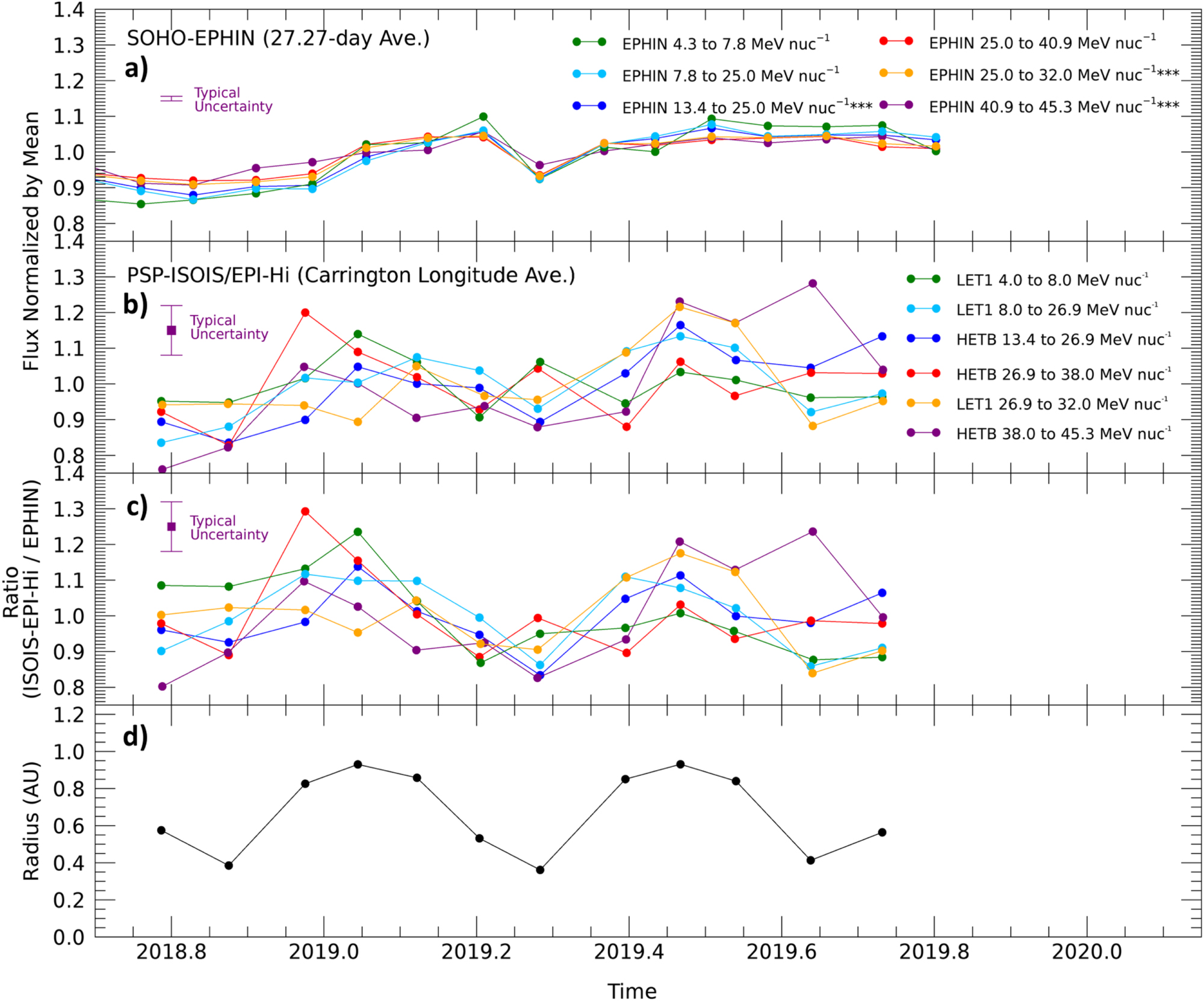}
   \caption{ACR normalized flux at (a) 1~AU averaged over 27.27 days and (b) {\emph{PSP}} averaged over Carrington longitude. The ratio of intensities (c) has a clear dependence on the radial distance of {\emph{PSP}} (d). Figure adapted from \citet{2021ApJ...912..139R}.}
              \label{Rankin_2021_2}
\end{figure*} 
        
\subsection{Open Questions and Future Collaborations}
\label{EPsRadOQ}

Over the first four years of the {\emph{PSP}} prime mission, large advances have been made regarding our understanding of inner heliospheric energetic particles and solar radio emissions. Looking forward, as the solar cycle ascends out of solar minimum, and as additional observatories such as {\emph{SolO}} enter into their science phase and provide robust energetic particle measurements \citep[see][]{2021A&A...656A..22W}, many new opportunities to study energetic particle populations and dynamics will present themselves. 

For example, while {\emph{PSP}} has begun exploring the radial evolution of SIRs and associated energization and transport of particles, future measurements will explore the cause of known solar cycle dependencies of the SIR/CIR-associated suprathermal ion composition \citep[{\emph{e.g.}},][]{ 2008ApJ...678.1458M, 2012ApJ...748L..31M, 2019ApJ...883L..10A}. Additionally, future {\emph{PSP}} observations of SIR/CIR-associated ions will be a crucial contribution to studies on the radial gradient of energetic ions \citep[{\emph{e.g.}},][]{1978JGR....83.4723V, Allen2021_CIR}. As {\emph{SolO}} begins to return off-ecliptic observations, the combination of {\emph{PSP}} and {\emph{SolO}} at different latitudes will enable needed insight into the latitudinal structuring of SIR/CIRs and associated particle acceleration.

As the solar cycle progresses, solar activity will increase. This will provide many new observations of CMEs and SEP events at various intensities and radial distances in the inner heliosphere, particularly for low energy SEP events that are not measurable at 1~AU \citep[{\emph{e.g.}},][]{2020ApJS..246...65H}. These {\emph{PSP}} observations will further our understanding of CME-associated shock acceleration and how the energetic content of CMEs evolves with heliographic distance. The current and future Heliophysics System Observatory (HSO) should also provide additional opportunities to not only study the radial evolution of CMEs \citep[{\emph{e.g.}},][]{2021AA...656A...1F}, but also the longitudinal variations of these structures, as was done for the 29 Nov. 2020 event \citep[{\emph{e.g.}},][]{2021A&A...656A..29C,2021A&A...656A..20K,2021A&A...656L..12M}.

As discussed in \S\ref{SEPs}, {\emph{PSP}} has already expanded our understanding of SEP events in the inner heliosphere. Because {\emph{SolO}}, which also returns observations of \textsuperscript{3}He-rich SEP events \citep[{\emph{e.g.}},][]{2021A&A...656L...1M,2021A&A...656L..11B}, will soon be taking measurements of SEP events at different latitudes than {\emph{PSP}}, the combination of these missions will enable exploration of latitudinal variations in SEP content. Similarly, energetic electron measurements on {\emph{SolO}} \citep[{\emph{e.g.}},][]{2021A&A...656L...3G}, soon to be taken off-ecliptic, will enable future studies into the latitudinal variations of electron events.

In addition to radio observations using multiple S/C, space-based and ground-based multi-wavelength observations enable new types of coordinated analysis of solar activity. \cite{2021A&A...650A...7H} combined {\emph{Hinode}}, {\emph{SDO}}/AIA, and RFS observations in a joint analysis of a non-flaring AR and a type III storm observed during {\emph{PSP}} Enc.~2, identifying the source of electron beams associated with the storm and using radio measurements to show the evolution of the peak emission height throughout the storm. \cite{2021A&A...650A...6C} studied a different storm occurring slightly after Enc.~2 using radio observations from {\emph{PSP}} and {\emph{Wind}}, and solar observations from {\emph{SDO}}/AIA, {\emph{SDO}}/HMI, and the Nuclear Spectroscopic Telescope ARray \citep[{\emph{NuSTAR}};][]{2013ApJ...770..103H}, finding correlated periodic oscillations in the EUV and radio data indicative of small impulsive electron acceleration.

Additionally, the continuation of the {\emph{PSP}} project science team’s close relationship with the Whole Heliosphere and Planetary Interactions (WHPI\footnote{https://whpi.hao.ucar.edu/}) international initiative, the successor of Whole Sun Month \citep{1999JGR...104.9673G} and Whole Heliosphere Interval \citep{2011SoPh..274....5G, 2011SoPh..274...29T, 2011SoPh..274....1B} will allow for multifaceted studies that incorporate ground-based and space-based observatories providing contextual information for the {\emph{PSP}} measurements. Many of these studies are beginning now, and should propel our fundamental understanding of the connection of the solar surface to interplanetary space and beyond.

\section{Dust}
\label{PSPDUST}

\subsection{Dust Populations in the Inner Heliosphere} 

The ZDC is one of the largest structures in the heliosphere. It is comprised of cosmic dust particles sourced from comets and asteroids, with most of the material located in the ecliptic plane where the majority of these dust sources reside. The orbits of grains gravitationally bound to the Sun, termed ``\amsn'', lose angular momentum from Poynting-Robertson and solar wind drag \citep[{\emph{e.g.}},][]{1979Icar...40....1B} and subsequently circularize and spiral toward the Sun. Due to the inward transport of zodiacal material, the dust spatial density increases as these grains get closer to the Sun \citep[{\emph{e.g.}},][]{1981A&A...103..177L}, until they are ultimately collisionally fragmented or sublimated into smaller grains \citep[{\emph{e.g.}},][]{2004SSRv..110..269M}.  Dust-dust collisions within the cloud are responsible for generating a significant portion of the population of smaller grains. Additionally, a local source for dust particles very near the Sun are the near-Sun comets, ``Sunskirters'', that pass the Sun within half of Mercury’s perihelion distance, and sungrazers that reach perihelion within the fluid Roche limit \citep{2018SSRv..214...20J}. Because these comets are in elongated orbits, their dust remains in the vicinity of the Sun only for short time \citep{2004SSRv..110..269M,2018A&A...617A..43C}. 

Sub-micron sized grains, with radii on the order of a few hundred nm, are most susceptible to outward radiation pressure. The orbital characteristics of these submicron-sized ``\bmsn'' are set by the ratio of solar radiation pressure to gravitational force, $\beta = F_{R}/F_{G}$, dependent on both grain size and composition \citep{1979Icar...40....1B}. Grains with $\beta$ above a critical value dependent on their orbital elements have positive orbital energy and follow hyperbolic trajectories escaping the heliosphere. This population of grains represents the highest number flux of micrometeoroids at 1~AU \citep{1985Icar...62..244G}. For the smallest nanograins ($\lesssim50$~nm), electromagnetic forces play an important role in their dynamics \citep[{\emph{e.g.}},][]{1986ASSL..123..455M}, where a certain population of grains can become electromagnetically trapped very near the Sun \citep{2010ApJ...714...89C}. Fig.~\ref{fig:dust_overview} summarizes these various processes and dust populations.

\begin{figure}
  \includegraphics[width=4.5in]{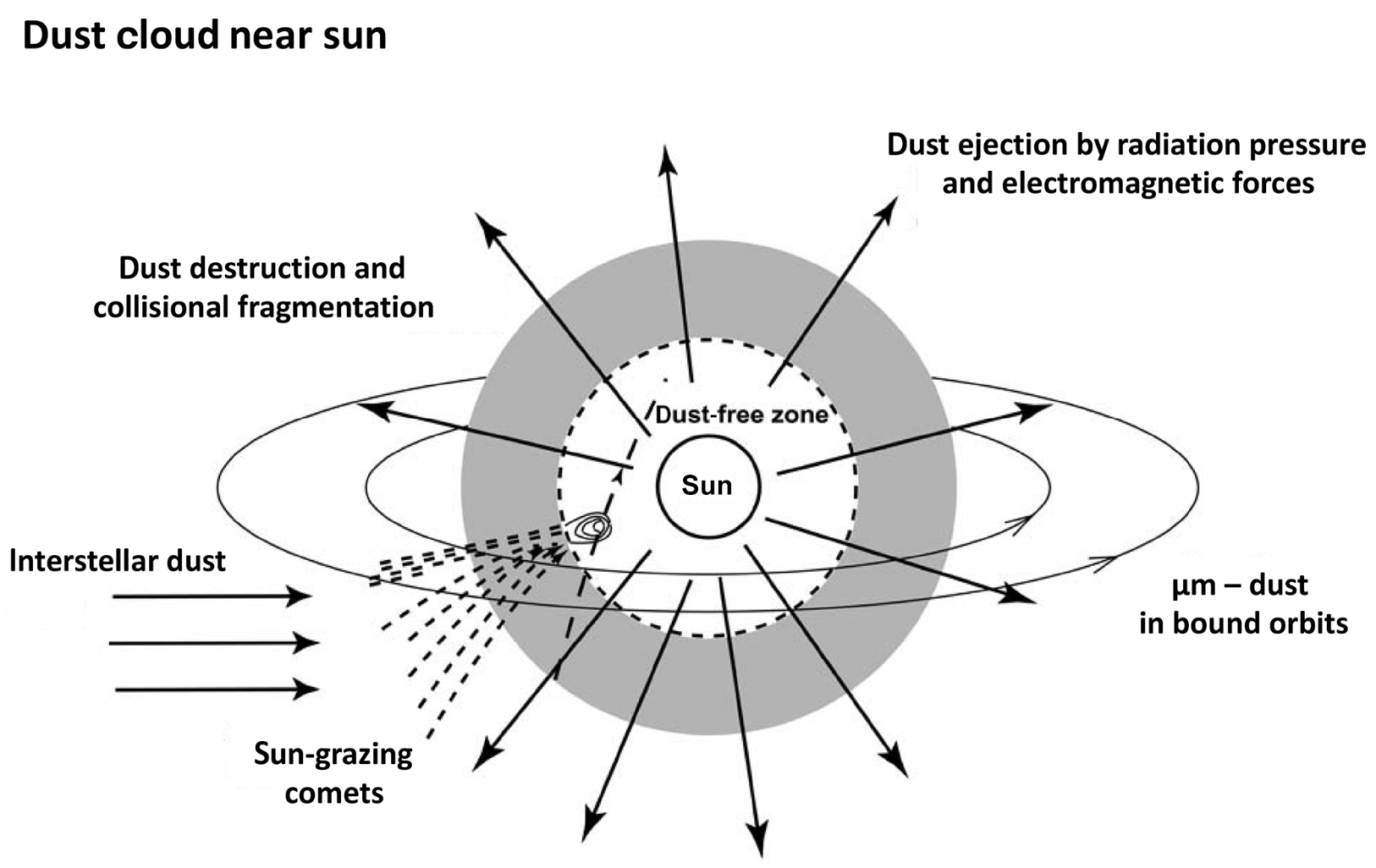}
\caption{The dust environment near the Sun \citep{2019AnGeo..37.1121M}.\label{fig:dust_overview}   }   
\end{figure}

When dust particles approach very near to the Sun, they can sublimate rapidly, leaving a region near the Sun relatively devoid of dust. Different estimates of this DFZ have been made based on Fraunhofer-corona (F-corona) observations and model calculations, predicting a DFZ within 2 to 20 solar radii and possible flattened radial profiles before its beginning \citep{2004SSRv..110..269M}. These estimates are based on dust sublimation; however,  an additional  destruction process recognized in the innermost parts of the solar system is sputtering by solar wind particles. \citet{2020AnGeo..38..919B} showed that sputtering is more effective during a CME event than during other solar wind conditions and suggested that multiple CMEs can lead to an extension of the DFZ. Dust destruction near the Sun releases molecules and atoms, where photoionization, electron-impact ionization, and charge exchange quickly ionize the atoms and molecules. This process contributes to a population of pickup-ions in the solar wind and provides a seed population for energetic particles \citep{2000JGR...105.7465S,2005ApJ...621L..73M}.

The inner heliosphere's dust populations within a few~AU have been observed both with {\emph{in~situ}} dust impact detections and remotely via scattered light observations. Due to their higher number densities, dust grains with radii on the order of $\sim\mu$m and smaller can be observed with {\emph{in~situ}} impact measurements. Dedicated dust measurements within this size range have been taken in the inner solar system with {\emph{Pioneers}} 8 and 9 \citep{1973spre.conf.1047B}, {\emph{HEOS-2}}\footnote{The Highly Eccentric Orbit Satellite 2} \citep{1975P&SS...23..985H, 1975P&SS...23..215H}, {\emph{Helios}} \citep{1978A&A....64..119L, 1981A&A...103..177L, 1980P&SS...28..333G, 2006A&A...448..243A, 2020A&A...643A..96K}, {\emph{Ulysses}} \citep{1999A&A...341..296W,2004A&A...419.1169W,2003JGRA..108.8030L, 2015ApJ...812..141S, 2019A&A...621A..54S}, and {\emph{Galileo}} \citep{1997Icar..129..270G}. These observations identified three populations of dust:  \amsn, \bmsn, and interstellar grains transiting the solar system \citep{1993Natur.362..428G}. Before {\emph{PSP}}, the innermost dust measurements were made by {\emph{Helios}} as close as 0.3~AU from the Sun.

For grains on the order of several $\mu$m and larger, astronomical observations of the F-corona and ZL \citep{1981A&A...103..177L} provide important constraints on their density distributions. Unlike the broader zodiacal cloud (ZC) structure, which is most concentrated near the ecliptic plane, the solar F-corona has a more spherical shape, with the transition from one to the other following a super-ellipsoidal shape according to the radial variation of a flattening index using observations from the {\emph{STEREO}}/SECCHI instrument \citep{2018ApJ...864...29S}. Measurements from {\emph{Helios}}~1 and {\emph{Helios}}~2 at locations between 0.3 to 1~AU showed that the brightness profile at the symmetry axis of the ZL falls oﬀ as a power law of solar distance, with exponent 2.3 \citep{1981A&A...103..177L}, which is consistent with a derived dust density profile of the form $n(r) = n_0\ r^{-1.3}$. This dust density dependence is well-reproduced by the dust produced by Jupiter-family comets \citep{2019ApJ...873L..16P}. Additionally, there were a number of discussions on the influence of excess dust in circumsolar rings near the Sun \citep[][]{1998EP&S...50..493K} on the observed F-corona brightness \citep[][]{1998P&SS...46..911K}. Later on, \cite[][]{2004SSRv..110..269M} showed that no prominent dust rings exist near the Sun.

More recently, \cite{2021SoPh..296...76L}. \citet{2021SoPh..296...76L} analyzed images obtained with the {\emph{SOHO}}/LASCO-C3 between 1996 and 2019. Based on a polarimetric analysis of the LASCO-C3 images, they separated the F- and K-corona components and derived the electron-density distribution. In addition, they reported the likely increasing polarization of the F-corona with increasing solar elongation. They do not discuss, however, the dust distribution near the Sun. They further discuss in detail the properties of the F-corona in \cite{2022SSRv..218...53L}.

To date, our understanding of the near-Sun dust environment is built on both {\emph{in~situ}} and remote measurements outside 0.3~AU, or 65 $\rs$. {\emph{PSP}}, with its eccentric and progressively low perihelion orbit, provides the only {\emph{in~situ}} measurements and remote sensing observations of interplanetary dust in the near-Sun environment inside 0.3~AU. In the first six orbits alone, {\emph{PSP}} has transited as close as 20 $\rs$ from the center of the Sun, offering an unprecedented opportunity to understand heliospheric dust in the densest part of the ZC and provide critical insight into long-standing fundamental science questions concerning the stellar processing of dust debris discs. Key questions {\emph{PSP}} is well-posed to address are: How is the ZC eroded in the very near-Sun environment?; which populations of material are able to survive in this intense environment?; how do the near-Sun dust populations interact with the solar wind?, among others.

\subsection{Dust Detection on {\emph{PSP}}} 

A number of sensors on {\emph{PSP}} are capable of detecting interplanetary dust in the inner heliosphere, each by a different mechanism. The FIELDS instrument detects perturbations to the S/C potential that result from transient plasma clouds formed when dust grains strike the S/C at high velocities, vaporizing and ionizing the impacting grain and some fraction of the S/C surface \citep{2020ApJS..246...27S, Page2020, Malaspina2020_dust}.  WISPR detects solar photons scattered by dust in the ZC \citep{2019Natur.576..232H, 2021A&A...650A..28S}, and IS$\odot$IS is sensitive to dust penetration of its collimator foils by dust \citep{2020ApJS..246...27S}.  Dust detection by these mechanisms has lead to advances in the our understanding of the structure and evolution of the ZDC, and we describe these observations in the context of {\emph{in~situ}} and remote-based measurements separately below.

\subsection{{\emph{in~situ}} impact detection}
\subsubsection{FIELDS} 

As {\emph{PSP}} traverses the inner heliosphere, its orbital trajectory results in high relative velocities between the S/C and interplanetary dust grains.  Relative velocities for $\alpha$-meteoroids can approach 100~km~s$^{-1}$ and exceed 100's~km~s$^{-1}$ for $\beta$-meteoroids and retrograde impactors \citep{2020ApJS..246...27S}.   The impact-generated plasma cloud perturbs the S/C surface potential, creating a signal detectable by the FIELDS electric field sensors.  This method of {\emph{in~situ}} dust detection was first demonstrated on the {\emph{Voyager}} S/C by \citet{Gurnett1983} and has subsequently been reported on numerous other S/C. See the review by \citet{2019AnGeo..37.1121M} and references therein.  

While there is agreement that electric field sensors detect impact ionization of dust, the physical mechanism by which potential perturbations are generated continues to be an active topic of research, with a range of competing theories \citep{Oberc1996, Zaslavsky2015, Kellogg2016, MeyerVernet2017, 2019AnGeo..37.1121M, 2021JGRA..12628965S}, and rapidly advancing lines of inquiry using controlled laboratory experiments \citep[{\emph{e.g.}},][]{2014JGRA..119.6019C, 2015JGRA..120.5298C, 2016JGRA..121.8182C, Nouzak2018, 2021JGRA..12629645S}.  

On {\emph{PSP}}, the vast majority of dust impacts ionization events produce high amplitude ($>10$~mV), brief ($\mu$s to ms) voltage spikes.  These can be detected in various FIELDS data products, including peak detector data, bandpass filter data, high cadence time-series burst data, and lower cadence continuous time-series data  \citep{2016SSRv..204...49B, Malaspina2016_DFB}.

Impact plasma clouds often produce asymmetric responses on electric field antennas \citep[{\emph{e.g.}},][]{Malaspina2014_dust}.  By comparing the relative dust signal amplitude on each FIELDS antenna for a given impact, the location of the impact on the S/C body can be inferred. From the impact location, and constraints imposed by dust population dynamics, one can deduce the pre-impact directionality of the dust that struck the S/C \citep{Malaspina2020_dust, Pusack2021}.  

{\emph{PSP}} data have revealed new physical processes active in the impact ionization of dust.  \citet{Dudok2022_scm} presented the first observations of magnetic signatures associated with the escape of electrons during dust impact ionization.  \citet{Malaspina2022_dust} demonstrated strong connection between the plasma signatures of dust impact ionization and subsequent debris clouds observed by WISPR and the {\emph{PSP}} star trackers. This study also demonstrated that long-duration S/C potential perturbations, which follow some dust impacts, are consistent with theoretical expectations for clouds of S/C debris that electrostatically charge in the solar wind \citep{2021JGRA..12629645S}.  These perturbations can persist up to 60 seconds, much longer than the brief ($\mu$s to ms) voltage spikes generated by the vast majority of dust impacts.

\subsubsection{Data-Model Comparisons}

\begin{figure}[ht]
\centering
\includegraphics[width=4.5in]{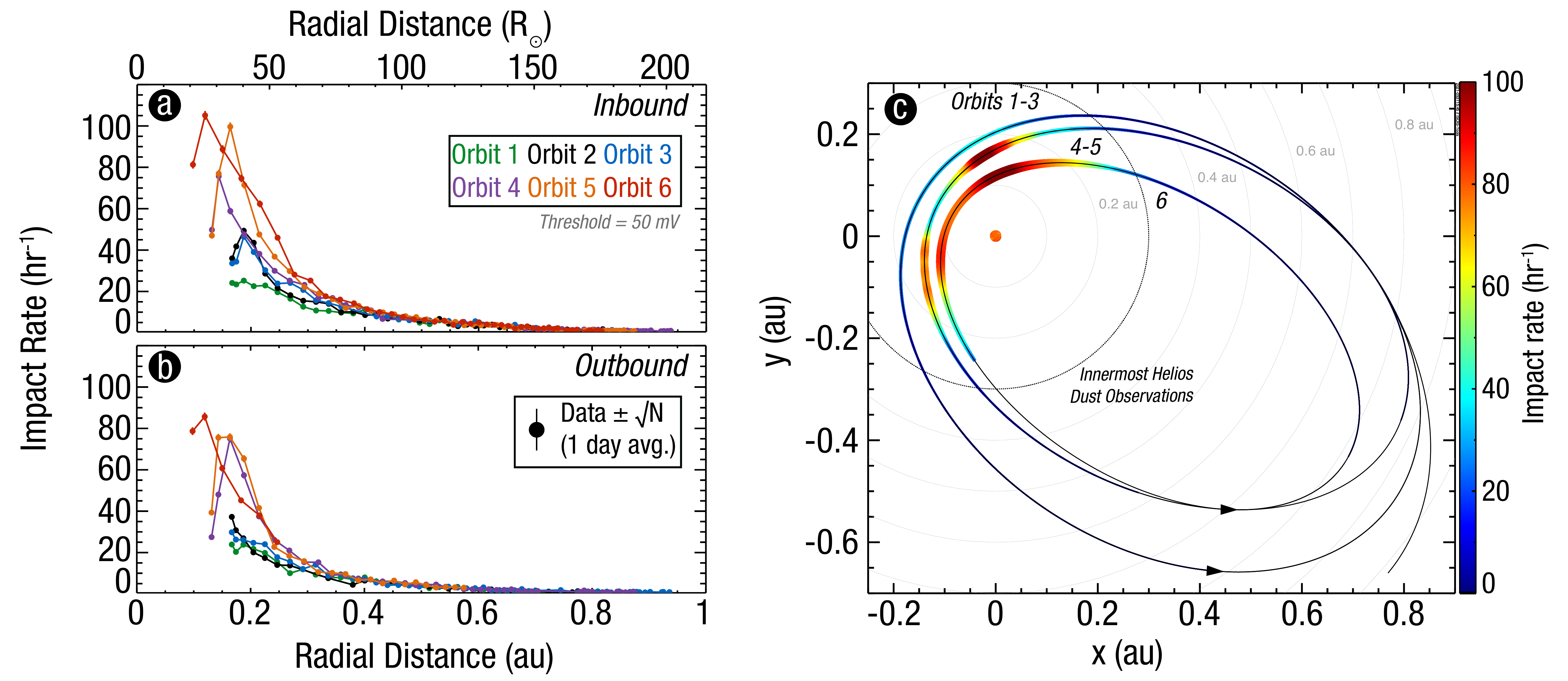}
\caption{Daily averaged impact rates as a function of radial distance for orbits $1-6$, separated by inbound (a) and outbound (b). (c) Impact rates overlaid on the {\emph{PSP}} trajectory in the ecliptic J2000 frame, averaged over orbits $1-3$, $4-5$, and individually shown for orbit 6. Color and width of the color strip represents the impact rate.  Figure adapted from \citet{szalay:21a}. \label{fig:dust_rates}}
\end{figure}

Since FIELDS can detect impacts over the entire S/C surface area, in the range of $4-7$ m$^2$ \citep{Page2020}, {\emph{PSP}} provides a robust observation of the total impact rate to the S/C. Fig.~\ref{fig:dust_rates} shows the impact rates as a function of heliocentric distance and in ecliptic J2000 coordinates \citep{szalay:21a}. There are a number of features that have been observed in the impact rate profiles. For the first three orbits, all with very similar orbits, a single pre-perihelion peak was observed. For the subsequent two orbit groups, orbits $4-5$, and orbit 6, a post-perihelion peak was also observed, where a local minimum in impact rate was present near perihelion. As shown in Fig.~\ref{fig:dust_rates}c, the substructure in observed impact rate occurs inside the previous inner limit of {\emph{in~situ}} dust detections by {\emph{Helios}}.

While {\emph{PSP}} registers a large number of impacts due to its effective area, determining impactor speed, mass, and directionality is not straightforward. To interpret these impact rates into meaningful conclusions about inner zodiacal dust requires data-model comparisons. Analysis of {\emph{PSP}} dust impact data from the first three orbits found the orbital variation in dust count rates detected by FIELDS during the first three solar Encs. were consistent with primarily sub-micron $\beta$-meteoroids \citep{2020ApJS..246...27S,Page2020,Malaspina2020_dust}. From the first three orbits, it was determined that the flux of \bms varies by approximately 50\%, suggesting the inner solar system's collisional environment varies on timescales of 100's of days \citep{Malaspina2020_dust}. Additionally, nanograins with radii below 100 nm were not found to appreciably contribute to the observed impact rates from these first orbits \citep{2021A&A...650A..29M}.

Subsequent analysis which included the first six orbits \citep{szalay:21a} compared {\emph{PSP}} data to a two-component analytic dust model to conclude {\emph{PSP}}'s dust impact rates are consistent with at least three distinct populations: ($\alpha$) bound zodiacal \ams on elliptic orbits, ($\beta$) unbound \bms on hyperbolic orbits, and a distinct third population of impactors. Unlike during the first three orbits of dust impact data, which were dominated by escaping \bmsn, larger grains have been inferred to dominate FIELDS detections for sections of each orbit \citep{szalay:21a} during orbits $4-6$. Data-model comparisons from the first six orbits have already provided important insight on the near-Sun dust environment. First, they placed quantitative constraints on the zodiacal collisional erosion rate of greater than 100 kg s$^{-1}$. This material, in the form of outgoing \bmsn, was found to be predominantly produced within $10-20~\rs$. It was also determined that \bms are unlikely to be the inner source of pickup ions, instead suggesting the population of trapped nanograins \citep{2010ApJ...714...89C} with radii $\lesssim 50$ nm is likely this source. The flux of \bms at 1~AU was also estimated to be in the range of $0.4-0.8 \times 10^{-4}$ m$^{-2}$ s$^{-1}$.

\begin{figure}[ht]
\centering
\includegraphics[width=4.5in]{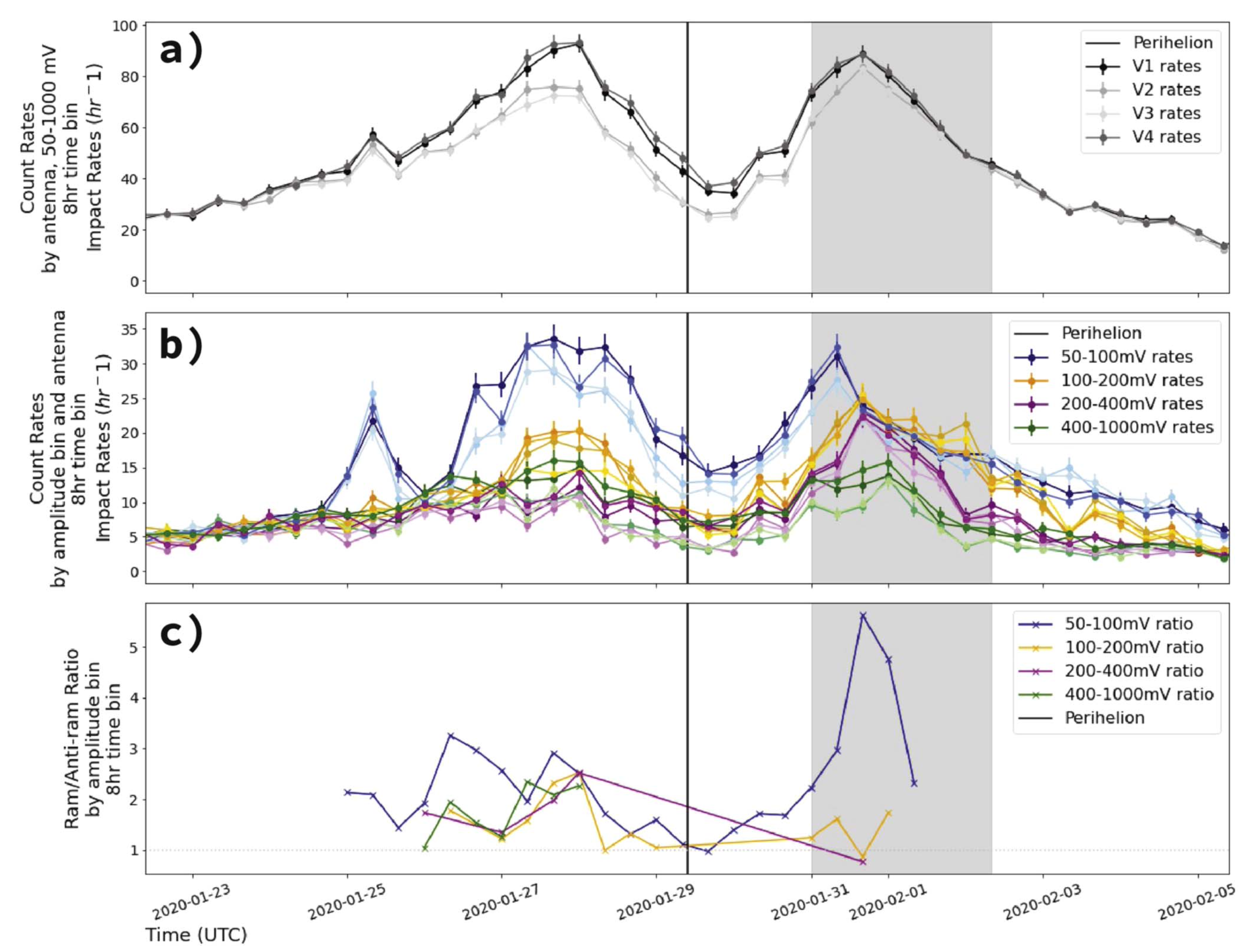}
\caption{(a) Orbit 4 count rates vs. time for antennas V1, V2, V3, and V4 using the $50–1000$ ~mV amplitude window with darker gray lines corresponding to the ram direction of the S/C and lighter gray lines corresponding to the anti-ram direction. (b) Count rates vs. time for each amplitude window on all four planar antennas. Gray shaded region indicates the anomaly duration. (b) carries with it the same gradation as (a) of tone to the various color families depicted, where each color family corresponds to a different amplitude window: blues for $50–100$~mV, orange–yellows for $100–200$~mV, pinks for $200–400$~mV, and greens for $400–1000$~mV. (c) Amplitude window ram/anti-ram rates vs. time with the same color families as (b). Figure adapted from \citet{Pusack2021}.  \label{fig:pusack}}
\end{figure}

From the data-model comparisons, orbits $4-6$ exhibited a post-perihelion peak in the impact rate profile that was not well-described using the two-component model of nominal \ams and \bms \citep{szalay:21a}. Two hypotheses were provided to explain this post-perihelion impact rate enhancements: (a) {\emph{PSP}} directly transited and observed grains within a meteoroid stream or (b) {\emph{PSP}} flew through the collisional by-products produced when a meteoroid stream collides with the nominal ZC, termed a \bsn. The timing and total flux observed during this time favors the latter explanation, and more specifically, a \bs from the Geminids meteoroid stream was suggested to be the most likely candidate \citep{szalay:21a}. A separate analysis focusing on the directionality and amplitude distribution during the orbit 4 post-perihelion impact rate enhancement also supports the Geminids \bs hypothesis \citep{Pusack2021}. Fig.~\ref{fig:pusack} shows the amplitude and directionality trends observed during orbit 4, where the two impact rate peaks exhibit different behaviors. For the pre-perihelion peak, predicted by the two-component model, impact rates for multiple separate amplitude ranges all peak at similar times (Fig.~\ref{fig:pusack}b) and impact the S/C from similar locations (Fig.~\ref{fig:pusack}c). The post-perihelion peak exhibits a clear amplitude dispersion, where impacts producing smaller amplitudes peak in rate ahead of the impacts that produce larger amplitudes. Additionally, the ram/anti-ram ratio is significantly different from the pre-perihelion peak. As further described in \citet{Pusack2021}, these differences are also suggestive of a Geminids \bsn. We note that grains that are smaller than the detected \bms and affected by electromagnetic forces have a much larger flux close to the orbital perihelia than at other parts of the orbit \citep{2021A&A...650A..29M}, yet their detection is difficult with {\emph{PSP}}/FIELDS due to the low expected impact charge generated by such small mass grains \citep{szalay:21a}.

Fig.~\ref{fig:dust_overview_PSP} summarizes the dust populations {\emph{PSP}} is likely directly encountering. From the data-model comparisons, the relative fluxes and densities of bound \ams and unbound \bms has been quantitatively constrained. {\emph{PSP}}'s dust impact measurements have been able to directly inform on the intense near-Sun dust environment. Furthermore, the existence of a third dust population suggests collisions between material along asteroid or cometary orbits can be a significant source of near-sun collisional grinding and \bmm production in the form of \bs \citep{szalay:21a}, and is a fundamental physical process occurring at all stellar dust clouds.  

\begin{figure}[ht]
\centering
\includegraphics[width=4.5in]{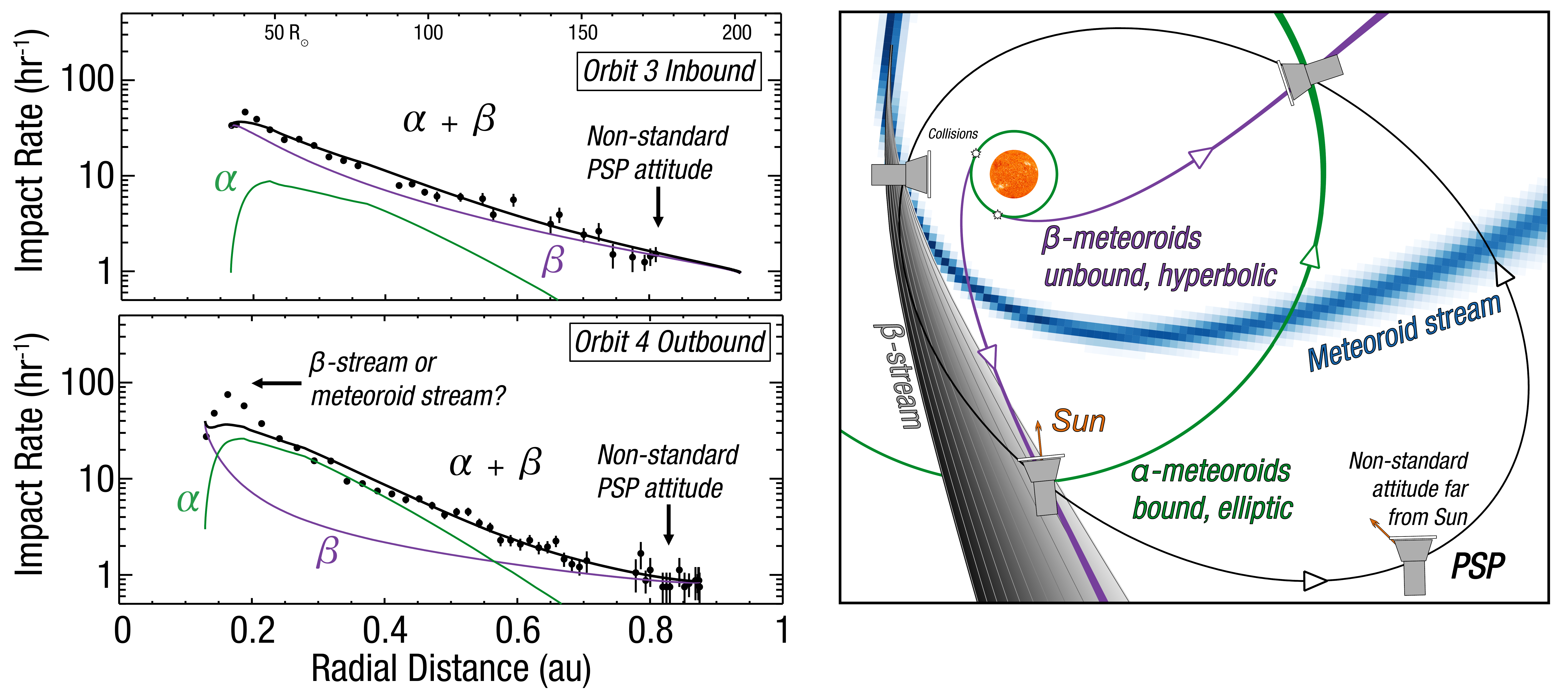}
\caption{{\emph{PSP}} detects impacts due to \amsn, \bmsn, and likely from discrete meteoroid streams. {\it Left}: Impact rates and model fits from orbit 3 (inbound) and orbit 4 (outbound). {\it Right}: Sources for the multiple populations observed by {\emph{PSP}}. Figure adapted from \citet{szalay:21a}.  \label{fig:dust_overview_PSP}}
\end{figure}

\subsection{Remote Sensing}

\subsubsection{Near-sun dust density radial dependence}
\label{dddZ}

\begin{figure}
  \includegraphics[scale=0.35, clip=true]{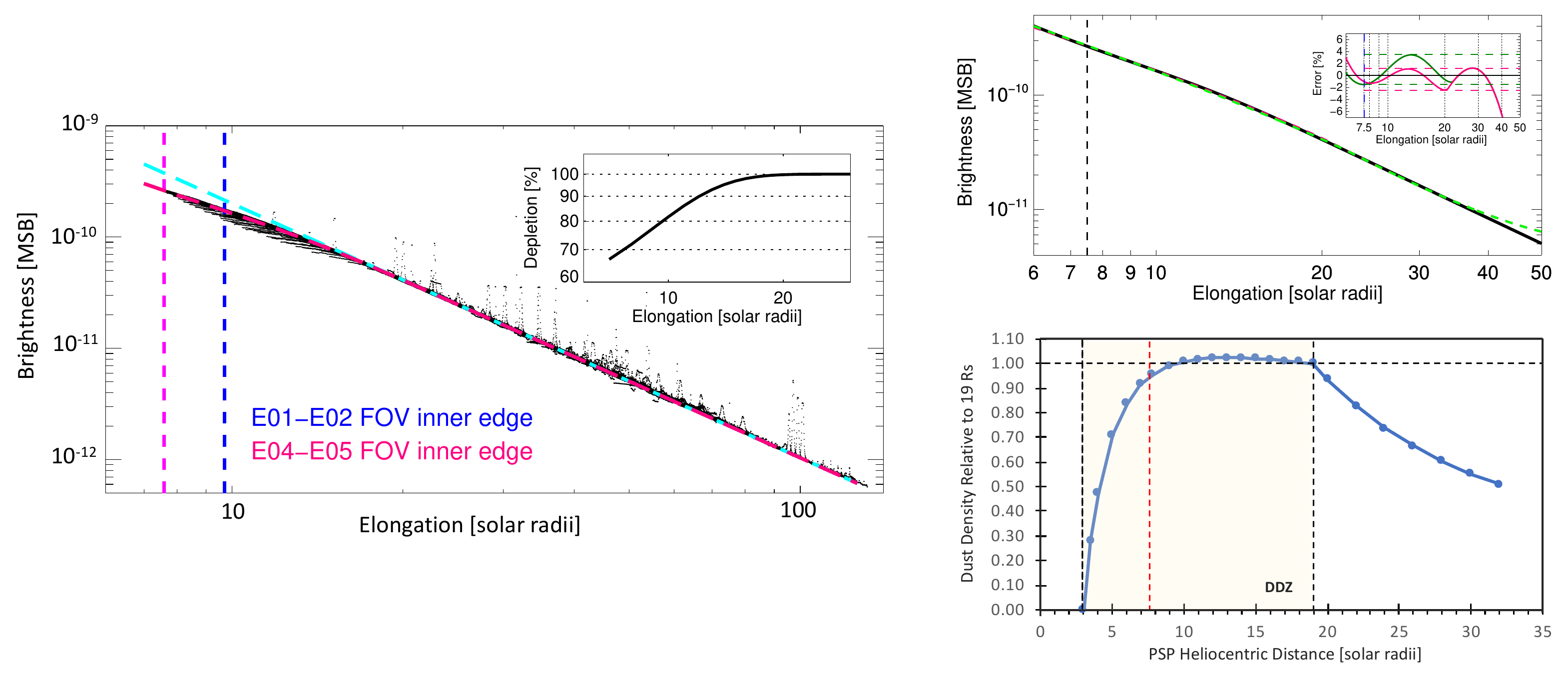}
\caption{(a) Left panel: Sample of radial brightness gradients along the symmetry axis of the F-corona (black) and data fit with an empirical model (red dashed line). The linear portion of the model is delineated with the light-blue dashed line. The inset shows the percentage departure of the empirical model from the linear trend. Upper Right panel: Comparison of the empirical model (in black color) and the forward modeling of the ZL brightness along the symmetry axis considering a DDZ between $2-20~R_\odot$ (in green color) and $3-19~R_\odot$ (in red color). The inset shows the relative error of the simulations compared to the empirical model (same color code). Bottom Right panel: Dust density model used in the simulations relative to the outermost edge of the DDZ between $3–19~R_\odot$ assuming a linear decrease in the multiplier of the nominal density. The dashed, vertical line in a red color indicates the innermost distance of the WISPR FOV in this study. Figure adapted from \cite{2021A&A...650A..28S}.}
\label{fig:depl}      
\end{figure}

In anticipation of the {\emph{PSP}} observations, several studies of the ZL/F-corona based on observations from the {\emph{STEREO}}/SECCHI instrument were carried out  \citep{2017ApJ...839...68S, 2017ApJ...848...57S, 2018ApJ...862..168S,2018ApJ...864...29S}. These studies established a baseline of F-corona properties from 1 AU to help identify any differences that may arise due to the varying heliocentric distance of the corresponding WISPR observations.

The question of whether a DFZ \citep[][]{1929ApJ....69...49R} exists close to the Sun is long-standing and has not been answered by pre-{\emph{PSP}} observations of the ZL/F-corona. White light observations obtained from distances between $0.3-1$~AU \citep[{\emph{e.g.}},][]{2018ApJ...862..168S,1981A&A...103..177L} do not reveal any break in the radial gradient of the brightness along of the symmetry plane of the ZDC, which was found to follow a power law $I(r) \sim r^{-2.3}$ to at least below the theoretically predicted start of the DFZ at $\approx 4-5 R_\odot$.

The WISPR instrument has recorded the intensities of the ZL/F-corona from ever decreasing observing distances, down to about 0.074~AU ($\sim15.9~R_\odot$) at the last executed perihelion (by the time of this writing). This unprecedented observer distance corresponds to an inner limit of the FOV of WISPR-i of about 3.7~$R_\odot$ (0.017~AU). A striking result from the WISPR observations obtained during the first five orbits, was the departure of the radial dependence of the F-corona brightness profile along the symmetry axis of the ZDC from the previously-established power law \citep[][hereafter referred to as HS]{2019Natur.576..232H,2021A&A...650A..28S}. In the left panel of Fig.~\ref{fig:depl}, we show a sample of WISPR brightness profiles along the symmetry axis obtained during orbits 1, 2, 4, and 5 (in black color), along with the fitting of an empirical model comprising a linear and an exponential function (red dashed line). The linear portion of the empirical model is delineated with the light-blue dashed line. We note that the linear behavior ({\emph{i.e.}}, constant gradient) continues down to $\sim20~R_\odot$ with the same slope as observed in former studies ({\emph{i.e.}}, $\propto r^{-2.3}$). Below that elongation distance, the radial brightness gradient becomes less steep. The modeled brightness measurements depart by about 35\% at the inner limit of $7.65~R_\odot$ (0.036~AU) from the extrapolation of the linear part the model (see inset in Fig.~\ref{fig:depl}). The brightness decrease is quite smooth down to $7.65~R_\odot$, {\emph{i.e.}}, it does not show discrete brightness enhancements due to sublimation effects of any particular dust species  \citep[][]{2009Icar..201..395K}.

The brightness profile was forward-modeled using RAYTRACE \citep{2006ApJ...642..523T}\footnote{RAYTRACE is available in the {\emph{SOHO}} Solarsoft library, http://www.lmsal.com/solarsoft/. }, which was adapted to integrate the product of an empirical volume scattering function (VSF) and a dust density at each point along any given LOS. The VSF was given by \cite{1986A&A...163..269L}, which condenses all the physics of the scattering process into an empirical function of a single parameter, the scattering angle. The dust density along the symmetry axis of the ZDC was taken from \cite{1981A&A...103..177L} ($n(r) \propto r^{-1.3}$). The intensity decrease observed in WISPR results was ascribed to the presence of a dust depletion zone (DDZ), which appears to begin somewhere between 19 and 20~$R_\odot$ and extends sunward to the beginning of the DFZ. To model the density decrease in the DDZ, we used a multiplier in the density model defined as a linear factor that varied between 1 at the outer boundary of the DDZ and 0 at the inner boundary. The extent of the DDZ was determined empirically by matching the RAYTRACE calculation with the empirical model of the radial brightness profile.  

In the upper right panel of Fig.~\ref{fig:depl} we show in green and red colors (the red is fully underneath the green) the forward modeling of the brightness with two different boundaries for the DDZ along with the empirical model (in black). The inner boundary was a free parameter to give the best match to the empirical model. The $3-19~R_\odot$ range (green) for the DDZ yields a slightly better fit to the observation than in the $2-20~R_\odot$ range (red). Note that the behavior below the observational limit of 7.65~$R_\odot$ is only an extrapolation. The inset shows the difference between the two forward models. We thus choose $19~R_\odot$ as the upper limit of the DDZ, although depletion could start beyond $19~R_\odot$ but doesn't cause a noticeable change in the intensities until about $19~R_\odot$.

In the bottom right panel of Fig.~\ref{fig:depl}, we show the radial profile of the dust density relative to the density at $19~R_\odot$ for the best fit to the intensity profile. Note that from $\sim10$ to $19~R_\odot$, the density appears to be approximately constant. In future orbits, WISPR will observe the corona down to 2.3~$R_\odot$, which will help establish more accurately the actual limit of the DFZ.

\subsubsection{Implications for collisions and/or sublimation:}

The smooth behavior of the radial brightness profile of the F-corona along its symmetry axis from 35~$R_\odot$ down to 7.65~$R_\odot$ is suggestive of a smooth and continuous process of dust removal. No evidence is seen of dust depletion at a particular distance due to the sublimation of a particular species. Thus the dust remaining at these distances is probably similar to quartz or obsidian, which are fairly resistant to sublimation  \citep[{\emph{e.g.}},][]{2004SSRv..110..269M}.

\subsubsection{Dust density enhancement along the inner planets' orbits}
\label{VdustRing}
 
In addition to measurements of the broad ZC structure, discrete dust structures have also been observed by WISPR. A dust density enhancement nearby Earth's orbit was theoretically predicted in the late eighties by \cite{1989Natur.337..629J} and observationally confirmed by \cite{1994Natur.369..719D} using observations from the Infrared Astronomy Satellite \citep[{\emph{IRAS}};][]{1984ApJ...278L...1N}. \cite{1995Natur.374..521R} confirmed the predicted structure of the dust ring near Earth using observations from the Diffuse Infrared Background Experiment \citep[{\emph{DIRBE}};][]{1993SPIE.2019..180S} on the Cosmic Background Explorer mission \citep[{\emph{COBE}};][]{1992ApJ...397..420B}. More recently, in a reanalysis of white light observations from the {\emph{Helios}} mission  \citep{1981ESASP.164...43P}, \cite{2007A&A...472..335L} found evidence of a brightness enhancement nearby Venus' orbit , which was later confirmed by \cite{2013Sci...342..960J,2017Icar..288..172J} using {\emph{STEREO}}/SECCHI observations. Finally, in spite of the lack of a theoretical prediction, a very faint, circumsolar dust ring associated with Mercury was indirectly inferred from 6+ years of white-light observations \citep{2018ApJ...868...74S} obtained with the {\emph{STEREO}}-Ahead/HI-1 instrument. In all the observational cases mentioned above, only particular viewing geometries allowed the detection of just a small portion of the dust rings.

The {\emph{Helios}} measurements were carried out with the $90^\circ$ photometer of the Zodiacal Light Experiment \citep[{\emph{ZLE}};][]{1975RF.....19..264L}, which looked perpendicular to the ecliptic plane. The observations reported a 2\% increase in brightness as {\emph{Helios}} crossed just outside of Venus’s orbit \citep{2007A&A...472..335L}. On the other hand, the {\emph{STEREO}} observations were obtained with the SECCHI/HI-2 telescopes, which image the interplanetary medium about $\pm20^\circ$ above and below the ecliptic plane. In the latter, the brightness enhancements observed were detected only when the viewing geometry was tangent to the orbit of Venus. The findings were interpreted, via theoretical modeling, as due to the presence of a resonant dust ring slightly beyond Venus' orbit \citep{2013Sci...342..960J,2017Icar..288..172J}. However, in a more recent work, the dust environment near Venus' orbit was modeled by coalescing the orbital paths of more than 10,000,000 dust particles of different provenance under the influence of gravitational and non-gravitational forces \citep{2019ApJ...873L..16P}. According to this model, an hypothetical population of dust particles released by Venus co-orbital asteroids could be stable enough to produce enough signal to match the observations. So far, twilight telescopic surveys have not found any long-term stable Venus co-orbital asteroids \citep{2020PSJ.....1...47P}; however, their existence cannot be ruled out. 

At visible wavelengths, the high density and scattering properties of the dust particles in the ZC \citep[{\emph{e.g.}},][]{1986A&A...163..269L}, makes it difficult to detect localized density structures embedded in it from 1~AU. However, as shown in \cite{2021ApJ...910..157S}, the {\emph{PSP}} mission traveling through regions never visited before by any man-made probe, allows the comprehensive visualization of discrete dust density enhancements in the ZDC. As with other white-light heliospheric imagers, the scene recorded in WISPR observations is dominated by the ZL \citep[or F-corona close to the Sun; see, {\emph{e.g.}},][]{2019Natur.576..232H}. To reveal discrete, stationary F-corona features in the FOV of the WISPR instrument, it is necessary to estimate the  F-corona background component (for its subsequent removal from the images) with images where the stationary feature is present at a different location in the FOV. By exploiting the different rolls while the S/C was between 0.5 and 0.25~AU, \cite{2021ApJ...910..157S} revealed the first comprehensive, white light observation of a brightness enhancement across a $345^\circ$ longitudinal extension along the Venus' orbit.  

\begin{figure}
  \includegraphics[scale=0.33, clip=true, trim=0.0cm -0.5cm 0.0cm 0.0cm]{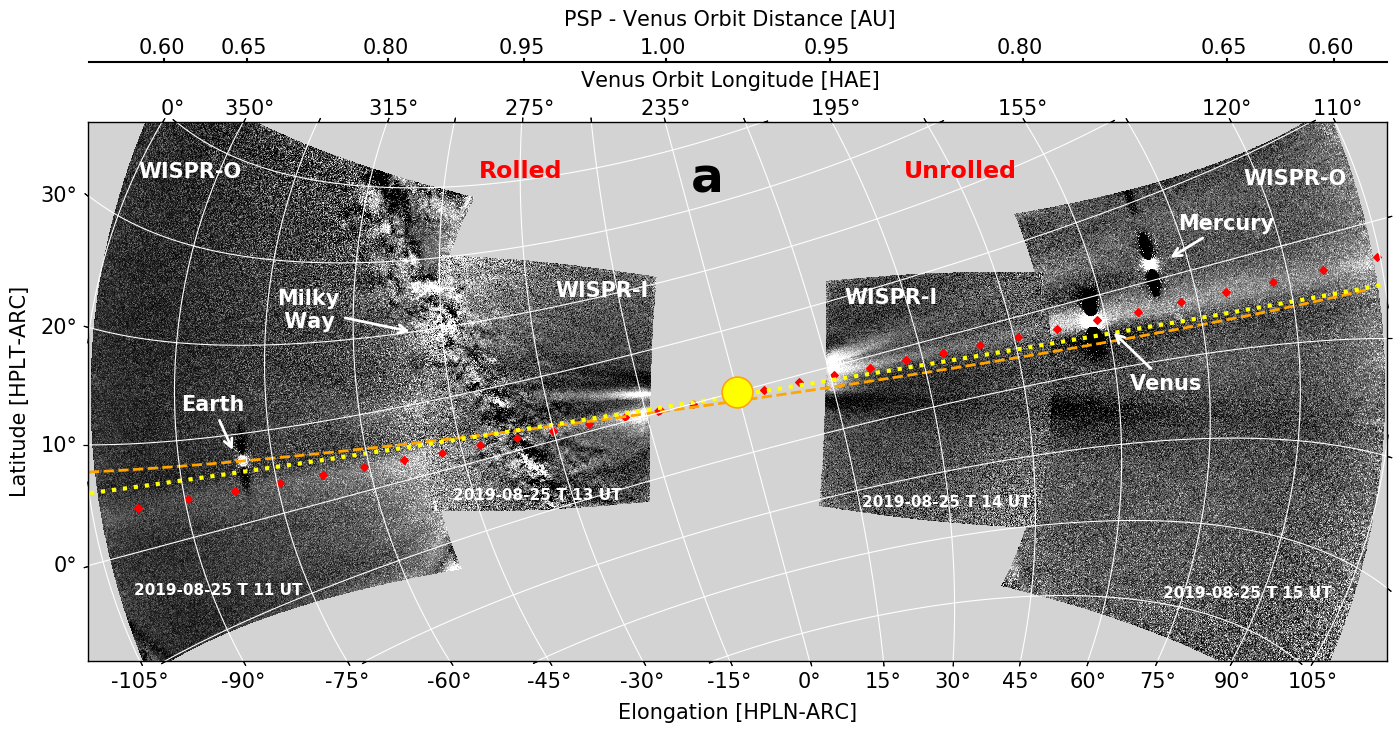}
\caption{Combined WISPR observations of a circumsolar dust ring near Venus’s orbit on 25 Aug. 2019. Images are projected onto the surface of a sphere with observer at the center ({\emph{PSP}} S/C) and radius equal to the heliocentric distance of the observer. The Sun is not to scale. The gray areas surrounding the bright point-like objects (Mercury, Venus, and Earth) are artifacts of the image processing due to the saturation caused by their excessive brightness. The odd oval-shaped object and its surrounding area are caused by reflections in the optics of the very bright Venus. The red dots delineate Venus’s orbital path, the dashed orange line the ecliptic, and the yellow dotted line the invariable plane. Figure adapted from \cite{2021ApJ...910..157S}.}
\label{fig:DustRing}
\end{figure}

Fig.~\ref{fig:DustRing} shows a composite panorama of the Venusian dust ring in WISPR images acquired during the inbound segment of orbit 3 while the {\emph{PSP}} S/C was performing roll maneuvers \citep[as extracted from][]{2021ApJ...910..157S}. The study showed that the latitudinal extension of the brightness enhancement corresponds to a dust ring extending 0.043~AU $\pm$ 0.004~AU, co-spatial with Venus' orbital path. Likewise, the median excess brightness of the band w.r.t. the background (of about 1\%), was shown to correspond to a dust density enhancement relative to the local density of the ZC of about 10\%. Both, the latitudinal extension and density estimates is in general agreement with the findings of \cite{2007A&A...472..335L} and  \cite{2013Sci...342..960J,2017Icar..288..172J}.  The viewing geometry only allowed a measure of the inclination and projected height of the ring, not of its radial position or extent. Therefore, no detailed information on the distance of the dust ring from the orbit of Venus could be extracted.

\subsubsection {Dust Trail of 3200 Phaethon}
\label{Phaethon}

Discovered in 1983 \citep{1983IAUC.3878....1G}, asteroid (3200) Phaethon is one of the most widely-studied inner solar system minor bodies, by virtue of a 1.434 year orbit, its large size for a near-Earth object \citep[6~km in diameter,][]{2019P&SS..167....1T}, and a low 0.0196~AU Earth minimum orbit intersection distance (MOID) favorable to ground-based optical and radar observations \citep{1991ASSL..167...19J,2010AJ....140.1519J}. Phaethon is recognized as the parent of the Geminid meteor shower and is associated with the Phaethon-Geminid meteoroid stream complex including likely relationships with asteroids 2005 UD and 1999 YC \citep[{\emph{e.g.}},][]{1983IAUC.3881....1W,1989A&A...225..533G,1993MNRAS.262..231W,2006A&A...450L..25O,2008M&PSA..43.5055O}. 
Due to a small 0.14~AU perihelion distance, observations of Phaethon near the Sun are impossible from traditional ground-based telescopes. The first detections of Phaethon at perihelion were made by {\emph{STEREO}}/SECCHI \citep{2013ApJ...771L..36J}. While Phaethon is active near perihelion and experiences an intense impact environment near the Sun, the mass-loss rates from cometary-like activity \citep{2013ApJ...771L..36J} and impact ejecta \citep{2019P&SS..165..194S} were both found to be orders of magnitude too low to sustain the Geminids.

\begin{figure}[ht!]
\centering
\includegraphics[scale=0.35]{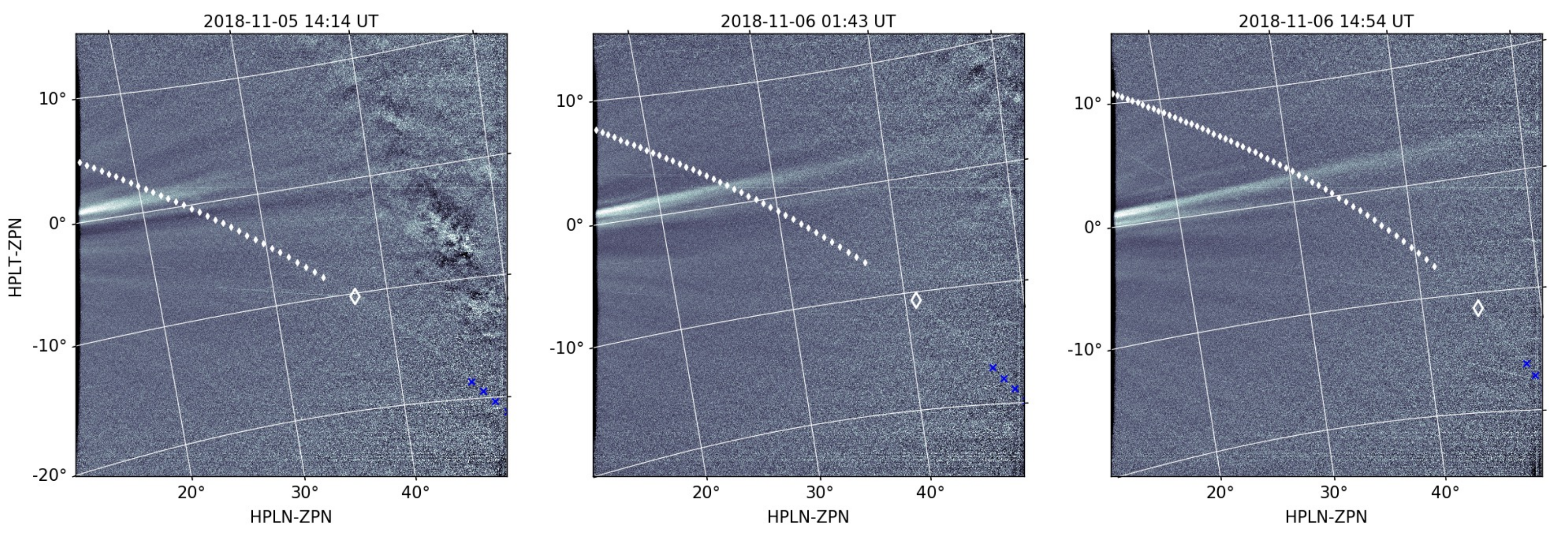}
\caption{WISPR-i observations recorded on 5 Nov. 2018, 14:14~UT, 6 Nov. 1:43~UT and 6 Nov. 14:54~UT. Plotted symbols indicate the imaginary position of Phaethon along the orbit in 60 minute increments, both pre-perihelion (blue) and post-perihelion (white). Symbols are excluded in the region where the trail is most easily visible. The white diamond indicates the perihelion position of the orbit in the FOV. Figure adapted from \cite{2020ApJS..246...64B}.
\label{fig:trail-evolution}}
\end{figure}

As presented in \cite{2020ApJS..246...64B}, an unexpected white-light feature revealed in the WISPR background-corrected data was the presence of a faint extended dust trail following the orbit of Phaethon. In Fig.~\ref{fig:trail-evolution}, we show three WISPR-i telescope observations that highlight the most visible portion of this dust trail as detected during {\emph{PSP}} Enc.~1, which is seen following the projection of  Phaethon's orbital path perfectly. 

Despite the dust trail being close to the instrument noise floor, the mean brightness along the trail was determined to be 8.2$\times10^{-15} B_\odot$ (where $B_\odot$ is the mean solar brightness), which equates to a visual magnitude of 15.8$\pm$0.3 per pixel. This result, coupled with the 70~arcsec per pixel resolution of WISPR-i, yields an estimated surface brightness of 25.0~mag~arcsec$^{-2}$ for the dust trail, which in turn is shown to yield a total mass of dust in the entire trail of $\sim(0.4-1.3){\times}10^{12}$~kg. This mass estimate is inconsistent with dust by Phaethon at perihelion, but is plausibly in-line (slightly below) mass estimates of the Geminids. The difference is attributed primarily to the faintness of the detection. 

This detection highlights the remarkable sensitivity of WISPR to white-light dust structures. Recent ground- and space-based surveys have failed to detect a dust trail in the orbit of 3200 Phaethon \citep{2018ApJ...864L...9Y}. The WISPR observation explains this as, when factors such as heliocentric distance and orbital spreading/clustering of the dust are considered, it can be shown that the surface brightness of the trail as seen from a terrestrial viewpoint is less than 30~mag~arcsec$^{-2}$, which constitutes an extremely challenging target even for deep sky surveys. 

The Phaethon dust trail continues to be clearly observed in the WISPR data in every {\emph{PSP}} orbit, and remains  under continued investigation. The dust trail of comet 2P/Encke is also quite clearly visible in the WISPR data, again highlighting the instrument's ability to detect faint dust features. The inner solar system is rich with fragmenting comets and comet families, yielding the potential for the discovery of additional dust features as the mission orbit evolves.

\subsubsection{Mass Loading of the Solar Wind by Charged Interplanetary Dust}

If charged dust grains reach sufficient density, they are theoretically capable of impacting solar wind plasma dynamics, primarily through mass-loading the wind \citep[{\emph{e.g.}},][]{Rasca2014a}.  As the solar wind flows over charged dust grains, the Lorentz force attempts to accelerate these grains up to the solar wind velocity.  The resulting momentum exchange can slow the solar wind and distort solar wind magnetic fields \citep{Lai2015}.  In practice, a high enough density of dust grains with sufficiently large charge-to-mass ratio to distort the solar wind flow is most likely to be found near localized dust sources, like comets \citep{Rasca2014b}.  The Solar Probe data so far have yielded one such potential comet-solar wind interaction, and a study of this event was inconclusive with regard to whether mass loading created an observable impact on the solar wind \citep{He2021_comet}.  

\subsection{Summary of Dust Observations and Future Prospects for {\emph{PSP}} Dust Measurements} 

Summarizing our understanding of the inner heliosphere's dust environment after four years of {\emph{PSP}} dust data:
\begin{itemize}
\item[1.] Impact rates from the first six orbits are produced by three dust sources: \ams on bound elliptic orbits, \bms on unbound, hyperbolic orbits, and a third dust source likely related to meteoroid streams.
\item[2.] The flux of \bms varies by at least 50\% on year-long timescales.
\item[3.] Directionality analysis and data-model comparisons suggests the third source detected during {\emph{PSP}}'s first six orbits is a \bsn.
\item[4.] A zodiacal erosion rate of at least $\sim100$~kg s$^{-1}$ is consistent with observed impact rates.
\item[5.] The flux of \bms at 1~AU is estimated to be in the range of $0.4-0.8 \times 10^{-4}$ m$^{-2}$ s$^{-1}$.
\item[6.] The majority of zodiacal collisions production \bms occur in a region from $\sim10-20~\rs$.
\item[7.] If the inner source of pickup ions is due to dust, it must be from nanograins with radii $\lesssim50$~nm.
\item[8.] The zodiacal dust density is expected to maintain a constant value in the range of $10-19~\rs$.
\item[9.] A dust ring along the orbit of Venus' orbit has been directly observed.
\item[10.] Multiple meteoroid streams have been directly observed, including the Geminids meteoroid stream.
\end{itemize}

There are a number of ongoing and recently open questions in the {\emph{PSP}}-era of ZC exploration. For example, it is not yet determined why the FIELDS dust count rate rises within each orbital group. Increases among orbital groups are expected because, as the S/C moves closer to the Sun, its relative velocity to zodiacal dust populations increases and the zodiacal dust density increases closer to the Sun \citep{2020ApJS..246...27S}.  While this effect is observed, it is also observed \citep{Pusack2021, szalay:21a} that successive orbits with the same perihelion distance show increasing dust count rates ({\emph{e.g.}}, high dust count rates on orbit 5 compared to 4).  Additionally, can FIELDS dust detections be used to differentiate between existing theories for the generation of voltage spikes by impact-generated plasma? {\emph{PSP}} traverses a wide range of thermal plasma, photon flux, magnetic field, and dust impact velocity conditions, enabling new tests of impact-plasma behavior as a function of these parameters.  

The WISPR remote measurements have provided an unparalleled look at the dust environment with a few 10's of $\rs$. Upcoming orbits will reveal whether the DFZ indicated by WISPR data \citep{2019Natur.576..232H, 2021A&A...650A..28S} will be directly observable {\emph{in~situ}} with FIELDS, and if the observed trend from WISPR for larger grains holds in the micron-sized regime. While {\emph{PSP}} will directly transit the region of constant radial density profile inferred by WISPR in the range of $10-19~\rs$, the decrease of this profile towards a DFZ occurs inside 10~$\rs$ where {\emph{PSP}} will not transit.

\begin{figure}[ht]
\centering
\includegraphics[width=\textwidth]{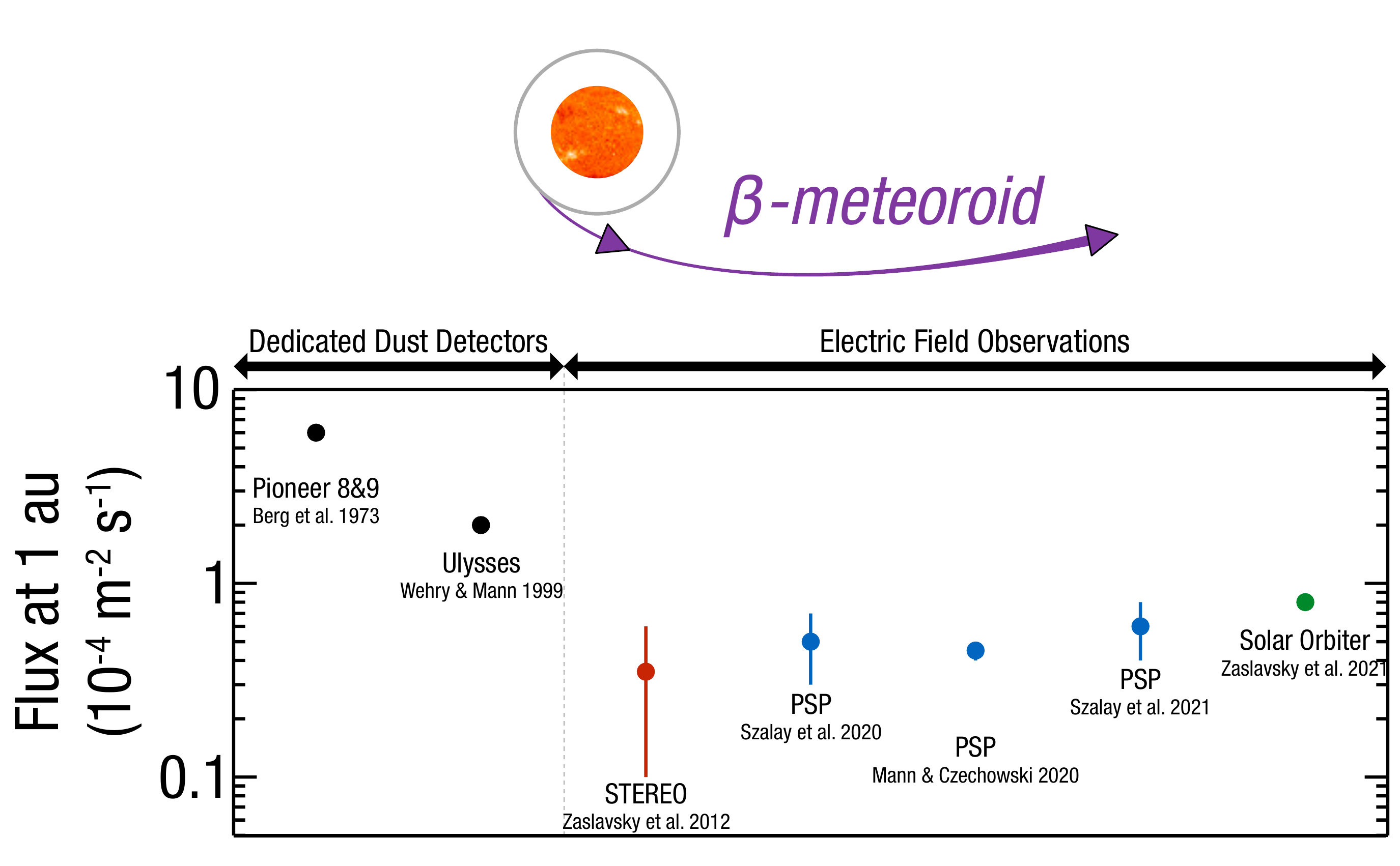}
\caption{\bmm fluxes observed by multiple S/C and detection schemes. \label{fig:beta_fluxes}}
\end{figure}

Finally, {\emph{PSP}}'s long mission duration will enable it to be a long-term observation platform for \bmm fluxes inside 1~AU. The flux of \bms directly encodes the collisional erosion occurring in the inner heliosphere, therefore a determination of their flux provides an important window into the dynamics and evolution of the ZC. Furthermore, the flux of \bms is the largest impactor source by number flux at 1~AU, and may be responsible for sustaining a significant portion of the Moon's impact ejecta cloud \citep{2020ApJ...890L..11S}. Hence, \bms may play a more important role in space weathering airless bodies than previously considered, and constraining their fluxes and variations with time can provide key insight on the space weathering of airless bodies which transit inside 1~AU. Fig.~\ref{fig:beta_fluxes} highlights the multiple \bmm flux estimates from dedicated dust instruments onboard {\emph{Pioneers}} 8 \& 9 \citep{1973spre.conf.1047B}, {\emph{Ulysses}} \citep{1999A&A...341..296W,2004A&A...419.1169W}, as well electric field-based observations from {\emph{STEREO}} \citep{2012JGRA..117.5102Z}, {\emph{SolO}} \citep{2021A&A...656A..30Z}, and {\emph{PSP}} \citep{2020ApJS..246...27S,szalay:21a,2021A&A...650A..29M}. As shown in this figure, the dedicated dust observations indicated a higher flux of \bms than the more recent estimates derived from electric field observations taken decades later. The extent to which the flux of \bms varies over time is a quantity {\emph{PSP}} will be uniquely posed to answer in its many years of upcoming operations.

\section{Venus}
\label{PSPVENUS}

Putting {\emph{PSP}} into an orbit that reaches within 10~R$_{\odot}$ of the Sun requires a series of VGA flybys to push the orbital perihelion closer and closer to the Sun. A total of seven such flybys are planned, five of which have already occurred as of this writing.  These visits to Venus naturally provide an opportunity for {\emph{PSP}} to study Venus and its interactions with the solar wind.  In this section, we review results of observations made during these flybys.

Direct images of Venus have been obtained by the WISPR imagers on board {\emph{PSP}}.  The first attempt to image Venus with WISPR was during the third flyby (VGA3) on 11 Jul. 2020.  The dayside of Venus is much too bright for WISPR to image.  With no shutter mechanism, the effective minimum exposure time with WISPR is the image readout time of about 2~s, much too long for Venus to be anything other than highly overexposed in WISPR images made close to the planet.  Furthermore, the VGA3 sequence of images demonstrated that if any part of dayside Venus is in the FOV, not only is the planet highly overexposed, but there are  scattered light artifacts that contaminate the entire image.

\begin{figure*}
   \centering
   \includegraphics[width=0.6\textwidth]{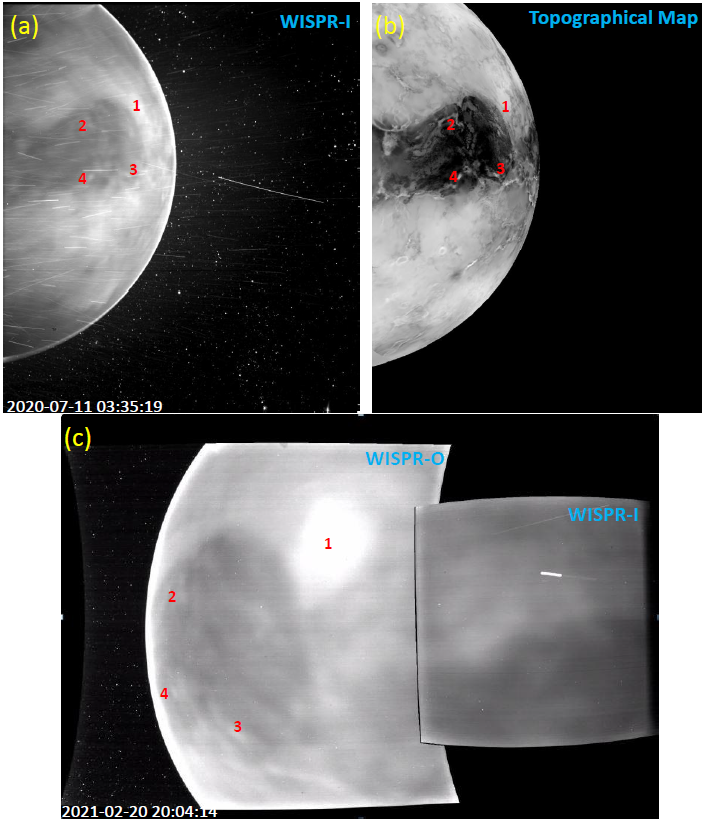}
   \caption{(a) WISPR-i image of the nightside of Venus from VGA3, showing thermal emission from the surface on the disk and O$_2$ nightglow emission at the limb. (b) Topographical map from Magellan, using an inverse black and white scale to match the WISPR image, with bright regions
     being low elevation and dark regions being high elevation. (c) WISPR-i and -o images of Venus from VGA4. The same part of the Venusian surface is observed as in (a). Red numbers in all panels mark common features for ease of reference. Figure adapted from \citet{2022GeoRL..4996302W}.}
              \label{Wood2022Fig1}
\end{figure*} 

Fortunately, there were a couple VGA3 images that only contained nightside Venus in the FOV, and these images proved surprisingly revelatory.  One of these images is shown in Fig.~\ref{Wood2022Fig1}a \citep{2022GeoRL..4996302W}. Structure is clearly seen on the disk.  Furthermore, comparison with a topographical map of Venus from the Magellan mission (see Fig.~\ref{Wood2022Fig1}b) makes it clear that we are actually seeing the surface of the planet.  This was unexpected, as the surface of Venus had never before been imaged at optical wavelengths.  Viewing the planetary surface is impossible on the dayside due to the blinding presence of scattered sunlight from the very thick Venusian atmosphere.

However, on the nightside there are windows in the near infrared (NIR) where the surface of the planet had been imaged before, particularly by the {\emph{Venus Express}} \citep[{\emph{VEX}};][]{2006CosRe..44..334T} and {\emph{AKATSUKI}} \citep{2011JIEEJ.131..220N} missions \citep{2008GeoRL..3511201H,nm08,ni18}.  This is not reflected light but thermal emission from the surface, which even on the nightside of Venus is about 735~K.  The WISPR imagers are sensitive enough to detect this thermal emission within their optical bandpass.  Because surface temperature decreases with altitude on Venus, as it does on Earth, dark areas in the {\emph{PSP}}/WISPR images correspond to cooler highland areas while bright areas correspond to hotter lowland regions. The dark oval-shaped region dominating the WISPR image near the equator is the Ovda Regio plateau at the western end of Aphrodite Terra, the largest highland region on Venus.

In addition to the thermal emission from the disk of the planet, a bright rim of emission is seen at the limb of the planet.  This is O$_2$ nightglow emission from the upper atmosphere of the planet, which had been observed by previous missions, particularly {\emph{VEX}} \citep{2009JGRE..11412002G,2013GeoRL..40.2539G}.  This emission is believed to be excited by winds of material flowing in the upper atmosphere from the dayside to the nightside.

The experience with the VGA3 images allowed for better planning for VGA4, and during this fourth flyby on 21 Feb. 2021 a much more extensive series of images was taken of the Venusian nightside, using both the WISPR-i and WISPR-o imagers. Fig.~\ref{Wood2022Fig1}c shows a view from VGA4, combining the WISPR-i and WISPR-o images.  It so happened that VGA4 occurred with essentially the same part of Venus on the nightside as VGA3, so the VGA3 and VGA4 images in Figs.~\ref{Wood2022Fig1}a and \ref{Wood2022Fig1}c are of roughly the same part of the planet, with the Ovda Regio area dominating both.

An initial analysis of the WISPR images has been presented by \citet{2022GeoRL..4996302W}. A model spectrum of the surface thermal emission was computed, propagated through a model Venusian atmosphere.  This model, assuming a 735~K surface temperature, was able to reproduce the count rates observed by WISPR.  A long-term goal will be to compare the WISPR observations with NIR images.  Ratios of the two could potentially provide
a diagnostic for surface composition.  However, before such mineralogy can be done, a more detailed analysis of the WISPR data must be performed to correct the images for scattered light, disk O$_2$ nightglow, and the effects of spatially variable atmospheric opacity.

Finally, additional WISPR observations should be coming in the future.  Although the Enc. geometry of VGA5 was not favorable for nightside imaging, and VGA6 will likewise be unfavorable, the final flyby (VGA7) on 6 Nov. 2024 should provide an opportunity for new images to be made.  Furthermore, for VGA7 we will be viewing the side of Venus not observed in VGA3 and VGA4.

{\emph{PSP}} also made extensive particle and fields measurements during the Venus flybys.  Such measurements are rare at Venus, particularly high cadence electric and magnetic field measurements \citep{Futaana2017}.  Therefore, {\emph{PSP}} data recorded near Venus has the potential to yield new physical insights. 

Several studies examined the interaction between the induced magnetosphere of Venus and the solar wind.  \citet{2021GeoRL..4890783B} explored kinetic-scale turbulence in the Venusian magnetosheath, quantifying properties of the shock and demonstrating developed sub-ion kinetic turbulence.  \citet{Malaspina2020_Venus} identified kinetic-scale electric field structures in the Venusian bow shock, including electron phase space holes and double layers.  The occurrence rate of double layers was suggested to be greater than at Earth's bow shock, hinting at a potential significant difference in the kinetic properties of bow shocks at induced magnetospheres vs. intrinsic magnetospheres.  \citet{Goodrich2020} identified subproton scale magnetic holes in the Venusian magnetosheath, one of the few observations of such structures beyond Earth's magnetosphere. 

Other studies used the closest portions of the flybys to examine the structure and properties of the Venusian ionosphere. \citet{Collinson2021} examined the ionospheric density at 1,100 km altitude, demonstrating consistency with solar cycle predictions for ionospheric variability. \citet{2022GeoRL..4996485C} used {\emph{PSP}} observations of cold plasma filaments extending from the Venus ionosphere (tail rays) to reconcile previously inconsistent observations of tail rays by {\emph{Pioneer}}~12 (also named Pioneer Venus Orbiter) and {\emph{VEX}}.  

Finally, \citet{Pulupa2021} examined radio frequency data recorded during {\emph{PSP}} Venus flybys, searching for evidence of lightning-generated plasma waves. No such waves were found, supporting results from earlier Cassini flyby observations \citep{2001Natur.409..313G}. 

\section{Summary and Conclusions}
\label{SUMCONC}

{\emph{PSP}} has completed 13 of its 24 scheduled orbits around the Sun over a 7-year nominal mission duration. The S/C flew by Venus for the fifth time on 16 Oct. 2021, followed by the closest perihelion of $13.28~R_\odot$. Generally, the S/C has performed well within the expectations. The science data returned is a true treasure trove, revealing new aspects of the young solar wind and phenomena that we did not know much about. The following is a summary of the findings of the {\emph{PSP}} mission during its four years of operations. We, however, refer the readers to the corresponding sections for more details.

\paragraph{\textbf{Switchbacks ---}}
The magnetic field switchbacks observed by the {\emph{PSP}} are a fundamental phenomenon of the young solar wind. SBs show an impressive effect; they turn the ambient slow solar wind into fast for the crossing duration without changing the connection to the source. These structures are Alfv\'enic, show little changes in the density, and display a slight preference to deflect in the tangential direction. The duration of the observed switchbacks is related to how S/C cross through the structure, which is in turn associated with the deflection, dimensions, orientation, and the S/C velocity. Most studies implied that these structures are long and thin along the flow direction. SB patches have shown the local modulation in the alpha fraction observed in-situ, which could be a direct signature of spatial modulation in solar sources. They also have shown large-scale heating of protons in the parallel direction to the magnetic field, indicating the preferential heating of the plasmas inside the switchbacks.
Observations provided a clue that switchbacks might have relevance to the heating and acceleration of the solar wind. Therefore it is essential to understand their generation and propagation mechanism. Some aspects of these features point toward ex-situ processes ({\emph{e.g.}}, interchange reconnection and other solar-surface processes) and others toward in-situ mechanisms (covering stream interactions, AWs, and turbulence) in which switchbacks result from processes within the solar wind as it propagates outwards. The various flavors of interchange-reconnection-based models have several attractive features, in particular their natural explanation of the likely preferred tangential deflections of large switchbacks, the bulk tangential flow, and the possible observed temperature enhancements. However, some important features remain unclear, such as the Alfv\'enicity of the structures and how they evolve as they propagate to {\emph{PSP}} altitudes.

 While, AW models naturally recover the Alfv\'enicity and radial elongation of switchbacks seen in {\emph{PSP}} observations, but can struggle with some other features. In particular, it remains unclear whether the preferred tangential deflections of large switchbacks can be recovered and also struggle to reproduce the high switchback fractions observed by {\emph{PSP}}. When radially stratified environment conditions are considered for AW models, studies showed that before propagating any significant distance, a switchback will have deformed significantly, either changing shape or unfolding depending on the background profile. This blurs the line between ex-situ and in-situ formation scenarios.
There are interrelationships and the coexistence of different mechanisms in some of the proposed models; moving forward, we must keep all the models in mind as we attempt to distinguish observationally between different mechanisms. Further understanding of switchback formation will require constant collaboration between observers and theorists.   

\paragraph{\textbf{Solar Wind Sources ---}}
A central question in heliophysics is connecting the solar wind to its sources. A broad range of coronal and heliospheric modeling efforts have supported all the {\emph{PSP}} Encs. {\emph{PSP}} has mainly observed only slow solar wind with a few exceptions. The first Enc. proved unique where all models pointed to a distinct equatorial coronal hole at perihelion as the dominant solar wind source. The flow was predominantly slow and highly Alfv\'enic. During the subsequent Encs., the S/C was connected to polar coronal hole boundaries and a flatter HCS. However, what has been a surprise is that the slow solar wind streams were seen to have turbulence and fluctuation properties, including the presence of the SBs, typical of Alfv\'enic fluctuations usually associated with HSSs. That slow wind interval appeared to have much of the same characteristics of the fast wind, including the presence of predominantly outwards Alfv\'enic fluctuations, except for the overall speed. The consensus is that the slow Alfv\'enic solar wind observed by {\emph{PSP}} originates from coronal holes or coronal hole boundaries. It is still unclear how the Alfv\'enic slow wind emerge: (1) does it always arise from small isolated coronal holes with large expansion factors within the subsonic/supersonic critical point? Or is it born at the boundaries of large, polar coronal holes? There is, however, one possible implication of the overall high Alfv\'enicity observed by {\emph{PSP}} in the deep inner heliosphere. All solar wind might be born Alfv\'enic, or rather that Alfv\'enic fluctuations be a universal initial condition of solar wind outflow. Whether this is borne out by {\emph{PSP}} measurements closer to the Sun remains to be seen.

Quiet periods typically separate the SBs-dominated patches. These quiet periods are at odds with theories relating to slow wind formation and continual reconfiguration of the coronal magnetic field lines due to footpoint exchange. This should drive strong wind variability continually \citep[{\emph{e.g.}},][]{1996JGR...10115547F}. Another interesting finding from the {\emph{PSP}} data is the well-known open flux problem persists down to 0.13~AU, suggesting there exist solar wind sources which are not yet captured accurately by modeling.

\paragraph{\textbf{Kinetic Physics ---}} {\emph{PSP}} measurements show interesting kinetic physics phenomena. The plasma data reveal the prevalence of electromagnetic ion-scale waves in the inner heliosphere for the first time. The statistical analysis of these waves shows that a near-radial magnetic field is favorable for their observation and that they mainly propagate anti-sunward. {\emph{PSP}} observed for the first time a series of proton beams with the so-called hammerhead velocity distributions that is an excessively broadened VDF in the direction perpendicular to the mean magnetic field vector. These distributions coincide with intense, circularly polarized, FM/W waves. These findings suggest that the hammerhead distributions arise when field-aligned proton beams excite FM/W waves and subsequently scatter off these waves. {\emph{PSP}} waveform data has also provided the first definitive evidence of sunward propagating whistler-mode waves. This is an important discovery because sunward-propagating waves can interact with the anti-sunward propagating strahl only if the wave vector is parallel to the background magnetic field.

\paragraph{\textbf{Turbulence ---}} Turbulence often refers to the energy cascade process that describes the energy transfer across scales. In solar wind turbulence, the energy is presumably injected at a very large scale ({\emph{e.g.}}, with a period of a few days). It cascades then down to smaller scales until it dissipates at scales near the ion and electron scales. The intermediate scale range between the injection scale and dissipation (or the kinetic) range is known as the inertial range. The {\emph{PSP}} observations shed light on the properties of the turbulence at various scales ({\emph{i.e.}}, outer scale, inertial-range scale, and kinetic scales) at the closest distances to the Sun. This includes the sub-Alfv\'enic region where the solar wind speed becomes smaller than the typical Alfv\'en speed. Several recent studies using {\emph{PSP}} data reveal the significance of solar wind turbulence on the overall heating and acceleration of the solar wind plasma. For instance, magnetic field switchbacks are associated with turbulent structures, which mainly follow the field's kink. Turbulence features such as the intermittency, the Alf\'venicity, and the compressibility have also been investigated. Overall, the data show that solar wind turbulence is mostly highly Alfv\'enic with less degree of compressibility even in the slow solar wind. Other studies used {\emph{PSP}} measurements to examine the typical plasma scale at which the energy spectrum breaks. However, it remains challenging to interpret the appropriate plasma scales corresponding to the empirical timescales using the standard frozen-in-flow Taylor hypothesis as the solar wind speed and the local Alfv\'en speed becomes comparable.

\paragraph{\textbf{Large Scale ---}}
Due to its low heliographic latitude orbit, {\emph{PSP}} crossed the HCS multiple times in each Enc. and observed many LFRs and SFRs. The observed locations of HCS crossings and PFSS model predictions were compared. An irregular source surface with a variable radius is utilized to minimize the timing and location differences. The internal structure of the HCS near the Sun is very complex, comprising structures with magnetic field magnitude depressions, increased solar wind proton bulk speeds, and associated suprathermal electron strahl dropouts, likely indicating magnetic disconnections. In addition, small flux ropes were also identified inside or just outside the HCS, often associated with plasma jets indicating recent magnetic reconnection. {\emph{PSP}} measurements also show that, despite being the site of frequent magnetic reconnection, the near-Sun HCS is much thicker than expected. HCS observations at 1 AU reveal significantly different magnetic and plasma signatures implying that the near-Sun HCS is the location of active evolution of the internal structures. In addition, our knowledge of the transition from CME to ICME has been limited to the in-situ data collected at 1 AU and remote-sensing observations from space-based observatories. {\emph{PSP}} provides a unique opportunity to link both views by providing valuable information that will allow us to distinguish the evidence of the early transition from CME to ICME. 

{\emph{PSP}} has also observed a multitude of events, both large- and small-scale, connected to flux ropes. For instance, at least one SBO event showed a flux rope characterized by changes that deviated from the expected smooth change in the magnetic field direction (flux rope-like configuration), low proton plasma beta, and a drop in the proton temperature. {\emph{PSP}} also observed a significant number of SFRs. Several tens of SFRs were analyzed, suggesting that the SFRs are primarily found in the slow solar wind and that their possible source is MHD turbulence. Other SFRs seem to be the result of magnetic reconnection.

From WISPR imaging data, the most striking features (in addition to CMEs) are the small-scale features observed when the S/C crosses the HCS. The imaging of the young solar wind plasma is revealing. The internal structure of CMEs is observed in ways not accessible before the {\emph{PSP}} era. Also, features such as the fine structure of coronal streamers indicate the highly-dynamic nature of the solar wind close to the Sun. An excellent example of the feature identified by WISPR are bright and narrow streamer rays located at the core of the streamer belt.

\paragraph{\textbf{Radio Emissions and Energetic Particles ---}}

The first four years of the {\emph{PSP}} mission enabled an essential understanding of the variability of solar radio emissions and provided critical insights into the acceleration and transport of energetic particles in the inner heliosphere. {\emph{PSP}} observed many solar radio emissions, SEP events, CMEs, CIRs and SIRs, inner heliospheric ACRs, and energetic electron events, which are critical to exploring the fundamental physics of particle acceleration and transport in the near-Sun environment.

The {\emph{PSP}}/FIELDS RFS measures electric fields from 10 kHz to 19.2 MHz, enabling radio observations. Only Enc.~2 featured multiple strong type III radio bursts and a type III storm during the first four Encs. As the solar activity began rising with Encs.~5 and beyond, the occurrence of radio bursts has also increased. The {\emph{PSP}} radio measurements enabled several critical studies, {\emph{e.g.}}: (1) Searching for evidence of heating of the corona by small-scale nanoflares; (2) Measurement of the circular polarization near the start of several type III bursts in Enc.~2; (3) Characterization of the decay times of type III radio bursts up to 10 MHz, observing increased decay times above 1 MHz compared to extrapolation using previous measurements from {\emph{STEREO}}; (4) Finding evidence for emission generated via the electron cyclotron maser instability over the several-MHz frequency range corresponding to solar distances where $f_{ce}>f_{pe}$; and (5) and determine the directivity of individual type III radio bursts using data from other missions, which was only possible using statistical analysis of large numbers of bursts.

{\emph{PSP}} observed many SEP events from different sources ({\emph{e.g.}}, SBOs, jets, surges, CMEs, flares, etc.) and with various properties that are key to characterizing the acceleration and transport of particles in the inner heliosphere. \citet{2021A&A...650A..23C} investigated the helium content of six SEP events from May to Jun. 2020 during the fifth orbit. At least three of these six events originated from the same AR. Yet, they have significantly different $^3$He/$^4$He and He/H ratios. In addition,  \citet{2021A&A...650A..26C} found that the path length of these events greatly exceeded that of the Parker spiral. They attributed that to the random walk of magnetic field lines. 

Most of the CMEs observed by {\emph{PSP}} were slow and did not produce clear events at 1~AU. They nonetheless produced particle events that were observed by {\emph{PSP}} closer to Sun. \citet{2020ApJS..246...29G} and \citet{2020ApJS..246...59M} reported on a particular CME event observed by {\emph{PSP}} shortly after the first perihelion pass that produced a significant enhancement in SEPs with energies below a few hundred keV/nuc, which also showed a clear velocity dispersion. The {\emph{PSP}} plasma measurement did not show any shock evidence, and the particle flux decayed before the CME crossed the S/C. Two different interpretations were proposed for this event. \citet{2020ApJS..246...29G} suggested diffusive transport of particles accelerated by the CME starting about the time it was 7.5~$R_\odot$ as observations suggest that very weak shocks, or even non-shock plasma compressions driven by a slow CME, are capable of accelerating particles. \citet{2020ApJS..246...59M} proposed an alternative based on the “pressure cooker” mechanism observed in the magnetosphere, where energetic particles are confined below the CME in the solar corona in a region bound by an electric potential above and strong magnetic fields below. The highest-energy particles overcome this barrier earlier and arrive at the S/C earlier than low-energy particles, presumably released much later when the CME has erupted from the Sun. The other interesting aspect is that the “pressure cooker” mechanism produces maximum energy that depends on the charge of the species. Although the event was relatively weak, there were sufficient counts of He, O, and Fe that, when combined with assumptions about the composition of these species in the corona, agreed with the observed high-energy cut-off as a function of particle species.

SIRs/CIRs are known to be regions where energetic particles are accelerated. Therefore, {\emph{PSP}} observations within 1~AU are particularly well suited to detangle these acceleration and transport effects as the SIR/CIR-associated suprathermal to energetic ion populations are further from shock-associated acceleration sites that are usually beyond 1~AU. Many of these nascent SIR/CIRs were associated with energetic particle enhancements offset from the SIR/CIR interface. At least one of these events had evidence of local compressive acceleration, suggesting that this event provides evidence that CIR-associated acceleration does not always require shock waves. 

{\emph{PSP}} also observed ACRs with intensities increasing over energies from $\sim5$ to $\sim40$ MeV/nuc, a characteristic feature of ACR spectra. However, the observed radial gradient is stronger ($\sim25\pm5$\%~AU) than observed beyond 1~AU. Understanding the radial gradients of ACRs in the inner heliosphere provides constraints on drift transport and cross-field diffusion models.

\paragraph{\textbf{Dust in the Inner Heliosphere ---}} 
The zodiacal dust cloud is one of the most significant structures in the heliosphere. To date, our understanding of the near-Sun dust environment is built on both in-situ and remote measurements outside 0.3~AU. {\emph{PSP}} provides the only in-situ measurements and remote sensing observations of interplanetary dust in the near-Sun environment inside 0.3~AU. {\emph{PSP}} provides the total dust impact rate to the S/C. The FIELDS instrument detects perturbations to the S/C potential that result from transient plasma clouds formed when dust grains strike the S/C at high velocities, vaporizing and ionizing the impacting grain and some fraction of the S/C surface. Several features have been observed in the impact rate profiles. For the first three orbits, a single pre-perihelion peak was observed. Another post-perihelion peak also marks the subsequent orbits. Between these two peaks, a local minimum in impact rate was present near the perihelion. 

Comparing the {\emph{PSP}} data to a two-component analytic dust model shows that {\emph{PSP}}’s dust impact rates are consistent with at least three distinct populations: a bound zodiacal $\alpha$-meteoroids on elliptic orbits; an unbound $\beta$-meteoroids on hyperbolic orbits; and a distinct third population of impactors. The data-model comparison indicates that the $\beta$-meteoroids are predominantly produced within $10-20~R_\odot$ and are unlikely to be the inner source of pickup ions, instead suggesting the population of trapped nanograins is likely this source. The post-perihelion peak is like the result of {\emph{PSP}} flying through the collisional by-products produced when a meteoroid stream ({\emph{i.e.}}, the Geminids meteoroid stream) collides with the nominal zodiacal cloud. 

At about 19~$R_\odot$, WISPR white-light observations revealed a lower increase of F-corona brightness compared to observations obtained between 0.3~AU and 1~AU. This marks the outer boundary of the DDZ. The radius of the DFZ itself is found to be about 4~$R_\odot$. The {\emph{PSP}} imaging observations confirm a nine-decade prediction of stellar DFZs by \citet{1929ApJ....69...49R}.

\paragraph{\textbf{Venus ---}} WISPR Images of the Venusian night side during VGAs 3 and 4 proved surprisingly revelatory, clearly showing structures on the disk. This was unexpected, as the surface of Venus had never before been imaged at optical wavelengths. The WISPR imagers are sensitive enough to detect this thermal emission within their optical bandpass. The WISPR images show the Ovda Regio plateau at the western end of Aphrodite Terra, the most extensive highland region on Venus. In addition to the thermal emission from the planet's disk, the data show an O\_2 night glow emission from the planet's upper atmosphere, which previous missions had observed. Another important planetary discovery is that of the dust ring along the orbit of Venus (see \S\ref{VdustRing}).

{\emph{PSP}} is over four years into its prime mission. It uncovered numerous phenomena that were unknown to us so far and which are about phenomena occurring during the solar cycle minimum, where the Sun is not very active. The activity level is rising as we approach the maximum of solar cycle 25. We will undoubtedly discover other aspects of the solar corona and inner heliosphere. For instance, we pine for the S/C to fly through many of the most violent solar explosions and tell us how particles are accelerated to extreme levels.

\section{List of Abreviations}

{\small
\begin{longtable}{ p{.10\textwidth} p{.90\textwidth} } 
\multicolumn{2}{l}{\textbf{Space Agencies}}  \\ 
ESA         &  European Space Agency   \\
JAXA        &  Japan Aerospace Exploration Agency   \\
NASA        &  National Aeronautics and Space Administration   \\
\end{longtable}
}

{\small
\begin{longtable}{ p{.15\textwidth} p{.5\textwidth} p{.35\textwidth}  } 
\multicolumn{3}{l}{\textbf{Missions, Observatories, and Instruments}}  \\ 
\multicolumn{1}{l}{\textbf{Acronym}}  &  \multicolumn{1}{l}{\textbf{Expanded Form}}   & \multicolumn{1}{l}{\textbf{References}}  \\ 
\endfirsthead

\multicolumn{3}{p{\textwidth}}{  --- {\it{continued from previous page}} --- } \\
 \multicolumn{1}{l}{\textbf{Acronym}}  &  \multicolumn{1}{l}{\textbf{Expanded Form}}   & \multicolumn{1}{l}{\textbf{References}} \\  
\endhead

 \multicolumn{3}{p{\textwidth}}{ --- {\it{continued on next page}} ---} \\ 
\endfoot

\endlastfoot

{\emph{\textbf{ACE}}}         &  The Advanced Composition Explorer mission  & \citet{1998SSRv...86....1S}  \\  
    \hspace{0.15cm}  {\emph{EPAM}}        &  The Electron, Proton, and Alpha Monitor instrument  & \citet{1998SSRv...86..541G}   \\ 
    \hspace{0.15cm}  {\emph{ULEIS}}         & The Ultra Low Energy Isotope Spectrometer instrument   & \citet{1998SSRv...86..409M}  \\ 

{\emph{\textbf{AKATSUKI}}}         &  The AKATSUKI/PLANET-C mission  & \citet{2011JIEEJ.131..220N}  \\ 

{\emph{\textbf{ARTEMIS}}}         &  The Acceleration, Reconnection, Turbulence and Electrodynamics of the Moon's Interaction with the Sun mission  & \citet{2011SSRv..165....3A}  \\ 

{\emph{\textbf{BepiColombo}}}         &  The BepiColombo mission  & \citet{2021SSRv..217...90B}  \\ 
{\emph{\textbf{Cluster}}}         &  The Cluster mission   & \citet{1997SSRv...79...11E}  \\ 
{\emph{\textbf{COBE}}}         & The Cosmic Background Explorer mission    & \citet{1992ApJ...397..420B}  \\ 
    \hspace{0.15cm}  DIRBE         & The Diffuse Infrared Background Experiment    & \citet{1993SPIE.2019..180S}  \\ 
{\emph{\textbf{Galileo}}}         &  The Galileo mission   & \citet{1992SSRv...60....3J}  \\ 
{\emph{\textbf{GOES}}}         &  The Geostationary Operational Environmental Satellite program  & \url{https://www.nasa.gov/content/goes-overview/index.html}  \\ 
{\emph{\textbf{GONG}}}         &  The Global Oscillation Network Group   & \citet{1988AdSpR...8k.117H}  \\   
{\emph{\textbf{Helios}}}         & The Helios (1 \& 2) mission   & \citet{Marsch1990}  \\    
    \hspace{0.15cm}  ZLE         & The Zodiacal Light Experiment    &  \citet{1975RF.....19..264L} \\   
{\emph{\textbf{HEOS-2}}}         & The Highly Eccentric Orbit Satellite-2   & \url{https://nssdc.gsfc.nasa.gov/nmc/spacecraft/display.action?id=1972-005A}  \\  
{\emph{\textbf{IMP-8}}}        &  The Interplanetary Monitoring Platform-8   & \url{https://science.nasa.gov/missions/imp-8}  \\   
{\emph{\textbf{IRAS}}}         &  The Infrared Astronomy Satellite   & \citet{1984ApJ...278L...1N}  \\  
{\emph{\textbf{ISEE}}}         & The International Sun-Earth Explorer    & \cite{1979NCimC...2..722D}  \\    
{\emph{\textbf{Mariner~2}}}         &  The Mariner~2 mission  & \url{https://www.jpl.nasa.gov/missions/mariner-2}  \\  
{\emph{\textbf{MMS}}}         &  The Magnetospheric Multiscale mission   & \citet{2014cosp...40E.433B}  \\ 
{\emph{\textbf{NuSTAR}}}         & The Nuclear Spectroscopic Telescope ARray    & \citet{2013ApJ...770..103H}  \\   
{\emph{\textbf{Pioneer}}}         &  The Pioneer mission  & \url{https://www.nasa.gov/centers/ames/missions/archive/pioneer.html}  \\   
{\emph{\textbf{PSP}}}         &  The Parker Solar Probe mission   & \citet{2016SSRv..204....7F}  \\  
                               &                                                 & \citet{doi:10.1063/PT.3.5120}  \\
    \hspace{0.15cm} FIELDS         & The FIELDS investigation   & \citet{2016SSRv..204...49B}  \\  
    \hspace{0.3cm}  AEB         & The Antenna Electronics Board    &  ---  \\  
    \hspace{0.3cm}  DFB         & The Digital Field Board    &  ---  \\  
    \hspace{0.3cm}  HFR         & The High Frequency Receiver    &  ---  \\   
    \hspace{0.3cm}  MAG(s)	&  The Fluxgate magnetometer(s)   &  ---  \\  
    \hspace{0.3cm}  RFS         & The Radio Frequency Spectrometer    & \citet{2017JGRA..122.2836P}  \\   
    \hspace{0.3cm}  SCM         & The Search Coil Magnetometer    & ---  \\  
    \hspace{0.3cm}  TDS         & The Time Domain Sampler    & ---  \\  

    \hspace{0.15cm}  SWEAP         & The Solar Wind Electrons Alphas and Protons Investigation    & \citet{2016SSRv..204..131K}  \\   
    \hspace{0.3cm}  SPAN         &  Solar Probe ANalzers (A \& B)   & \citet{2020ApJS..246...74W}  \\  
    \hspace{0.3cm}  SPAN-e         &  Solar Probe ANalzers-electrons   & \citet{2020ApJS..246...74W}  \\  
    \hspace{0.3cm}  SPAN-i         &  Solar Probe ANalzers-ions   & \citet{10.1002/essoar.10508651.1}  \\  
    \hspace{0.3cm}  SPC            &  Solar Probe Cup   & \citet{2020ApJS..246...43C}  \\  
    \hspace{0.3cm}  SWEM            &  SWEAP Electronics Module   & ---  \\  

    \hspace{0.15cm}  {\emph{IS$\odot$IS}}         & The Integrated Science Investigation of the Sun    & \citet{2016SSRv..204..187M}  \\  
    \hspace{0.3cm}  EPI-Hi         &  Energetic Particle Instrument-High   &  --- \\  
    \hspace{0.45cm}  HET            &  High Energy Telescope   & ---  \\  
    \hspace{0.45cm}  LET            &  Low Energy Telescope   & ---  \\  
    \hspace{0.3cm}  EPI-Lo         &  Energetic Particle Instrument-Low    & ---  \\  
      
    \hspace{0.15cm}  {\emph{WISPR}}       &  The Wide-field Imager for Solar PRobe   & \citet{2016SSRv..204...83V}  \\  
    \hspace{0.3cm}  {\emph{WISPR-i}}       &  WISPR inner telescope   & \citet{2016SSRv..204...83V}  \\  
    \hspace{0.3cm}  {\emph{WISPR-o}}      &  WISPR outer telescope   & \citet{2016SSRv..204...83V}  \\  

    \hspace{0.15cm}  {\emph{TPS}}       &  the Thermal Protection System   & ---  \\  

{\emph{\textbf{SDO}}}         &  The Solar Dynamic Observatory   & \citet{2012SoPh..275....3P}  \\   
    \hspace{0.15cm}  AIA         & The Advanced Imaging Assembly    & \citet{2012SoPh..275...17L}  \\  
    \hspace{0.15cm}  HMI         &  The Helioseismic and Magnetic Imager   & \citet{2012SoPh..275..207S}  \\  

{\emph{\textbf{SOHO}}}         &  The Solar and Heliospheric Observatory   & \citet{1995SSRv...72...81D}  \\  
    \hspace{0.15cm}  {\emph{EPHIN}}        & Electron Proton Helium INstrument    & \citet[EPHIN;][]{1988sohi.rept...75K}   \\  
    \hspace{0.15cm}  {\emph{LASCO}}        & Large Angle and Spectrometric COronagraph    & \citet{1995SoPh..162..357B}   \\  

{\emph{\textbf{SolO}}}         &  The Solar Orbiter mission  & \citet{2020AA...642A...1M}  \\   

{\emph{\textbf{STEREO}}}        &  The Solar TErrestrial RElations Observaory   & \citet{2008SSRv..136....5K}  \\  
    \hspace{0.15cm}  SECCHI         & Sun-Earth Connection Coronal and Heliospheric Investigation    &  \citet{2008SSRv..136...67H} \\ 
    \hspace{0.3cm}  COR2        &   Coronagraph 2  & ---  \\
    \hspace{0.3cm}  EUVI         &  Extreme Ultraviolet Imager   & \citet{2004SPIE.5171..111W}  \\  
    \hspace{0.3cm}  HI        & Heliospheric Imager  (1 \& 2)   & \citet{2009SoPh..254..387E}  \\ 

{\emph{\textbf{Ulysses}}}         &  Ulysses   & \citet{1992AAS...92..207W}  \\  
{\emph{\textbf{VEX}}}         &  Venus Express   & \citet{2006CosRe..44..334T}  \\   
{\emph{\textbf{Voyager}}}    & Voyager   (1 \& 2)  & \citet{1977SSRv...21...77K}  \\ 
{\emph{\textbf{Wind}}}         &  Wind   & \url{https://wind.nasa.gov}  \\
    \hspace{0.15cm}  {\emph{WAVES}}         & WAVES   & \citet{1995SSRv...71..231B}  

\end{longtable}
}

{\small
\begin{longtable}{ p{.15\textwidth} p{.4\textwidth} p{.35\textwidth} } 
\multicolumn{3}{l}{\textbf{Models}}  \\ 
\multicolumn{1}{l}{\textbf{Acronym}}  &  \multicolumn{1}{l}{\textbf{Expanded Form}}   & \multicolumn{1}{l}{\textbf{References}}  \\ 
\endfirsthead
\multicolumn{3}{p{\textwidth}}{  --- {\it{continued from previous page}} --- } \\
 \multicolumn{1}{l}{\textbf{Acronym}}  &  \multicolumn{1}{l}{\textbf{Expanded Form}}   & \multicolumn{1}{l}{\textbf{References}} \\  
\endhead
 \multicolumn{3}{p{\textwidth}}{ --- {\it{continued on next page}} ---} \\ 
\endfoot
\endlastfoot
3DCORE		&  3D Coronal Rope Ejection model									& \citet{2021ApJS..252....9W}	\\
ADAPT		&  Air Force Data Assimilative Photospheric Flux Transport model			& \citet{2004JASTP..66.1295A}	\\
CC        		&  Circular Cylindrical model										&  --- \\
EC			&  Elliptical-Cylindrical model										&  --- \\
EUHFORIA	&  European Heliopheric FORecasting Information Asset					& \citet{2018JSWSC...8A..35P}	\\
GCS         		&  Graduated Cylindrical Shell model									&  \citet{2011ApJS..194...33T} \\
HELCATS		&   Heliospheric Cataloguing, Analysis and Techniques Service model		& \citet{2014AGUFMSH43B4214B}	\\
HelMOD		&   Heliospheric Modulation model									& \citet{2018AdSpR..62.2859B}	\\
OSPREI		&  Open Solar Physics Rapid Ensemble Information model				& \citet{2022SpWea..2002914K}	\\
PARADISE	&  Particle Radiation Asset Directed at Interplanetary Space Exploration model	& \citep{2019AA...622A..28W,2020AA...634A..82W}	\\
PFSS		&  Potential-Field Source-Surface model								& \citet{1969SoPh....9..131A,1969SoPh....6..442S}	\\  
PIC			&  Particle-In-Cell												& ---	\\
SSEF30		&  The Self-Similar Expansion Fitting (30) model						& \citet{2012ApJ...750...23D}     \\ 
WSA			&  Wang-Sheeley-Arge (PFSS) model								& \citet{2000JGR...10510465A}	\\   
WSA/THUX	&  WSA/Tunable HUX model										& \citet{2020ApJ...891..165R}	
\end{longtable}
}

{\small
\begin{longtable}{ p{.20\textwidth} p{.80\textwidth}  } 
\multicolumn{2}{l}{\textbf{Acronyms and Symbols}}  \\ 
\multicolumn{1}{l}{\textbf{Acronym}}  &  \multicolumn{1}{l}{\textbf{Expanded Form}}     \\ 
\endfirsthead
\multicolumn{2}{p{\textwidth}}{  --- {\it{continued from previous page}} --- } \\
 \multicolumn{1}{l}{\textbf{Acronym}}  &  \multicolumn{1}{l}{\textbf{Expanded Form}}    \\  
\endhead
 \multicolumn{2}{p{\textwidth}}{ --- {\it{continued on next page}} ---} \\ 
\endfoot
\endlastfoot
1D			&  One-dimensional   \\  
2D			&  Two-dimensional   \\  
2PL			&  Two spectral range continuous power-law fit   \\   
3D			&  Three-dimensional   \\  
3PL			&  Three spectral range continuous power-law fit   \\   

ACR(s)		&  Anomalous cosmic ray(s)   \\ 
ACW(s)      	&  Alfv\'en ion cyclotron wave(s)   \\   
AR(s)		&  Active region(s)   \\   
AU          		&  Astronomical unit   \\  
AW(s)		&  Alfv\'en wave(s)   \\  


CIR(s)      		  &  Corotating interaction region(s)   \\  
CME(s)    		  &  Coronal mass ejection(s)   \\   
cobpoint		&   “Connecting with the OBserving” point \\   
CR        		  &  Carrington rotation   \\   

DC          		&  Direct current   \\  
DDZ        		 &  Dust depletion zone   \\  
DFZ       		  &  Dust-free zone   \\  
dHT        		 &  de Hoffman-Teller frame   \\  
DOY       		  &  Day of the year   \\  

ED         		&  ``Either" discontinuity   \\  
Enc. / Encs.      	&  Encounter(s)   \\  
ES (waves)	&  Electrostatic waves    \\  
EUV       		&  Extreme ultraviolet   \\  

$f_{ce}$   		&  Electron gyrofrequency  or electron cyclotron frequency \\   
$f_{cp}$   		&  Proton gyrofrequency  or proton cyclotron frequency \\   
$f_{pe}$   		&  Plasma frequency   \\   
$f_{LH}$    	&  Lower hybrid frequency   \\  
F-corona  		&  Fraunhofer-corona   \\   
$F_\mathrm{A}$	&  Alfv\'enic energy flux   \\  
$F_\mathrm{K}$	&  Bulk kinetic energy flux   \\  
FITS      		&  Flexible Image Transport System   \\   
FM/W        	&  Fast-magnetosonic/whistler   \\   
FOV        		&  Field of view   \\  

GCR(s)         	&  Galactic cosmic ray(s)   \\  

HCI         		&  Heliocentric inertial coordinate system   \\     
HCS         		&  Heliospheric current sheet   \\  
HEE        		&  Heliocentric Earth Ecliptic system  \\   
HEEQ      		&   Heliocentric Earth Equatorial system  \\   
HFR         		&   High frequency receiver   \\   
HHT         		&   Hilbert-Huang transform   \\  
HPC        		&   Helioprojective cartesian system  \\   
HPS         		&   Heliospheric plasma sheet   \\   
HSO			&   Heliophysics System Observatory  \\   
HSS(s)         	&   High-speed stream(s) \\   

ICME(s)     	&   Interplanetary coronal mass ejection(s)   \\   
ICW         		&   Ion cyclotron wave   \\   
ID(s)			&   Interplanetary discontinuity(ies)   \\   
KAW(s)      	&   Kinetic Alfv\'en wave(s)   \\   

LFR(s)         	&   Large-scale flux rope(s)   \\   
LTE         		&   Local thermal equilibrium   \\   
LOS         		&   Line of sight   \\  

$M_{A}$		&  Alfv\'enic Mach number   \\   
MAG(s)         	&   Fluxgate magnetometer(s)   \\  
MC(s)       		&   Magnetic cloud(s)   \\   
MFR(s)      	&  Magnetic flux rope(s)   \\   
MHD         	&   Magneto-HydroDynamic   \\  
MOID        	&  Earth Minimum Orbit Intersection Distance   \\   
MVA         		&   Minimum variance analysis   \\  

ND         		&  ``Neither" discontinuity   \\  
NIR         		&  Near Infrared   \\ 

PAD(s)         	&  Pitch angle distribution(s)   \\   
PDF(s)		&  Probability distribution function(s)   \\   
PIC			& Particle-in-cell   \\  
PIL(s)         	&  Polarity inversion line(s)   \\   
PVI         		&  Partial variance of increments   \\   

QTN         		&  Quasi-thermal noise   \\   

RD(s)          	&  Rotational discontinuity(ies)   \\  
RLO         		&  Reconnection/Loop-Opening   \\  
RTN         		&  Radial-Tangential-Normal frame   \\   
$R_\odot$		&  Solar radius    \\  

SBO(s)          	&  Streamer blowout(s)   \\   
SBO-CME(s)	&  Streamer blowout CME(s)   \\   
S/C			& Spacecraft    \\  
SEP(s)          	&  Solar energetic particle event(s)   \\  
SFR(s)          	&  Small-scale flux rope(s)   \\   
SIR(s)          	&  Stream interaction region(s)   \\   

TD(s)              	&  Tangential discontinuity(ies)   \\  
TH              	&  Taylor's hypothesis    \\  
TOF             	&  Time of flight   \\  
TPS             	&  Thermal Protection System   \\  

UT              	&  Universal Time   \\   

VDF(s)          	&  Velocity distribution function(s)   \\   
VGA(s)          	&  Venus gravity assist(s)   \\   
VSF			&  Volume scattering function     \\ 

WCS            	& World Coordinate System   \\   
WHPI            	& Whole Heliosphere and Planetary Interactions   \\   
WKB             	&  The Wentzel, Kramers, and  Brillouin approximation   \\  
w.r.t.             	&  With respect to   \\  
WTD             	&  Wave/Turbulence-Driven   \\  

ZC             	& Zodiacal cloud   \\ 
ZDC             	& Zodiacal dust cloud   \\  
ZL              	& Zodiacal light   \\  
\end{longtable}

\begin{acknowledgements}
Parker Solar Probe was designed, built, and is now operated by the Johns Hopkins Applied Physics Laboratory as part of NASA’s Living with a Star (LWS) program (contract NNN06AA01C). Support from the LWS management and technical team has played a critical role in the success of the Parker Solar Probe mission.

\smallskip
The FIELDS instrument suite was designed and built and is operated by a consortium of institutions including the University of California, Berkeley, University of Minnesota, University of Colorado, Boulder, NASA/GSFC, CNRS/LPC2E, University of New Hampshire, University of Maryland, UCLA, IFRU, Observatoire de Meudon, Imperial College, London and Queen Mary University London.

\smallskip
The SWEAP Investigation is a multi-institution project led by the Smithsonian Astrophysical Observatory in Cambridge, Massachusetts. Other members of the SWEAP team come from the University of Michigan, University of California, Berkeley Space Sciences Laboratory, The NASA Marshall Space Flight Center, The University of Alabama Huntsville, the Massachusetts Institute of Technology, Los Alamos National Laboratory, Draper Laboratory, JHU's Applied Physics Laboratory, and NASA Goddard Space Flight Center. 

\smallskip
The Integrated Science Investigation of the Sun (IS$\odot$IS) Investigation is a multi-institution project led by Princeton University with contributions from Johns Hopkins/APL, Caltech, GSFC, JPL, SwRI, University of New Hampshire, University of Delaware, and University of Arizona.

\smallskip
The Wide-Field Imager for Parker Solar Probe (WISPR) instrument was designed, built, and is now operated by the US Naval Research Laboratory in collaboration with Johns Hopkins University/Applied Physics Laboratory, California Institute of Technology/Jet Propulsion Laboratory, University of Gottingen, Germany, Centre Spatiale de Liege, Belgium and University of Toulouse/Research Institute in Astrophysics and Planetology. 

\smallskip
IM is supported by the Research Council of Norway (grant number 262941). OVA was supported by NASA grants 80NNSC19K0848, 80NSSC22K0417, 80NSSC21K1770, and NSF grant 1914670. 
\end{acknowledgements}

%
%

\section*{Compliance with Ethical Standards}
{\bf{Disclosure of potential conflicts of interest:}} There are no conflicts of interest (financial or non-financial) for any of the co-authors of this article. \\
{\bf{Research involving Human Participants and/or Animals:}} The results reported in this article do not involve Human Participants and/or Animals in any way. \\
{\bf{Informed consent:}} The authors agree with sharing the information reported in this article with whoever needs to access it.

\bibliographystyle{aa}      		
\bibliography{0_PSP_FirstFourYears_SpaceScienceReviews}    

\end{document}